\documentclass [11pt, proquest] {uwthesis}[2020/02/24]
 
\usepackage{alltt}  %
\usepackage{subfigure} %
\usepackage{graphicx}
\usepackage{caption}
\usepackage{amssymb}
\usepackage{amsmath}
\usepackage[T1]{fontenc}

\begin{document}
 
%
%

\prelimpages
 
%
%
\Title{Superconducting Resonator Development for the Axion Dark Matter eXperiment}
\Author{Thomas Braine}
\Year{2024}
\Program{Physics}

\Chair{Gray Rybka}{}{Physics}
\Signature{Gianpaolo Carosi}
\Signature{Alejandro Garcia}

\copyrightpage

\titlepage

%
%

%
%

\setcounter{page}{-1}
\abstract{%
The Axion Dark Matter eXperiment (ADMX) is the first axion haloscope search to reach DFSZ sensitivity in any mass range for the QCD axion. The QCD axion is a well-motivated dark matter candidate that additionally solves the strong CP problem in nuclear physics. A haloscope has three necessary components; a very strong external magnet, a high Q cavity resonator embedded in this field, and an ultra-sensitive RF read-out system. This dissertation mainly reports on the future development of the resonators for ADMX, specifically superconducting RF (SRF) cavities. It starts with an overview of axions, haloscopes, and the current ADMX experiment, followed by the work done at Lawrence Livermore National Laboratory (LLNL) on SRF cavities, and finally the preliminary results of the ADMX "hybrid" superconducting-copper Sidecar cavity in run 1D.
}
 
%
%
\tableofcontents
\listoffigures
\listoftables
 
%
%
\acknowledgments{
First and foremost, I would like to thank my University of Washington advisor, Gray Rybka, and my LLNL advisor, Gianpaolo Carosi, who are also both co-spokespersons for ADMX. It was a great opportunity to research at Lawrence Livermore National Laboratory with GP on future resonator development, while also getting to travel back to UW to help with the current experimental operations at CENPA.

Thank you to the other members of my committee: Alejandro Garcia, David Kaplan, and Ben Marwick.

Thank you to Leslie Rosenberg for being my first contact with ADMX, and giving me the initial opportunity join the collaboration as first-year graduate student.

I'd also like to thank the senior graduate students and postdocs that mentored me along the way: Nick Du, Raphael Cervantes, Chelsea Bartram, Nicole Crisosto, Nathan Woollett, Stephan Knirck, and Dan Zhang. Additionally, the ADMX graduate students that came after me, Michaela Guzzetti and Jimmy Sinnis, who helped with on-site Sidecar operations and analysis questions towards the end of my graduate tenure.

Thank you to Sam Posen and the others at the SQMS center at Fermilab for not only producing the superconducting tuning rod for Sidecar, but providing SRF accelerator cavity expertise.

Thank you to Lance Cooley, Wenura Winthrage, and Andre Juliao for the superconducting samples and material science expertise.

Thank you to Alex Baker for not only getting us time on the PPMS system but all the help with mounting samples. 

Thank you to the LLNL technicians that machined and fixed the many parts that made this research possible: Sean Durham and Nathan 'Eric' Robertson.

Thank you to the many intrepid helium liquefier and dilution refrigerator operators that kept ADMX running at UW: Nick Force, Seth Kimes, Grant Leum, Charles Hanretty, and Cyrus Goodman.

Thank you to the members of the Rare Event Detection group at LLNL for welcoming me in at the lab as well.

Thank you to my good friends, Kian Alden, Trevor Thomas, and Chance Pryor for the well-needed phone calls and trips along the way.

Thank you to my wonderful cats, Higgs and Boson, who stayed right by my side throughout writing this dissertation.

Finally, I'd like to thank my parents, Tim Braine and Judith Ivey, and my grandfather, Dr. Nathan Ivey, who've supported me to the fullest in pursuit of this degree. 
}

%
%
\dedication{\begin{center}To my grandmother, Dorothy Lee Lewis Ivey, 1922-2023. 
\par  "I finally got it done Grammy!" \end{center}.}

%

%
%

\textpages
 
 
\chapter {Introduction}
 One of the bigger ironies when reflecting on the immense progress in science and physics of the last century is that physicists actually know less about what the majority of the universe is comprised of than what it was thought to comprise of a century earlier. What I allude to is the perplexing discovery of dark matter from astronomical and cosmological measurements. The dark matter problem arises when one tries to account for all the matter that can be seen by interactions with the electromagnetic force, and compare it to the amount of matter that can be "seen" by interactions with the gravitational force; it becomes apparent that the vast majority of the matter in our galaxy, 85\% at current measurements, only interacts via gravitation appreciably, and is otherwise 'dark' and non-luminous when trying to detect it electromagnetically. Furthermore, scientists are confident this matter is non-baryonic and unlike any particles we know \cite{ARBEY,Bertone2018}. The crown jewel of physics theories, the standard model, seems to only explain the interactions of a small fraction of the universes' constituents; an experimental observation pointing not to just a small hole in the model, but a sea of darkness around it. However, it isn't to say that this dark matter isn't interacting with the electromagnetic force entirely, but perhaps it is such a weak interaction as to avoid detection by current technologies. It is the job of physicists to ask what new particle could have the right exotic properties if a few adaptations to the standard model could be made?
 \par Another lesser known mystery surrounding the standard model of particle physics is known as the "Strong CP Problem." Although its easy to explain dark matter in plain terms as "missing" matter to a laymen, the strong CP problem remains lesser known because it requires understanding physical law in the language of symmetries. In essence, the standard model describes the interactions of the electromagnetic, weak nuclear, and strong nuclear forces in terms of the symmetries they preserve and violate. A force that is symmetrical under a given transformation, such as charge conjugation (C), will produce the same phenomenon after such a transformation (changing the signs of all the charges in the case of charge conjugation). Quantum chromodynamics (QCD), the theory that describes the nature of the strong nuclear force, contains a term in its Lagrangian that is explicitly "CP"-violating, meaning a transformation of charge conjugation and parity (changing the charge and spin signs of all the particles in the system) should not produce the same system. The problem, however, is that experiments show that the strong nuclear force doesn't seem to be CP-violating at all; a classic piece of evidence for this is the measurement of the neutron's electric dipole moment (nEDM) that is consistent with zero. Original estimates based on QCD would have the nEDM be many orders of magnitude larger than what its upper value is measured to be. This has forced theorists to lower the strength parameter of the CP-violation, $\theta$, to a value of below $10^{-10}$. This is what constitutes a 'fine-tuning' problem; one is prescribing an incredibly specific number to make the theory work without any mechanism as to why that number is that specific value.
 \par In 1977, Helen Quinn and Roberto Peccei proposed a mechanism that would explain the near zero value of $\theta$, known now as Peccei-Quinn theory \cite{PecceiQuinn1}. If true, it implied the existence of a new particle dubbed the axion, named after a popular detergent because it cleaned up the strong CP problem. At the time, it seemed like new particles were being discovered everyday; it was only a matter of time until this axion was found in an accelerator. And yet, this never happened, which pushed the energy scales higher, and lowered the expected mass of the axion. It was then realized that this lighter axion had a lot of properties that a dark matter particle would have; it would have mass, feebly interact with photons, and would be created in abundance during the early universe. They are produced non-thermally so that they are naturally cold (unlike neutrinos), which allows them to easily cluster around galaxies early in their formation. What makes them exotic, however, is that these particles would be so light that they would have de Broglie wavelengths on the order of kilometers, and would behave more like waves than particles. Because of these properties, physicists initially thought that these particles could never be detected, and coined them "invisible axions."
 \par An experimental apparatus was then proposed by Pierre Sikvie in 1983, to detect this axion field within our galaxies' dark matter halo \cite{Sikivie}: a resonant cavity embedded in a strong static magnetic field with an ultra-low noise receiver. The static magnetic field provides a large density of virtual photons to stimulate the axion to photon conversion. If the resonant cavity is tuned to the frequency of these photons, the build up of power could be enough to detect the signal over the noise via the ultra-low noise receiver. This was dubbed an axion \textit{halo}scope, and plans were made to one day construct such a device.
 \par Just over 40 years later, the world now has several axion haloscopes, and many more planned to be built. As large WIMP dark matter experiments continue to show no events, the experimental axion field is gaining momentum. However, there is no doubt who were the pioneers in this field: the Axion Dark Matter eXperiment (ADMX). Initially formed in the 1990s, for over thirty years, ADMX has paved the way to creating the first discovery-sensitive haloscope. This has involved the development of several novel technologies to be possible.
 \par For the past 6 years, it has been my privilege to work alongside the small team of scientists that make up the ADMX collaboration, developing yet another novel axion haloscope technology. My graduate work has been focused on the development and testing of the current and future cavity resonators used in ADMX. More specifically, this dissertation focuses on efforts to replace the current copper cavities used with cavities that are made of superconductors; this could potentially increase the haloscope's sensitivity to axions. As part of this project, I spent about 4 years of my graduate career stationed at Lawrence Livermore National Laboratory (LLNL), working under Dr. Gianpaolo Carosi, testing superconducting materials, developing cavity measurement techniques, and designing prototype cavity experiments. This work has culminated in the first axion search that uses a superconducting tuning rod within the resonant cavity; This is an upgrade to an in-situ R\&D testbed cavity system known as ADMX Sidecar, which runs in parallel with the main ADMX cavity. 
 \par The structure of this dissertation is a follows: Chapter \ref{chap:axiontheory} will formally introduce the axion theory through the solution to the strong CP problem, and ultimately how the axion would act as the dark matter particle within our universe. Chapter \ref{chap:Haloscopes} will introduce the experimental detector concept, the haloscope. Chapter \ref{chap:Sitelayout} gives the specifics of the main ADMX haloscope and experimental site. Chapter \ref{chap:analysis} describes the analysis techniques that we use to report axion detection events or lack thereof. Chapter \ref{chap:Cavities} is a detailed description of resonant cavities and the techniques used to characterize them. Chapter \ref{chap:SRF} is a formal introduction to the theory of superconductors. Chapter \ref{chap:LLNL} reviews my work on superconducting cavities at LLNL. Finally, Chapter \ref{chap:Sidecar1D} reports performance of and the preliminary axion search analysis of the hybrid superconducting Sidecar cavity during the currently on-going ADMX data run.

\chapter{Axion theory and cosmology}
\label{chap:axiontheory}
In this chapter, I will cover the mathematical foundation needed to understand why the axion may be the invisible dark matter particle. This will start with introducing the strong CP problem, the Peccei-Quinn solution, and introduction of the axion. I will then move toward the initial experimental attempts to find the axion and the subsequent implications of its characteristics. Finally, I will motivate the axion as a dark matter candidate, and describe theories of how axions would have become the dark matter, and which axions would be preferred for dark matter production.
\section{The strong CP problem}
As mentioned in the introduction, the strong CP problem originates in the quantum chromodynamics' (QCD) prediction of CP-violation. The minimal form of the QCD Lagrangian can be written:
\begin{equation}
    \mathcal{L}_{QCD}=\sum_{f=u,d,s,c,t,b} \bar{q_f} (i\gamma^{\mu}D_{\mu}-m_f)q_f-\frac{1}{4}G^a_{\mu\nu}G^{\mu\nu}_a
    \label{eqn:QCDlagrangian}
\end{equation}
where the sum of f is over the 6 flavors of quarks, up (u), down (d), charm (c), strange (s), top (t) and bottom (b), $q_f$ is the quark field of the given flavor, $\gamma^{\mu}$ are the gamma matrices, $D^{\mu}$ is the covariant derivative, $m_f$ is the given quark mass, and $G^a_{\mu\nu}$ is the gluon field strength tensor.
\par This Lagrangian poses a problem because it is symmetric under $U(1)_A$ global axial rotations, which would imply the existence of hadron parity doublets, meaning there would be another form of the proton with opposite parity for instance; this particle has not been observed in nature. One can propose that this symmetry is broken in the early universe, but that implies the existence of a pseudo-goldstone boson similar to $\eta$, but with pion mass order; no particle like this has been discovered either, when it should have already appeared in accelerator experiments. This means the $U(1)_A$ symmetry must be broken by other means; it can be avoided if it is anomalously broken by an axial current. This is what the Alder-Bell-Jackiw (ABJ) anomaly does, originally introduced to explain the observed decay of pions. It, however, introduces the CP-violating term to the QCD Lagrangian:
\begin{equation}
    \mathcal{L}_{\bar{\theta}}=\frac{\bar{\theta}}{32 \pi^2}G^a_{\mu\nu}\tilde{G}^{\mu\nu a}
    \label{eqn:QCDlagrangianCPterm}
\end{equation}
The $\bar{\theta}$ term can be thought of as the strength parameter of the CP-violation, but it is also made up of two sub-terms, one that describes the QCD vacuum, and the other based on the quark masses; $\bar{\theta}=\theta+argdet(M)$, where $\theta$ describes the vacuum violation, and M is the quark mass matrix. This could also be written in terms of the electric and magnetic components of the gluon field:
\begin{equation}
    \mathcal{L}_{\bar{\theta}}=\frac{\bar{\theta}}{16 \pi^2}E^a \cdot B^a
    \label{eqn:QCDlagrangianCPterm2}
\end{equation}
\par Taking a step back, this term points to the non-abelian nature of the vacuum gluon field, and indicates that it contains degenerate, yet topologically distinct vacuum states \cite{POLYAKOV}. Each topologically distinct state can be labeled by an integer $n$, a topological winding number. The actual QCD vacuum then can be described as a superposition of these states:
\begin{equation}
    |\Theta>=\sum_n e^{-in\theta}|n>
    \label{eqn:thetastates}
\end{equation}
In this way, there are an infinite number of vacuum states in QCD, and the real vacuum is continuously 'quantum-tunneling' through the energy barriers between the various vacuum states. $\theta$ is an arbitrary phase-offset that describes the strength of the quantum-tunneling between vacuum states; hence it is an angle between 0 and $2 \pi$.
\par
\begin{figure*}[htb!]
    \centering
    \includegraphics[width=0.4\linewidth]{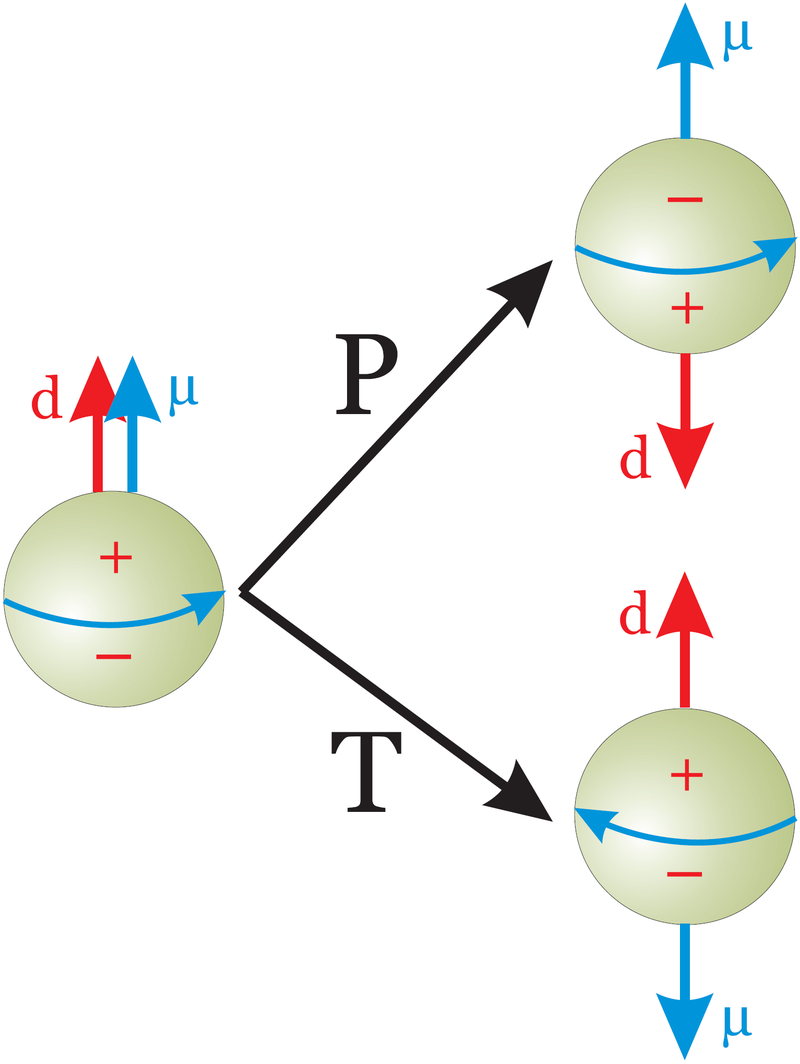}
    \caption{A diagram of the neutron with its charge distribution and spin resulting in aligned electric and magnetic dipole moments. Applying a time-reversal symmetry (T), preserves the charge distribution but reverses the spin, causing the magnetic moment to reverse directions as well; By CPT theorem, this is the equivalent of a CP operation. Under just a parity operation, the charge distribution can be thought of as flipping with the respect to the spin axis, resulting in the electric dipole changing direction; a subsequent (C) operation will lead to the same result as the time reversal operation. One can see in the end, the neutron now has electric and magnetic dipole moments that are anti-aligned, meaning CP symmetry is not preserved in this system. Credit: public domain work from Andreas Knecht.}
    \label{fig:nEDM}
\end{figure*}

 There is no reason why $\theta$ or $argdet(M)$ should be zero or should precisely cancel each other out; one expects the value to be order 1 as they both vary from 0 to $2\pi$. Yet if it is order 1, this would imply several observable effects, such as the existence of electric dipole moments in neutral particles with non-degenerate ground states, such as the neutron. This is outlined in Figure \ref{fig:nEDM}, where a time-reversal operation results in the dipole moments being anti-aligned, a clear sign of CP-violation. However, the neutron EDM has been measured extensively over decades only to further constrain it closer to a zero value, with a currently accepted value of $d_n=(0.0 \pm 1.1_{stat} \pm 0.2_{sys} \times 10^{-26} e\,$cm), which further constrains $\bar{\theta} < 10^{-10}$ \cite{nEDM}. 
\par The heart of the strong CP problem is why this number $\bar{\theta}$ would be so close to zero. There is no reason $\theta$, originating from strong force vacuum, and $argdet(M)$, originating from the weak interaction, should cancel each other out as they are unrelated quantities. They seem to be 'fine tuned' to the precise value we see in nature. One way to solve this quickly is to introduce a massless quark, which would cause the $argdet(M)$ term to become undefined, and therefore make $\bar{\theta}$ not observable; such a quark has not been discovered, and has other conflicting implications for lattice QCD. There are also other proposed solutions that can make $\theta$ and $argdet(M)$ both go to zero \cite{NELSON1984387,Barr}, but this is beyond the scope of this dissertation. We will focus on the Peccei-Quinn solution to the strong CP problem, because it is a minimal extension to the standard model and produces the axions we hope to find.
\section{Peccei-Quinn solution}
\begin{figure*}[htb!]
    \centering
    \includegraphics[width=0.8\linewidth]{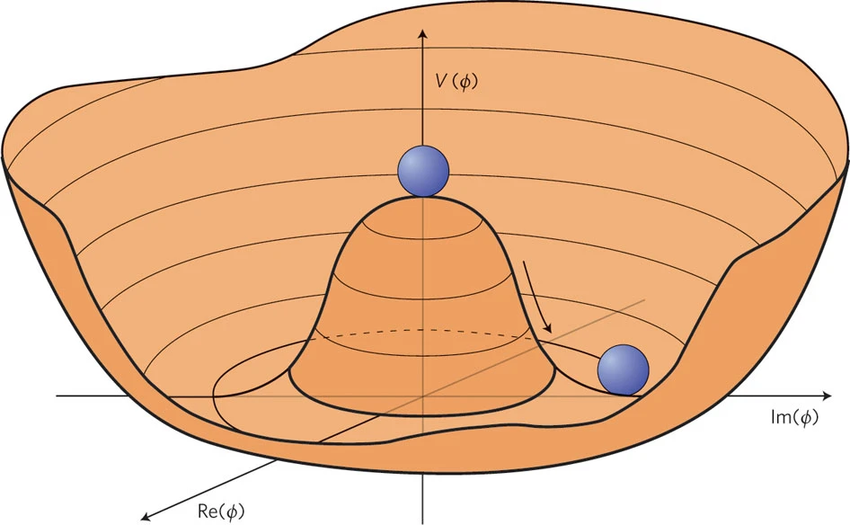}
    \caption{The "wine-bottle" potential of the field $\phi$ introduced by Peccei-Quinn. Credit: CERN}
    \label{fig:winebottle}
\end{figure*}

In 1977, Roberto Peccei and Helen Quinn published the first paper on what came to be known as the Peccei-Quinn mechanism of solution to the strong CP problem \cite{PecceiQuinn1}. It is popular because it is quite simple and sensible, and most of all, produces the axion as a by-product for physicists to look for. Their solution was to add a new global $U(1)$ chiral symmetry known as the Peccei-Quinn symmetry $U(1)_{PQ}$ that is broken in the early universe. This would introduce an new complex, pseudoscalar field, $\phi$. $\bar{\theta}$ is promoted from a constant to a variable in this theory equal to $arg(\phi)$. For temperatures greater than $\Lambda_{QCD}$, this symmetry takes on a classic "wine bottle" potential:
\begin{equation}
    V(\phi)= \mu^2 (|\phi|^2-f_{PQ}^2/2)^2
    \label{eqn:PQpotential}
\end{equation}
where $\mu$ is a dimensionless parameter and $f_{PQ}$ is the symmetry-breaking energy scale above which CP is not conserved. This potential is shown in Figure \ref{fig:winebottle}. Above the symmetry breaking scale, $f_{PQ}$, the vacuum expectation of $\phi$ is zero, and all values of $\bar{\theta}$ are equally probable, and strong CP violation exists. This means the neutron would have had an EDM in the early universe. However, as the universe cools below this energy, the field is forced to the minima of the potential, which is $\phi=f_{PQ}/\sqrt{2}$. In a classical potential, $\bar{\theta}$ could continue to take on a variety of values around the minima 'rim' of the wine bottle, but when $T\propto \Lambda_{QCD}$, quantum instanton effects tip the wine bottle potential off axis (an anomaly not described here), forcing $\bar{\theta}$ to a global minima. A new PQ Lagrangian term arises:
\begin{equation}
    \mathcal{L}_{PQ}=\xi \frac{\phi}{f_{PQ}}\frac{g_s^2}{32 \pi^2} \tilde{G}^{\mu \nu}_b G_{b \mu \nu} + \mathcal{L}_{interactions}
    \label{eqn:PQlagrange}
\end{equation}
where $\xi$ is a model dependent parameter, and $g_s$ is the strong force coupling constant.
\par This additional CP-violating term can be grouped with the others, and $\bar{\theta}$ can be redefined as:
\begin{equation}
    \bar{\theta}=\theta+argdet(M)-\xi\frac{\phi}{f_{PQ}}
    \label{eqn:PQtheta}
\end{equation}
One can then see that $\bar{\theta}$ naturally relaxes to zero when the potential is minimized at:
\begin{equation}
    <\phi>= \frac{f_{PQ}}{\xi}(\theta+ argdet(M))
    \label{eqn:PQphi}
\end{equation}
They had found a mechanism that naturally gives $\bar{\theta}$ a zero value! The strong CP problem is solved, but what are the side-effects? Within a year after the Peccei-Quinn paper being published, Steven Weinberg \cite{WeinbergAxion} and Frank Wilczek \cite{WilczekAxion} had published papers showing that the PQ mechanism postulates the existence of a new pseudo-Nambu-Goldstone boson that, because of the anomalous axial symmetry breaking in the potential, would have non-zero mass. Weinberg called this particle the higglet, because of the similarities between the PQ mechanism and the Higgs mechanism. Frank Wilczek dubbed it the 'axion', after the laundry detergent, because it had cleaned up the strong CP problem; the name also works because it arises from the axial symmetry breaking in the wine-bottle potential. Wilczek's name caught on, and the hunt for the axion began!
\section{The original and invisible axions}
The original axion, sometimes called the Peccei-Quinn-Weinberg-Wilczek (PQWW) axion, had an associated symmetry breaking scale with the electroweak scale, that is $f_{PQ} \approx 10^2-10^3 GeV$. This would place the mass of the axion around 10-100 keV. These axions should have been detected in a variety of accelerator experiments via several particle channels: $a\rightarrow\gamma\gamma$,$a\rightarrow e^- e^+$, $J/\Psi \rightarrow \gamma a$, $K^+ \rightarrow \pi^+ a$, $\Upsilon \rightarrow \gamma a$, and $N* \rightarrow Na$. When they didn't appear in experiments, this ruled out the PQWW axion, and it pushed the $f_{PQ}$ to higher energy values \cite{PQWWaxion1,PQWWaxion2}. The result of this was a much lighter axion given by the equation:
\begin{equation}
    m_a=\frac{(m_um_d)^{1/2}}{m_u+m_d}\frac{f_{\pi}}{f_{PQ}/N}m_{\pi} \approx 0.62 eV \frac{10^7 GeV}{f_{PQ}/N}
    \label{eqn:m_axion1}
\end{equation}
Where $m_u$ and $m_d$ are the up and down quark masses respectively, $f_{\pi}$ is the pion decay constant, N is the PQ color anomaly (related to the $\xi$ term earlier), and $m_{\pi}$ is the pion mass. This can be even more simply written for the experimenter:
\begin{equation}
    m_a \approx 5.691 \mu eV \frac{10^12 GeV}{f_{PQ}}
    \label{eqn:m_axion2}
\end{equation}
\par
\begin{figure*}[htb!]
    \centering
    \includegraphics[width=0.8\linewidth]{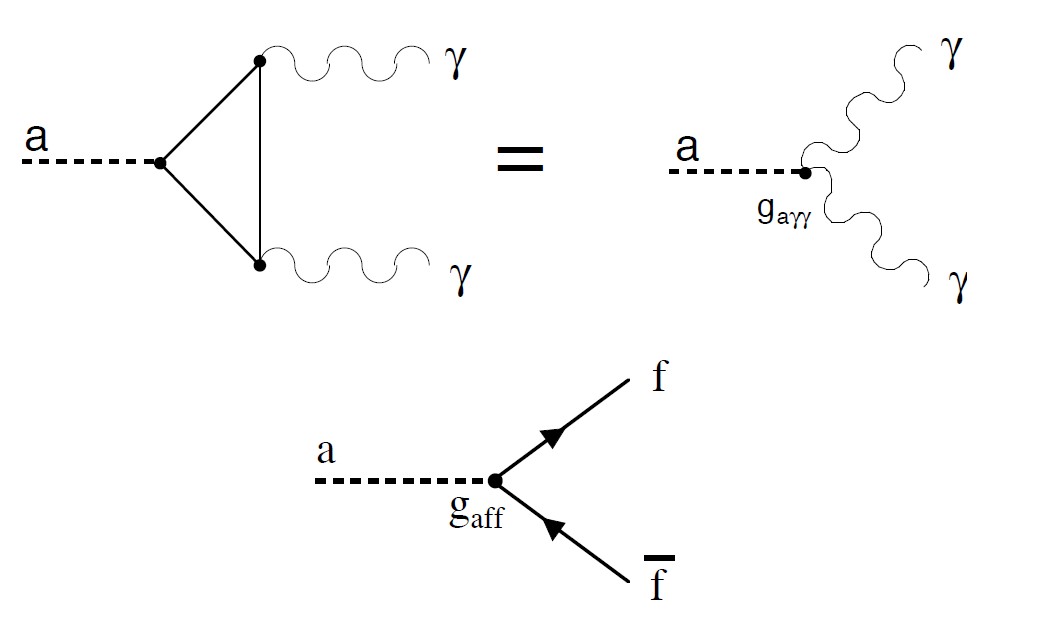}
    \caption{Feynman diagrams of axion coupling to photons and fermions.}
    \label{fig:axionfeynmandiagram}
\end{figure*}

 Because the axion would now have an extremely light mass, trying to detect and measure the mass of axions via meson decay channels ($K^+ \rightarrow \pi^+ a$, $\Upsilon \rightarrow \gamma a$) or nuclear de-excitations ($N* \rightarrow Na$) would be a similar to, but even harder than, measuring the neutrino mass; the axion would be a tiny drop of energy and momentum in comparison with its partner in the interaction, being probably a far lighter particle than a neutrino. The $a\rightarrow e^- e^+$ channel would only occur in some axion models, as discussed later in this section, so it is usually not of interest. The decay channel,  $a\rightarrow\gamma\gamma$, is of interest, pictured in Figure \ref{fig:axionfeynmandiagram}, because it would be directly correlated with the axion energy, and would most likely produce measurable microwave/RF photons based on its predicted mass. 
\par The interaction Lagrangian describing this decay of the axion into two photons is given by,
\begin{equation}
    \mathcal{L}_{a\gamma\gamma}=-g_{a\gamma\gamma}\phi_a\vec{E}\cdot\vec{B}
    \label{eqn:interactionlagrangianAgammagamma}
\end{equation}
where $\phi_a$ is the axion field, $\vec{E}$ and $\vec{B}$ are the electric and magnetic fields, and $g_{a \gamma \gamma}$ is the coupling constant of the decay. This constant is given by 
\begin{equation}
    g_{a \gamma \gamma}=\frac{\alpha}{2 \pi f_{PQ}} (\frac{E}{N}-1.92)=\alpha \frac{g_{\gamma}}{\pi f_{PQ}}
    \label{eqn:gagammagamma}
\end{equation}
where $\alpha$ is the fine structure constant, $g_{\gamma}$ is a model-dependent coupling, and $E$ and $N$ are the electromagnetic and color anomalies of the axial symmetry. The model dependent parameter, $g_{\gamma}$, arises from the fact that there is a fermionic loop in this interaction; which fermions have interactions with the axion in this loop are dependent on the model.
\par They are two benchmark "invisible" axion models referred to as the Kim-Shifman-Vainshtein-Zakharov (KSVZ) model \cite{KSVZ1,KSVZ2}, and the Dine-Fischler-Srednicki-Zhitnitsky (DFSZ) model \cite{DFSZ1,DFSZ2}. The KSVZ model only permits the axion to couple to heavy quarks, whereas the DFSZ model allows coupling to all fermions, quarks and leptons. The KSVZ model has $g_{\gamma}=-0.97$, whereas the DFSZ has coupling, $g_{\gamma}=+0.36$. DFSZ axions can also be easily implemented into grand unified theories that contain the $SU(5)$ group.
\section{Axions as dark matter}
The invisible axion quickly became recognized as a potential dark matter candidate. For the purposes of keeping this dissertation to a reasonable length, I will assume the reader is familiar with the dark matter problem and its origins, and has accepted the numerous pieces of evidence for its existence; galactic rotation curves, gravitational lensing measurements, and signatures in the cosmic microwave background to name a few. I will point to two previous ADMX dissertations that go over the supporting evidence well \cite{Boutan,DuThesis}. What is important for the purposes of ADMX is what we know about dark matter's characteristics based on the supporting evidence. It must be:
\begin{enumerate}
    \item Very stable
    \item Very weakly coupled to photons
    \item Non-baryonic
    \item Mostly cold or non-relativistic
    \item Numerous enough to make up 85\% of the mass in the universe
\end{enumerate}
From what we know about the invisible axion already, it matches the first three criteria pretty well; a non-baryonic particle with an extremely light mass that would very weakly couple to photons and have few other decay channels once created. The real question is how do enough axions get produced in the early universe and cool off in time to make up the dark matter.
\par There are three primary ways in which axions could have been produced in the early universe: thermally in quark and nucleon interactions when the universe was a hot, dense plasma, string or domain wall decays, and a process called the "vacuum realignment" mechanism. In the case of thermal axions, the mass would need to be greater than 1 m$e$V for its cross-sections to be large enough to make a significant amount of matter, and the axions produced in these interactions would be very hot, which is not in agreement with the structure of dark matter we observe. Furthermore, measurements of red giant stars and supernova rule out most axions above 1 m$e$V, as is discussed later in the chapter. The axions produced in string or domain wall decay are outside the scope of this dissertation, but also have several constraints that indicate they are not the dominant production mechanism for dark matter. In the case of domain wall decay, over production of axions is very likely unless symmetry-breaking parameters are finely tuned to get the desired axion energy density \cite{HARIGAYA20181}. String decays have similar dependencies to become a dominant dark matter production mechanism \cite{HAGMANN199981}. The vacuum realignment (VR) mechanism is, however, a very compelling process for the creation of dark matter.
\par This mechanism originates from the fact that before PQ symmetry is broken at its energy scale, $\theta$ can take on any value from 0 to $2 \pi$ at any point in space. At the point where it cools below the $f_{PQ}$ scale, there will be some initial mis-alignment value, $\theta_i$, and the value will begin to drop down the wine-bottle potential towards its minimum, following the equation of a damped harmonic oscillator:
\begin{equation}
    \ddot{\bar{\theta}}+3H(t)\dot{\bar{\theta}}+m_a^2sin(\bar{\theta})=0
    \label{eqn:VReqn1}
\end{equation}
where $H(t)$ is the Hubble parameter and $m_a$ is the axion mass. For $H<\theta$, the oscillations will be under-damped and lose energy via "Hubble Friction." These oscillations correspond to a zero-momentum, pressureless condensate of axions that would be cold immediately. This would allow them to then fall into the gravitational wells produced by the anisotropies in the hot baryonic matter, seeding the large scale structure we see today as the dark matter. The fractional cosmic mass density from this VR mechanism is:
\begin{equation}
    \Omega_A^{VR}h^2=0.12(\frac{6 \, \mu eV}{m_A})^{1.165} F \bar{\theta_i}^{2}
    \label{eqn:VReqn2}
\end{equation}
where h is the Hubble expansion parameter, $\bar{\theta_i}$ is the initial misalignment angle relative to the CP-conserving value between $-\pi$ and $\pi$, and F is a factor accounting for the anharmonicities in the axion potential \cite{VR1,VR2,VR3}. What is important to note here is that the leading value, if the rest is kept to $\mathcal{O}(1)$, is at the energy density value expected for cold dark matter, $\Omega _{cdm}h^2=0.12$, from cosmological measurements. This means if $F \bar{\theta_i}$ was kept to $\mathcal{O}(1)$, then an axion of order a $\mu eV$ would account for 100\% of the dark matter observed in the universe.
\par This is the point where one must consider the cosmological event known as inflation, and its effect on this production process. Inflation is another well accounted for cosmological event in the early universe, whereby the universe expanded rapidly, 'faster' than the speed of light, expanding its volume by at least a factor of $10^{78}$ \cite{inflation}. This event resulted in separate casual volumes within the universe, and our casual volume (a light-sphere) is often called our visible universe. This means if PQ symmetry is broken after inflation, $\theta_i$ would take on different values in different causal volumes, and would average out to the root-mean-square average of all the volumes, giving a value of $\bar{\theta}= \pi/\sqrt{3}$. This simplifies the fractional cosmic mass density of the axions:
\begin{equation}
    \Omega_A^{VR}h^2=0.12(\frac{30 \mu eV}{m_A})^{1.165}
    \label{eqn:VReqn3}
\end{equation}
One can then see a plausible range of axion masses is around $25\, \mu eV < m_a < 1000 \,\mu eV$ for a post-inflationary scenario.
\par In the pre-inflationary scenario, the symmetry would break first at a random value of $\theta_i$, and from that tiny volume expand rapidly with inflation, making $\theta_i$ the same random value for all casual volumes. Nonetheless, it is still constrained to be $\mathcal{O}(1)$, such that axions much lighter than $\mu eV$ are produced too abundantly in the early universe, making too much dark matter, and causing the universe to over-close. There is one argument, using the pre-inflationary scenario, where the universe randomly gets a near zero $\theta_i$, which allows for much lighter axions; the anthropic argument is that only these small $\theta_i$ scenarios allow for life to form, and therefore we must be in a small $\theta_i$ universe. Whether or not one believes this argument, there are other reasons to want ultra light axions, as they can help explain some discrepancies astronomers see in the large-scale structure of dark matter, by use of 'fuzzy' dark matter, where the axion has de Broglie wavelengths on the order of light years. There are also several grand unified theories that are preferential to it. There are several axion experiments targeting this ultralight mass range as will be discussed at the end of this chapter.
\section{Axion energy spectrum lineshapes}
In this section I will cover the velocity distribution axions take on after they are generated by the VR mechanism. Since these axions are generated initially as a pressureless zero-momentum gas, it is fair to assume that the axions around us fell into the gravitational well that became our galaxy, and became thermalized in the process. This simple model is called the "isothermal sphere" model and would take on the classical Maxwell-Boltzmann thermal velocity distribution. The axion line shape based on a Maxwell -Boltzmann distribution, following the standard halo model would be \cite{MaxwellBoltzmannLineshape}:
\begin{equation}
        g(f)=\frac{2}{\sqrt{\pi}}\sqrt{f-f_a}\left(\frac{3}{f_a\frac{c^2}{\langle v^2 \rangle}}\right)^{3/2} e^{\frac{-3(f-f_a)}{f_a}\frac{c^2}{\langle v^2 \rangle}}
        \label{eqn:MBlinshape}
\end{equation}
 where $f$ is the measured frequency, $f_a$ is the axion rest mass frequency, and the rms velocity of the dark matter halo is $\langle v^2 \rangle\approx (270 \,km/s)^2$. Note that this equation averages out the orbital motion of the Earth around the Sun and rotational motion of the detector about the earth's axis; these would produce a small spectral modulation in the line shape on a scale smaller than what's needed to just detect the signal. High resolution analysis of this modulation, once an axion is found, would be highly beneficial for characterizing the local distribution of dark matter in our solar system. As it stands, this would result in an axion virial velocity of $\beta=v/c=10^{-3}$, and a fractional line width of $10^{-6}$. 
\par The isothermal sphere model is a bit naive however, and unable to explain features of our galaxy. A primary mechanism that can produce errors in the model is galactic mergers; two galaxies that pass between each other can have a tendency to pull some dark matter on to the more massive partner. This results in a 'dark disc' on top of the isothermal sphere for some galaxies \cite{DarkDisk}. It is also possible that axions have entered our halo and not fully thermalized, resulting in "cold flows" as they fall towards the galactic core. These flows support the formation of "caustic rings" at points where the number of flows present change value \cite{CausticRings}. The rings can be thought as stationary points in the velocity phase space, producing a very high density region of axions, with diminishing velocity dispersion, making an extremely small energy line width of the axion photon signal locally.
\par
\begin{figure*}[htb!]
    \centering
    \includegraphics[width=0.8\linewidth]{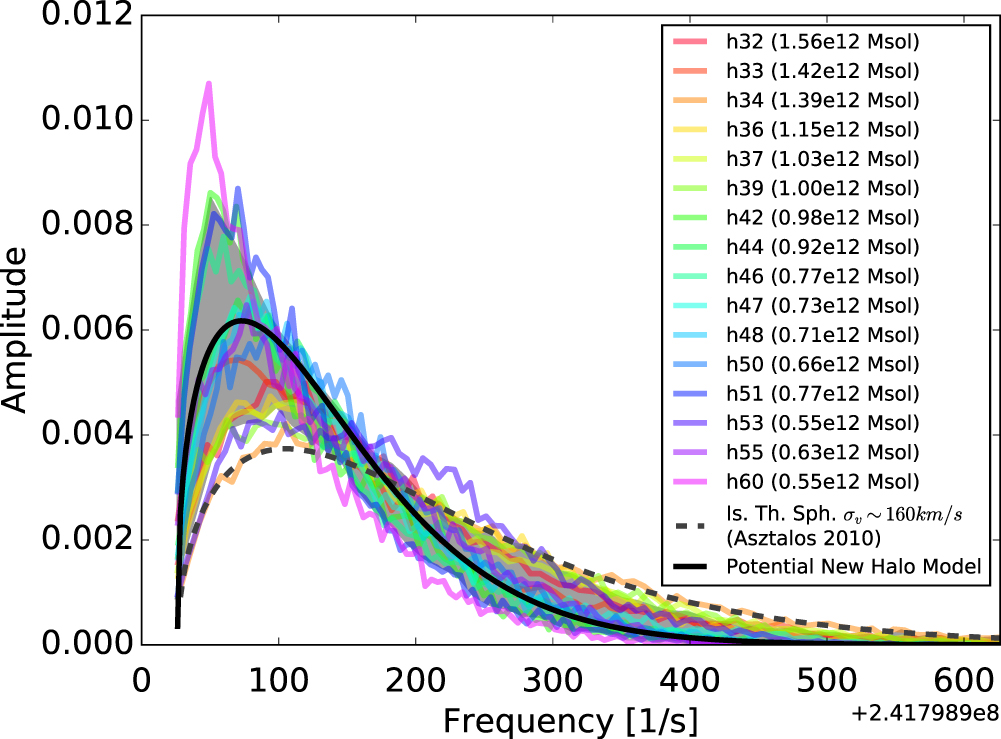}
    \caption{N-Body Simulations from Romulus25 showing the axion frequency spectra for various Milky-Way-like galaxies (color lines), containing an axion of $m_a=1 \mu eV$. The solid black line shows a fit to the new line width model shown in Equation \ref{eqn:N-bodylinshape}, whereas the dotted line shows the Mawell-Boltzmann distribution (Equation \ref{eqn:MBlinshape}) \cite{Lentz_2017}.}
    \label{fig:Nbody}
\end{figure*}

 Developments in galaxy formation simulations have enabled calculation of alternative velocity distributions for the axion that can account for these unique phenomenon. Ref. \cite{Lentz_2017} outlines simulation work done at the University of Washington to better approximate the axion linewidth for ADMX. ADMX now reports its exclusion limits (more on this later), using both the Maxwell-Boltzmann shape, and the predicted "N-body" shape from these simulations. These simulations describe galaxies using the N-body+smooth-particle-hydrodynamics (N-body+SPH) method, in lieu of the standard halo model (SHM). The axion line shape based on these N-body simulations would be:
        \begin{equation}
        g(f)\approx \left(\frac{(f-f_a)}{m_a \kappa}\right)^\alpha e^{-\left(\frac{(f-f_a)}{m_a \kappa}\right)^\beta}
        \label{eqn:N-bodylinshape}
        \end{equation}
Again, $f$ is the measured frequency, $f_a$ is the axion rest frequency, and $m_a$ is the axion rest mass. The best fit parameters were computed from simulation to be $\alpha \approx 0.36 \pm 0.13$, $\beta \approx 1.39 \pm 0.28$, and $\kappa \approx (4.7 \pm 1.9)\times10^{-7}$ \cite{Lentz_2017}. Figure \ref{fig:Nbody} shows that the axion linewidth could be narrower than the Maxwell-Boltzmann prediction. This however doesn't effect detection very much, because one is simply looking for any peak in the beginning and looking for an energy excess in a wider shape is a more conservative strategy. Once the axion signal is found in a wider bin, it is relatively easy to narrow in on the specifics of its line shape. Throughout the rest of this dissertation, I will assume the Maxwell-Boltzmann shape unless stated otherwise.
\section{The current axion parameter space}
\begin{figure*}[htb!]
    \centering
    \includegraphics[width=1.0\linewidth]{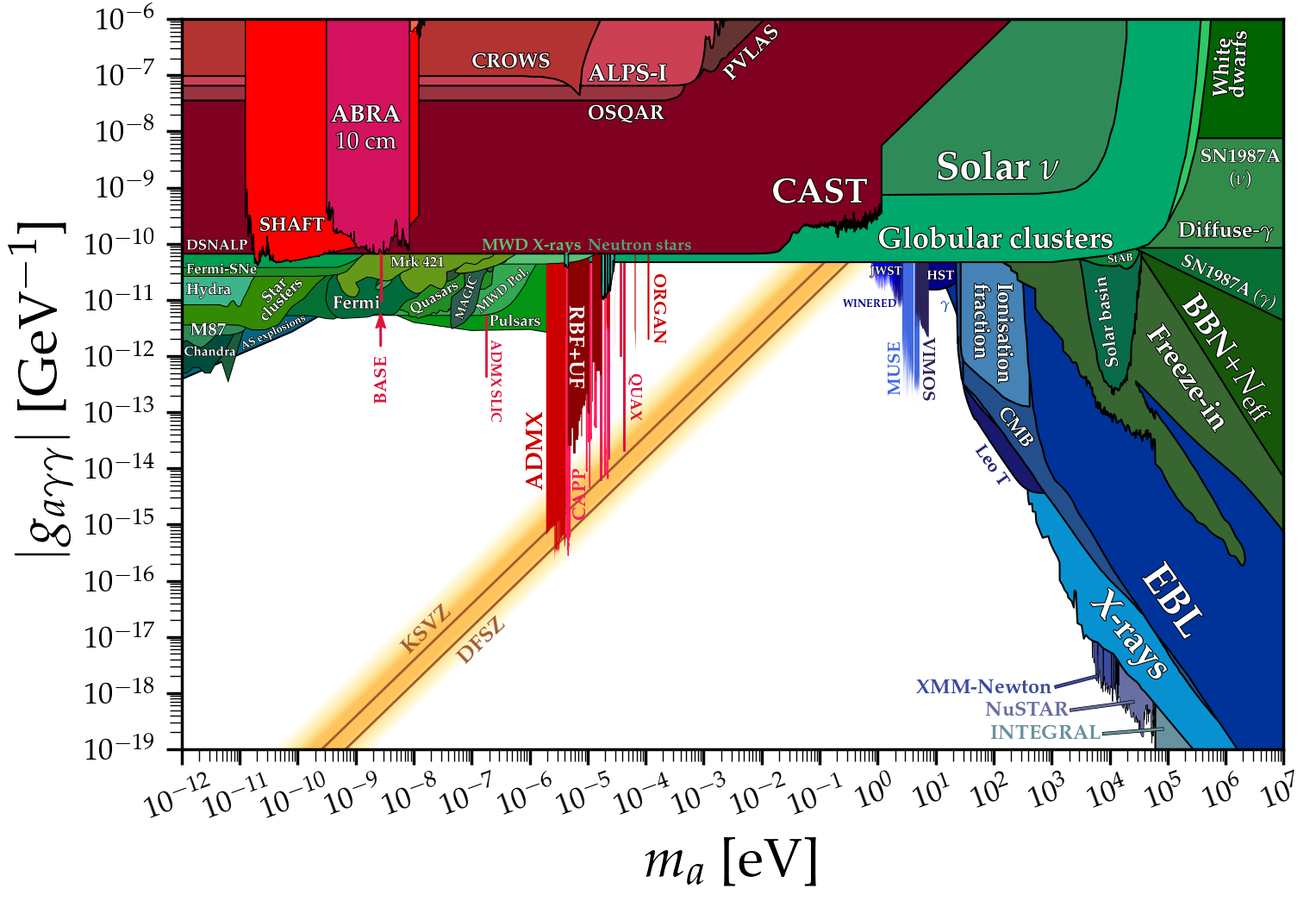}
    \caption{The axion-photon coupling exclusion limit plot for a variety of experimental techniques. The KSVZ and DFSZ theory lines are plotted in orange, with a yellow band around them marking the "discovery" range of where the invisible axion is expected to be. Laboratory/dedicated axion searches are highlighted in red colors, with ADMX and other haloscopes being the most sensitive and able to exclude axion within the yellow "discovery" range. Astrophysical constraints using measurements of various stars, galaxies etc. are highlighted in green; These measurements do not assume the axion is dark matter. More astrophysical measurements such as constraints set by the CMB are highlighted in blue; these assume the axion is the entirety of dark matter.}
    \label{fig:Axionlimitstheory}
\end{figure*}

\begin{figure*}[htb!]
    \centering
    \includegraphics[width=1.0\linewidth]{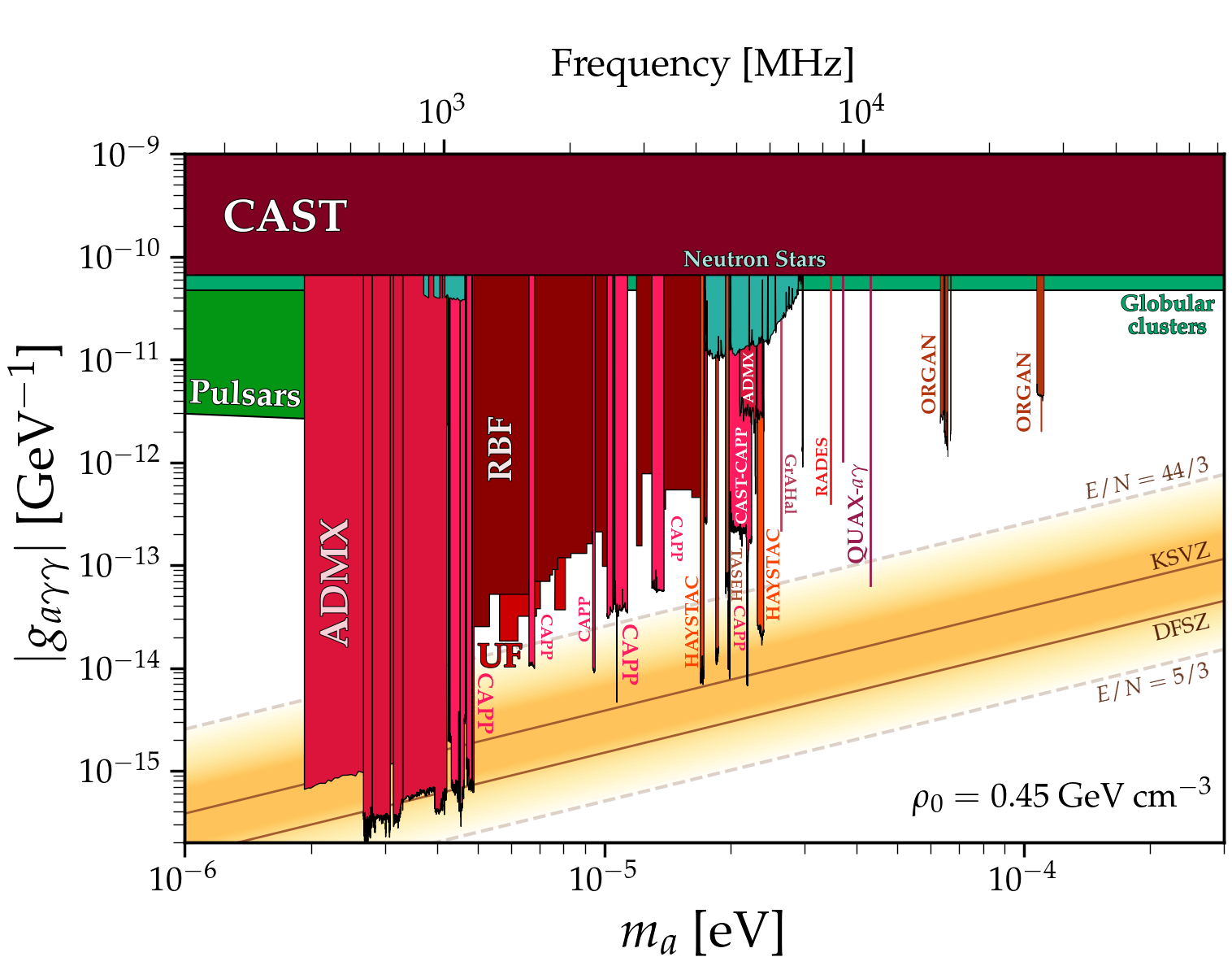}
    \caption{The axion parameter space centered around the haloscope exclusion region with various haloscope experiments, including ADMX, highlighted in red. The yellow region highlights the invisible axion discovery region, centered around the predicted DFSZ and KSVZ mass-coupling lines.}
    \label{fig:ADMXlimitstheory}
\end{figure*}

As of now, no axions have ever been detected, but through a variety of experimental techniques, areas of the axion parameter space have been able to be excluded, meaning an axion with those properties could not exist. As the axion theory has been outlined, one can see that there are only two open theory parameters that define the axion: $f_{PQ}$ and $g_{\gamma}$. $f_{PQ}$ determines the axion mass, $m_a$, by Equation \ref{eqn:m_axion2}, and both $f_{PQ}$ and $g_{\gamma}$ determine $g_{a \gamma \gamma}$ by Equation \ref{eqn:gagammagamma}. For most experimenters dealing with measuring the power of a given axion interaction coupling, it is better to report experimental sensitivity in terms of $g$ (y-axis) vs. $m_a$ (x-axis), thus we will be reporting the haloscope sensitivities in axion-photon coupling, $g_{a \gamma \gamma}$, although this could also be for coupling with any fermion, $g_{af}$. One can see that, in the case of the KSVZ and DFSZ invisible axion theories, $g_{a \gamma \gamma} \propto m_a$, and each theory produces a predicted coupling-mass line through a plot of $g_{a \gamma \gamma}$ vs $m_a$ based on its value of $g_{\gamma}$. Figure \ref{fig:Axionlimitstheory} shows such a plot, with the KSVZ and DFSZ predicted lines overlay in a yellow/orange band; this band marks the 'discovery' range where one expects to find invisible axions. Any axions found outside of this yellow band would be considered 'Axion-Like Particles' or ALPs, and would not solve the strong CP problem. The rest of the plot contains exclusion limits from three major categories: astrophysical measurements where axions aren't the dark matter, astrophysical measurements where they are the dark matter, and dedicated laboratory searches. One can see that the upper end of the mass range has been well excluded by astronomical and cosmological measurements, making KSVZ and DFSZ axions impossible above 100 meV; as we already discussed, however, these higher mass axions from 1-100 meV are more likely made thermally, and therefore 'hot', so they would most likely not be the dark matter. Another important contribution to this plot, with a lot of real estate in the axion parameter space, is the CAST experiment in dark red; It is an axion helioscope that looks for solar axions, and is able to set limits over a wide range of masses, but only to a limited value of $g_{a \gamma \gamma}$. Finally, one should notice the series of red rectangles reaching down to the KSVZ/DFSZ yellow region starting at about $m_a=1 \mu eV$ that start to get smaller and sparser above $10 \mu eV$ ; These are the haloscope experiments including CAPP, HAYSTAC, ORGAN, RBF+UF, and of course ADMX! These are dedicated instruments designed to reach this discovery region, and the goal for haloscopes is to fully exclude the discovery region up to a 1 meV axion mass. Figure \ref{fig:ADMXlimitstheory} highlights this haloscope discovery region, from 1 $\mu eV$ to $0.1 meV$, more clearly and shows all the various experiments. I will briefly outline how each of these techniques work in the following sections, with the exception of the haloscope as that will be covered in the next chapter in detail.
    \subsection{Astrophysical constraints}
    If axions exist, depending on their mass and coupling, they would be a source of energy dissipation in astrophysical objects. Similar to a neutrino, if axions couple strongly enough to regular matter, they would be produced enough in stars and supernovae to affect their energy loss rates and lifetimes. However, at the same time, if they are too strongly coupled, they would be trapped within the star endlessly colliding off the dense matter within. By looking at the lifetime of stars and their energy budgets, constraints on the axions mass and coupling can be made.
    \par In the case of our sun, which is only burning hydrogen currently, axions would be primarily generated through a Primakoff process, $\gamma + Ze \rightarrow a + Ze$, The additional energy loss from this extra channel would cause the sun to contract faster, in turn raising its temperature, and thereby increasing the fusion rate in its core. This increase in fusion would result in increased neutrino production, which can be measured by experiments on Earth. The Sudbury Neutrino Observatory (SNO) was able to set limits in this way, citing $g_{a \gamma \gamma} < 7 \times 10^{-10} GeV^{-1}$ \cite{SNOaxion}.
    \par A stronger limit than the Solar $\nu$ is the constraints set by globular clusters in Figure \ref{fig:Axionlimitstheory}. Globular clusters are not full-form galaxies, but spheroidal conglomerations of stars caused by gravity. Most importantly, they do not have active, appreciable star formation taking place within them. This means most stars within them are all nearly the same age, and vary by their initial masses. This means the population sizes of the various phases of stars within them is proportional to the time the star is spending in such a phase. Because axions would have the effect of shortening the helium burning phase of stellar evolution, the horizontal branch stars (HB), one can compare the populations of these stars to the hydrogen-burning phase stars, the Red Giant Branch (RGB), within globular clusters, and set limits on axions accordingly \cite{RGBlimit1}. This has ruled axions out over a very wide mass range to a coupling of $g_{a \gamma \gamma} < 6.6 \times 10^{-11} GeV^{-1}$ \cite{RGBlimit2}.
    \par Limits based on supernovae observations can also be set. If axions exist, the primary means of axionic cooling would be from nucleon-nucleon axionic brehmsstrahlung ($NN \rightarrow a NN$). Its important to note however, that axion masses $<1\, meV$ would have not taken away enough energy to be observable, and axions with mass $> 2\,eV$ would not have escaped the supernova, giving their energy back to the event. But if the axion has mass in the intermediate range, this would have resulted in a significantly shortened neutrino burst during the SN1987A event, which was not observed \cite{SN1987A1,SN1987A2}. 
    \par There are many other astrophysical constraints that I do not have time to discuss here; neutron star lifetimes, black-hole super radiance, and optical telescope measurements for instance \cite{neutronStars,Blackholesuperradiance,Telscopes}. If one assumes axions are the majority of dark matter, constraints on the CMB, Active Galactic Nuclei (AGN), and other phenomena can also be made. Exclusion limits from the James Webb Telescope have even been made recently \cite{JWST}. In general, these other astrophysical methods do not exclude axions in the KSVZ and DFSZ coupling ranges, either by not being sensitive enough to the low couplings, or by looking at axion masses much less than 1 $\mu eV$. 
    \subsection{Laboratory Searches}
    Axion helioscopes are the high-energy sibling to the axion haloscope \cite{Sikivie}. They aim to convert solar axions originally produced in the sun back into X-ray photons via a large magnet. A large-bore, long, high-field magnet is pointed at the sun where solar axions would enter and along its length decay into photons. An X-ray optics package focuses these potential photons onto a detector. Because of this setup, axions can be a variety of masses below a limiting upper mass which is constrained by other astrophysical backgrounds. The most famous of these axion telescopes is the CERN Axion Solar Telescope (CAST), which has set a limit of $g_{a \gamma \gamma} < 6.6 \times 10^{-11} GeV^{-1}$ for $m_a < 0.02\, eV$ \cite{CAST}. The physics world is greatly anticipating the follow-up experiment, the International Axion Observatory (IAXO), which has a target sensitivity of $g_{a \gamma \gamma} < 5 \times 10^{-12} GeV^{-1}$.
    \par Another major category of laboratory searches are "light-shining-through-walls" experiments. These experiments contain two Fabry-Perot cavities; an axion generator cavity and receiver. The generator cavity is embedded in a dipole magnetic field and contains a laser that is pumping photons within it. The laser photons are stimulated to convert to axions, with an enhancement factor based on the cavity optical path length. Once axions are generated, they could travel through an optical barrier (the "wall") and end up in the receiver cavity, which is also embedded in a magnet. The axion in the receiver cavity then decays back into a photon that would be detected by a CCD detector. These experiments include ALPS, ALPS-II, CROWS, PVLAS, and OSQAR. OSQAR has set the most stringent limit at $g_{a \gamma \gamma} < 3.5 \times 10^{-8} GeV^{-1}$ for $m_a < 0.3 \,meV$ \cite{OSQAR}.
    \par Another technique gaining popularity looks for axions in the low mass range via LC-resonant circuits, such as the ADMX-SLIC, ABRACADABRA and SHAFT experiments \cite{SHAFT,ABRA}. They consist of a toroidal magnet that causes a very tiny oscillating current within the loop of the toroid due to axions; a superconducting sheath acts as the wire for this current to travel on. This oscillating current would produce a magnetic flux in the center of the toroid, which could be detected by an ultra-sensitive magnetometer. This inductor at the center of the toroid is usually a tunable SQUID sensor. These can be thought of as haloscopes that simply replace the resonant cavity with lumped circuit elements. ADMX-SLIC, sited at the University of Florida, uses this technique and was able to exclude axions with coupling $g_{a \gamma \gamma} < 10^{-12} GeV^{-1}$ in the $174.98-175.19\, neV$ and $177.34-177.38\, neV$ mass ranges \cite{ADMXSLIC}. ABRACADABRA was able to recently set a new limit of $g_{a \gamma \gamma} < 3.2 \times 10^{-11} GeV^{-1}$ for axions in the $0.41-8.27\, neV$ mass range \cite{ABRA}. The DM-radio collaboration will soon push the limits of this technique, with construction of DM Radio 50 L underway at Stanford, and DM-radio $m^3$ in the design phase.
    \par As you can see from Figure \ref{fig:Axionlimitstheory}, the through-line for all these other techniques, is none of them have been able to exclude significant parts of the DFSZ and KSVZ region except the haloscopes; I will cover how these work in the next chapter.
 
 
\chapter{Overview of haloscopes}
\label{chap:Haloscopes}
 This chapter will cover what an axion haloscope is, its main components, and how these components affect the sensitivity of the detector system. The concept of an axion haloscope was first proposed by Pierre Sikivie in 1983 \cite{Sikivie,Sikivie2}. Invisible axions are made visible by exploiting the coupling of axions to the electromagnetic field with a large static magnetic field. In this arrangement, the magnetic field will act as a virtual photon, and axions from the galactic halo will convert to photons within the high static magnetic field via the two photon coupling illustrated in Figure \ref{fig:axionfeynmandiagram} (hence the name 'halo'-scope). The experimental apparatus must then have a way of detecting the resultant photons and distinguish them from any background sources; this is done with a resonant cavity, low-noise readout, and cooling system.
    \section{Components}
    \label{sec:components}
    \begin{figure*}[htb!]
    \centering
    \includegraphics[width=1.0\linewidth]{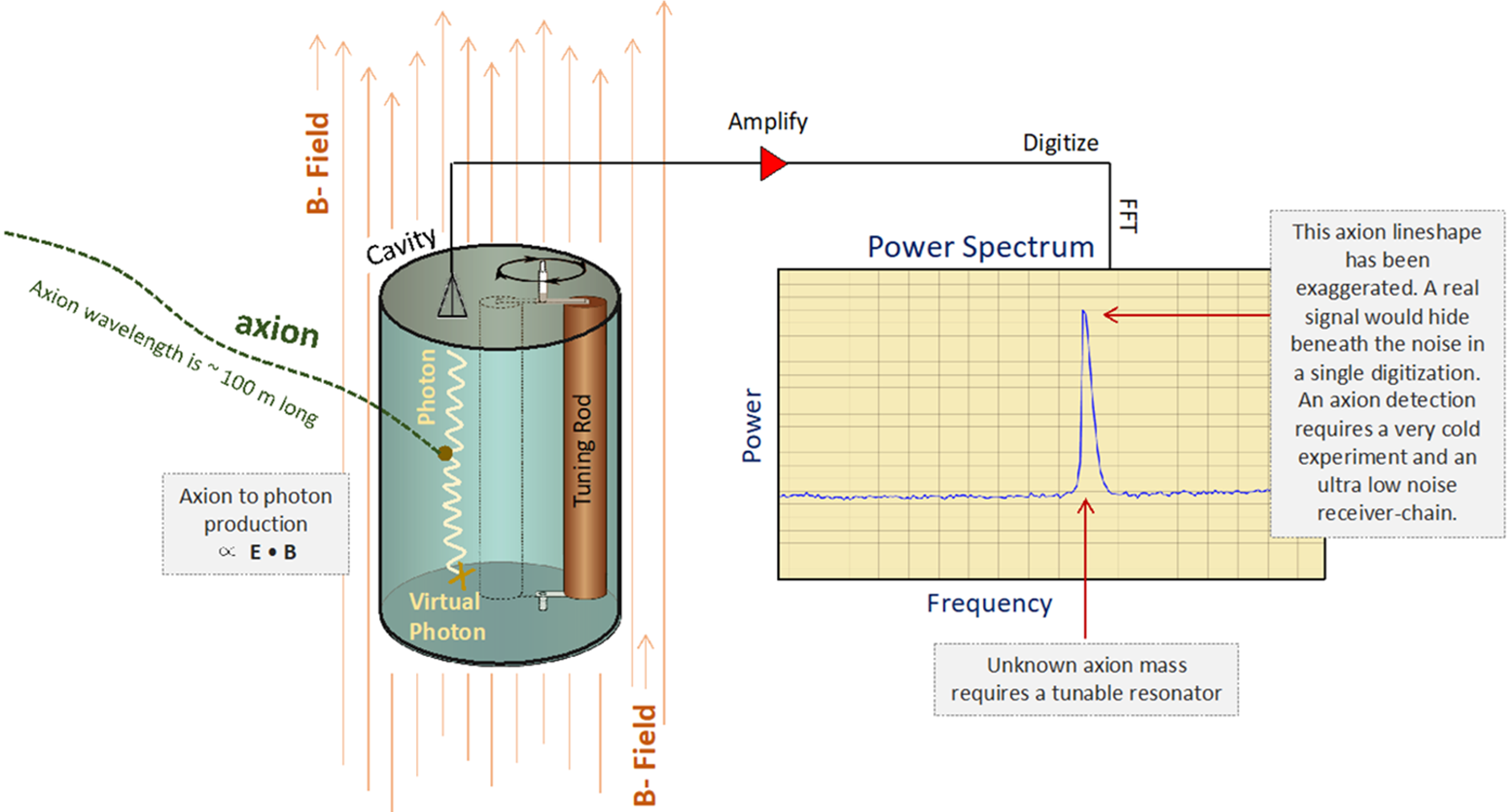}
    \caption{A toy model of an axion haloscope's main components and detection process \cite{Boutan}.}
    \label{fig:HaloscopeDiagram}
    \end{figure*}
    The 4 main components of any axion haloscope as alluded to above are:
    \begin{enumerate}
        \item High field strength magnet to stimulate axion to photon conversion
        \item An embedded, frequency tunable, resonant cavity or cavities within that magnet's field to collect axion photon power
        \item A readout system for the resonant cavity system to detect resultant photon power
        \item A cooling system to minimize external photon noise (primarily thermal photons) within the cavity and readout systems
    \end{enumerate}
    \par These components are pictured in Figure \ref{fig:HaloscopeDiagram} and will be used as a visual guide for this section. The specifics of these components build on one another and dictate what kind of axion masses and couplings can be searched for. From a fiscal standpoint, the magnet, because it is usually the most expensive equipment of these components, is the starting point for what kind of axion search is possible within a given budget. 
        \subsection{Magnet}
        \par The magnet and the static field it produces creates the potential detection volume for axions as pictured by the field lines in Figure \ref{fig:HaloscopeDiagram}. A higher peak field strength will stimulate more axion to photon production, specifically proportional to the interaction Lagrangian as shown in Equation \ref{eqn:interactionlagrangianAgammagamma}. Therefore the axion signal power is proportional to the magnet field strength, specifically by $B^2$ as will become clear in the axion power equation section. Additionally, the volume of the external field will cast a larger detection footprint. It is important to note, however, because axions have a de Broglie wavelength much larger than the length scale of most magnet volumes, a classical particle picture isn't quite accurate; more volume isn't casting a larger target to 'catch' more axions like in many WIMP experiments, rather it is covering a larger portion of the axion field and that will drive more power into the embedded resonant cavity system. Simply said, a magnet's figure of merit for a search would be $\oint\vec{B}  \,dV$ where $B$ is the magnet field strength and $V$ is the field volume.
        \par Superconducting solenoid magnets are the standard for generating the highest static magnetic fields in any laboratory, and thus have dictated the geometry and design of axion haloscope experiments so far. The detection volume in this case is the internal cylindrical bore dimensions, which subsequently dictates the size of a resonant cavity system that can be inserted within it. These magnets are commonly wound with NbTi or NbSn superconducting wires, with the latter being used for higher strength magnets that exceed the critical field of NbTi ($\approx15\,\mathrm{T}$). NbSn magnets, because of their winding materials' critical fields and flux creep, are limited to $\approx10-20\,{\rm T}$ depending on construction details (these superconducting physics terms are discussed in Chapter \ref{chap:SRF}).
        \par These superconducting magnets commonly make use of a closed loop design with a persistent switch, where the power supply is only connected when ramping the magnet current. This saves on electricity costs when operating, leaving cooling of the magnet the principle expense. These magnets are cooled with a pumped liquid helium system, and if operating for a long duration like axion searches do, often have an associated helium liquefaction plant and recovery system.
        \par This limit can be exceeded by the use of a Bitter electromagnet design, with the record being $37.5\,{\rm T}$ at Radboud University High Magnetic Field laboratory in the Netherlands \cite{RadboudMagnet}. The strongest continuous, static magnet field ever generated in a laboratory was of a hybrid design: a Bitter magnet inside a superconducting solenoid. This was constructed at the National High Magnetic field laboratory (NHMFL) in Tallahassee, FL, USA. The Bitter magnet generated $33.5\,{\rm T}$ and the outer solenoid generated the remaining $11.5\,{\rm T}$ for a total peak field of $45\,{\rm T}$ \cite{NHMFL45T}. Although incredible feats of engineering, these magnets do not make sense for current axion searches because of their prohibitive energy and cooling expenses. In the case of NHMFL's record magnet, it consumed around 33 megawatts of power for its liquid helium system and surrounding infrastructure, which at the time cost approximately \$1500 per hour; An axion search must operate for months at a time, making use of these ultra-high field magnets prohibitively expensive at current budgets. Because of the requirement of a large bore, most current and planned axion searches stick to conventional NbTi or NbSn magnets with very few experiments having magnets over $10\,{\rm T}$.
        \subsection{Cavity}
        \par Similar to the magnet, there is no axion signal to be detected without a cavity resonator to be driven by the axion field. As previously mentioned the resonant conversion of halo axions to photons is proportional to $\vec{E}\cdot\vec{B}$, see Equation \ref{eqn:interactionlagrangianAgammagamma}, where $\vec{E}$, is the electric field of the cavity mode, and $\vec{B}$ is the external magnetic field. This means for a solenoid magnet with a cylindrical cavity, having an axial electric field for as much of the internal cavity volume as possible, maximizing the $\vec{E}\cdot\vec{B}$. The $TM_{010}$ resonant cavity mode does this, with its electric field entirely along the cylindrical cavities' axial direction (see Figure \ref{fig:HaloscopeDiagram}). The specifics of resonant cavity modes will be discussed in detail during Chapter \ref{chap:Cavities}. In order for this cavity mode resonance to be excited, it would have to be tuned precisely to the resultant photons frequency:
        \begin{equation}
        f_a\approx\frac{m_ac^2}{h}
        \label{eqn:axionfreq}
        \end{equation}
        \par For axion rest masses of $10^{-5}\,{\rm eV}$, this corresponds to a frequency of $2.4\,{\rm GHz}$ or wavelength of 12 cm. As mentioned in Chapter \ref{chap:axiontheory}, the local cold dark matter density is approximately $0.45\,{\rm Gev/cc}$, therefore, if axions make up the entirety of the dark matter, the number density will be over $10^{13}\,{\rm cm^{-3}}$. Also discussed earlier, these axions will be cold, with $\beta\approx10^{-3}\,c$, corresponding to a de Broglie wavelength of 1000 m (its also of note that this is why the approximation in Equation \ref{eqn:axionfreq} works). This means that on the length scale of resonant cavity, $\approx 12\,{\rm cm}$, an axion halo will behave like a classical field, coherent over a much larger scale than the cavity, $~100-1000\,{\rm m}$. Thus, one can view the axion field as a coherent classical source of radiation that is driving the resonant cavity mode. Because of the very small velocity dispersion of the axion field, the fractional width of the excitation peak will be around $10^{-6}$ within the cavity power spectrum. Therefore a resonator with a high quality factor, as close to $10^{+6}$ as possible, is ideal for initial detection (more on this in Chapter \ref{chap:Cavities}). 
        \par This is all to say that the aim of any haloscope experiment is to detect this resonant peak by monitoring the power spectrum of the cavity near the $TM_{010}$ mode or other search mode of interest (see the exaggerated resonant peak pictured in Figure \ref{fig:HaloscopeDiagram}). To do this, antennae are carefully inserted into the cavity space to sample this power spectrum precisely. Because one does not know the axion rest mass precisely, this resonant mode frequency must also be adjustable, or "tunable" in order to line it up with a range of potential axion photon frequencies. This is why tuning rods are used as pictured in Figure \ref{fig:HaloscopeDiagram}.
        \par There are several factors that go into the design of an optimum cavity resonator. One wants to maximize the volume of axion field driving the resonant system, therefore maximizing $V_{cav}$ within the magnet bore volume is ideal. However, this effects the resonant frequency of the $TM_{010}$ mode, $f_{010}^{TM}$, which must be set according to the desired axion search frequency range. For an empty right cylindrical cavity the analytical solution to this mode frequency is:
        \begin{equation}
        f_{010}^{TM}=\frac{x_{01}}{2\pi\sqrt{\mu\epsilon}R}
        \label{eqn:TM010Freq}
        \end{equation}
        where $\mu$ and $\epsilon$ are the interior permeability and permittivity, $x_{01}$ is the first root of the Bessel function $J_0(x)$, and $R$ is the radius of the cavity (the derivation of this expression is covered in Chapter \ref{chap:Cavities}). Therefore, at least for an empty cavity, the radius sets the resonant frequency, and the length can be maximized to the length of the magnet bore. When tuning rods are introduced, the axial symmetry is broken, and there are no analytical solutions; one is forced to use numerical modelling methods to calculate frequency. The cavity length in this case will have a perturbative effect on the frequency. Although it will be discussed more in Chapter \ref{chap:Cavities}, this connection between volume and frequency can be circumvented by the use of multi-cavity systems that maximize detection volume while still targeting a specific resonant frequency; this is especially useful for higher axion mass searches.
        \par The factor that includes the axion-photon interaction Lagrangian, $\propto\vec{E}\cdot\vec{B}$, we refer to as the cavity form factor. It is dimensionless with a range of 0 to 1, and is maximized for the $TM_{010}$ mode, being 0.7 for an empty cavity, and typically between 0.3-0.5 for a cavity with tuning rods. Similar to frequency, it can be calculated explicitly for an empty cavity, but only through numerical modelling for tunable cavities. Although it will be derived explicitly in the next section, I leave it here for reference:
        \begin{equation}
        C_{lmn}=\frac{(\int\vec{E}(\vec{x},t)\cdot \vec{B}_{ext}(\vec{x})dV)^2}{\int\vec{B}_{ext}^2dV\int\epsilon_r\vec{E}^2dV}
        \label{eqn:CavityFormFactor}
        \end{equation}
        where $\vec{E}(\vec{x},t)$ is the electric field of the $lmn$ indexed cavity mode (such as 010 in the case of $TM_{010}$), $V$ is over the interior cavity volume, $\epsilon_r$ is the relative permittivity of the interior cavity, and $\vec{B}_{ext}(\vec{x})$ is the externally applied magnetic field. One can see that only modes of the form $TM_{0n0}$ will have axial electric fields that overlap with the axial magnetic field in order for form factor to be appreciably non-zero. 
        \par Finally, the quality factor of the cavity, the primary parameter of this dissertation, should be maximized in order to achieve unity with the very narrow axion line shape and maximize resonant power exchange into the mode. Although it will be discussed extensively in Chapter \ref{chap:Cavities}, the quality factor is a measure of how many cycles the resonator mode can hold before energy is dissipated through the cavity's interior surfaces, antennae, and other radiated losses. This definition can be written as:
        \begin{equation}
        Q=\omega_0\frac{U}{P_d}
        \label{eqn:CavityQualityFactor1}
        \end{equation}
        where $\omega_0$ is the angular frequency of the mode ($2\pi f_0$), $U$ is the energy stored in the mode, and $P_d$ is the power dissipated via both material and radiated losses. 
        \par In summation, a cavity system must be designed to target a certain axion mass frequency range, and its figure of merit for sensitivity to axions would be $\approx C^2V^2Q$, where $C$ is the form factor, $Q$ is the quality factor, and $V$ is cavity system volume (the powers of two for $C$ and $V$ will become apparent later in this chapter).
        \subsection{Readout System and Cooling System}
        The readout and cooling system is designed to detect an ultra-low power resonant signal generated by the cavity over the pre-dominant thermal noise background. I combine these two systems together, because in terms of operations and analysis they are very closely connected. This is because the dominant background within the cavity as well as the readout system itself is Johnson-Nyquist thermal noise, due to the non-zero temperature of these components. Therefore, the cavity is cooled to the lowest cryogenic temperatures feasible and the lowest noise electronics available are employed to readout the very cold power spectrum. This includes low-loss superconducting lines, and quantum amplifiers to boost the power signal as it is sent to room temperature electronic components, also known as the 'warm receiver'. According to Nyquist theorem, the noise power, $P_{Noise}$ deposited over a given bandwidth of frequency space, $\Delta f$, from a system with a total system noise temperature, $T_{sys}$ is:
        \begin{equation}
        P_{Noise}=k_BT_{sys}\Delta f
        \label{eqn:NoisePower}
        \end{equation}
        where $k_B$ is the Boltzmann constant. The system noise temperature is not only the physical temperature of the cavity, but takes into account the temperature of the electronic components along the chain. Since the cavity readout signal is passed through a series of amplifiers, each with an associated gain that multiplies the input signal, the subsequent up-stream added noise is suppressed by the gains of the down-stream components. In this arrangement, it is easier to associate noise introduced by cables and other electronic components with the amplifier stage it is contained within by a single noise temperature. For a chain of N amplifiers this translates to the total system noise temperature expression being:
        \begin{equation}
        T_{sys}=T_{phys}+T_{1}+\frac{T_2}{G_1}+\frac{T_3}{G_1G_2}+...+\frac{T_N}{G_1G_2G_3...G_{N-1}}
        \label{eqn:NoiseTemp}
        \end{equation}
        Where $T_{phys}$ is the physical temperature of the cavity, $T_1,...,T_N$ are the noise temperatures of each amplifier stage including the down-stream cable component of each stage, and $G_1,...,G_N$ are the linear gains of each amplifier in the chain. This means even if a cavity is cooled to a $100\,{\rm mK}$, a typical solid state amplifier with a noise temperature of $2\,{\rm K}$ if used as a primary amplifier ($n=1$), will add its full noise temperature to your system. However, if it is used just a stage above an amplifier with $10\,{\rm dB}$ gain, it only adds $200\,{\rm mK}$ to the system. Attenuation of the coaxial line between the cavity and primary amplifier can be made near-zero by the use of superconducting cables. For this reason, the primary amplifier is the most critical component to any haloscope RF chain. 
        \par Most current experiments, including ADMX, make use of Superconducting Quantum Interference Device (SQUID) type amplifiers where the noise temperature scales with its physical temperature, and is only limited by the uncertainty principle where,
        \begin{equation}
        T_{SQUID}\geq\frac{hf}{k_B}
        \label{eqn:QuantumLimit}
        \end{equation}
        This is typically referred to as the standard quantum limit. For an amplifier operating at $2\, {\rm GHz}$, this is about $100\,{\rm mK}$. 
        \par However, this is much easier said than done, especially since these SQUIDs are sensitive to even a single magnetic flux quanta ($h/2e$), yet must sit not far from an $8\,{\rm T}$ solenoid magnet so it is still cooled by the same fridge system as the cavity. It's for this reason that another application of SQUIDS is extremely sensitive magnetometers for condensed matter physics and material science. For our purposes, we have to protect this amplifier as best as possible from magnetic overload. To do this, ADMX employs a 'bucking' magnet that is counter-wound to have zero mutual inductance with the main magnet, but cancels the main magnet's field to $<100$ Gauss within its bore. Within this "bucking" magnet bore, the quantum-sensitive electronics are protected (a citadel for the SQUIDs, or 'SQUIDadel' so to speak). However, to maintain the generality of this section, the quantum amplifier package could be stored in a far-field region and cooled separately, or in some other way, to ensure optimal performance.
        \par These SQUID amplifiers are a field of research and development in themselves, and the axion haloscope community likes to think we are driving the need for new emerging SQUID amplifier technologies. In the case of ADMX, this includes implementing Microstrip SQUID Amplifiers (MSA) initially, Josephson Parametric Amplifiers (JPA) on the current main experiment, and most recently a Travelling Wave Parametric Amplifier (TWPA) on the ADMX pathfinder experiment, Sidecar. Increasing the operational bandwidth of these amplifiers is the main incremental improvement, which allows the haloscope to scan a larger frequency range before needing to be tuned or even replaced.
        \par This leaves the physical temperature of the cavity, $T_{phys}$, in Equation \ref{eqn:NoiseTemp} as the main driver of the system noise. In its simplest form, the cavity is already being inserted into a superconducting magnet cooled by liquid helium, which will bring the bore temperature to $4\, {\rm K}$, but this can be improved upon. The first generation of ADMX implemented a pumped liquid helium system which could cool the cavity to $1.3\,{\rm K}$. In its current Gen 2 configuration today, it uses a $He_3/He_4$ dilution refrigerator that can achieve a base temperature of $80\, {\rm mK}$, but it typically operates in the $100\, {\rm mK}$ range. This seems to be the most popular, cost effective technology of the day driven by the demand of quantum computing research. These refrigerators can go as low as $2\, {\rm mK}$,  just not for a large thermal mass with moving components like the ADMX insert. Beyond the $2\, {\rm mK}$ limit, magnetic cooling technologies would be necessary. However, as stated above, $2\,{\rm mK}$ is already below the quantum limit for the typical axion haloscope frequency ranges, thus would only be relevant to low-mass axion searches below $\approx 42\,{\rm MHz}$; it would lower $T_{sys}$, but the primary amplifier would continue to sit at the quantum limit. There are ways of circumventing the quantum thermal limit, such as quantum squeezing, but this is out of the scope of this dissertation. 
        \par In summary, a haloscopes' dominant background is thermal noise, therefore the cavity followed by a high gain, quantum-limited, primary amplifier are cooled as close to the quantum limit as possible for a low system noise temperature. Up-stream amplifiers can be significantly warmer because they will be suppressed by the gains of down-stream amplifiers.
    \section{Axion power in a haloscope}
    In this section, I will review an equivalent circuit derivation of the axion signal power in a haloscope, however, there are several ways to arrive at this equation, and this derivation is not at all original, but follows ref. \cite{DawThesis,RogersThesis,KinionThesis,YuThesis}. It can also be derived by introducing an axion source term to Maxwell's equations, which is done in ref. \cite{PhysRevLett.55.1797,Hagmann,MalagonThesis,brubakerThesis}. What I would call a combination of these methods can be found in ref. \cite{HotzThesis,LyapustinThesis}. I have chosen to include this derivation here because it introduces the equivalent circuit picture and methods that will be useful in Chapter \ref{chap:Cavities} for thinking about resonant cavity characterization. 
    \begin{figure*}[htb!]
    \centering
    \includegraphics[width=0.4\linewidth]{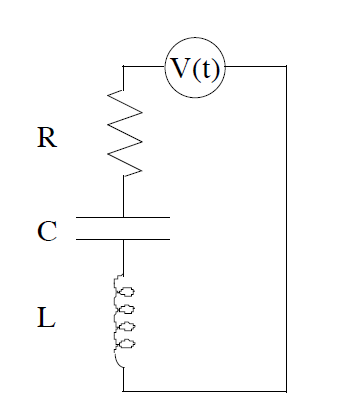}
    \caption{An equivalent circuit for a resonant cavity being driven by axion photonic power.}
    \label{fig:RLCCircuit}
    \end{figure*}
    \par The resonant cavity being driven by the axion field decaying into photons can be written as an RLC circuit with a time dependent voltage source term, as pictured in Figure \ref{fig:RLCCircuit}. A resonant cavity stores both electrical and magnetic field energies, giving it an inherent capacitance and inductance, while also dissipating this energy through its associated surface resistances and radiation losses, acting as the resistor. The time-dependent voltage source is the axion photon power being deposited in the cavity. This circuit has a classical Lagrangian of:
    \begin{equation}
    \mathcal{L}=\frac{1}{2}L\dot{q}^2-\frac{1}{2C}q^2+qV(t)
    \label{eqn:CircuitLagrangian}
    \end{equation}
    As well as the dissipative power for the resistor:
    \begin{equation}
    P_{d}=\frac{1}{2}R\dot{q}^2
    \label{eqn:dissipationL}
    \end{equation}
    We can relate the inductive and capacitive energy terms in this equivalent circuit Lagrangian back to Maxwell's equations by the magnetic and electric field Poynting energy integrals for a given cavity mode:
    \begin{equation}
    \frac{1}{2}L\dot{q}^2=\frac{\mu}{2}\int |\vec{H}|^2 \,dV \
    \label{eqn:InductiveEnergy}
    \end{equation}
    \begin{equation}
    \frac{1}{2C}q^2=\frac{\epsilon}{2}\int |\vec{E}|^2 \,dV \
    \label{eqn:CapacitiveEnergy}
    \end{equation}
    In the non-dissipative and non-driven case ($P_{d}=0,\; V(t)=0$), the Euler-Lagrange equation will simply result in simple harmonic oscillations as expected:
    \begin{equation}
    \frac{d}{dt}(\frac{\partial\mathcal{L}}{\partial\dot{q}})-\frac{\partial\mathcal{L}}{\partial q}=0
    \label{eqn:Euler-Lagrange-Nondissapative}
    \end{equation}
    \begin{equation}
    L\ddot{q}+\frac{q}{C}=0 \rightarrow \; \ddot{q}=-\omega_0 q , \; \omega_0=\frac{1}{\sqrt{LC}}
    \label{eqn:LCHarmonicMotion}
    \end{equation}
    The Euler-Lagrange equation in the driven, dissipative case is:
    \begin{equation}
    \frac{d}{dt}(\frac{\partial\mathcal{L}}{\partial\dot{q}})-\frac{\partial\mathcal{L}}{\partial q}+\frac{\partial P_{d}}{\partial \dot{q}}=0
    \label{eqn:Euler-Lagrange-dissapative}
    \end{equation}
    with equation of motion:
    \begin{equation}
    L\ddot{q}+R\dot{q}+\frac{q}{C}=V(t)
    \label{eqn:RLCmotion}
    \end{equation}
    We expect for the dissipative, driven cavity case to behave like an under-damped, continuously driven oscillation; the mode will be excited for more than a single cycle from the deposited axion energy. In this way, we can assume that the mode energy will be transferred back and forth between the electric and magnetic fields over the course of a single oscillation cycle, making the time-averaged energy stored in the mode for the initial oscillation equal to either Equations \ref{eqn:CapacitiveEnergy} or \ref{eqn:InductiveEnergy}:
    \begin{equation}
    U_{0}=\frac{1}{2}L\dot{q}^2=\frac{\mu}{2}\int |\vec{H}|^2 \,dV \
    =\frac{1}{2C}q^2=\frac{\epsilon}{2}\int |\vec{E}|^2 \,dV \
    \label{eqn:ModeEnergy}
    \end{equation}
    Next, we can relate the power dissipated in the circuit picture to the power dissipated in the field equation picture:
    \begin{equation}
    P_{d}=\frac{1}{2}R\dot{q}^2=\frac{R}{2}\int |\vec{H}|^2 \,dS \
    \label{eqn:Powerdissipated}
    \end{equation}
    Combining Equations \ref{eqn:CavityQualityFactor1}, \ref{eqn:ModeEnergy}, and \ref{eqn:Powerdissipated} together, we can arrive at a circuit expression for quality factor in the circuit picture (note this assumes no antenna or radiation losses):
    \begin{equation}
    Q=\frac{1}{R}\sqrt{\frac{L}{C}}
    \label{eqn:QualityfactorCircuitDef}
    \end{equation}
    \par Next we must relate the axion-photon coupling to the voltage source, $V(t)$. To do this, we take the Lagrangian term describing the axion fields coupling to the electromagnetic fields (Equation \ref{eqn:interactionlagrangianAgammagamma}) and integrate this interaction of the cavity volume ($V_{cav}$). This result should be equal to the energy of the voltage source $q(t)V(t)$:
    \begin{equation}
    q(t)V(t)=-g_{a\gamma\gamma}c\epsilon_0a(t)\int \vec{E}(\vec{x},t)\cdot\vec{B}(\vec{x},t) \,dV_{cav}
    \label{eqn:axionvoltagesourcelagrangian}
    \end{equation}
    We can then substitute the form factor expression from the last section, Equation \ref{eqn:CavityFormFactor}, as well as Equation \ref{eqn:CapacitiveEnergy} into the expression for form factor. Finally, I will assume a perfectly axial magnetic field, $\vec{B}(\vec{x},t)=B_0\hat{z}$ which simplifies the expression for voltage source to:
    \begin{equation}
    V(t)=-g_{a\gamma\gamma}cB_0\sqrt{\frac{\epsilon_0C_{lmn}V_{cav}}{C}}a(t)
    \label{eqn:axionvoltagesourcesimplified}
    \end{equation}
    Note that $C_{lmn}$ is cavity form factor, $C$ is capacitance, $V(t)$ is voltage, and $V_{cav}$ is cavity volume. Next, sticking to the circuit picture, the power, $P$, deposited in the cavity due to this axion-photon conversion voltage source would be given by:
    \begin{equation}
    P=\frac{\langle V^2(t)\rangle}{R}=g_{a\gamma\gamma}^2\epsilon_0c^2B_0^2V_{cav}C_{lmn}\frac{1}{RC}\langle a^2(t)\rangle
    \label{eqn:axionpower1}
    \end{equation}
    We can then use the circuit picture expression for quality factor, Equation \ref{eqn:QualityfactorCircuitDef}, and the circuit resonant frequency, Equation \ref{eqn:LCHarmonicMotion}, to get rid of the circuit parameters:
    \begin{equation}
    P_{a \rightarrow \gamma} =g_{a\gamma\gamma}^2\epsilon_0c^2B_0^2V_{cav}C_{lmn}\omega_0 Q\langle a^2(t)\rangle
    \label{eqn:axionpower2}
    \end{equation}
    The square amplitude of the axion field can be related to the mass density of axions in the region:
    \begin{equation}
    \langle a^2(t) \rangle = \frac{\rho_a \hbar^2}{m_a^2c}
    \label{eqn:axionfieldamplitude}
    \end{equation}
    This gives the expression:
    \begin{equation}
    P_{a \rightarrow \gamma} =\frac{2\pi \epsilon_0\hbar^2c}{m_a^2}g_{a\gamma\gamma}^2 \rho_af_aB_0^2V_{cav}C_{lmn}Q
    \label{eqn:axionpower3}
    \end{equation}
    Using the Chapter \ref{chap:axiontheory} theory expressions for $m_a$ and $g_{a\gamma\gamma}$, Equations \ref{eqn:m_axion2} and \ref{eqn:gagammagamma}, $m_a$ can be eliminated and $g_{\gamma}$ can be used in place of $g_{a\gamma\gamma}$. Additionally, we have ignored the antenna that is sampling this cavity power; this introduces a factor of $\frac{\beta}{1+\beta}$, where $\beta$ is the coupling coefficient. This coefficient will be covered in Chapter \ref{chap:Cavities}, and technically varies from 0 to $\infty$, but is typically kept at a value of 1-2 for axion searches. Substituting in the appropriate constants and normalizing the equation to typical values seen in the laboratory, one gets an expression more useful to the experimenter:
    \begin{equation}
    \begin{split}
     P_{a \rightarrow \gamma} =
     6.94 \times 10^{-23} W &\left[\left(\frac{g_{\gamma}}{0.36}\right)^2\left(\frac{\rho_a}{0.45 \mathrm{GeV/cm^3}}\right)\left(\frac{f}{1 \mathrm{GHz}}\right)\right]\times\\
     &\left[\left(\frac{B}{7.6\, \mathrm{T}}\right)^2\left(\frac{V_{cav}}{136\,\mathrm{L}}\right)\left(\frac{C_{lmn}}{0.4}\right)\left(\frac{\beta}{1+\beta}\right)\left(\frac{Q_0}{70000}\right)\right]   
    \end{split}
    \label{eqn:axionpower4}
    \end{equation}
    where values of the terms grouped in the left brackets are fixed by nature (the experiment is designed around the frequency range) and terms in the right brackets are parameters that can be controlled by the scientist. The value $g_{\gamma}$ here is the DFSZ model value, and the axion mass density is assumed to be the local halo dark matter density; that is to say axions make up the entirety of dark matter. One can adjust these values based on their assumed axion model and dark matter halo composition. By far the most striking feature of this expression is the incredibly feeble size of this signal power: $10^{-23} W$ or a 10 'yotta' watts order. Because of this, one needs to 'listen' at each frequency long enough to ensure they could detect such a weak power excess over the thermal background before moving onto the next. That is to say, there is a proper frequency scan rate to exclude axions to a certain sensitivity. Next I will combine this expression with the expected thermal background noise, obtain a signal-to-noise ratio, and derive the instantaneous frequency scan rate.
    \section{Scan rate}
     As illustrated in Figure \ref{fig:HaloscopeDiagram}, a power spectrum of the cavity near its mode resonance is taken and searches for a power excess over the noise floor that could be from an axion. Once no power excess is seen, the tuning rods are moved to change the mode resonant frequency, and a new power spectrum is taken near that new resonance; this is the 'scanning' process of the experiment. 
    \par More precisely, a time-domain digitization is taken over a period of time of the cavity antennas' field power response. This time-series data is then Fourier transformed into a frequency domain power spectrum. Power excesses above the mean noise floor can then be characterized by a signal-to-noise ratio (SNR) according to the Dicke Radiometer Equation \cite{DickeRadiometer}:
    \begin{equation}
    SNR=\frac{P_{a\rightarrow\gamma}}{k_B T_{sys}}\sqrt{\frac{t}{B}}
    \label{eqn:DickeRadiometer}
    \end{equation}
    where $P_{a\rightarrow\gamma}$ is the axion power deposited according to Equation \ref{eqn:axionpower3}, $T_{sys}$ is the system noise temperature described by Equation \ref{eqn:NoiseTemp}, $t$ is the integration time over which the Fourier transform is taken, and $B$ is the integrated bandwidth of the Fourier transform. That is to say, for a given noise temperature, by taking a longer time sitting on a given cavity resonance, or using a smaller frequency bandwidth for the transform, the SNR can be increased. However, spending more time on one frequency limits how much frequency space can be scanned in a given amount of time. ADMX typically tries to keep an SNR of about 3.5 across its scanning range, so I will assume that to be the default value going forward. 
    \par It is important to make a note about bandwidth $B$ here. Effectively it is the width of our frequency 'bins' as we take multiple transforms centered at different frequencies. Yet, the cavity resonant mode peak has an effective  frequency width itself $\Delta f_c$, and so does the axion signal lineshape, $\Delta f_a$. This a good time to introduce the next definition of quality factor:
    \begin{equation}
        Q=\frac{f_0}{\Delta f}
    \label{eqn:QualityfactorDeltaF}
    \end{equation}
    where $f_0$ is the resonant frequency and $\Delta f$ is the full-width at half maximum (FWHM) of the resonant peak (more on this in Chapter \ref{chap:Cavities}). As discussed in Chapter \ref{chap:axiontheory}, the axion lineshape is dictated by its expected velocity dispersion, and a safe assumption is $\Delta f_a  \leq 1 \,{\rm kHz}$ for an axion at $1 \,{\rm GHz}$, producing an associated axion quality factor, $Q_a \approx 10^6$. In the case of an axion search, where one's goal is to just find the axion signal, and not try to measure the line-shape structure itself, one would not want to take an integration bandwidth smaller than $\Delta f_a$, because it would not capture the full axion signal power; our purposes here are just to detect the axion and not capture the signal structure. ADMX does perform high resolution analysis of its data that uses smaller integration bandwidths to search for small signal modulations, but this is outside the scope of this proof. Additionally, any axion signal that is outside the cavity bandwidth $\Delta f_c$ will not have the resonant amplification and therefore won't be detectable. In fact, having a larger bandwidth will contribute extra thermal noise from those other included frequencies to the transformed power, according to Equation \ref{eqn:NoisePower}. Therefore, it is clear that one wants $\Delta f_c \gtrsim B \gtrsim \Delta f_a$, and in the best case scenario, $B=\Delta f_a$, where extraneous frequency noise is excluded. Therefore I will assume for the rest of this section that $B=\Delta f_a$, is the ideal case; it is important to note that $\Delta f_a$ is an estimate based on the velocity dispersion of the axion field, but it turns out this value will cancel out later in the proof. The rest of this scan rate proof is heavily inspired from Ref. \cite{brubakerThesis}.
    \par Since we are searching over multiple cavity resonant frequencies, another vary important parameter is the optimal tuning step size, $\delta f_c$. One can imagine the following search scheme: the cavity starts at a center mode frequency, $f_c=f_0$, time-series data is taken for a fixed $t$, a single scan, then the center cavity frequency is tuned up by $\delta f_c$ and the cycle is repeated, such that the entire frequency range is covered ($B=\Delta f_a$, but we aren't performing the Fourier transforms yet, just taking time-series data). One can see in this scheme, assuming no down time between scans, that the frequency scan rate would be $\delta f_c/t$. For simplicity, assume the other major signal and noise parameters such as $T_{sys}$, $C_{lmn}$, and $Q_L$ are frequency independent for now. The cavity is sensitive to axions between $f_c+\Delta f_c> f_a > f_c-\Delta f_c$ for a given tuning arrangement, therefore if $\delta f_c > \Delta f_c$, there will be gaps, or at least significant drops, in axion sensitivity between scans; in truth, our analysis procedure accounts for the drop off in sensitivity outside $\Delta f_c$ based on the Lorentzian shape of the cavity resonance, and it is actually the bandwidth of power spectrum that determines this, but if this bandwidth is significantly larger than $\Delta f_c$, one will see proportional drop-offs in sensitivity between scans. If $\delta f_c \leq \Delta f_c$ then there will be no gaps, but some overlap even. I will therefore define $\delta f_c=\Delta f_c/F$, where $F\geq 1$. In this regime, measurements at multiple consecutive tuning steps will contribute to the SNR at each potential axion frequency $f_a$ in the tuning range, because cavity bandwidths at each step overlap with each other to some degree by a factor of $F$. That is to say, there is some $K>F$ value of tuning step scans in either direction of a given $f_a$, which itself is within a central $f_c$ scan, that contribute to the SNR at $f_a$. To complicate things just a little bit more, we know that if $f_a$ is not on the central cavity frequency, $f_c$, for a given scan, the signal power will degrade slightly according to the cavity resonance peak shape, a Lorentzian. How far off the $f_a$ is from the central frequency in a given scan is $\delta f_a=|f_c-f_a|$.
    \par One can see that adding the SNR of multiple integrations of the same frequency would add in quadrature. This can be shown by splitting the integration time of a single scan into two scans. The original scan has $SNR=R_0$ with $t_0$, and it will become two scans with times $t_1$ and $t_2=t_0-t_1$. Although the SNR will obviously go down for each individual scan, The SNR should remain $R_0$ when we combine these two scans again. Using the Dicke Radiometer equation, Equation \ref{eqn:DickeRadiometer}, one can derive the expression $R_i=R_0\sqrt{t_i/t_0}$ for $i=1,2$. Therefore, if we want to combine a set of $K$ scans in each direction of a central scan centered at $f_0$, we can write assuming $f_a=f_0$:
    \begin{equation}
     SNR(f_0)^2=\frac{1}{(k_BT_{sys})^2}\frac{t}{\Delta f_a}\sum_{k=-K}^{K} P_{sig}^2(\delta f_a =k\delta f_c)
    \label{eqn:MultiscanSNR}
    \end{equation}
    Where $P_{sig}(\delta f_a =k\delta f_c)$ would represent the signal power that is off from the central peak cavity frequency by $k\delta f_c$. This signal power would follow the Lorentzian shape $P_{sig}=P_0/(1+x^2)$ where $x=\delta f_a/\Delta f_c$ and $P_0$ is the peak power (more on Lorentzians in Chapter \ref{chap:Cavities}). This expression can be converted to the scan rate by dividing out the SNR, and multiplying by $\delta f_c/t$:
    \begin{equation}
     \frac{df}{dt}=\frac{\delta f_c}{t}=\frac{1}{SNR(f_0)^2}\frac{1}{(k_BT_{sys})^2}\frac{\Delta f_c}{\Delta f_a}P_{sig}^2(\delta f_a=0)\frac{1}{F}\sum_{k=-K}^{K} (\frac{1}{1+(2k/F)^2})^2)
    \label{eqn:ScanRate1}
    \end{equation}
    This sum at the end of the scan rate expression can be solved numerically for a variety of values, but I will use the convenient approximation:
    \begin{equation}
    Z(F,K)=\frac{1}{F}\sum_{k=-K}^{K} (\frac{1}{1+(2k/F)^2})^2)\approx\frac{4}{5}
    \label{eqn:ScanRateNumSum}
    \end{equation}
    for any $F \geq 2$ and any $K \gtrsim F$, which we already required for $K$, and can easily be satisfied for $F$. Considering antenna losses again, $\delta f_c$ will degrade by a factor $1+\beta$ when the Q is loaded. Finally, if we substitute Equation \ref{eqn:axionpower4} in for signal power, pull constants to the front (and remember $Q_a=f_a/\Delta f_a \approx 10^6$), and then normalize to some base experimental values, we arrive at a much more useful expression for experimenter:
    \begin{equation}
    \begin{split}
     \frac{df}{dt}=
     & \left(\frac{3.43\,\mathrm{GHz}}{\mathrm{year}}\right)\left(\frac{0.36}{g_{\gamma}}\right)^4\left(\frac{\rho_0}{0.45\,\mathrm{GeV/cm^3}}\right)^2\left(\frac{f}{1\, \mathrm{GHz}}\right)^2\left(\frac{3.5}{\mathrm{SNR}}\right)^2\left(\frac{0.2 K}{T_{sys}}\right)^2 \\
     & \left(\frac{B}{8\,\mathrm{T}}\right)^4\left(\frac{V}{100\,\mathrm{L}}\right)^2\left(\frac{C_{lmn}}{0.5}\right)^2\left(\frac{\beta^2}{(1+\beta)^3}\right)\left(\frac{Q_0}{10^5}\right)
    \end{split}
    \label{eqn:ScanRate2}
    \end{equation}
    As you can see from this expression, $g_{\gamma}$ and $B$ have the very strongest dependence on scan rate, nonetheless any improvement in $Q$ would directly translate to speeding up the experiment by the same factor.
    \par In the above derivation, it was assumed many parameters were frequency independent, while in reality, the experimental parameters, mainly $T_{sys}$ $f$, $C_{lmn}$, and $Q$, are changing as one tunes the cavity. This is okay as long as they are being measured periodically in between scans (or simulated beforehand in the case of $C_{lmn}$), as they all are relatively frequency independent on the scale of a single scan width. Therefore, this scan rate expression can be used to determine the frequency "speed limit" to tune the cavity at any given time for a given sensitivity. These periodic measurements, calibrations, and repairs all introduce dead time to this data taking cycle, and this expression is sometimes adapted to include a dead-time fraction to account for that additional time. In actual data-taking, SNR based on the expected DFSZ power is assigned to every frequency bin in the "grand spectrum" where individual overlapping spectra have been added together. Areas of power excess with high SNR can be flagged as axion candidates, and are subsequently rescanned to search for persistent axion signals. In the absence of persistent signals, this SNR is ultimately translated into a minimum value of $g_{a \gamma \gamma}$ that can be set for the frequency bin, creating the exclusion limits shown in Figure \ref{fig:ADMXlimitstheory}. 
    \par In conclusion, scan rate is often the key parameter looked at when designing a haloscope experiment; based on the expected experimental values, one can estimate how fast the targeted axion frequency space can be searched at a given $g_{a \gamma \gamma}$, and therefore how long the experimental run would be. It is also a key operations parameter because it signals to the operator if tuning needs to be slowed or if it can be sped up through certain regions of the cavity tuning range. Scan rate is also one of the best figures of merit for comparing different haloscopes, because although one haloscope may have a much higher signal power, another may have a much lower noise; both are accounted for in the scan rate. The haloscope scan rate expression is also indicative of a big problem with higher frequency haloscopes, which will be discussed in Chapter \ref{chap:Cavities}.
    

\chapter{Current ADMX experimental site}
\label{chap:Sitelayout}
The Axion Dark Matter eXperiment (ADMX) is the worlds' premier axion haloscope experiment, and, at the time of writing, has performed the highest sensitivity axion searches published so far. This chapter will outline the specifics of the ADMX haloscope and experimental site at the Center for Nuclear, Particle, and Astrophysics (CENPA) at the University of Washington as of the time of writing, spring 2024. The haloscope is currently taking data for our run 1D search. This "run" nomenclature will be explained in this section. It will start with a brief history of the experiment and outlines the key steps and equipment that lead to the current experimental configuration and site. The following subsections will briefly outline the technical specifications of each sub-system.
    \section{Brief history of ADMX}
    \label{sec:ADMXHistory}
    ADMX was originally conceived to be the first Sikivie haloscope with enough sensitivity to detect axions of masses and couplings predicted by theoretical invisible axion models, initially the KSVZ model, and subsequently the DFSZ model after several upgrades (see Chapter \ref{chap:axiontheory}). It was known that this would involve multiple developmental stages, as technology improved, and be a multi-decade project. The collaboration was formed and work began on the phase "0" detector in 1993 (the same year that the writer of this dissertation was born!).
    \par The experiment was sited initially at Lawrence Livermore National Laboratory (LLNL), in Livermore, CA, where the superconducting magnet was sourced and located. It operated with a pumped helium refrigeration system with a $1.3\,{\rm K}$ base temperature, and used 2 cryogenic HEMT amplifiers supplied by the National Radio Astronomy Observatory (NRAO). The copper resonant cavity was 50 cm in diameter and 1 meter length, using two 5 cm diameter tuning rods. The first results of the phase "0" experiment were reported in 1998, excluding axions to the KSVZ limit between $2.9-3.3\,{\rm \mu eV}$; it was the first haloscope experiment to reach the theoretical benchmark model \cite{ADMX1998, DawThesis}. Subsequent searches ending in 2005 expanded this range to $1.9-3.3\,{\rm \mu eV}$ with this system.
    \par The next major upgrade phase of the experiment was to replace the HEMT amplifiers with Superconducting Quantum Interference Devices (SQUIDs) that would greatly lower the noise temperature of the primary amplifier stage. This required development of SQUID amplifiers that could operate at the higher microwave frequency range of the cavity, which was significantly above the resonant range of SQUIDs at the time. These amplifiers would be developed at UC Berkeley by John Clarke and become known as DC micro-strip SQUID amplifiers (MSAs). Because SQUIDs operate via transfer of single magnetic flux quanta, they would be overwhelmed in the magnetic field of ADMX; therefore a field-free region had to be developed to store these sensitive electronics. This was solved with the installation of a bucking magnet into the insert. This culminated with the first SQUID search from 2008-2010 setting new limits KSVZ axions between $3.36-3.69\,{\rm \mu eV}$ \cite{ADMX2010,HotzThesis}.
    \par
\begin{figure*}[htb!]
    \centering
    \includegraphics[width=0.5\linewidth]{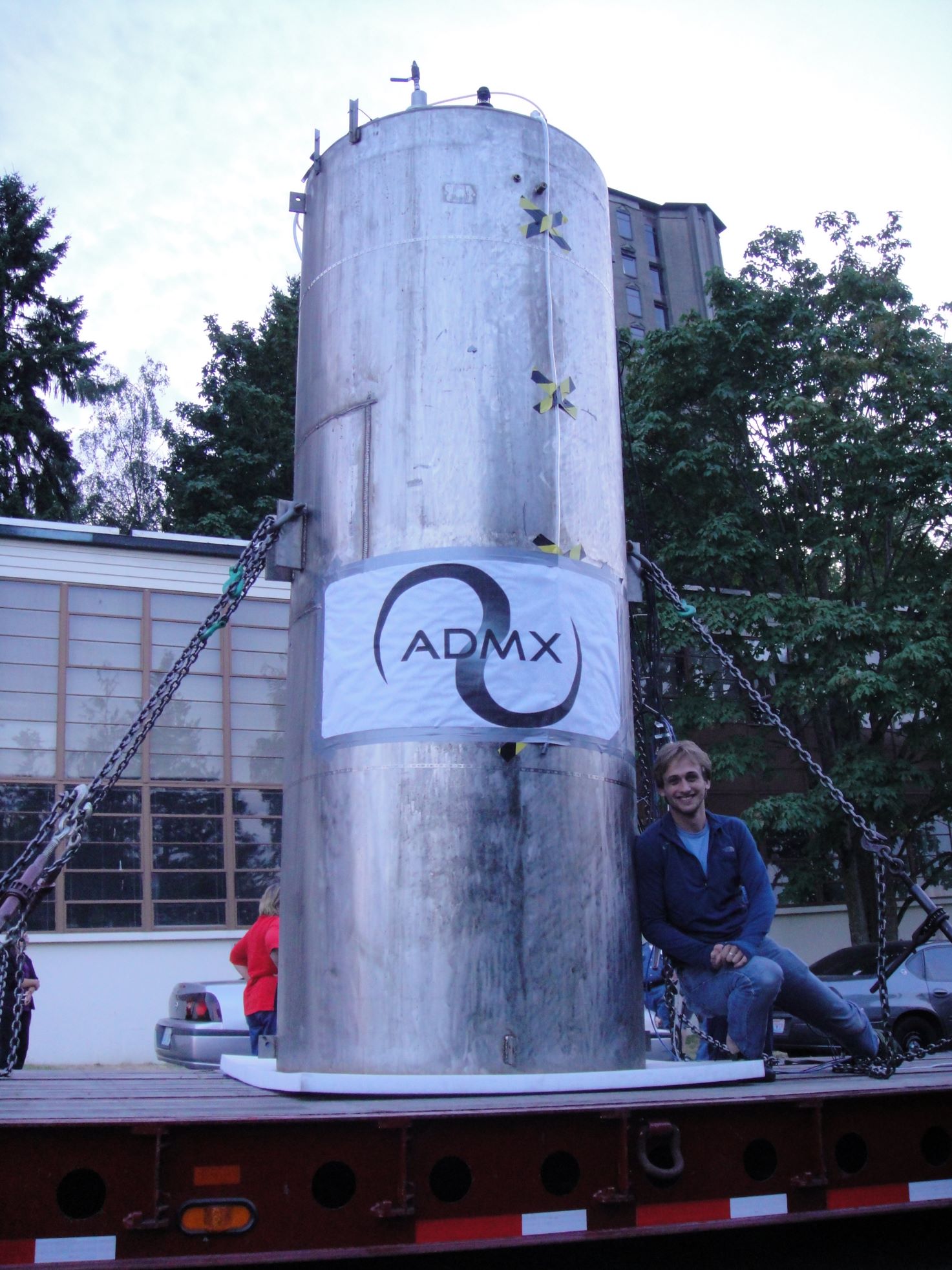}
    \caption{The ADMX superconducting solenoid magnet arriving at UW CENPA, Seattle, WA, from LLNL, Livermore, CA.}
    \label{fig:MagnetOnTruck}
    \end{figure*}

 In 2010, the 9 Ton, $8.5\,{\rm T}$ magnet was transported to CENPA at the University of Washington in Seattle, WA from LLNL in California (see Figure \ref{fig:MagnetOnTruck}). Around this time, DOE and NSF were conceptualizing "Generation 2" dark matter detectors with a significant increase in sensitivities, and, although they primarily had WIMP searches in mind, ADMX was slotted in as the single experiment to search for axions. The primary ADMX "Gen-2" upgrade was the installation of the helium dilution refrigerator, which lowered the physical temperature to $\approx 150\,{\rm mK}$, significantly speeding up the scan rate, and finally opening the possibility of searching for the much more elusive DFSZ model axions.
    \par This goal was accomplished in 2016, when ADMX became the first axion haloscope to exclude DFSZ axions, with limits set in the $2.66-2.81\,{\rm \mu eV}$ mass range; this run within the collaboration is referred to as the Gen-2 run 1A \cite{ADMXRun1A}. This run was limited to a small mass range because of the gain bandwidth of the MSA used. The subsequent run in 2017, Run 1B, implemented a Josephson Parametric Amplifier (JPA), with significantly larger bandwidth, allowing for DFSZ exclusion of axions between $2.81-3.31\,{\rm \mu eV}$ \cite{ADMXRun1B}.
    \par Running parallel to these main experimental searches, the ADMX "Sidecar" experiment was installed in 2016 in what was otherwise wasted free space in the insert. The Sidecar is a high frequency ($\approx4-7\,{\rm GHz}$), piezoelectrically tuned cavity, meant to test budding technologies for future high mass axion detectors. It isn't meant to achieve KSVZ or DFSZ sensitivity, but does occupy a mass range that hasn't entirely been excluded to either sensitivity, meaning it can produce new science limits as a by-product. The run 1A-C Sidecar papers can be found in Refs. \cite{Sidecar1A,Sidecar1C}, and the preparations and results of the Sidecar run 1D will be the primary topic of Chapter \ref{chap:Sidecar1D}. 
    \par Run 1C started in January of 2020, and required installing larger tuning rods to increase the frequency range of the copper resonator, among other RF component exchanges to accommodate the new mass range. In March 2020, the COVID-19 pandemic caused a halt to all non-essential on-site operations. It was then observed that the system noise temperature was too high to exclude DFSZ axions in a reasonable amount of time, and extraction, inspection, and reparation of the insert was needed. Unable to do so during the pandemic, the initial report of run 1C includes both a DFSZ range and KSVZ exclusion limits between $3.3-4.2\,{\rm \mu eV}$. This entire range was subsequently excluded to DFSZ level concluding in December 2022. Run 1D began in mid-December 2023 and is currently on-going, with target mass exclusion range of $4.2-5.75\,{\rm \mu eV}$ or $1.02-1.39\,{\rm GHz}$. To achieve this frequency range with the same diameter cavity, a single large diameter tuning rod was implemented. Increasing the rod's diameter further than this will cause significant losses in sensitivity via the volume loss, and therefore it will be the last configuration with the current single copper cavity system. The future haloscopes, ADMX Run 2 and ADMX-EFR, will use multi-cavity arrays that will be discussed at the end of Chapter \ref{chap:Cavities}. 
    \section{Site infrastructure}
\begin{figure*}[htb!]
    \centering
    \includegraphics[width=1\linewidth]{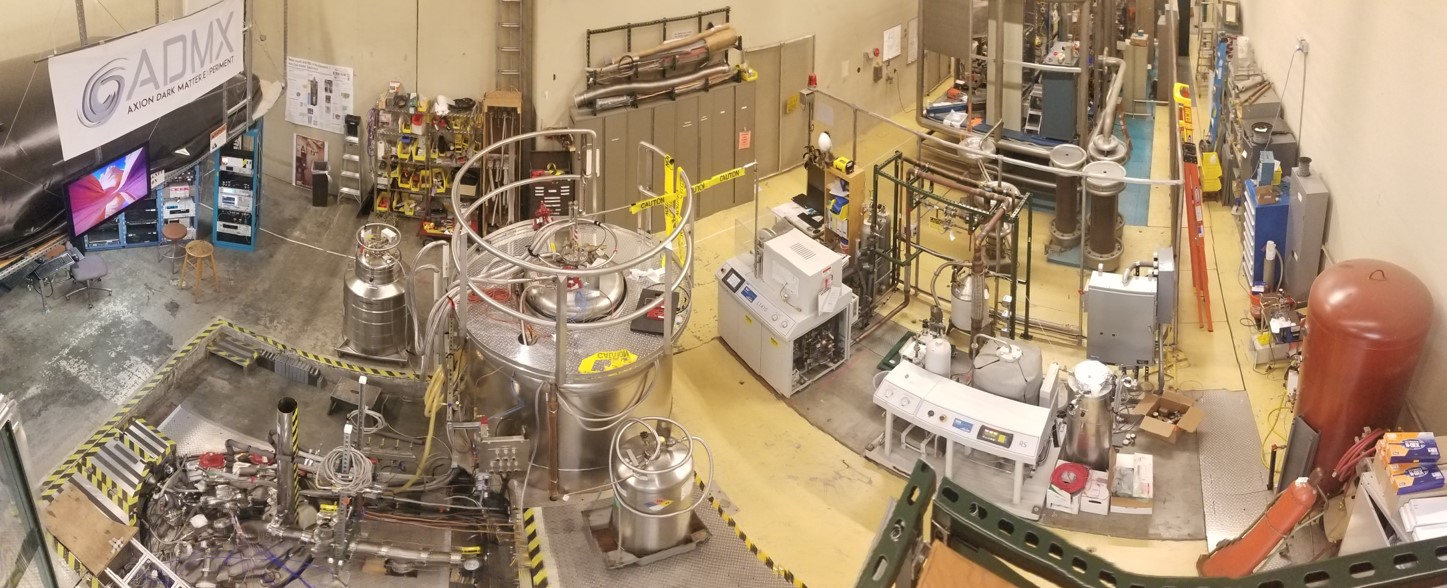}
    \caption{The ADMX experimental site.}
    \label{fig:ADMXSite}
    \end{figure*}

\begin{figure*}[htb!]
    \centering
    \includegraphics[width=0.7\linewidth]{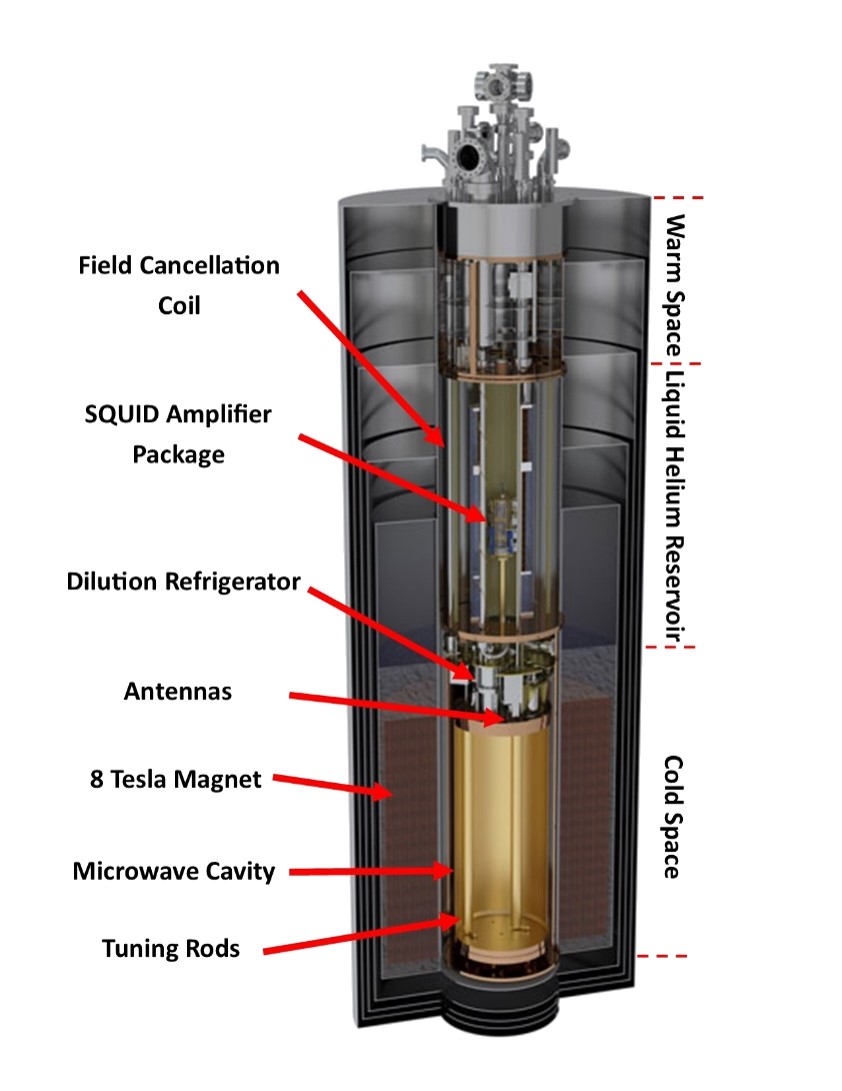}
    \caption{A cut-away diagram of the ADMX insert within the solenoid magnet.}
    \label{fig:ADMXInsert}
    \end{figure*}

\begin{figure*}[htb!]
    \centering
    \includegraphics[angle=270, width=0.7\linewidth]{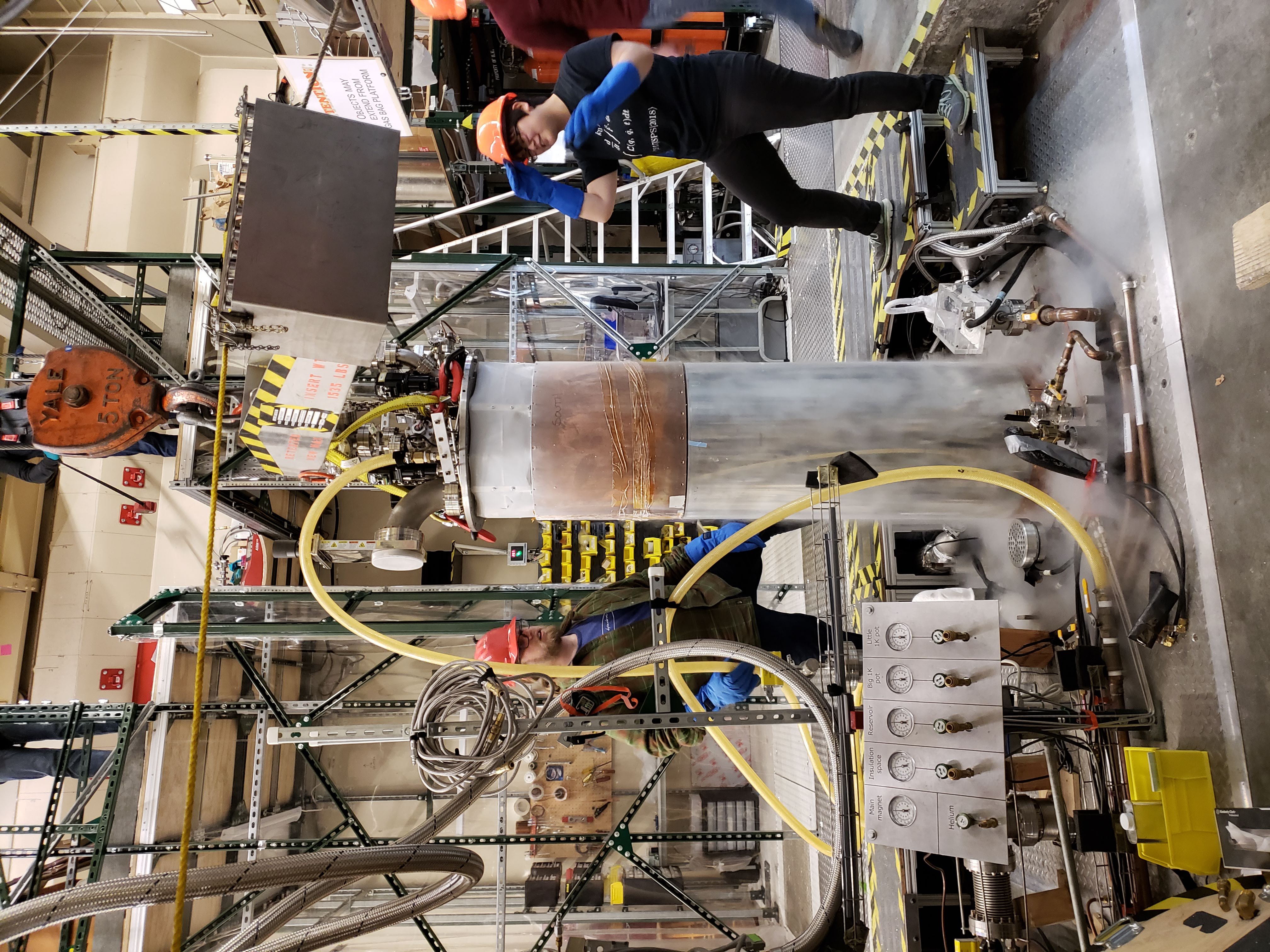}
    \caption{Extraction of the ADMX insert from magnet after run 1B.}
    \label{fig:ADMXInsert_Extraction}
    \end{figure*}

\begin{figure*}[htb!]
    \centering
    \includegraphics[width=0.6\linewidth]{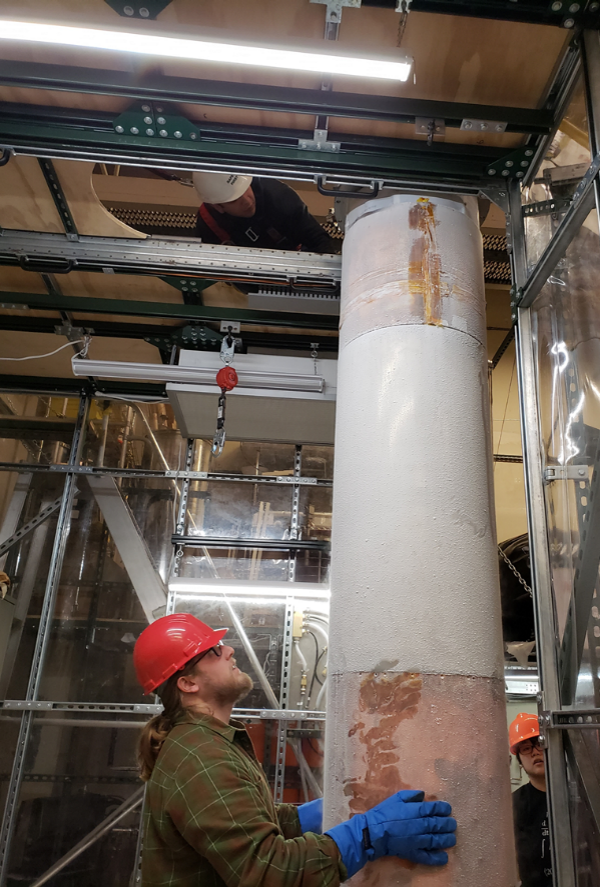}
    \caption{Mounting ADMX insert to clean room ceiling after run 1B.}
    \label{fig:ADMXInsert_Extraction1}
    \end{figure*}

    The ADMX Gen 2 experiment resides at the end of the CENPA accelerator hall shown in Figure \ref{fig:ADMXSite}. At the center of the site, within a 11' concrete pit, putting its top at ground level, is the solenoid magnet. The heart of the experiment is referred to as the insert, pictured in Figure \ref{fig:ADMXInsert}, which contains the main resonant cavity, dilution refrigerator, 'SQUIDadel' cold electronics package within its bucking coil, Sidecar cavity, and a myriad of cabling and sensors. Next to the magnet is a scaffold clean room capable of mounting the insert to its ceiling for hands-on repairs and upgrades (see Figure \ref{fig:ADMXInsert_Extraction1}). At the beginning of a run, the insert is moved by crane and inserted into the magnet to cool down before ramping up the magnet (see Figure \ref{fig:ADMXInsert_Extraction}). At the conclusion of each run, the magnet is ramped down, refrigerator helium mixture collected, and the insert is extracted from the magnet to be mounted in the clean room. The rest of the site is almost entirely made up of the various helium gas handling systems, including a helium liquefier. There are two separate closed-loop cryogenic systems: the $^4He$ magnet cooling system that the liquefier is providing its output to, and the  $^4He/^3He$ dilution system. 
    \section{Magnet}
\begin{figure*}[htb!]
    \centering
    \includegraphics[width=1\linewidth]{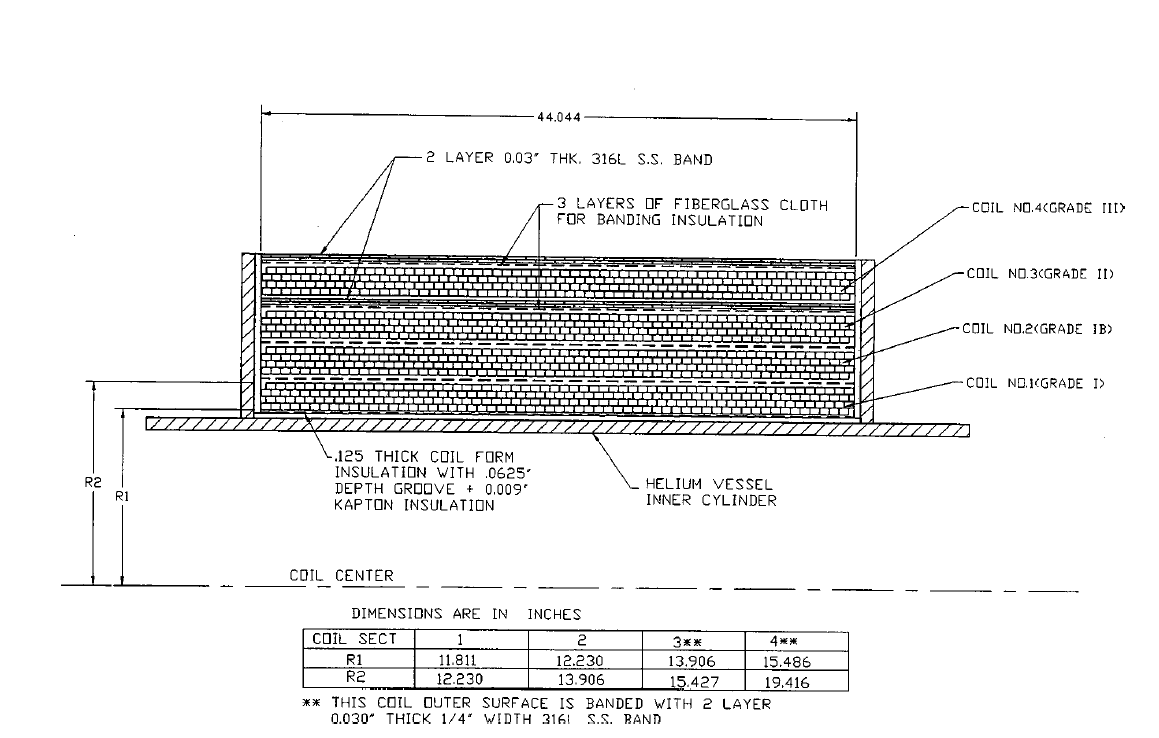}
    \caption{A cut-away diagram of the ADMX main magnet windings.}
    \label{fig:MagnetWindings}
    \end{figure*}

    The main magnet and its cryostat was manufactured by WANG NMR in Livermore, CA in 1993. It has niobium titanium (NbTi) windings forming a $1.12\,{\rm m}$ tall solenoid coil with a $60\,{\rm cm}$ inner diameter. It is actually made up of four concentric solenoids connected in series with progressively higher critical current ratings for the wire as pictured in Figure \ref{fig:MagnetWindings}. In total, there is $99\,{\rm km}$ of copper coated NbTi wire that is wound around the stainless steel spool frame. The magnet itself weighs $~6000\,{\rm kg}$ and the cryostat is about $~3000\,{\rm kg}$. Its inductance is $~534\,{\rm H}$ and is rated to have a peak field strength of $8.5\,{\rm T}$, corresponding to $~249\,{\rm A}$. In terms of stored energy, this is $\approx 16.5\,{\rm MJ}$, about the explosive energy of sixteen and half sticks of dynamite. However it is normally operated at $7.6\,{\rm T}$ during data-taking in order to mitigate any risk of quenching (only 13 sticks of dynamite stored!). This field strength drops off quickly along the length of the coil, with only about $70\%$ of the center field strength at either end of the coil. After 30 years, it still typically consumes 65 liquid liters of helium every day under normal operations, necessitating the helium liquefier.
    \par Part of the reason for running below its rated strength is because its persistent switch is no longer operational increasing Ohmic heating and a risk of quench. A persistent switch usually works by having the power supply connected in parallel to a small portion of the superconducting winding, and the switch operates by a heating source to that winding section. When in "non-persistent" operation or a ramping phase of the magnet, the winding section is forced into a normal state by the heat source raising it above its critical temperature, and current flows to the power supply instead, allowing the power supply to adjust the current within the windings. When ramping operations are done, it can be set to be persistent by turning off the heat source, allowing that winding portion between the leads to go superconducting, creating a closed current loop. This cuts down on helium consumption by lowering Ohmic heating from the power supply leads. The ADMX magnet's switch failed at Wang NMR during commissioning and was delivered as is. LLNL engineers decided the cost-effective repair was to permanently install vapor-cooled leads instead of the removable configuration.    
    \par
\begin{figure*}[htb!]
    \centering
    \includegraphics[width=0.5\linewidth]{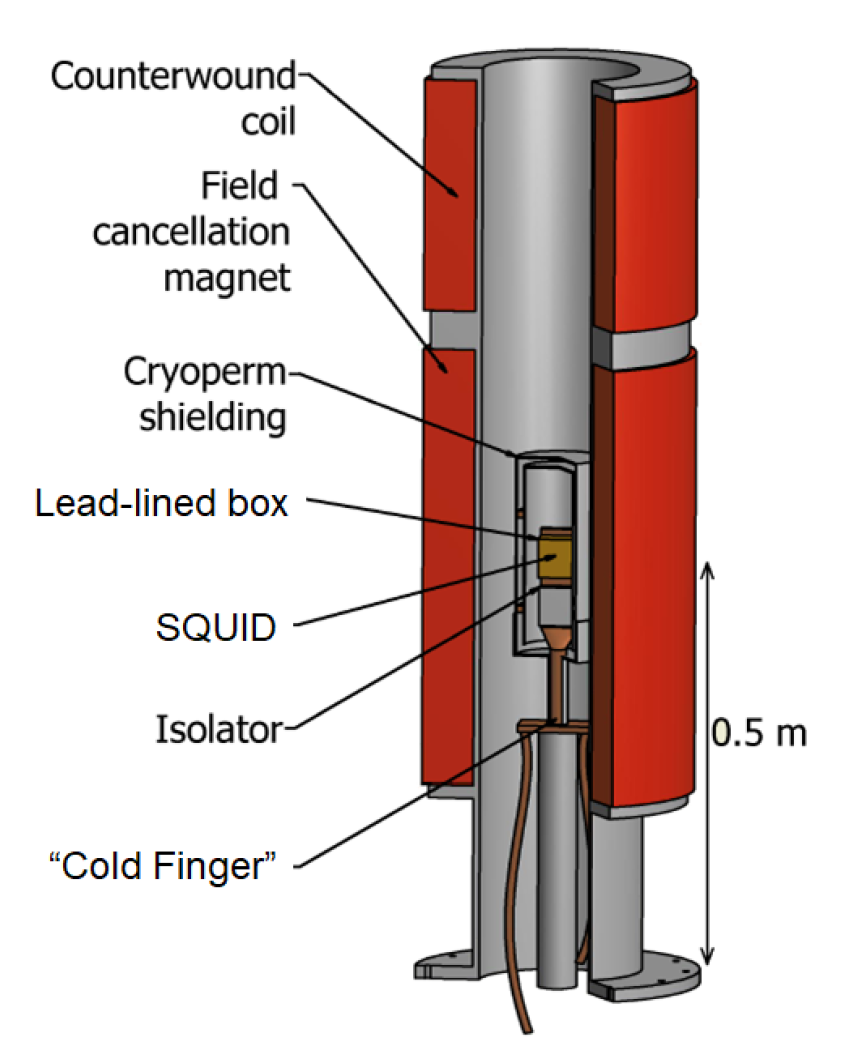}
    \caption{A sketch diagram of the bucking coil.}
    \label{fig:BuckingCoil}
    \end{figure*}

 The second 'bucking-coil' magnet resides within the insert (see Figure \ref{fig:ADMXInsert}) and generates a field opposite the main magnet, creating a net-zero field region for the field-sensitive electronics. Figure \ref{fig:BuckingCoil} outlines the magnet's construction, consisting of one coil to cancel the main magnet's field and another counter-wound coil that prevents a non-zero mutual inductance between the main magnet and the first coil. This prevents any significant net force between the main magnet and bucking coil. The bore of the bucking coil has net field of less than $100\,{\rm Gauss}$, and the sensitive electronics within can be passively shielded by mu-metal for further field reduction from there.
    \section{Cavity}
\begin{figure*}[htb!]
    \centering
    \includegraphics[width=.5\linewidth]{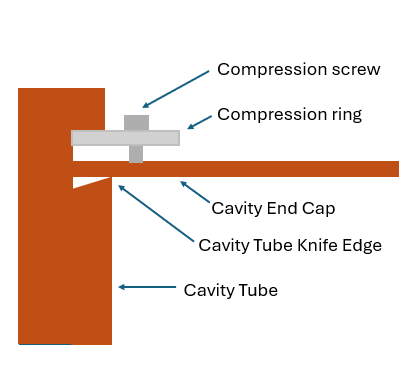}
    \caption{A diagram of the Knife-Edge clamp mechanism for ADMX run 1 cavities. A sharp knife edge is constructed off of an inset cut a couple of inches within the end of the tube. This inset will have the knife edge on the interior edge with a perpendicular face above it. The copper-plated end cap is placed on top of the knife edge, and the stainless steel compression ring inserts into the space above it within the inset of the cavity tube end. Compression screws create separation between the ring and end cap, causing the end cap to compress against the copper knife edge, and the compression ring against the opposing surface. This creates a strong seal between the copper end cap and tube, creating a solid current path for a good $TM_{010}$ resonance.}
    \label{fig:EndCapKnifeEdge}
    \end{figure*}

    The ADMX Gen-2 run 1 cavity has been the same base design through all the runs 1A-C, with only the tuning rod system changing between. It consists of a stainless steel tube that is $~101\,{\rm cm}$ and has an inner diameter of $~42\,{\rm cm}$, giving it an empty volume of $136\,{\rm L}$. Within the tube on top and bottom is a sharp, knife-edge lip, as well as a blunt lip above for applying counter-pressure against the end plates (or end-caps) seating them firmly against the knife-edge (see Fig \ref{fig:EndCapKnifeEdge}). The top and bottom end plates are also stainless steel and contain port fixtures for the antennae and tuning rods. These components are all copper plated with oxygen free copper. Upon assembly, the end-caps are placed to rest onto the knife edge, and a stainless steel mounting ring is placed on top. Several stainless section pieces are inserted under the blunt lip and affixed to the mounting ring, then 96 bolts pass through clearance holes in the section pieces and thread into the mounting ring. As the bolts are torqued it creates a separation between the blunt lip and mounting ring, pushing the end-cap firmly against the knife edge (see Fig \ref{fig:EndCapKnifeEdge}). This is because currents for the $TM_{010}$ mode run across this joint, so it is very important that the knife-edge to end cap electrical connection is solid.
    \par
\begin{figure*}[htb!]
    \centering
    \includegraphics[width=1\linewidth]{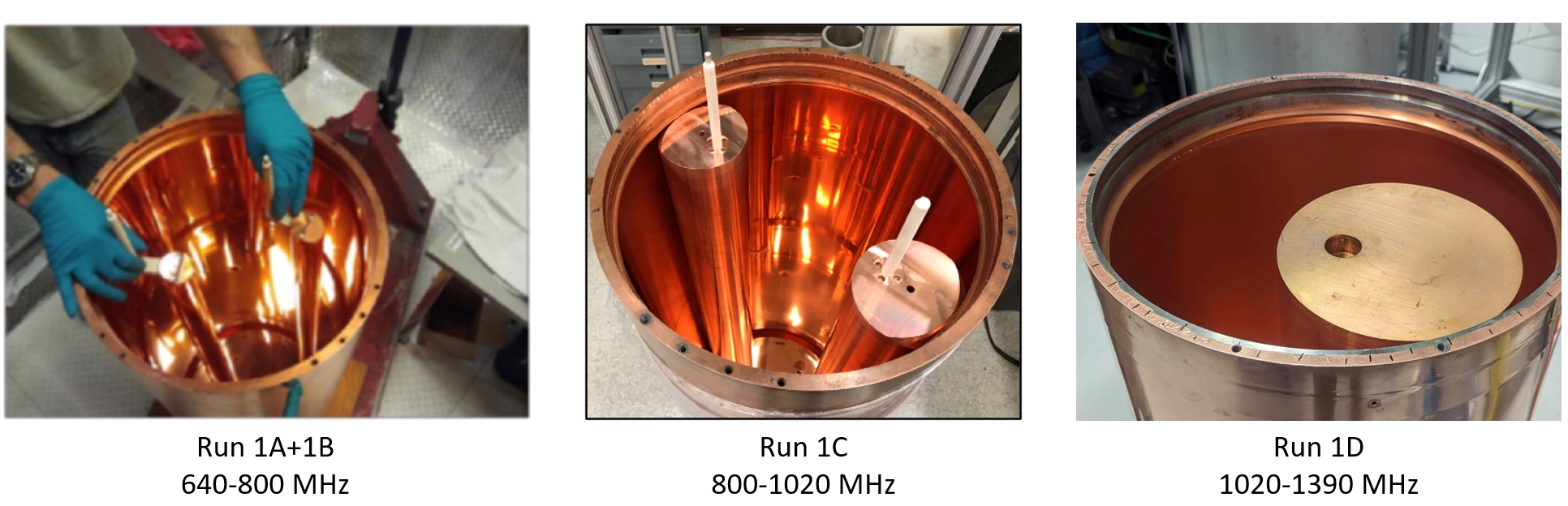}
    \caption{The various ADMX tuning rod arrangements within the run 1 cavity.}
    \label{fig:TuningRodsinCavity}
    \end{figure*}

 The run 1A-C systems used two different double tuning rod systems. Having two rods allows for different tuning arrangements that can be advantageous for shifting the frequencies of mode crossings, which will be discussed further in Chapter \ref{chap:Cavities}. They are made of hollow OHFC copper tubes with the end-caps welded on. They run the length of the interior cavity, with a $\approx0.020"$ clearance gap on top and bottom. This is because they must electrically disconnect from the cavity itself, in order for the mode structure to be maintained and tuned (see Chapter \ref{chap:Cavities}). The axles are connected off-center of the rod such that they will turn the rod center to various radii within the cavity, shifting the resonant frequency of the $TM$ modes. The axles extend through axle ports in the end-caps where they attach to a gearbox system above the cavity top. This gearbox system runs the length of the insert where it is powered by room temperature servo motors with a gear ratio of 19,600 to 1. Again, to maintain electric disconnect from the cavity, and minimize power losses out of the cavity, the axles were made of alumina in earlier runs, and, more recently, of sapphire. Sapphire and alumina are low-loss dielectric material that prevent the axles from acting like antennae themselves, dissipating power out of the cavity and spoiling the quality factor. Sapphire is preferred over alumina because it has a higher thermal conductivity, making cooling of the tuning rods faster and more homogeneous with the cavity. The run 1A-B rods were $5\,{\rm cm}$ in diameter, whereas the run 1C tuning rods were $11.1\,{\rm cm}$. The run 1D system implements a single $~20.5\,{\rm cm}$ diameter rod and thus necessitating a different end cap design with only 1 tuning rod axle port. This rod weighs $~51.2\,{\rm lbs}$, and therefore is installed via crane unlike the much lighter predecessors that could be installed by hand. These are all pictured in Figure \ref{fig:TuningRodsinCavity}.
    \par The cavity has 3 antenna ports: 2 "weak" or fixed coupling ports, and 1 "strong" or variable coupling port. The strong port antenna is the data-taking antenna for power time-series data, and has its own gearbox and room temperature stepper motor that allows the operator to adjust its insertion depth into the cavity, which allows for variable coupling to the cavity (coupling will be discussed more in depth in Chapter \ref{chap:Cavities}). The other two antenna ports have fixed antennae that are set to be very weakly coupled to the cavity. These weak ports are only used for periodic cavity characterization periods: monitoring the quality factor, resonant frequency, and coupling between tuning cycles. For this reason, the coupling is kept weak to prevent it from dissipating potential axion power. In practice, only one of these weak ports was actively used, and the other port was kept as a back-up. The strong port, being the axion data antenna, is kept critically coupled (much stronger coupling) to maximize the potential axion power going into the receiver system. This data-taking cadence will be discussed more in depth in the next section of this chapter.
    \section{Helium dilution refrigerator}
    The helium dilution refrigerator is a custom high-flow design developed and manufactured by Janis Research (part of which has more recently been bought by Lakeshore Cryogenics) with collaborative help from the ADMX group at University of Florida. It is a 'wet' dilution system and has a cooling power of $800\,{\rm \mu W}$ at an operating temperature of $100\,{\rm mK}$; this is the physical temperature, not to be confused with system noise temperature that determines the noise floor of the experiment. 
    \par A helium dilution refrigerator makes use of a phase separation between $^{3}He$ and $^{4}He$ for mixtures below $870\,{\rm mK}$: The concentration of $^{3}He$ can either be highly concentrated (essentially pure $^{3}He$) or a dilute mixture of about 6.6\% (this ratio is fixed by Van Der Waals forces in a stable equilibrium). This mixture exists in the part of the refrigerator called the mixing chamber, with the pure phase on the top of this chamber, while the fixed dilution is below. By pumping from the dilution side of the mixing chamber, $^{3}He$ is drawn into the dilution to maintain the fixed concentration ratio. Because this transition is going from an orderly to more disorderly phase, entropy increases, and heat must be absorbed from the system to maintain the dilution ratio. This endothermic reaction is what removes heat from the cavity and surrounding components. 
    \par
\begin{figure*}[htb!]
    \centering
    \includegraphics[width=0.6\linewidth]{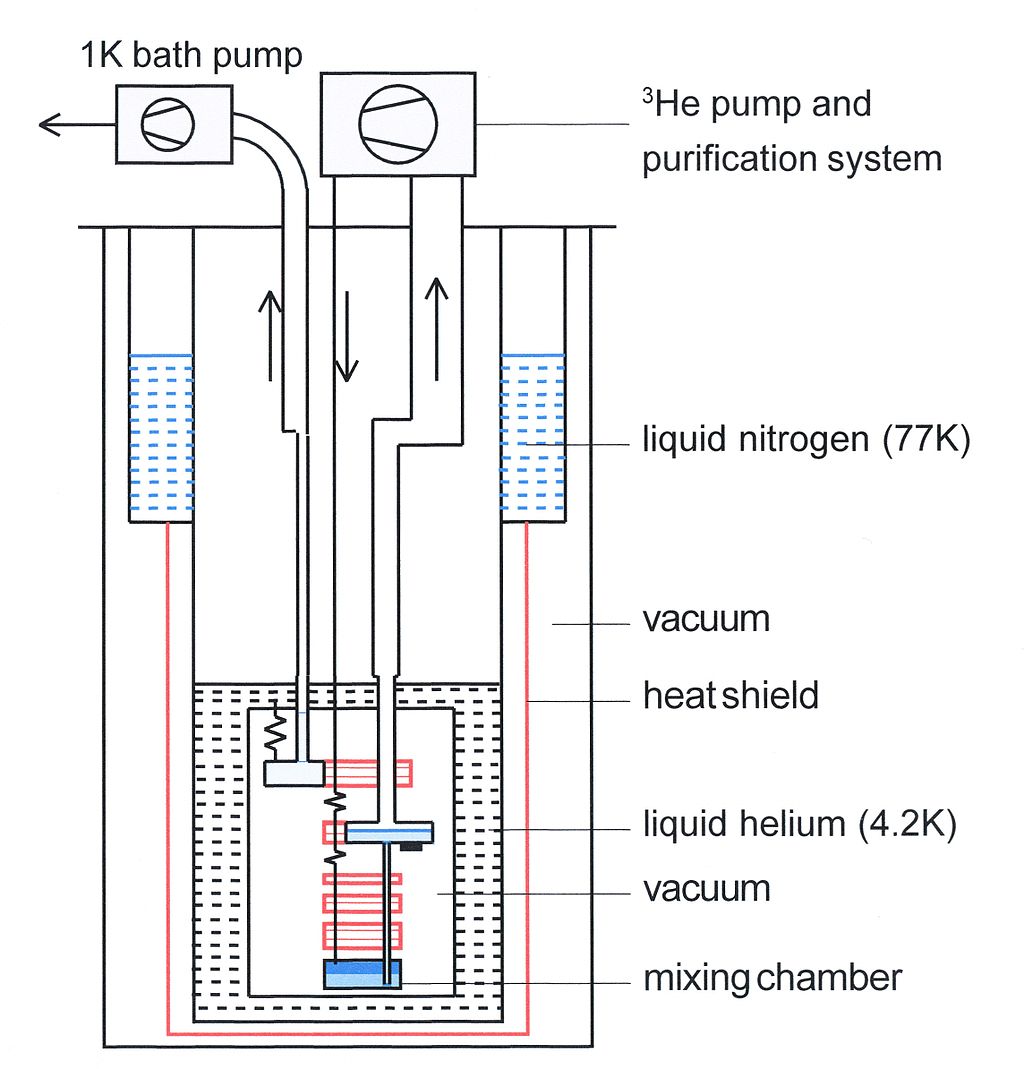}
    \caption{A simple schematic of a generic 'wet' helium dilution refrigerator \cite{DilutionWiki}.}
    \label{fig:DilutionSchematic}
    \end{figure*}

 As outlined in Figure \ref{fig:DilutionSchematic}, there are many other important components to the standard fridge system that pre-cool the helium to make this interaction possible. First of all, typically a liquid nitrogen bath surrounds the system to provide base cooling to $77\,{\rm K}$, connected to some heat shielding system, as well as a liquid $^{4}He$ bath to then cool to $4\,{\rm K}$ . In the case of ADMX, the magnet cryostat serves this purpose. The "4-K stage" consists of a helium reservoir that was filled about once a day. This reservoir supplies helium to the 1-K "pots" in the experiment. As shown in Figure \ref{fig:DilutionSchematic}, the 1-K pot is connected to the helium reservoir, as well as a 1-K bath pump, and by pumping in the vapor above the $^{4}He$ in the pot, the boiling point is lowered to $1\,{\rm K}$. This pot is then connected via a heat exchanger to the pure $^{3}He$ input line for pre-cooling before it enters the mixing chamber. In the ADMX system, this $^{3}He$ pre-cooling pot is referred to as the "small 1-K pot", while another, "big 1-K pot", is an auxiliary pot to provide cooling for shielding and other components to $1\,{\rm K}$. After pre-cooling the pure helium via the small 1-K pot, the pure $^{3}He$ continues through the main impedance line that has a high flow resistance and is connected via heat exchangers to the "still", described below, which cools it below the phase transition point to $500-700\,{\rm mK}$. This is where the nearly pure $^{3}He$ then enters the mixing chamber, where it mixes and energy is pulled from the system as it dilutes, cooling the system to the $130\,{\rm mK}$ cavity operating temperature. As the pumped dilution moves out of the mixing chamber, it provides cooling to the incoming $^{3}He$, and then enters the "still". The still is the area where $^{3}He$ is then separated again from the $^{4}He$. Pressure is still kept low (about $10\,{\rm Pa}$) via a $^{3}He$ pump, and because $^{3}He$ has a lower boiling point than $^{4}He$, the vapor in the still is basically pure $^{3}He$, whereas the $^{4}He$ remains liquid and stationary. A still heater controls this distillation process, keeping a steady flow of $^{3}He$ out, where it is then subsequently compressed again and fed back into the system on the input end, thus completing the fridge cycle. 
    \par
\begin{figure*}[htb!]
    \centering
    \includegraphics[width=0.6\linewidth]{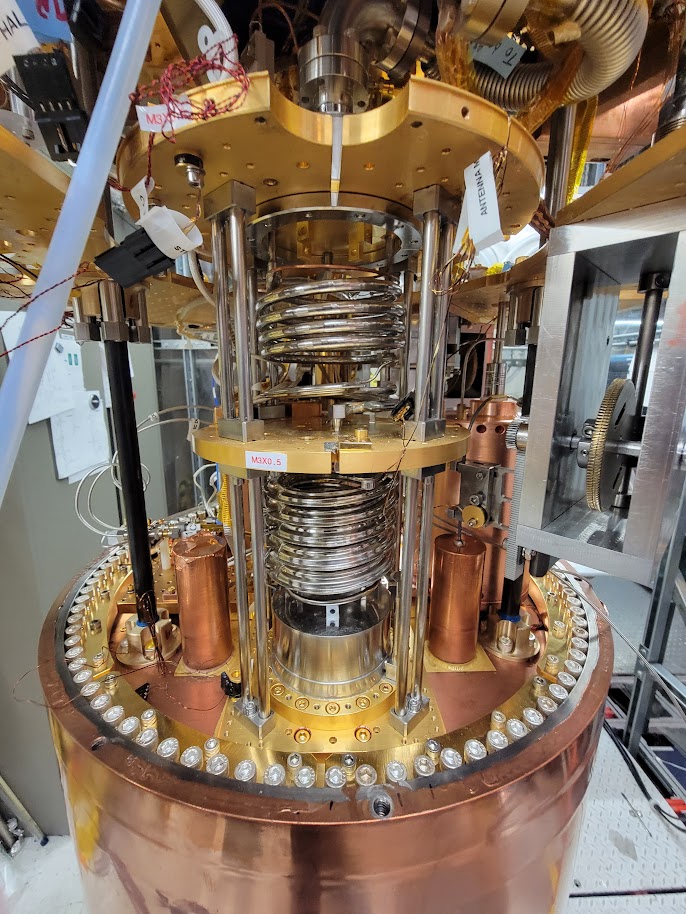}
    \caption{The ADMX dilution refrigerator mixing chamber mounted on top of cavity with heat exchangers (the coils) above providing pre-cooling to the  $^{4}He$/$^{3}He$ mixture input.}
    \label{fig:ADMXmixingchamber}
    \end{figure*}

\begin{figure}
    \centering
    \subfigure[Fridge control.]{\label{fig:fridgecontrol}\includegraphics[ width=0.4\linewidth]{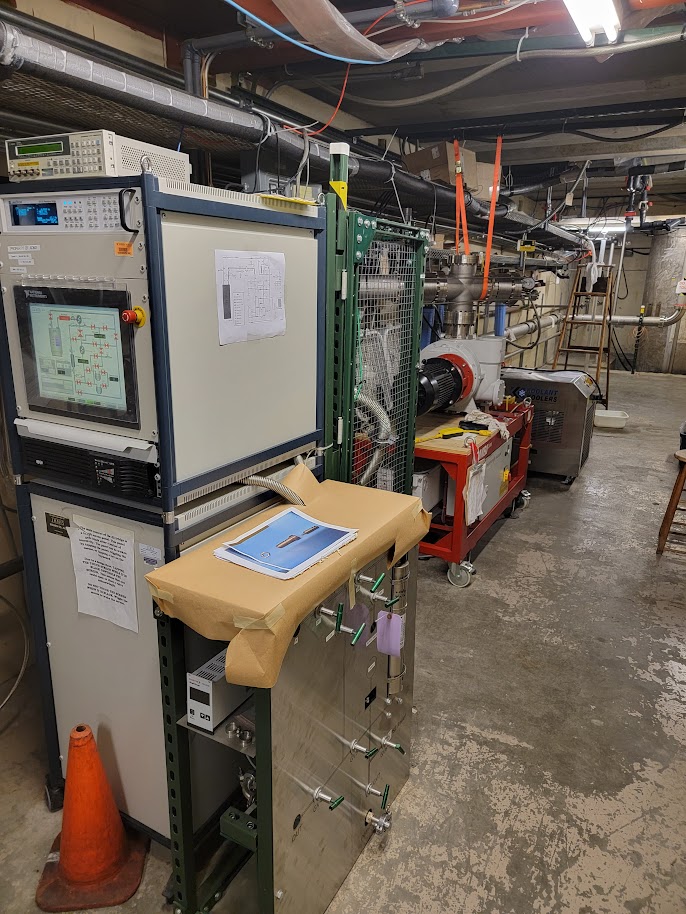}}
    \subfigure[Fridge pump.]{\label{fig:fridgepump}\includegraphics[ width=0.4\linewidth]{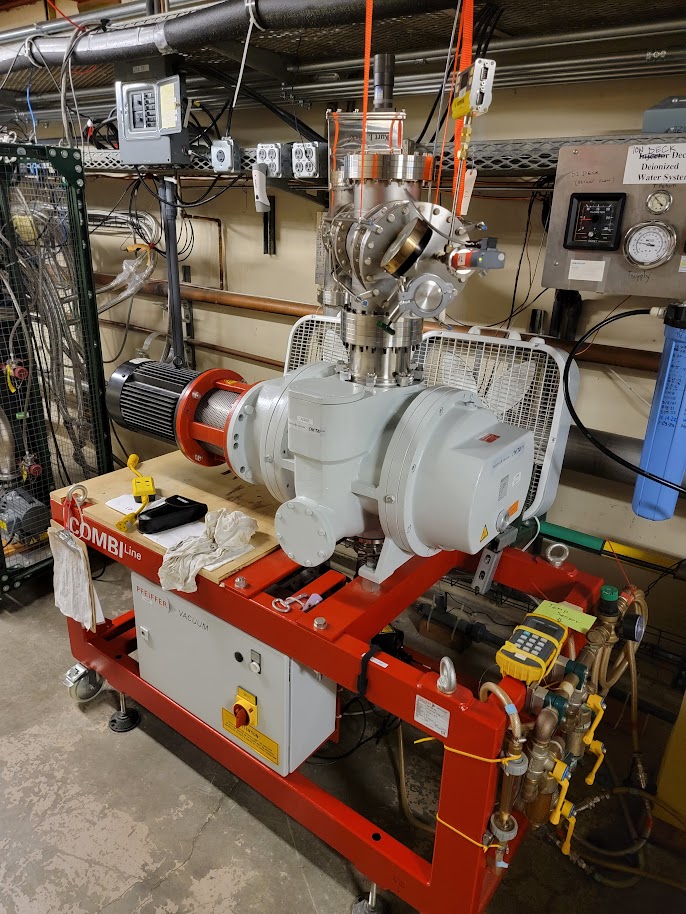}}
    \subfigure[Helium traps.]{\label{fig:fridgetrap}\includegraphics[width=0.5\linewidth]{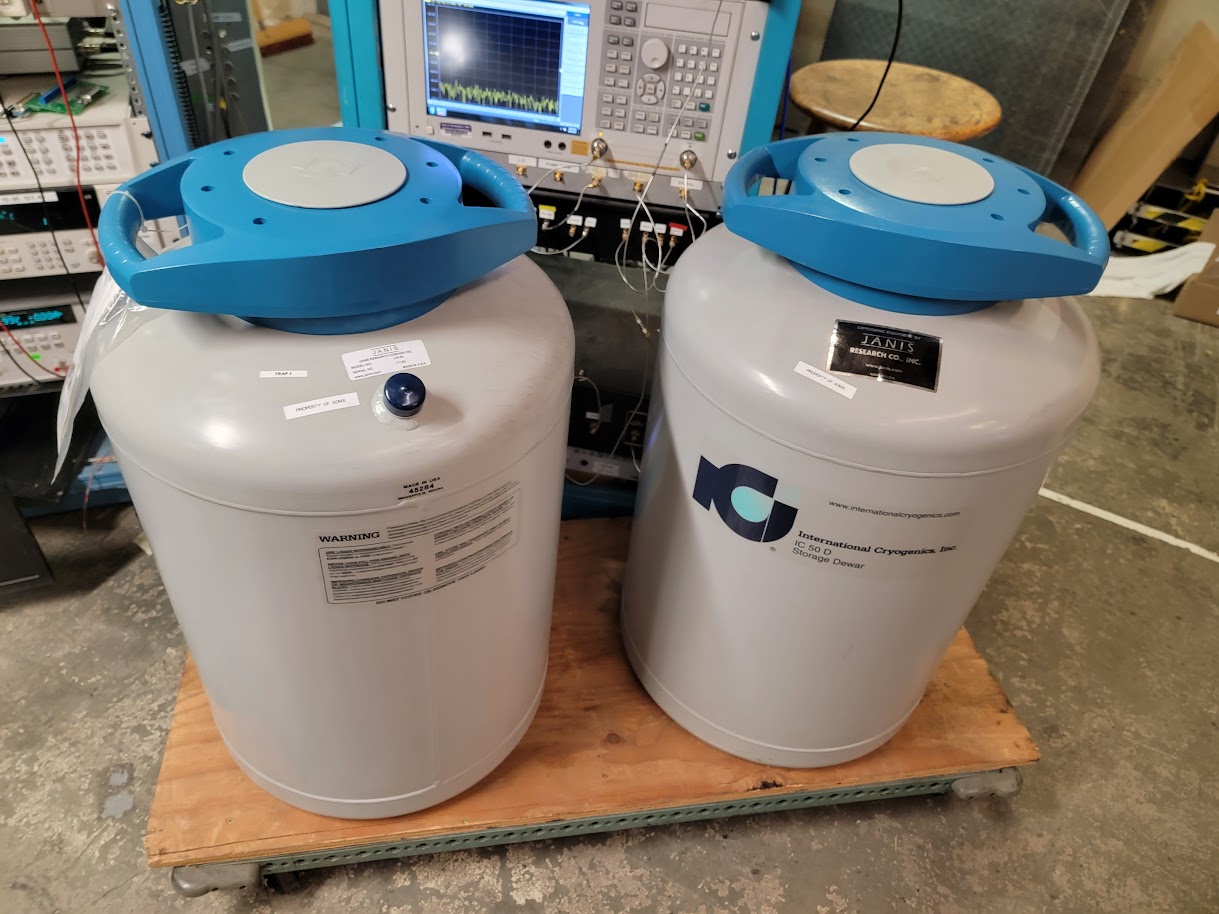}}
    \caption{The exterior ADMX dilution refrigerator components.}
    \label{fig:Fridgesystems}
    \end{figure}

 As pictured in Figure \ref{fig:ADMXmixingchamber}, The ADMX dilution fridge mixing chamber is attached to the "T-plate" or "mK stage" which is directly on top of the top cavity end cap. This is also where the bottom of the Squidadel arm (more on this later) is attached to, as well as the Sidecar cavity that will be discussed in Chapter \ref{chap:Sidecar1D}. The pots and still sit above this stage, and are not pictured here. The output plumbing of all this comes out of the top of the insert, and is then fed back down into an area known as "the pit" where the magnet itself sits. This is where the pumps and control systems are kept for the dilution fridge, as well as traps for storing the $^{4}He$/$^{3}He$ mixture when performing maintenance (see Figure \ref{fig:Fridgesystems}).
    \section{Cryogenic receiver}
    \begin{figure*}[htb!]
    \centering
    \includegraphics[width=0.8\linewidth]{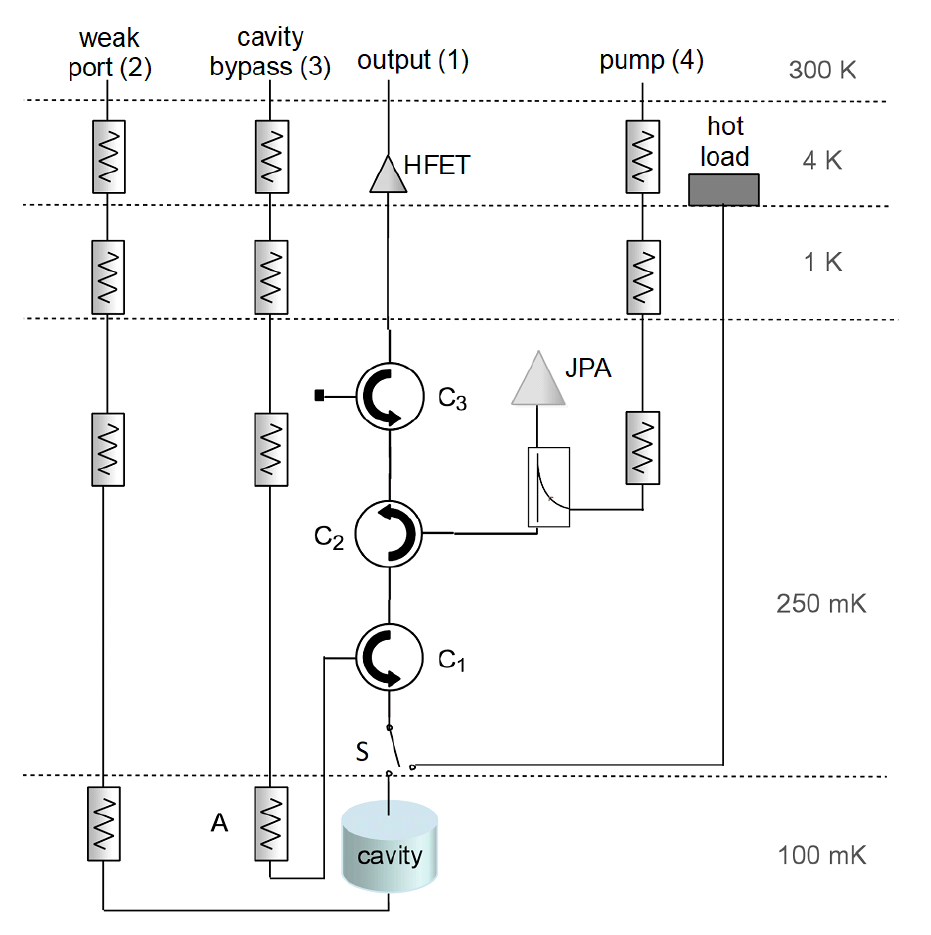}
    \caption{The schematic of the ADMX cryogenic receiver. "$C_n$" represents the circulators. "S" is a solenoid switch. "A" represents the attenuators. "JPA" is the Josephson parametric amplifier. "HFET" is a heterostructure field-effect transistor amplifier. The pump line is used for supplying a pump tone to the JPA. The weak port and cavity bypass port are used for transmission and reflection measurements respectively, which allow for in situ cavity characterization and calibration.}
    \label{fig:RecieverSchematic}
    \end{figure*}
    The RF circuit layout of the signal receiver within the insert is picture in Figure \ref{fig:RecieverSchematic} with the various temperature stages outlined.
    \par As mentioned in the cavity section, the cavity has two antennae, one strongly coupled antenna for extracting axion data, and one weak antenna for inputting power during calibration measurements. For instance, this weak port is what allows for the in-situ measurement of the transmission cavity quality factor. Within the schematic, the weak antenna is the connection on the bottom, and the strong antenna is on top of the cavity. In addition to calibration, some "synthetic axion" signals are injected into the weak port periodically for testing the haloscope and its operators ability to detect signals, although this sub-system won't be discussed in depth in this dissertation.
    \par In addition to the transmission measurements through the weak port, the cavity bypass line allows for reflection measurements off the strong port: Power is injected through the bypass line. It then travels through the circulator to bounce off the strong port antenna, back into the circulator and out the output line (ignoring the upstream amplification components). These reflection measurements determine the coupling coefficient ($\beta$) of the strong port, which is a measure of the fraction of cavity mode power being absorbed by the antenna (more on this in Chapter \ref{chap:Cavities}). The cavity bypass line can also be used to assess the overall health of the receiver and amplification components separate from the cavity as discussed in the next chapter. 
    \par As shown in Figure \ref{fig:RecieverSchematic}, both of these lines have several attenuators along the various thermal stages of the insert for two reasons: These act as heat sinks for the connecting coaxial cables themselves, as well as act as noise suppression for the warm thermal noise coming into the cold space. These coaxial cables for the two lines are made of stainless steel flashed with copper, in order to minimize the thermal conduction between the temperature stages. 
    \par The circulators along the output line are critical for directing the RF signals through the right components and stopping them from entering the wrong components. Circulators are non-reciprocal devices that act as a one-way roundabout or rotary for RF signals. In our case, they have 3 ports. The key difference between these circulators and a roundabout you'd find driving, is that the RF signals have to exit at the next port from when they entered. For example, a signal entering on port 1 will exit via port 2 and never from port 3 ($1\rightarrow2$, $2\rightarrow3$, $3\rightarrow1$). This one-way directionality is represented by the counter-clockwise arrow in the schematic. Its important to note, however, these are real-world devices, so there is some signal loss between ports, and isolation isn't perfect, but it is characterized. Being that this is the output line, coaxial cables used here were also superconducting NbTi, in order to minimize all RF loss, up until the HFET.
    \par Each circulator has a specific purpose. $C_1$ is primarily for directing the cavity bypass reflection signal as described above; the input signal on the bypass line can only travel to the strong port antenna first, and, vice-versa, the reflected power can only travel out the output line, not the bypass line. $C_2$ directs the raw cavity power signal to the primary amplifier, the JPA, where it is amplified by being reflected off the JPA. This amplified reflected power is then directed only up the output line by $C_2$ and not back down to the cavity. Additionally, any thermal noise from upstream output line is then dumped into $C_1$ where it is directed into the by-pass line, preventing any noise getting into the cavity. $C_3$ serves a similar isolation purpose: the amplified signal coming out of $C_2$ is directed to the HFET, secondary amplifier, and any noise from the HFET side is directed to a terminator rather than back down the output line.
    \par The JPA itself requires whats called pump tone for operation, hence the 4th RF line labeled "pump" in Figure \ref{fig:RecieverSchematic}. Josephson parametric amplifiers work similarly to how you swing on a swing set; you pump your legs at a given frequency back and forth, which drives the amplitude of the swinging to increase. This only happens because the frequency of your leg pumping motion is in the vicinity of the resonant frequency of that swing (thinking of it as a pendulum), and the resonance makes each leg pump more efficiently transfer energy into increasing the amplitude to your swing. However, if you pump too fast (or too slow) the swing amplitude stays relatively low. The JPA pump tone acts as the legs in this case, and instead of the pendulum frequency of the swing, you have your cavity output power spectrum. In electronics terms, the JPA acts as a resonant LC circuit here, where the inductance is adjusted by varying the critical current of a Josephson junction, which is done by applying a magnetic flux via a separate squid loop threaded through the junction. The pump tone is a modulation of bias current supplied to this threaded squid loop which then adjusts the resonant frequency of the JPA. As in the swing set example, the gain of the JPA is optimal when the pump tone matches the frequency of the incoming cavity signal (in the case of current-biased amplifiers. It is twice this frequency for flux-pumped amplifiers). However, this isn't optimal because it puts a separate signal right over where one wants to look for axions, therefore to decouple the pump tone from the cavity signal, it is offset from the signal frequency by $375\,{\rm kHz}$. This spacing was chosen to allow use of band pass filters in the warm space to remove the pump tone before digitizing the power from receiver. 
    \par The JPA provides at least $20\,{\rm dB}$ of gain in most runs over a $10-20\,{\rm MHz}$ bandwidth. Since this bandwidth is considerably smaller than the frequency space covered in a run, it is adjusted by changing the DC bias current and power of the pump tone signal roughly every 10 minutes of operation. 
    \par In Figure \ref{fig:RecieverSchematic}, the switch, S, just above the cavity strong port is a solenoid operated coaxial switch that allows for switching between the cavity and a "hot-load" on the 4-K stage. The hot-load acts as a variable temperature noise source, which allows for measurement of the JPA noise temperature, a process that will be discussed later in Chapter \ref{chap:analysis} . The switch was actuated by mA order DC current applied for about 1 second originally, which supplied a considerable amount of heat to the system. In the next subsection, I will talk about a small project I did to fix this issue, and minimize the heat to actuate.
    \par
\begin{figure*}[htb!]
    \centering
    \includegraphics[width=0.9\linewidth]{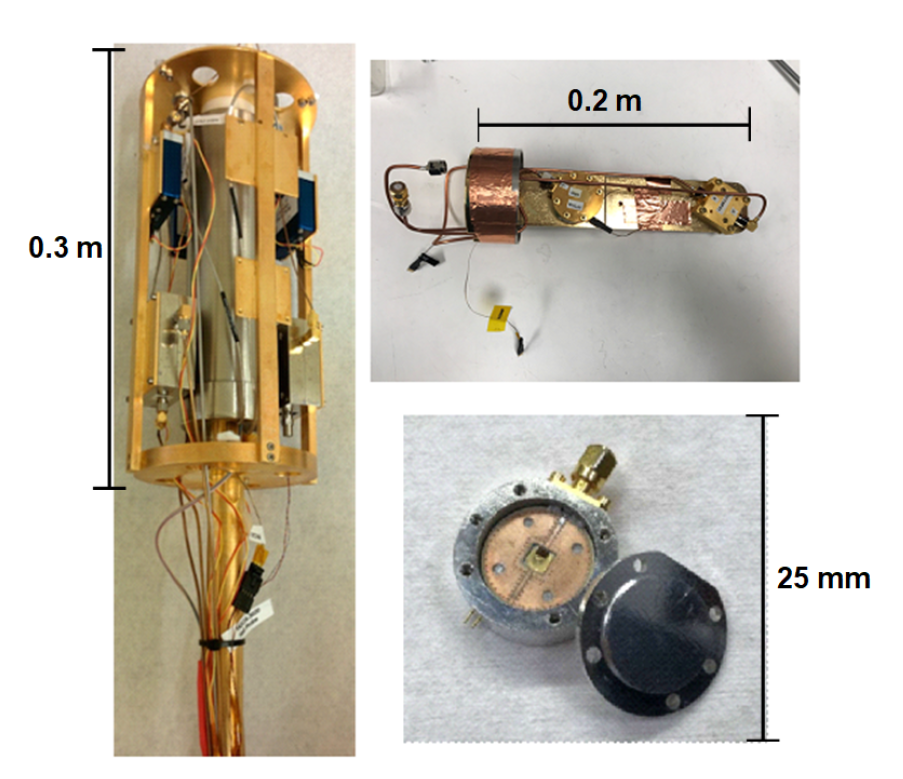}
    \caption{Left: The SQUIDadel chassis with circulators and coaxial switches and amplifier core. Top right: The amplifier core with amplifiers mounted and shielding removed. Bottom right: the JPA itself used in Run 1B.}
    \label{fig:SquidadelPic}
    \end{figure*}

 What Figure \ref{fig:RecieverSchematic} doesn't show you is what we call the "SQUIDadel", which houses all the magnetic-field sensitive components within the bucking coil: The ferromagnetic circulators, the SQUID amplifiers, and the solenoid switches. It consists of a long gold-plated copper shaft, called the "cold-finger", with a bolt head on one end for bolting it to the T-plate, thermally sinking it to the dilution fridge mixing chamber for maximum cooling. The other end is a chassis for the package itself containing switches and circulators on the outside and in the center a $\mu$-metal shielded core for the ultra-sensitive amplifiers as shown in Figure \ref{fig:SquidadelPic}. This shaft raises the SQUIDadel up into the field-free region, while also allowing it to be cooled by the mixing chamber. The coaxial cables are wrapped around the cold finger down to the mK-stage where they then break out and move up the insert in thermal stages. In order to prevent a thermal touch and minimize radiation imposed on the package from the relatively warm Bucking coil space (4 K), nylon spacers have been added to prevent any leaning of the cold finger into the bucking coil. I will also note that although this section only covered the cold receiver for the main experiment, a parallel receiver exists for the Sidecar system and is also housed in the same SQUIDadel cold finger. The other components, mainly HFET, attenuators, and hot-load, are mounted in the field region, on their appropriate temperature stages.
        \subsection{A pulsed RF switch box}
        \par As mentioned in the last section, the coaxial switch, used for hot-load calibrations, was hand-actuated by an operator with a DC power supply by applying around $160\,{\rm mA}$ for one second, which could overheat the refrigerator. One of my first projects on ADMX was designing a pulsed power supply system that would remove the operator from turning the direct connection to the power supply on and off. The idea was to implement a simple RC circuit that would be charged up by the power supply, disconnected, and then reconnected to the coaxial switch where it would discharge, sending a pulse to actuate the switch.
        \par
\begin{figure*}[htb!]
        \centering
        \includegraphics[width=0.7\linewidth]{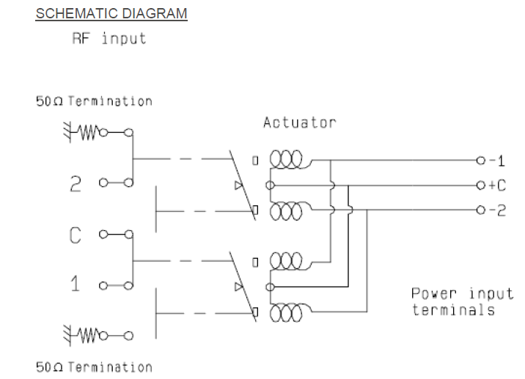}
        \caption{The circuit schematic of the Radiall coaxial switch.}
        \label{fig:SwitchSchematic}
        \end{figure*}

 The solenoid coaxial switch used is from Radiall, model R585433210, and operates from $0-18\,{\rm GHz}$, with a nominal switching voltage of $28\,{\rm V}$ at $25\,{\rm C}$. The schematic diagram of this switch is shown Figure \ref{fig:SwitchSchematic}. It is a Single-Pole Double-Throw (SPDT) switch, where 'C' in the diagram refers to common pole and ports 1 and 2 are each of the "throw" ports. The left side of the diagram is the coaxial connections and the right side is the DC power input terminals. In the diagram, the switch is currently configured with the common pole connected to port 1, and port 2 is connected to a $50\,{\rm \Omega}$ terminator. By applying a positive voltage to the common input and the port 2 power input to its ground, the temporary solenoid magnetic field exerts a force on a ferromagnet that tilts the angled  sections into the reciprocal configuration where the common coax is connected to the port 2 coax, and port 1 is terminated by the other load. What was discovered about this switch configuration is one cannot simply reverse the polarity of the voltage applied to switch back, but one needs to connect the port 1 input to the negative connection and disconnect the port 2 input. If not, simple reversing the polarity can result in a incomplete throw where all 3 coaxial ports are weakly connected with high loss, until a voltage with the correct polarity is applied (or worse an absurdly high negative voltage forces it into place).
        \begin{figure*}[htb!]
        \centering
        \includegraphics[width=1\linewidth]{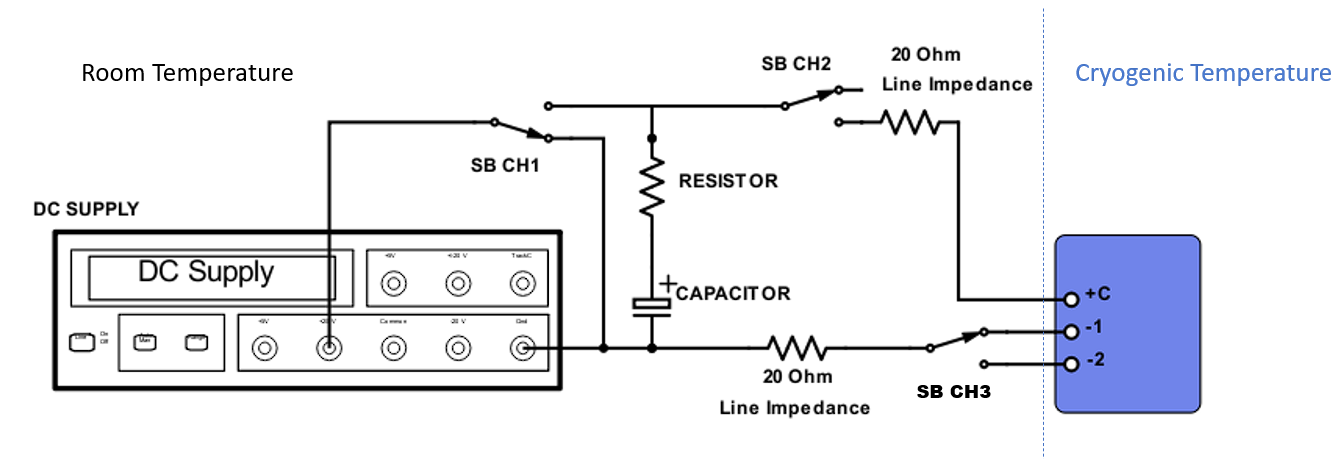}
        \caption{The pulsed power supply circuit design for a single coaxial switch.}
        \label{fig:PulsedSystem}
        \end{figure*}
        \par The pulsed switch system designed is shown in Figure \ref{fig:PulsedSystem}, and was made using an Agilent switch box module (model 349980A) that consisted of an array of DC SPDT switches. The CH1 switch connected the power supply to the RC circuit during the charging phase, while the CH2 switch disconnected the common input of the Radiall switch. Once the capacitor had charged, CH1 would disconnect from the power supply and CH2 would connect to the common pole, discharging the capacitor into the switch and actuating it. The CH3 switch allowed for going back and forth between the port 1 and 2 ground connections, depending on which port one needed to switch to. The main parameters in this circuit then were the capacitance, $C$, resistance, $R$, and the applied voltage $V_0$. These determined the key parameters of stored energy and time constant for the discharge:
        \begin{equation}
        E=\frac{1}{2}CV_0^2 , \tau=RC
        \label{eqn:RCenergy}
        \end{equation}
        The voltage supplied then to the common pole as a function of time during a discharge was thus:
        \begin{equation}
        V(t)=V_0e^{-t/RC}
        \label{eqn:dischargevoltage}
        \end{equation}
        The key was to supply the nominal $28\,{\rm V}$ switching voltage with the correct $RC$ time constant to actuate the switch, while still minimizing the energy deposited as heat in the system. 
        \par Test were first done at room temperature to figure out the optimal resistance and capacitance for a variety of operating voltage conditions. Initially, the thought was to match the pulse time length to the cited actuation time of the switch from Radiall, $\tau_{pulse}\approx3RC\approx10 ms$, and use the cited nominal voltage of $28\,{\rm V}$. A $47\,{\rm \Omega}$ resistor and $100\,{\rm \mu F}$ electrolytic capacitor gave us an $RC= 4.7\,{\rm ms}$. This did work, but one couldn't lower the voltage anymore to minimize the energy. Although one could lower the energy stored by lowering the capacitance while increasing the resistance accordingly, capacitors are sold with capacitance varying in factors of 10, which lowered the stored energy beyond the minimum to fire. Therefore, it was settled to lower the time constant by a factor of 10, by lowering the resistance, to increase the power supplied to the switch for the same energy stored. With this strategy, the nominal switching voltage was lowered to $20\,{\rm V}$, for an $RC=0.47\,{\rm ms}$ and $E= 20\,{\rm mJ}$. It was this resistance, $RC= 4.7\,{\rm \Omega}$, and capacitance, $C= 100\,{\rm \mu F}$, that was settled on for cryogenic testing.
        \par
\begin{figure*}[htb!]
        \centering
        \includegraphics[width=0.5\linewidth]{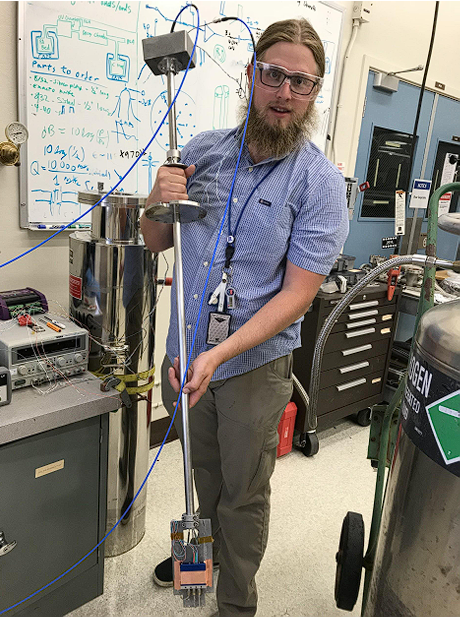}
        \caption{The author holding the coaxial switch mounted to paddle for cryogenic testing. The cryogenic vessel located behind.}
        \label{fig:SwitchPaddle}
        \end{figure*}

 Cryogenic testing was done in both liquid nitrogen and liquid helium. The switch was mounted to a paddle that could be inserted into a cryogenic vessel holding either liquid nitrogen or helium, pictured in Figure \ref{fig:SwitchPaddle}. Coaxial cables were connected to the switch and went up and out the top of the paddle so that transmission through the switch could be verified on a network analyzer. The DC wiring for actuating the switch also came out of the top of the paddle and connected to the RC circuit kept at room temperature. Because the conductivity of the wiring and the switch's internal components goes up at cryogenic temperatures, unsurprisingly, the minimum energy to actuate the switch goes down. The results are reported in Table \ref{tab:switchresults}.
        \begin{table}
        \centering
        \renewcommand{\arraystretch}{1.3}
        \begin{tabular}{@{}lcr@{}}
        Temperature ($K$) & Minimum Pulse Energy ($mJ$) & Initial Pulse Power (W)  \\
        \hline
        $300$ & $20$ & $85$ \\
        \hline
        $77$ & $12$ & $31$ \\
        \hline
        $4$ & $3$ & $2$ \\  
        \hline
        \end{tabular}
        \caption{A summary of the minimum energy and initial pulse power to fire the Radiall coaxial switch at various temperatures. The $RC$ time constant was fixed at $4.7\,{\rm ms}$.}
        \label{tab:switchresults}
        \end{table}
        \par The switch box was then finalized and installed on the ADMX site for use during Run 1C. A python program handled the switching process by automating the various channel switches in the pulse firing process. Unfortunately, the operations team blew out the polar electrolytic capacitor by running the circuit with reversed polarity very early in the run, rather than using the CH3 switch. This was due to poor communication on how to operate the system and the complexity of the sequence of switches that need to be made to fire a pulse. If CH1 is not disconnected before CH2 or vice versa, the system fails. If the voltage leads are switched, the circuit destroys itself. Although frustrating as the designer of this system it is to see, I understand now that these are very key weak points for its usability. The system has been fixed since this incident, and now employs software that prevents such a situation from occurring as easily again.
    \section{Helium liquefaction plant}
    \begin{figure*}[htb!]
    \centering
    \includegraphics[width=1\linewidth]{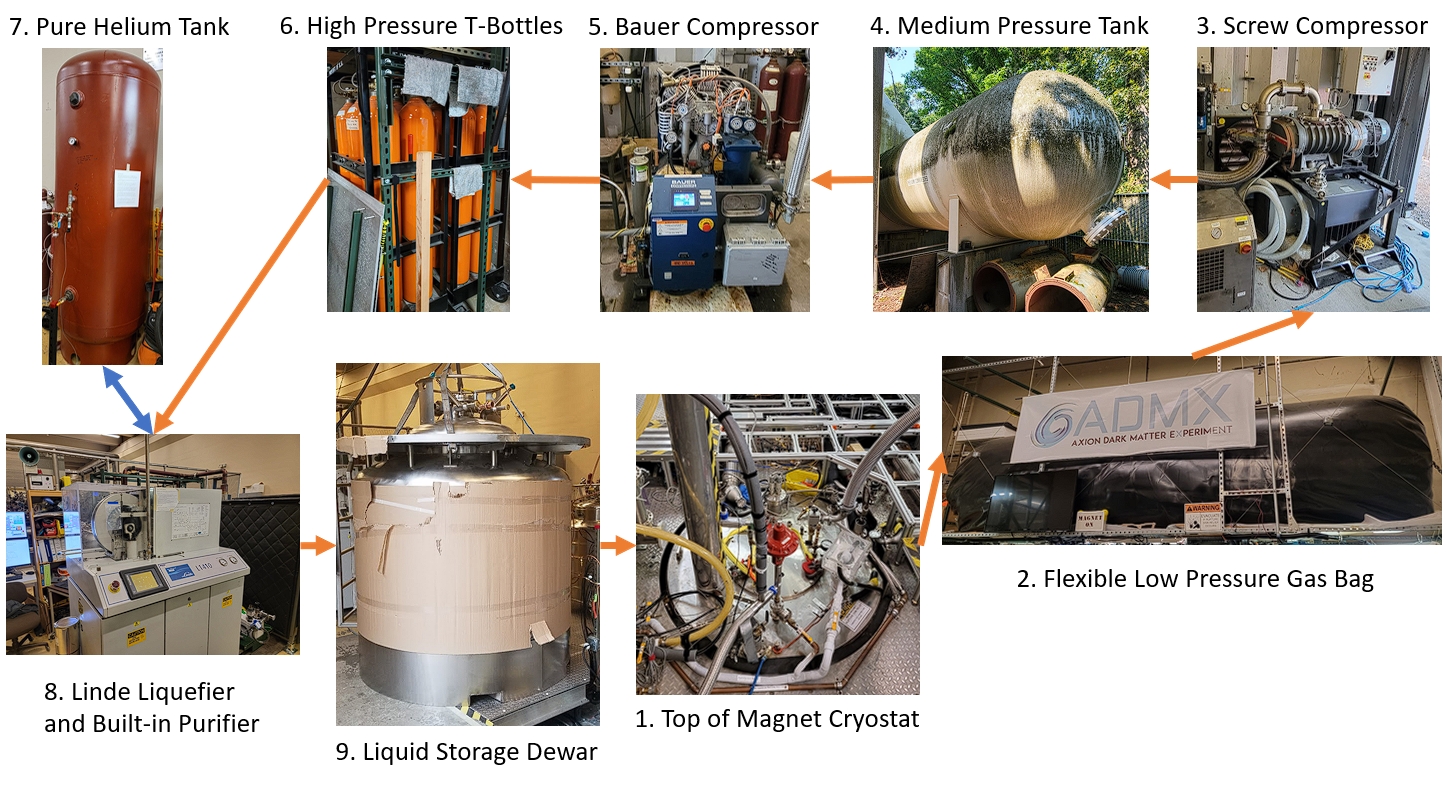}
    \caption{An outline of the ADMX liquid helium plant.}
    \label{fig:HePlant}
    \end{figure*}
    The helium liquefaction plant at the experimental site provides cost-effective cooling for the superconducting magnet, as well as cooling of the insert components to $4\,{\rm K}$. As pictured in Figure \ref{fig:HePlant}, the magnet cryostat reservoir is a good starting point for understanding the cooling cycle: this is filled with liquid helium and supplies liquid helium to the magnet superconducting coils with periodic magnet fills, keeping the coils fully submerged at all times. Helium vapor boils off over time in the magnet cryostat (about 20\% of its capacity per day) and is then transferred to a flexible gas bag (Figure \ref{fig:HePlant}-2) that is at roughly atmospheric pressure; this bag slowly inflates over the period of about 12 hours, and upon 92\% capacity is pumped down by the Kaeser Quantum Designs screw compressor (Figure \ref{fig:HePlant}-3) into the medium pressure helium gas tank that is kept outside the accelerator hall (Figure \ref{fig:HePlant}-4). This tank fluctuates between $2-12\:{\rm Bar}$ with a period of about 12 days, averaging out to about $5\:{\rm Bar}$ or $72\:{\rm PSI}$. Two Bauer compressors (Figure \ref{fig:HePlant}-5) transfer helium gas from the medium pressure tank to a set of high pressure T-bottles (Figure \ref{fig:HePlant}-6) kept at $120\,{\rm Bar}$. The Linde helium liquefier (Figure \ref{fig:HePlant}-8) requires certain purity level of helium to operate, therefore it contains a purifier on its gas input. However, it needs a certain amount of pure helium to start the liquefier and the purification process; therefore a separate high purity helium tank is connected to the inlet as well (Figure \ref{fig:HePlant}-7). Once the liquefier and purifier have started, it then starts to pull helium from the high pressure T-bottles to liquefy (it will also fill the pure helium tank back up off the outlet of the purifier). The produced liquid helium is then transferred via a Remote Delivery Tube (RDT) to $2500\,{\rm LL}$ (Liquid Liter) storage dewar, often called the 'Mother Dewar'(Figure \ref{fig:HePlant}-9). This dewar then contains automated liquid helium pumps and transfer lines that periodically fill the magnet cryostat reservoir to maintain liquid level, thus completing the cycle.
        \subsection{2500 LL mother dewar installation at CENPA}
\begin{figure*}[htb!]
    \centering
    \includegraphics[angle=180, width=0.7\linewidth]{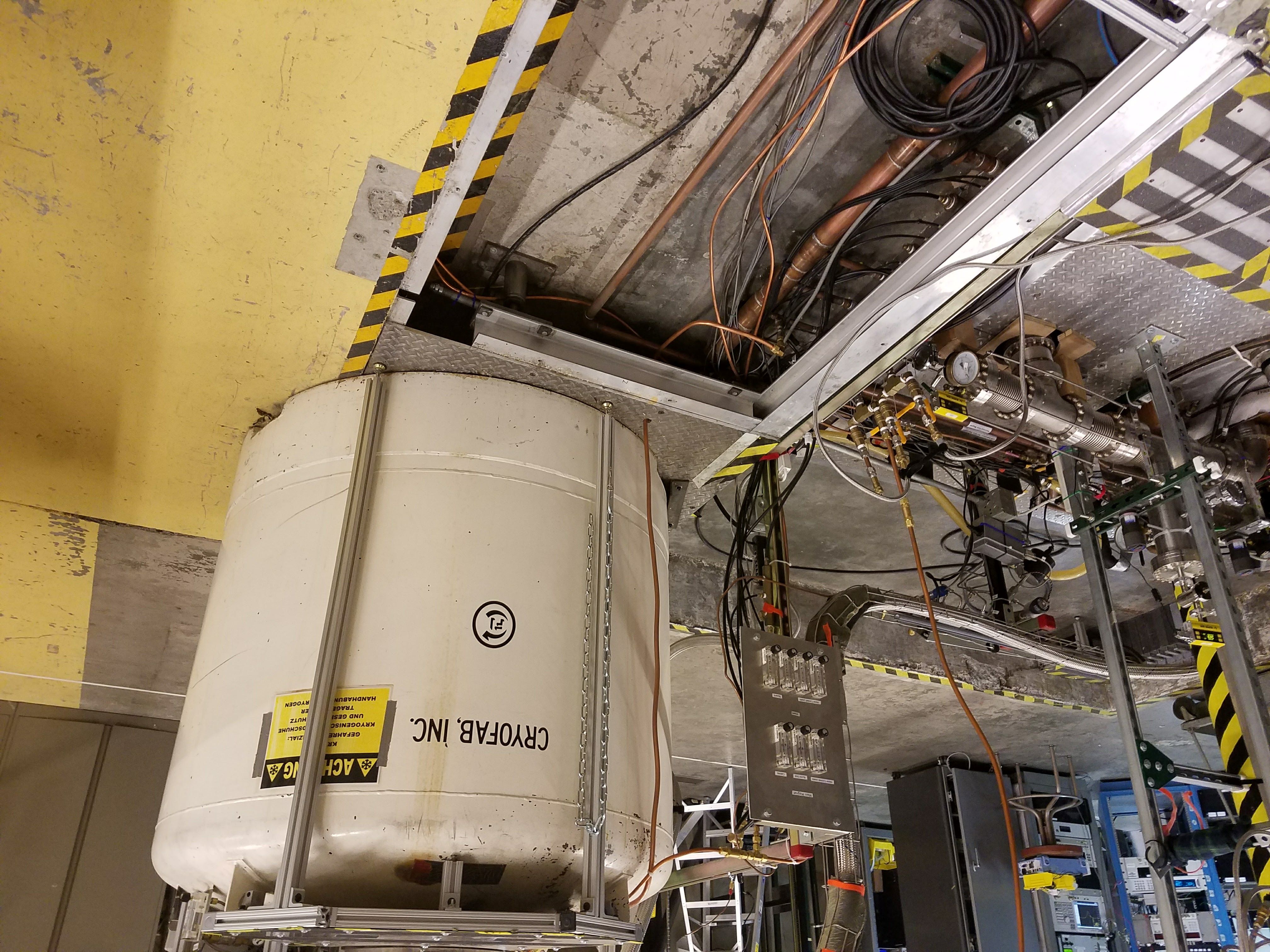}
    \caption{The old 500 LL ADMX helium dewar.}
    \label{fig:OldDewar}
    \end{figure*}

\begin{figure}[!t]
    \centering
    \begin{minipage}{0.5\textwidth}
        \captionsetup{width={.9\linewidth},font={small}}
        \centering
        \includegraphics[angle=270, width=0.9\linewidth]{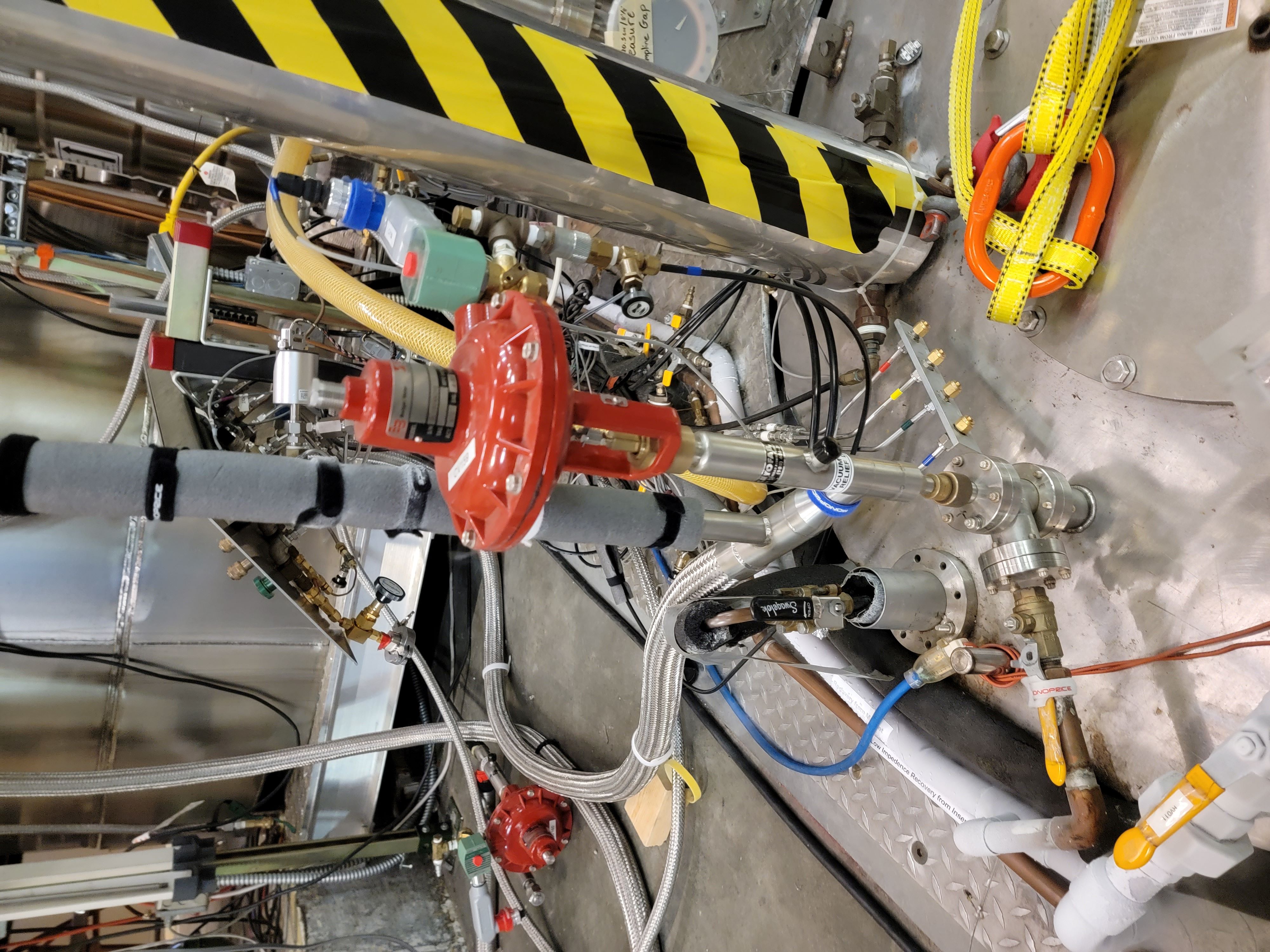}
        \caption{An electronic liquid helium fill pump. 4 of these pumps allow for automated fills of the magnet cryostat and its reservoir from the mother dewar.}
        \label{fig:FillPump}
    \end{minipage}%
    \hfill
    \begin{minipage}{0.5\textwidth}
        \centering
        \captionsetup{width={.9\linewidth},font={small}}
        \includegraphics[angle=270, width=0.9\linewidth]{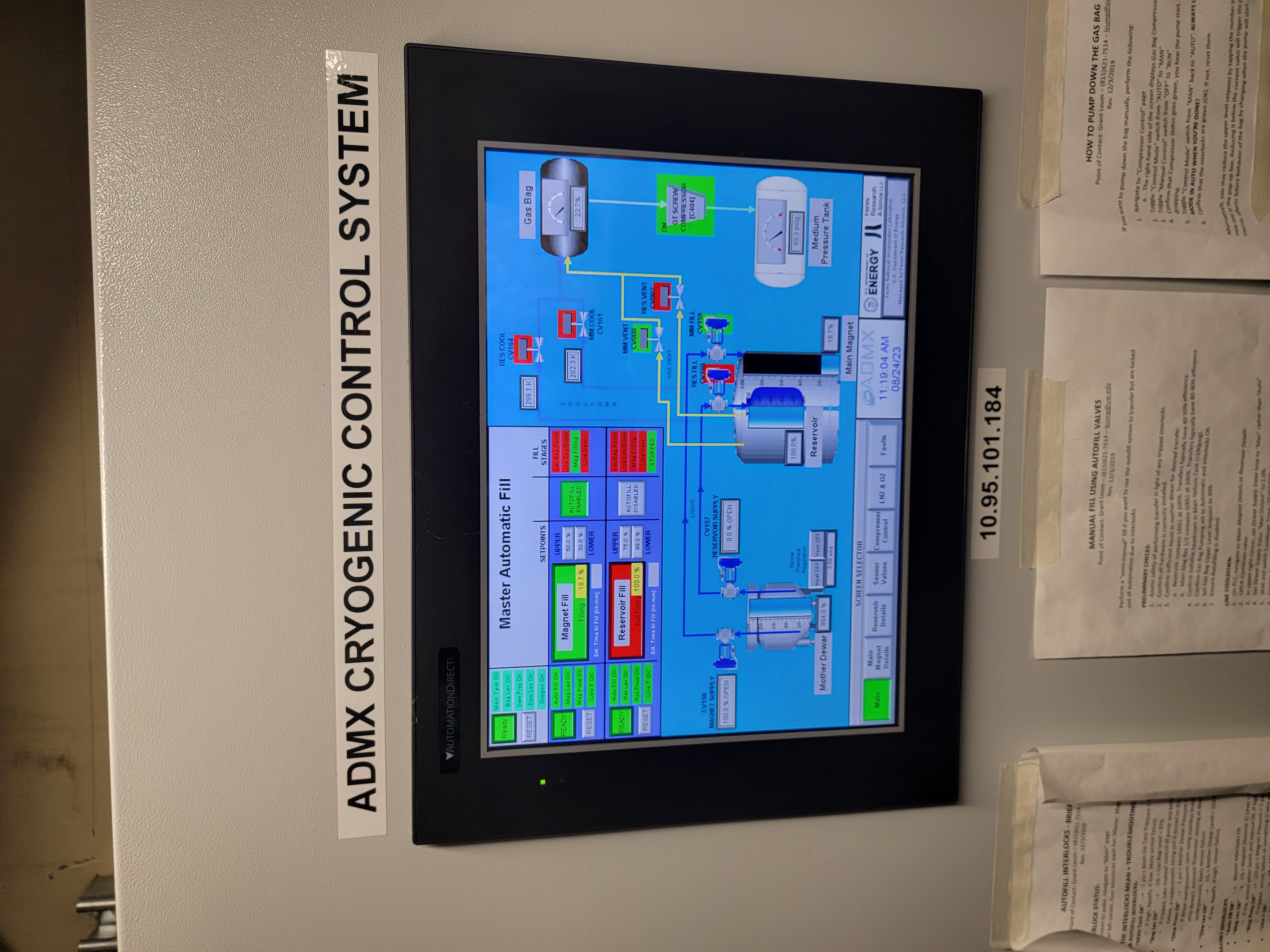}
        \caption{The ADMX cryogenic control system display. This allows for operators to adjust automatic helium fill parameters and monitor helium inventory.}
        \label{fig:CryoControl}
    \end{minipage}
    \end{figure}

        During my first year as a research assistant for ADMX, the liquid helium plant had several key upgrades to smooth out operations that I assisted with. The primary upgrade was replacing the older and much smaller $500\,{\rm LL}$ mother dewar (Figure \ref{fig:OldDewar}) with the current $2500\,{\rm LL}$ model. The other two upgrades included installation of the automated liquid fill system from this dewar to the magnet cryostat (Figure \ref{fig:FillPump}) , as well as a new computerized cryogenic control system for the entire plant (Figure \ref{fig:CryoControl}).
    \par The primary motivation for this upgrade was to increase the time between liquefaction runs by increasing storage capacity, and lightening the weekly duties of the liquefier operators. Previously, the liquefier had to be run 1-2 times per week to maintain liquid inventory for the magnet, as well as manual magnet fills nearly every day and reservoir fills every 3 days. This meant a liquefier operator had to be monitoring and controlling the system around the clock with very little room for error or contingency time for maintenance. With the increase in liquid storage by a factor of 5, it reduces liquefier runs to once every 10 days approximately, and the automatic fills simply need to be monitored via a control website, rather than an in-person operation. 
    \par
\begin{figure}[!b]
    \centering
    \begin{subfigure}{}
        \includegraphics[angle=270, width=0.375\linewidth]{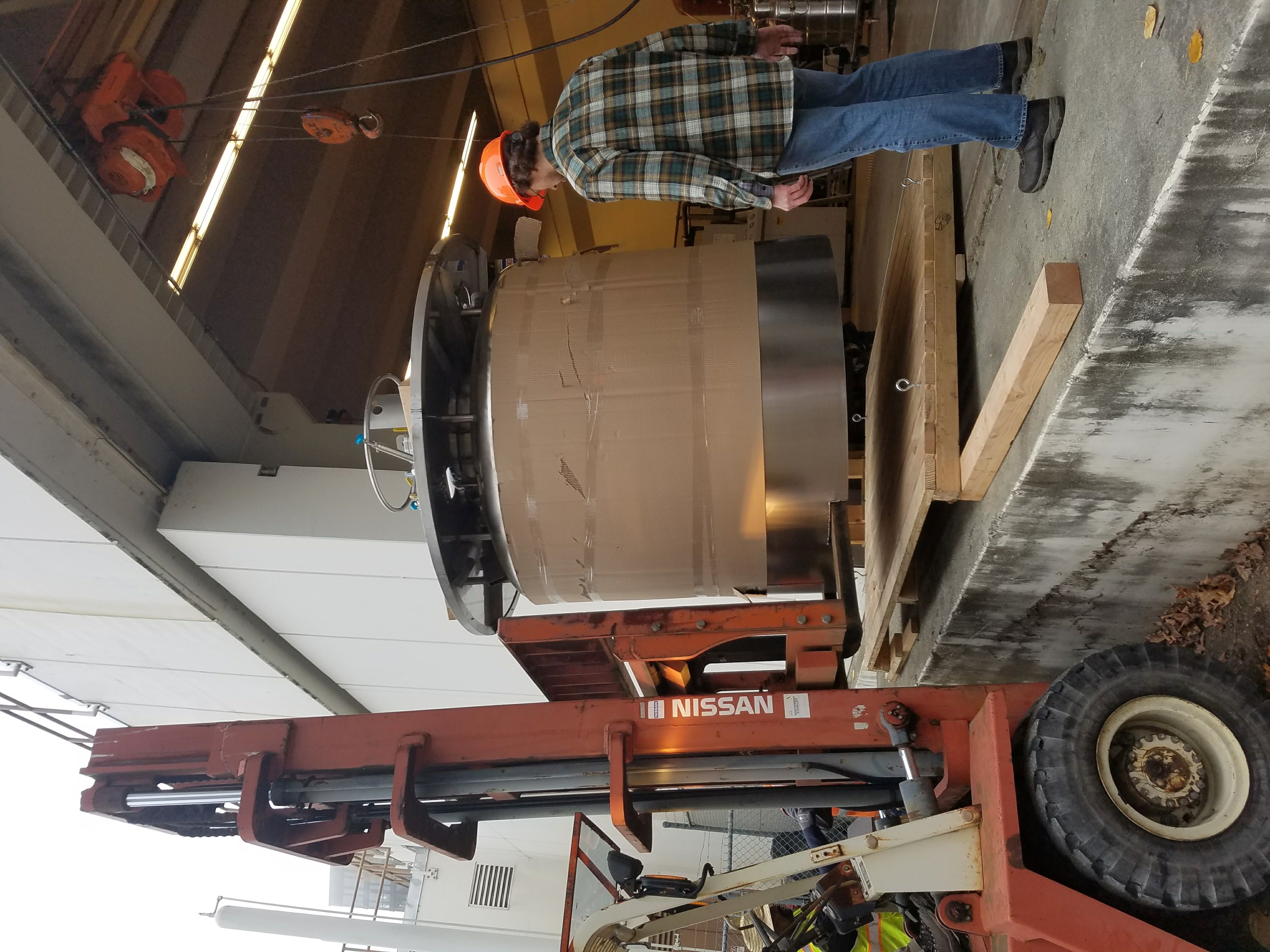}
    \end{subfigure}%
    \begin{subfigure}{}
        \includegraphics[angle=270, width=0.375\linewidth]{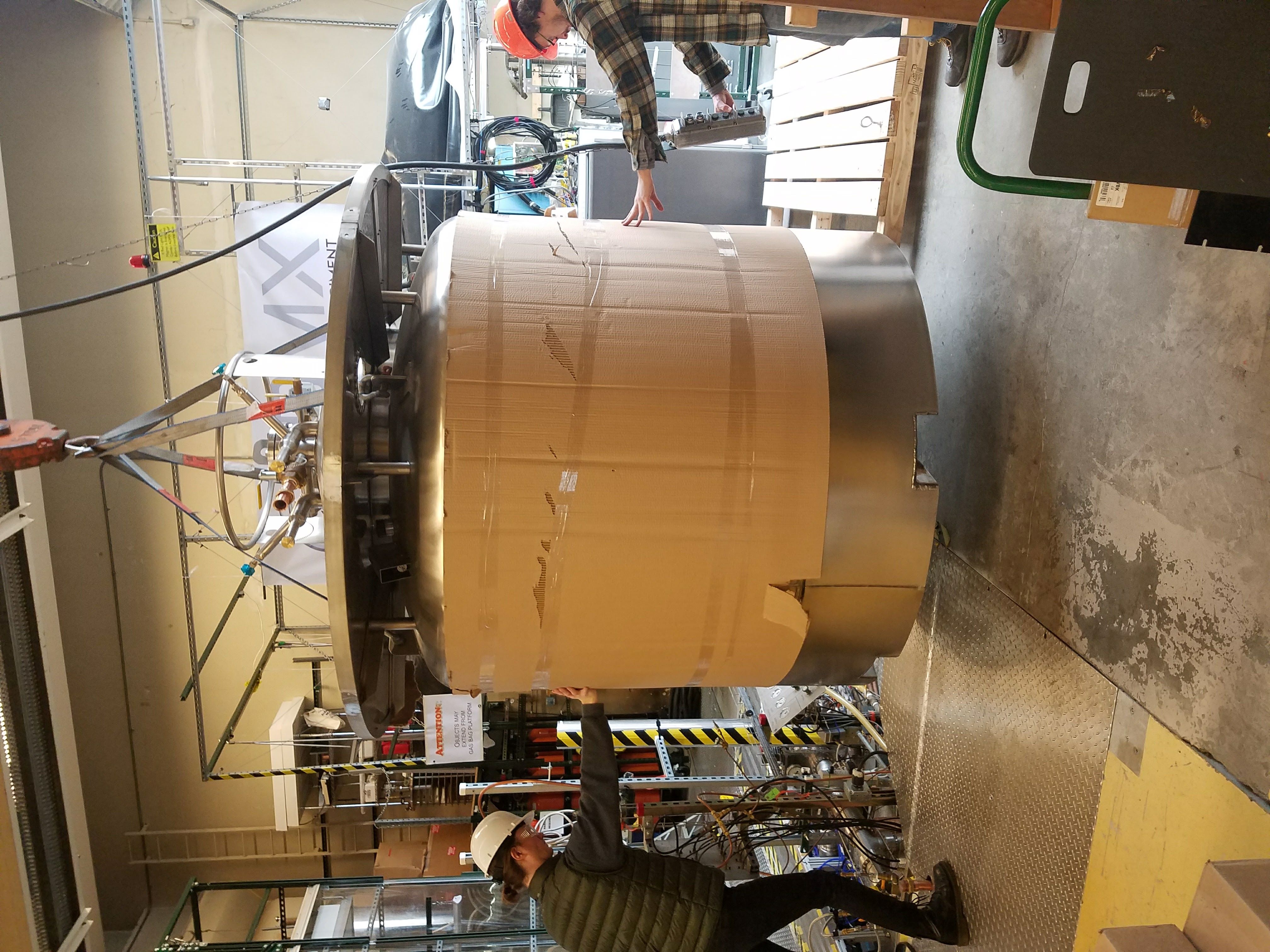}
    \end{subfigure}%
    \begin{subfigure}{}
        \includegraphics[width=0.375\linewidth]{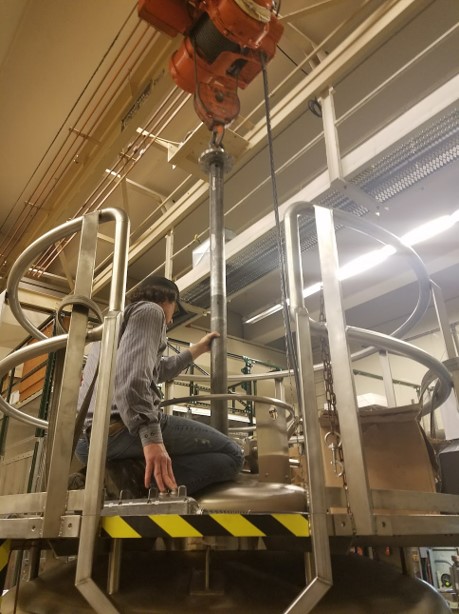}
    \end{subfigure}%
    \caption{Installation of 2500 LL helium dewar.}
    \label{fig:dewarinstall}
    \end{figure}

 The installation process of the dewar is outlined in Figure \ref{fig:dewarinstall}. The installation of this new dewar required use of a forklift to remove it from the shipping trailer onto the deck of the accelerator hall. From there, it was transferred by crane to its current location next to the helium liquefier. A ladder and railing was installed on top of it for safety when working on top of it. The dip tube was installed by crane, followed by the RDT that connected to the liquefier. Finally, the automatic pumps were installed for transferring liquid to the magnet cryostat.

    
\chapter{Data-taking and analysis techniques for haloscopes}
\label{chap:analysis}
This chapter will cover the necessary data-taking operations of a haloscope including the data-taking cycle as well as less periodic calibrations and operations. It will then lead into an analysis overview, focused on how the data could translate into an axion signal or the more common exclusion limit plot that is the key figure reported in every ADMX run paper so far (in the absence of a positive axion detection). Chapter \ref{chap:Sidecar1D} will cover the analysis results of the run 1D Sidecar, which will provide more details of what a real analysis looks like. The bulk of this chapter follows Refs. \cite{Run1BAnalysis,Sidecar1C}. 
    \section{Data-taking cadence}
    As discussed in Chapter \ref{chap:Haloscopes}, an axion haloscope scans through cavity resonant frequencies looking for power fluctuations above an average thermal noise background that could be due to an axion-photon decay. These power fluctuations can often be false axion signals such as RF interference, statistical fluctuations, and even intentionally injected "synthetic" axion signals. Therefore, certain frequency regions must be re-scanned to eliminate non-persistent signals. A given data-taking run covers the larger tuning range of the cavity, usually on the order of $100\,{\rm MHz}$ for the main experiment, while individual raw spectral centers of data were usually $2\,{\rm kHz}$ apart. To deal with this large separation in frequency span and organize the re-scanning process, runs are broken up into "nibbles", sections of the run on the order of $10\,{\rm MHz}$ wide, that take roughly a week to initially scan through.
    \par A given nibble is first scanned through at the fastest rate possible while still achieving the desired sensitivity (this is DFSZ sensitivity for the main experiment today), but if run conditions are not ideal, this might not be the case over the entire nibble. The goal of this first pass is to identify candidates for the re-scanning process. A data-taking cycle in this first pass starts with tracking and characterizing the resonant frequency of the $TM_{010}$ mode after the cavity has been tuned by moving the rods at the end of the last cycle. This involves a transmission measurement using the cavity weak port, which provides the resonant frequency of the cavity and its quality factor, and then a reflection measurement via the cavity bypass line, which provides the antenna coupling coefficient for the given tuning arrangement (these network analyzer measurements will be discussed more in depth in Chapter \ref{chap:Cavities}). After this, a primary amplifier re-biasing procedure might take place, which would optimize the gain for the target frequency, but this is only done every 5-7 cycle iterations as it is quite time-intensive. In the case of 'un-blinded' Synthetic Axion Generated injected signals (SAG injections), there is a check if any such injections are in the target frequency band. The actual axion data taking then commences with a 100 second power digitization. Finally, the cavity is tuned to the next frequency by stepping the rods to a new position, and the next cycle begins. An additional process of manually re-coupling the strong port antenna could occur if the last coupling coefficient was not ideal. However, this requires stopping the automated data-taking software, so it is not usually included in the typical cycle. The frequency and fraction of time these processes take in a given cycle are outlined in Table \ref{tab:DataCycle}.
    \begin{table}
    \centering
    \renewcommand{\arraystretch}{1.3}
    \begin{tabular}{@{}lcr@{}}
    Process ($K$) & Frequency ($mJ$) & Fraction of Time per Cycle  \\
    \hline
    Transmission Measurement & Every Cycle & $<1\%$ \\
    \hline
    Reflection Measurement & Every Cycle & $<1\%$ \\
    \hline
    JPA Re-bias & 5-7 Cycles & $25\%$ \\  
    \hline
    Check for SAG injecion & Every Cycle & $<1\%$ \\  
    \hline
    Data Digitization & Every Cycle & $98\%$ \\  
    \hline
    Move Rods & Every Cycle & $<1\%$ \\  
        \end{tabular}
        \caption{A Summary of the frequency and fraction of time each process takes in a data-taking cycle of ADMX based on data from run 1B. Axion search data is only acquired during the data digitization cycle. Since the JPA bias procedure only occurs every 5-7 cycles the average fraction of time spent digitizing is kept at $98\%$. SAG stands for Synthetic Axion Generator.}
        \label{tab:DataCycle}
        \end{table}
    \par
\begin{figure*}[htb!]
    \centering
    \includegraphics[width=0.97\linewidth]{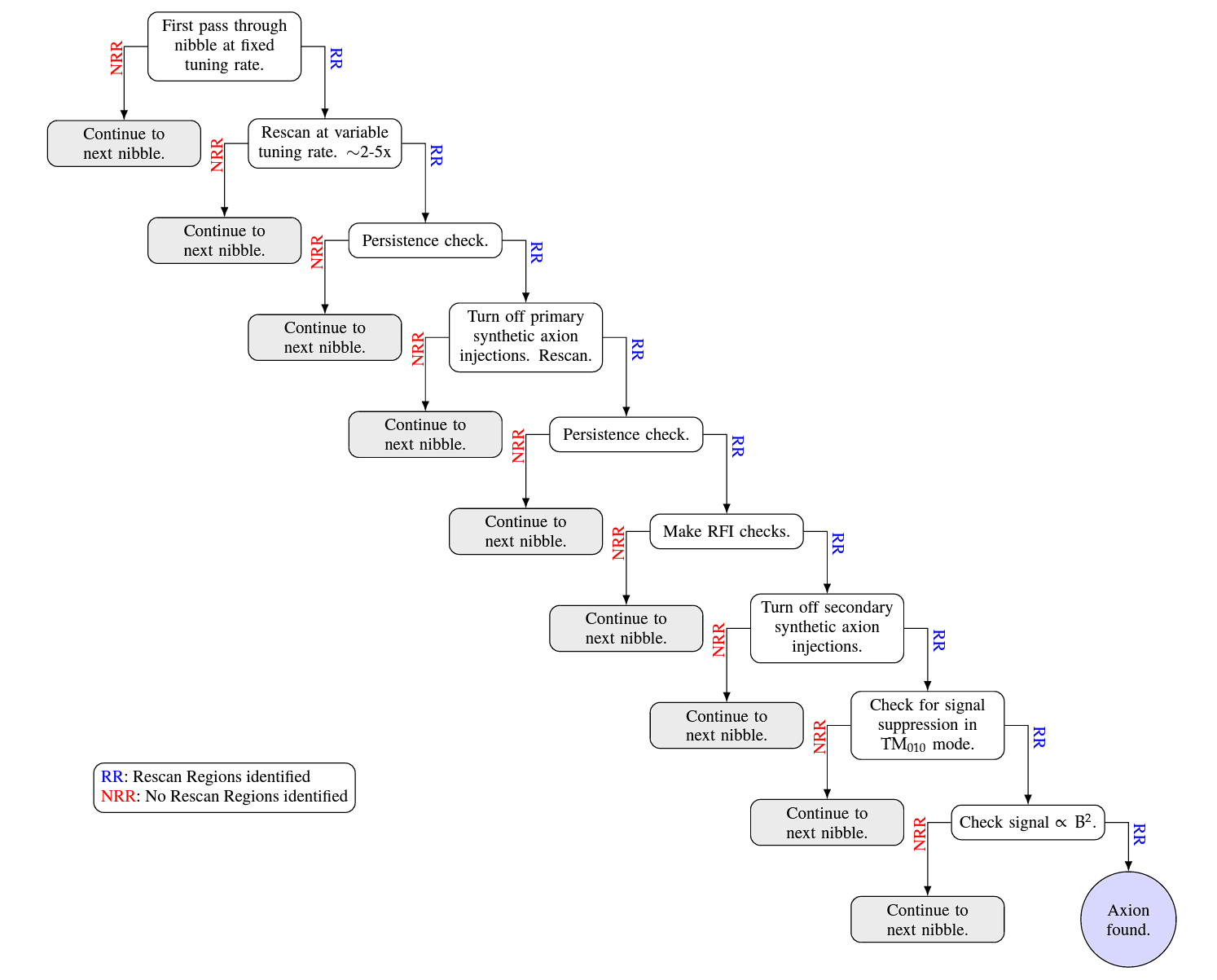}
    \caption{Data-taking decision tree. Typically a first nibble pass-through always results in several re-scan triggers in the analysis due to the statistics associated with the chosen tuning rate. A persistence check means verifying that a potential signal appears in all the spectra acquired for it so far, and it is not intermittent. Primary synthetic axion injections are un-blinded by the operations team, whereas secondary injections are only known by an off-site SAG operator to check the operations teams' detection efficiency. RFI checks are made with a signal analyzer outside the cavity, ruling out any external signals such as local radio stations. Final checks include trying to modulate the signal power according to Equation \ref{eqn:axionpower3} by looking for suppression of the signal off-resonance and modulating the magnet field strength. Figure credit from Ref. \cite{Run1BAnalysis}.}
    \label{fig:DecisionTree}
    \end{figure*}

 Ideally, the first nibble pass-through can run with relative autonomy, after which the nibble re-scan procedure begins. Re-scan regions are flagged by the first pass through in the analysis software. The cavity is tuned back to these regions to repeat the same data-taking cycle, and skipping the non-flagged regions entirely. After this re-scan, any candidate signals that are persistent through both digitization cycles are then subjected to other tests that would evaluate if it had an "axionic" nature. These tests include RFI checks outside the cavity, checking for dependence on the $TM_{010}$ mode, and ultimately, seeing if the signal power scaled with $B^2$ via ramping the magnet. Once all the candidates in a nibble are eliminated, the experiment then moves on to the next nibble, repeating the process until all nibbles in the run are complete. This decision tree is outlined in Figure \ref{fig:DecisionTree}.
    \section{Hot load measurements/Y-factor method 1}
    \label{hotload1}
    The "hot load" shown in Figure \ref{fig:RecieverSchematic} is critical to measuring the system noise temperature of the haloscope. As alluded to before, the system noise temperature is critical to understanding the noise level in the receiver and quantifying the SNR of potential signals. Critical to the noise temperature calculation, is the receiver temperature as a whole. These measurements are done only a few times throughout the run, as they tend to produce stable results; in run 1B, this procedure was only performed 4 times. These are also done sparingly because it requires halting the data-taking process for several hours at least. They are referred to as "Hot-Cold" measurements (hence the "hot load" name) or "y-factor" measurements \cite{wilson2011techniques}.
    \par In abstract terms, this is done by connecting the receiver to a known variable temperature  resistive source at its input and as its incremented in temperature by some $\Delta T$, the receiver output power will change by some $\Delta P$ (in the case of coherent receivers, with a "DC", linear response). If we label the temperature of the hot and cold states of the load, $T_H$ and $T_L$ respectively, then we can write down the power output of the receiver in each case, $P_H$ and $P_L$:
    \begin{equation}
    \begin{split}
         P_L= (T_L+T_{rx})G \\
        P_H= (T_H+T_{rx})G
    \end{split}
    \label{eqn:hotload1}
    \end{equation}
    defining:
    \begin{equation}
    y=\frac{P_H}{P_L}
    \label{eqn:yfactor}
    \end{equation}
    the resultant receiver temperature would then be:
    \begin{equation}
    T_{rx}=\frac{T_H-T_Ly}{y-1}
    \label{eqn:hotload2}
    \end{equation}
    This measurement assumes, however, that this variable temperature source doesn't change the operation of the receiver components, which is easier said than done for ADMX.
    In the case of ADMX, an ideal noise measurement would be done with the JPA pump enabled allowing for full characterization along the entire receiver chain. However, since the JPA can easily be saturated by thermal noise and the hot load was kept on the $4-K$ stage during run 1B, the JPA was kept disabled (the hot load has since been moved to a 175 mK arrangement for run 1C-D). When its not enabled, the JPA conveniently acts as a passive RF mirror, reflecting signals with minimal attenuation back down the output line. During a measurement, the RF switch is set to point to the hot load in this arrangement (see Figure \ref{fig:RecieverSchematic}) and a heater on the hot-load raises the temperature as a temperature sensor monitors this warming process. A wide-bandwidth power measurement is then taken, which has the expected power output per unit bandwidth:
    \begin{equation}
    P=G_{HFET}k_B[T_{JPA}(1-\epsilon)+\epsilon T_{load}+T_{HFET}]
    \label{eqn:hotload3}
    \end{equation}
    where $T_{JPA}$ and $T_{load}$ are the physical JPA and hot load temperatures measured by sensors, $G_{HFET}$ is the gain of the HFET secondary amplifier, and $T_{HFET}$ is the noise temperature of the HFET and all down stream electronics. The transmission efficiency of the quantum amplifier package is given by $\epsilon$, and is simply the linear version of the attenuation, $\alpha$, in dB ($\epsilon = 10^{-\alpha/10}$. It was quantified ex-situ by measuring the dominant loss source, the two up-stream circulators from the JPA, but also in-situ by a more complicated Y-factor measurement, or using a JPA signal-to-noise ratio improvement (SNRI) measurement (more on SNRI in the next section). In short, Equation \ref{eqn:hotload3} was used to fit for $G_{HFET}$ and $T_{HFET}$ by holding the hot load (and as a result, the JPA too) at a variety of temperatures. 
    \section{Y-factor method 2}
    \label{hotload2}
    The second method for determining the receiver noise temperature essentially replaces the hot-load with the final cavity bypass line attenuator (labeled 'A' in Figure \ref{fig:RecieverSchematic}). In this method, the RF switch is still connected to cavity. Off-resonant thermal photons originating from the bypass line will travel through the first circulator (labeled 'C1' in Figure \ref{fig:RecieverSchematic}), reflect off the cavity, then reflect off the disabled primary amplifier package, and out through the output line through the post-amplifiers. Since these thermal photons pass through an extra length of the first circulator, they will be slightly more attenuated than a photon originating from the hot load; this was approximated 0.5 dB based on ex-situ measurements of the circulator, hence $\epsilon_c=10^{-(\alpha+0.5\,dB)/10}$. The form of Equation \ref{eqn:hotload3} still applies here, however one replaces $\epsilon$ with $\epsilon_c$, and $T_{load}$ with $T_{cav}$ since the bypass attenuator is mounted directly to the cavity top (there is an assumption that $T_{attenuator}=T_{cav}$). Additionally, it was approximated that $T_{cav}$ and $T_{JPA}$ were the same, therefore Equation \ref{eqn:hotload3} can be re-written as: 
    \begin{equation}
    P=G_{HFET}k_B[T_{JPA}+T_{HFET}]
    \label{eqn:hotload4}
    \end{equation}
    In this case, the fit parameters were $G_{HFET}$ and $T_{HFET}$ whereas $T_{JPA}$/$T_{cav}$ was the varied and measured quantity. Because this temperature must still be varied, there must be a way of heating the JPA. This was done in run 1B by applying a small current to the RF switch such that it heated, but did not actuate. As will be discussed in Chapter \ref{chap:Sidecar1D}, the run 1D Sidecar HFET was characterized using the magnet ramp as a heating source; The ramping of the magnet heats the insert through eddy currents, and therefore the mK attenuator. Operators in anticipation of this heating and cooling process, turn on wide-bandwidth power measurements. These wide-bandwidth power measurements are done by taking a series of digitizations across the tuning range of the cavity; This is because the bandwidth of the digitizer is limited to 2 MHz. The idea here is to capture off-resonant photons across the search frequency range (on-resonance thermal photons will be absorbed by the cavity), so that one can give an associated HFET noise temperature for each target search frequency during normal data-taking. Gaps in between these power spectra are okay, as the noise temperature can be interpolated as a function of frequency (it should follow a black-body distribution). 
    \par There are several advantages to this method. Primarily, this gives a separate confirmation of $T_{HFET}$ independent of the attenuation through the SQUIDadel, which improves confidence in the analysis. In terms of operations, this method can be less intensive and time-consuming, or even the only option; the RF switch can fail to actuate, rendering a hot-load measurement impossible. Conveniently, an RF switch is usually said to fail because it is heating the system towards the limits of the dilution fridge before actually actuating; one can then use this heat to perform this back-up method. It is also adaptable; any anomalous heating or cooling of the cavity and amplifiers can be a moment to perform this operation. Finally, in some cases, this can be done remotely, whereas a hot-load measurement requires on-site personnel to flip switches and monitor. See Chapter \ref{chap:Sidecar1D} for details of the Sidecar HFET characterization.
        \subsection{Designing the run 1C part 2 "hot-cold" load}
        One of my projects during my tenure as a graduate student was redesigning the hot-load for run 1C, where it would be moved from the 4-K stage to the intermediate cold plate that normally operated at 150 mK. This move would bring the y-factor fit measurement closer to the operating temperature regime of the JPA. The hot load would have to be heated from this base temperature up to about 500 mK without overwhelming the dilution fridge, but also cool down in a reasonable amount of time. This is done by making the hot load thermally disconnected from the mounting plate except for a well-known cooling wire connection that controls the cooling power of the load.
        \par
\begin{figure*}[htb!]
        \centering
        \includegraphics[angle=0, width=0.4\linewidth]{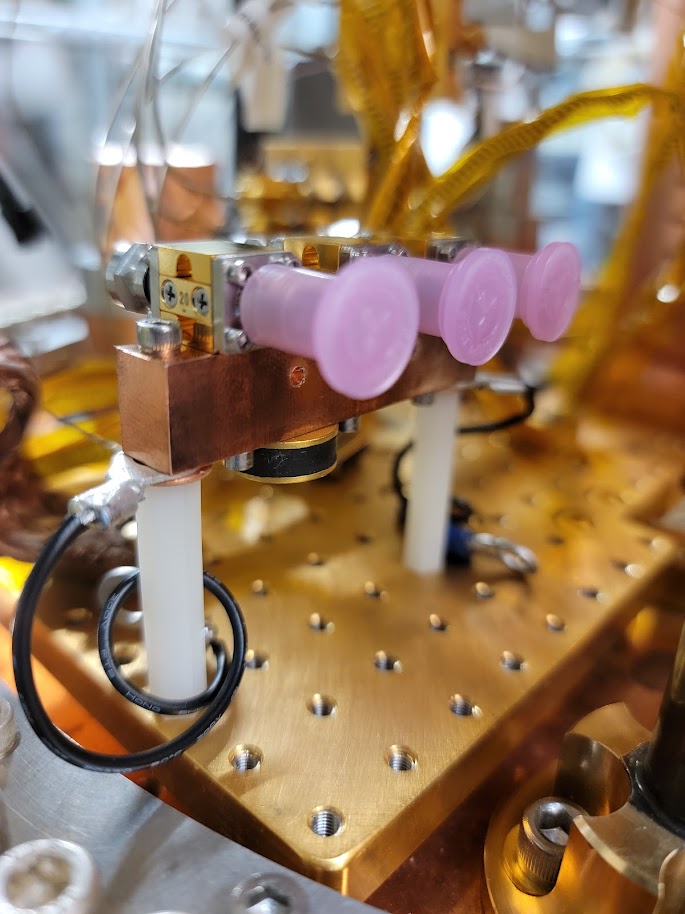}
        \caption{The run 1C hot load.}
        \label{fig:Run1Chotload}
        \end{figure*}

 The original run 1C hot load is pictured in Figure \ref{fig:Run1Chotload}. Two nylon standoffs thermally disconnect the main OHFC copper body, on which 3 attenuators are mounted to each by a single screw. One attenuator is for the main experiment receiver, one for Sidecar, and one extra was budgeted in. Within the copper body, 3 through-holes are made where 10 k$\Omega$ resistors are embedded and epoxied in place; these resistors act as the heaters for the load. The resistors are connected to superconducting wire that leads to power supplies, thermally disconnecting them from the warm space above by the superconductor, but electrically connected to act as a heater when in use. A temperature sensor is mounted on the underside to monitor the heating and cooling process. Finally, two copper cooling wires were connected underneath each nylon standoff, and attached to the intermediate cold plate below.
        \par
\begin{figure*}[htb!]
        \centering
        \includegraphics[angle=0, width=0.4\linewidth]{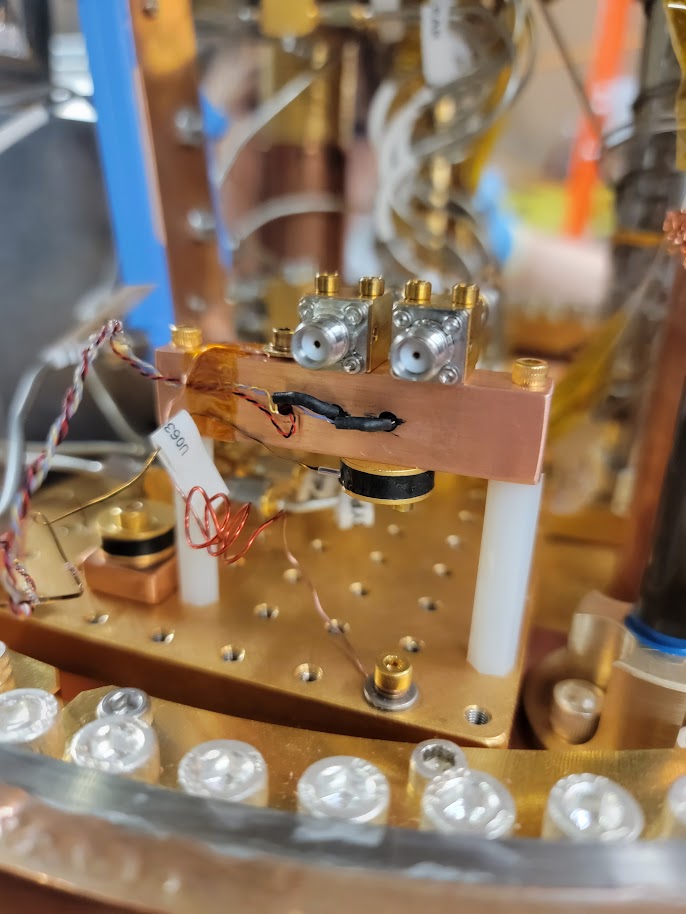}
        \caption{The run 1D hot load.}
        \label{fig:Run1Dhotload}
        \end{figure*}

 This original design has been iterated since, and the current configuration is pictured in Figure \ref{fig:Run1Dhotload}. First, the extra attenuator was removed because it was an unnecessary, unknown thermal load on the piece. The nylon standoffs, being plastic threading, did not provide enough torque to properly connect the cooling wire to the body, and thus it was given its own threaded connection with a metal screw in place of the 3rd attenuator. Additionally, the attenuators were mounted with two screws from the top, to ensure a more solid thermal connection to the body. This was motivated by the hot load being hotter than predicted, and cooling slower than predicted, which is indicative of bad thermal contacts.
        Key to this design is understanding how to calculate the thermal conductive power of each component touching the hot load, as well as the heat capacity of the load itself. The conductive power of a wire/cylinder connecting two bodies can be written:
        \begin{equation}
        G=kT\frac{r^2}{L}
        \label{eqn:wirethermalpower}
        \end{equation}
        where $k$ is the thermal conductivity of the wire material, $T$ is the temperature of the wire, $r$ is the radius of the wire, and $L$ is the length of wire between the two bodies. In the case of the hot load, we calculate the conductive power of the two nylon stand offs, cooling wire, heater wire, 4 coaxial cable connections, and temperature sensor connection. Next one wants to know the heat capacity of the hot load body, which is simply $C=cM$ where $c$ is the specific heat and $M$ is the mass of the body of copper. The screws and attenuator cubes also add to the heat capacity, but are not included in this calculation. From this we calculate the time constant for cooling as:
        \begin{equation}
        \tau_{cool}=\frac{C}{G_{cool}}
        \label{eqn:thermaltimeconstant}
        \end{equation}
        where $G_{cool}$ is just the combined power of all connections to the intermediate cold plate, not the warm space connections like the heater wire and temperature sensor wire; these, if anything, should be subtracted from the cooling power, but being superconducting, they are minimal sources of heat. The heating time constant can simply be written based on the current supplied to the resistor:
        \begin{equation}
        \tau_{heat}=\frac{C}{I^2R}
        \label{eqn:heatingtime}
        \end{equation}
        where $R$ is the total resistance of the heaters and $I$ is the current from the power supplies. The other warm and cold space connections are negligible compared to this supplied heater power, so they are not included in this calculation. 
    \section{SNRI}
    As mentioned earlier, signal-to-noise improvement (SNRI), commonly used to quantify performance of amplifiers, can be used to circumvent directly measuring the attenuation of the quantum amplifier package. SNRI can be defined as:
    \begin{equation}
    SNRI=\frac{T_{sys,off}}{T_{sys,on}}=\frac{G_{on}P_{off}}{G_{off}P_{on}}
    \label{eqn:SNRI}
    \end{equation}
    where $T_{sys,on/off}$ refers to the noise temperature with the JPA powered and off, $G_{on/off}$ is the gain of the JPA powered and off, and $P_{on/off}$ is the measured noise power when powered and off. Essentially, it gives the ratio of improvement expected from having the primary amplifier operating in the receiver. Hot load measurements have given us $T_{HFET}$ with the amplifier powered off. We can write down the total system noise temperature with the amplifier off as:
    \begin{equation}
    T_{sys,off}=T_{cav}+T_{JPA,off}+\frac{T_{HFET}}{G_{off}}
    \label{eqn:Tsysoff}
    \end{equation}
    Since $T_{HFET}\approx 11 K$ while $T_{cav/JPA}\approx 0.1 K$, we can make the assumption $T_{HFET}>>T_{cav}+T_{JPA,off}$. Additionally, an amplifier that isn't powered shouldn't have significant gain, and therefore can be approximated as just 1. This means we can write the total system noise temperature with the JPA powered ON just in terms of $T_{HFET}$ and SNRI :
    \begin{equation}
    T_{sys,on}=\frac{T_{HFET}}{\epsilon SNRI}
    \label{eqn:Tsyson}
    \end{equation}
    where $\epsilon$ is the transmission efficiency from the cavity to the primary amplifier. Since it has been determined that $T_{HFET}$ is pretty stable throughout the run, this means periodic SNRI measurements are all that is needed to update the system noise temperature measurements. This can be done when re-biasing the amplifier, and uses the latter half of Equation \ref{eqn:SNRI}, measuring the noise power and gain both on and off. This can be done using the cavity bypass line; a known power off-resonant signal inputted through the bypass will reflect off the cavity, then be amplified by JPA, giving the gain both on and off. The noise power level can just be sampled from the output with the amplifier on and off. This makes periodic checks of the noise temperature much less time-intensive using SNRI.
    \section{Axion search analysis procedure}
    This section is an overview of the ADMX "medium" resolution analysis procedure, and how it ultimately produces limits on the axion parameter space when there are no persistent candidates left. As described in the data-taking cadence section, after time-series digitization data is recorded, "medium"-resolution analysis is performed on the data, separate from the data-taking cycle, to look for the presence of axion candidates, determining re-scan regions for a given nibble. Once all these candidates have been reviewed and eliminated, a limit on $g_{a\gamma\gamma}$ over the frequency range can be set. This process can be divided into spectra background processing, analysis cuts, and grand spectrum construction.
    \subsection{Spectra Background Processing}
\begin{figure}
    \centering
    \subfigure[Example of a raw spectrum from run 1B.]{\label{fig:RawSpectrum}\includegraphics[ width=0.7\linewidth]{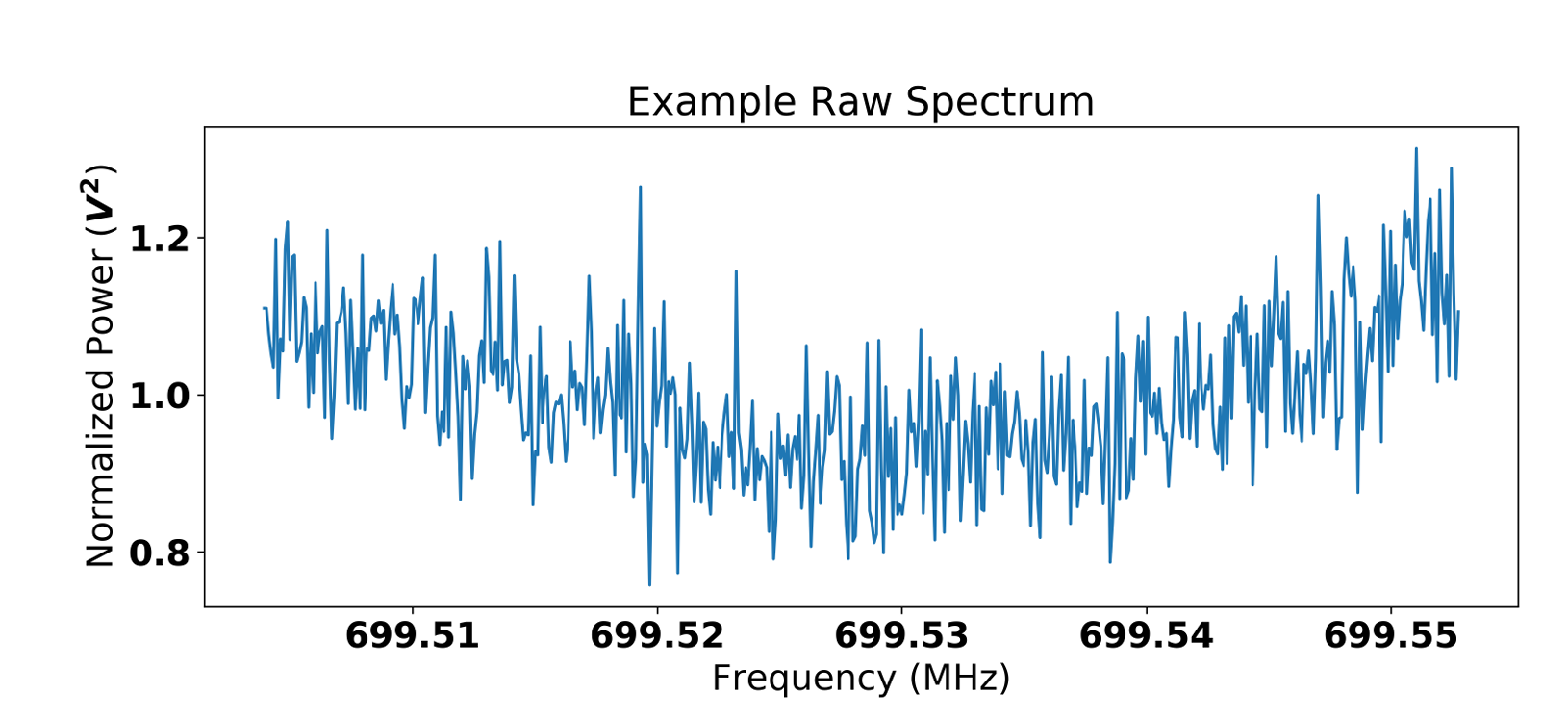}}
    \subfigure[Example of a filtered spectrum from run 1B.]{\label{fig:FilterSpectrum}\includegraphics[ width=0.7\linewidth]{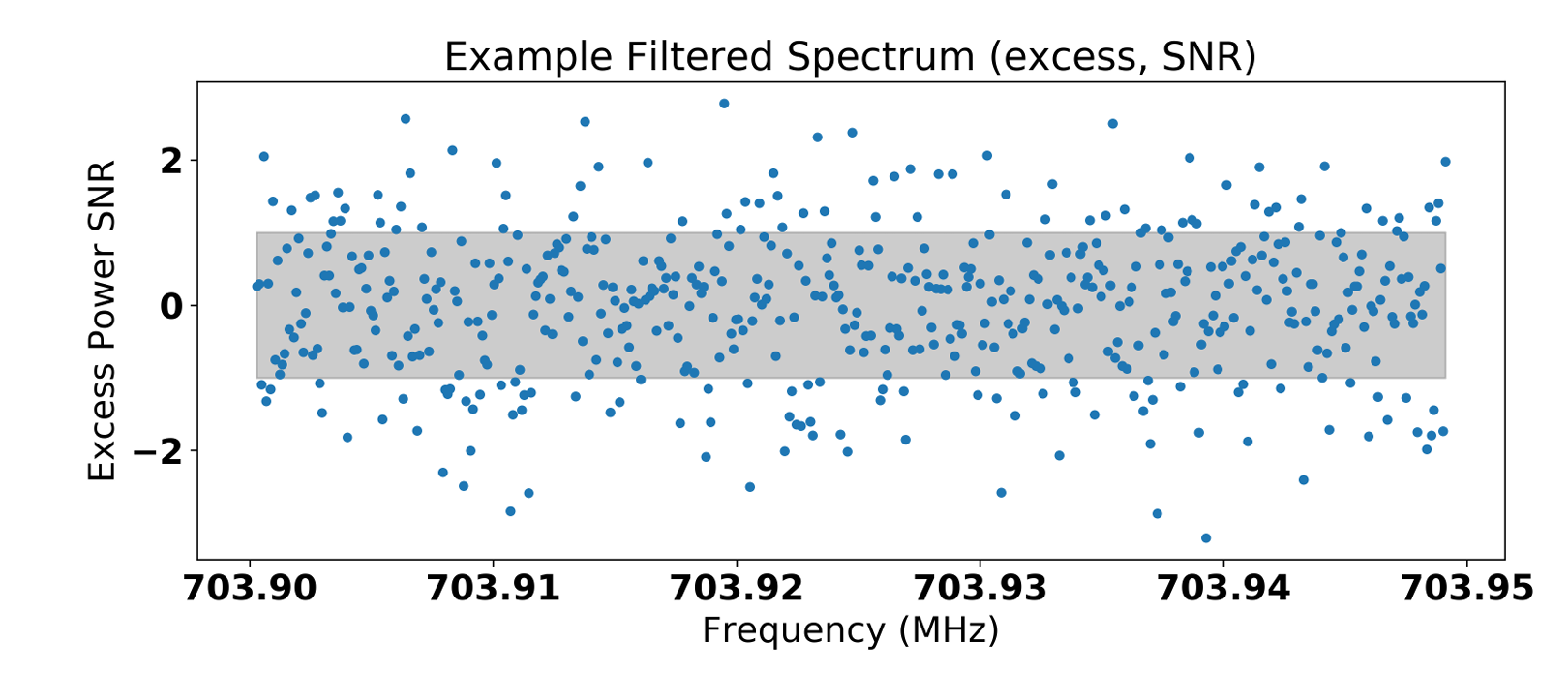}}
    \subfigure[Example of a Lorentzian weighted spectrum from run 1B.]{\label{fig:LorentzSpectrum}\includegraphics[width=0.7\linewidth]{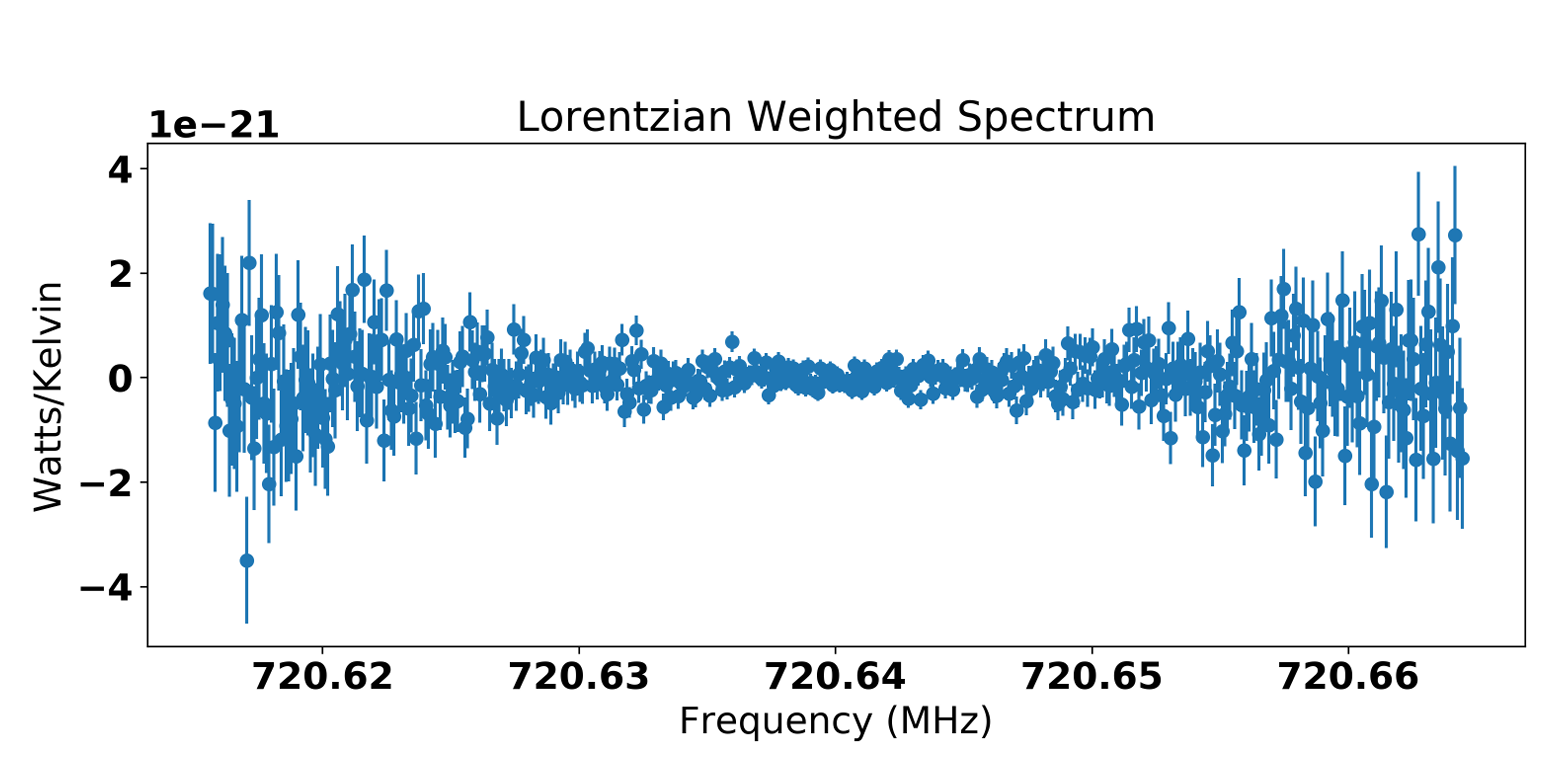}}
    \caption{The raw spectrum preparation process.}
    \label{fig:Spectras}
    \end{figure}

    Each spectrum in the main experiment during run 1B consisted of 512 bins each with a width of 95 Hz, for a total spectrum width of 48.8 kHz; an example spectrum is shown in Figure \ref{fig:RawSpectrum}. A single spectrum is the result of 100 seconds of time-series data being chopped up into $10^4$ Fourier transforms of each 10 millisecond segment. There is frequency overlap in the spectra, as was mentioned in the scan rate section, meaning we will have to account for multiple spectra having the same frequency bins. Therefore it's easiest to denote the power in the "j"th spectra within the "i"th bin as $P^j_i$.
    \par
\begin{figure*}[htb!]
    \centering
    \includegraphics[angle=0, width=0.7\linewidth]{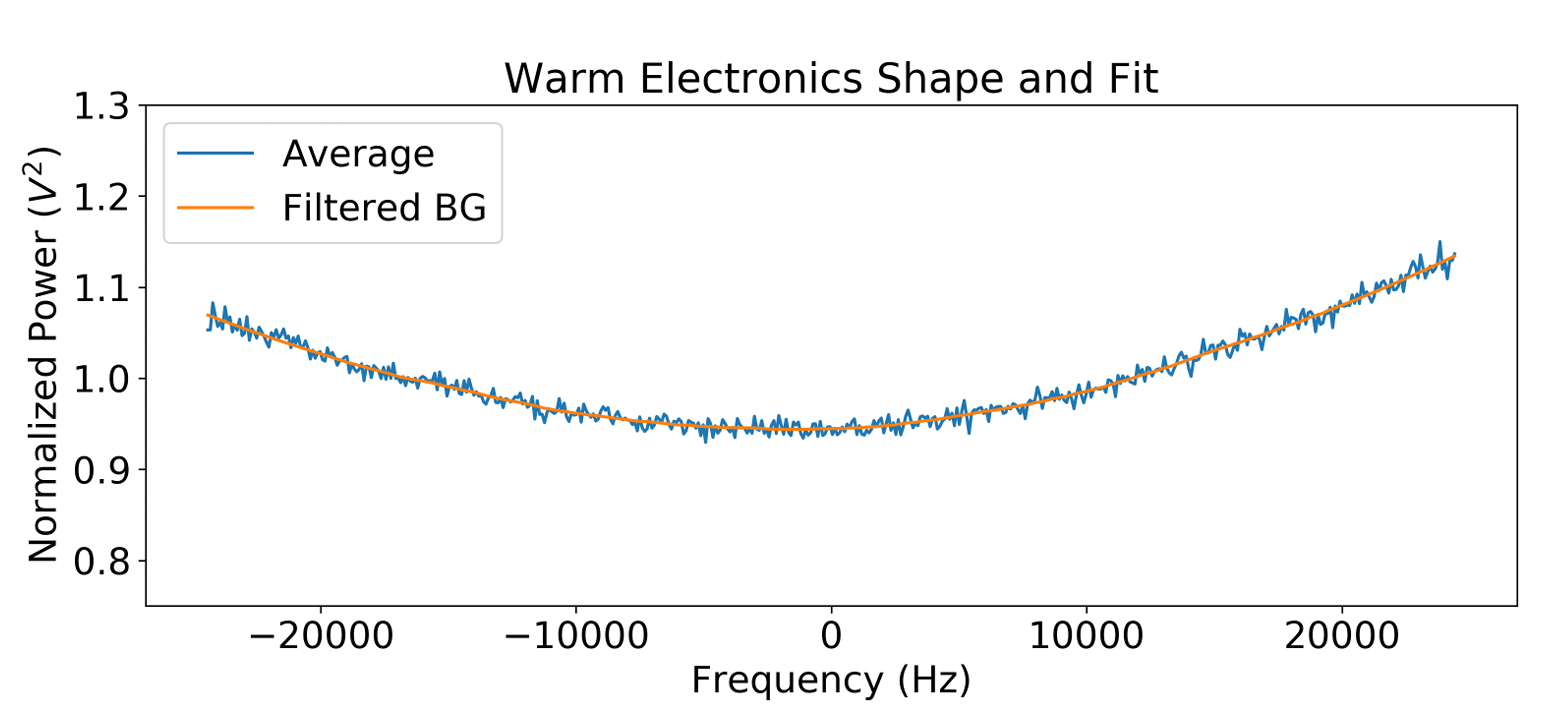}
    \caption{The filtered background shape ("Filtered BG") and average base-line fitted shape from the warm electronics during run 1B \cite{Run1BAnalysis}.}
    \label{fig:WarmElectronicsBG}
    \end{figure*}

 The first step in processing the raw spectra is removing the fixed background shape due to the warm electronics from all the raw spectra. This is primarily due to frequency dependent gain variations after mixing. The shape of these variations is somewhat predictable by the filters in the receiver chain. Subdominant contributions are gain variations before mixing that originate in the cold space, as well as frequency dependent noise variations. The background shape is pictured in Figure \ref{fig:WarmElectronicsBG}, alongside its fit which was averaged and smoothed using a Savitzsky-Golay filter \cite{SavitskyGolay1964,MalagonThesis}. This baseline filtered shape is then divided from all the raw spectra.
    \par After the baseline shape is removed, there is still more filtering and normalization to be done on each raw spectra. A six-order Pad\'e filter was convoluted on each spectra to remove the residual shape from the cryogenic receiver transfer function. This is motivated by a derivation of the power spectrum shape at the output of the last-stage cold amplifier \cite{DawThesis}. The power in each bin is then normalized by dividing by the mean power of the whole spectrum. This means in the absence of an axion signal, one would expect a Gaussian distribution around a mean of one, which is the case if one creates a histogram of the data. Since we are looking for power excesses above one, one is then subtracted from all the power bins, so it is more intuitively looking for excesses above zero. This filtered result is pictured in Figure \ref{fig:FilterSpectrum}. 
    \par Another feature that must be accounted for in this raw data is the Lorentzian shape of the cavity resonance. If an axion signal is towards the edges of the spectrum it will not be enhanced fully by the Q of the cavity, whereas a signal that is directly centered will be. This Lorentzian shape is determined by the cavity resonant frequency and loaded quality factor at the time of the scan. Each bin is then further normalized based on its distance from the cavity resonant frequency, and where it would lay on this shape. Evident after this processing is the increase in the excess on the edges of a Lorentzian weighted spectrum, as pictured in Figure \ref{fig:LorentzSpectrum}.
    \subsection{Analysis Cuts}
    After the spectra are processed, there are a series of cuts to the data to be made for quality control. The vast majority of the cuts made are referred to as "timestamp" cuts. These are simply scans taken during known, aberrant run conditions. This could result from manually adjusting the dilution fridge, manual biasing of the JPA, digitizer failures, software malfunctions, and various other engineering studies that could compromise the data quality. During run 1B, these made up about $57 \%$ of all the analysis cuts made. The 2nd most common analysis cut, about $36 \%$ of cuts in run 1B, are scans that have an associated system noise temperature that is either less than $0.1 K$, which is below the quantum limit and nonphysical, or greater than $2\, K$; these both likely result from incorrectly measuring the SNRI and therefore not trusted to be included in the data set. The remaining $6\%$ of cuts are similarly on associated data values being badly measured; quality factors less than 10,000 or greater than 120,000 are omitted because they are likely from a bad transmission fit measurement. The six-order Pad\'e fit had to have a $ \chi ^2$ per degree of freedom less than 2 to be included, as well as a few additional cuts on the warm receiver filter shape. All in all, in run 1B, these cuts reduced the total run scans from 197,680 spectra to 185,188, or about $7\%$ of the data set. 
    \subsection{Grand Spectrum Construction}
    \label{GrandSpectrum}
    The final step in the analysis process to finding a candidate axion is to construct a 'grand spectrum' that combines all of the processed, normalized raw spectra together, usually for just a single nibble. To do this, we must account for the varying potential axion signal height as well as noise floor across spectra, based on their associated run conditions at the time of digitization; scaling the spectra based on their measured $Q$,$f$,$C_{010}$,and $T_{sys}$. Each filtered $P_{i}^j$ is converted to a $P_{i_{scaled}}^j$ based on this. To further increase sensitivity, a filtering process based on the axion line shape is also performed, by convolving each bin power with a filter associated with either the Maxwell-Boltzmann line shape (Equation \ref{eqn:MBlinshape}) or N-body line shape (Equation \ref{eqn:N-bodylinshape}) discussed in Chapter \ref{chap:axiontheory}. 
    \par The combining process uses a well-established 'optimal weighting procedure' outlined in Ref. \cite{ADMX2001,HaystacAnalysis}. This process finds weights for the individual power excesses that results in the optimal SNR for the grand spectrum. The weights are chosen such that the maximum likelihood estimation of the true mean value, $\mu$, is the same for all contributing bins; this can be outline mathematically on a bin-by-bin basis using the equation:
    \begin{equation}
    P_{w}=\frac{\sum_{j=0}^{N} \frac{P_{scaled}^j}{\sigma^{j^2}}}{\sum_{j=0}^{N} \frac{1}{\sigma^{j^2}}}
    \label{eqn:WeightedPower}
    \end{equation}
    where $N$ is the total number of spectra for a given frequency bin, and $P_w$ is the weighted power for an individual RF bin in the grand spectrum. The standard deviation for each bin in the grand spectrum is then calculated with
    \begin{equation}
    \sigma_{w}=\sqrt{\frac{1}{\sum_{j=0}^{N} \frac{1}{\sigma^{j^2}}}}
    \label{eqn:WeightedSTDEV}
    \end{equation}
    The grand spectrum is then completely defined by power excess $P_w$ and its standard deviation, $\sigma_w$. It is expected that this grand spectrum will be relatively level and flat with a Gaussian distribution if the background has been properly subtracted. From this grand spectrum, sharp power excesses above the noise can by flagged as axion candidates; more precisely, when the weighted power at that frequency is in excess of $3\sigma$. A re-scan would also be triggered if a candidate's power exceeded that of a DFSZ axion by $0.5\sigma$. Regions in the grand spectrum that didn't meet the target SNR or target DFSZ sensitivity are also flagged to be re-scanned. Once all candidates are deemed not to be axion-like, and the desired sensitivity is met over the grand spectrum, an exclusion limit can be set.
    \subsection{Setting Limits}
\begin{figure*}[htb!]
    \centering
    \includegraphics[angle=0, width=1\linewidth]{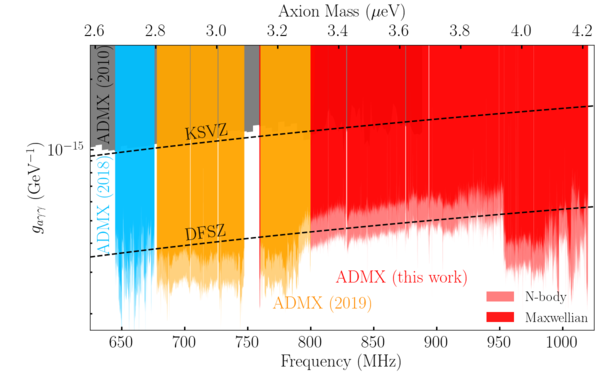}
    \caption{The ADMX Gen 2 exclusion limit plot from October 2021. The light blue is from the run 1A 2018 paper, Orange is run 1B concluded in 2019, and in red is run 1C part 1 that concluded in 2021. Run 1C continued into 2022, to exclude the full range to DFSZ level, but these results have not been reported at the time of writing yet.}
    \label{fig:2021ADMXlimits}
    \end{figure*}

    Once all data-taking is completed, a final limit is set on $g_{a\gamma\gamma}$ over the frequency range that was scanned. A frequency bin that contained an axion signal, if scanned multiple times, would have a Gaussian distribution about a mean, $\mu=g_{\gamma}^2\eta$, where $\eta$ is the SNR for the given measurement. Whereas, a frequency bin not containing a signal would be Gaussian about a mean $\mu=0$. A limit is set by computing $\mu$ for each bin that gave a 90\% confidence limit that the measurement did not contain an axion. To deal with negative power values, a cumulative distribution function for a truncated normal distribution is used when determining the 90\% confidence $\mu$ value \cite{PhysRevD.57.3873}. For each bin, a normal distribution is generated using the weighted power as the mean and the uncertainty as the standard deviation. The generated distribution is then randomly sampled 100 times, clipping to zero if a negative value is generated; this list of generated $\mu$ values is then sorted, and the 90\% confidence limit is value closest to 90\% to the top of the sorted list. This results in a very jagged limit plot, with each bin value being around 100 Hz wide, thus some combining and smoothing is also done; A small number of bins, around 200 representing a single plot pixel, were combined into a single limit value. The most recent limit plot for the main ADMX experiment is shown in Figure \ref{fig:2021ADMXlimits}. These limits are also shown, albeit zoomed out, in Figure \ref{fig:ADMXlimitstheory}.
    

\chapter{Cavity Resonators}
\label{chap:Cavities}
This chapter will cover the physics and techniques of resonant cavities. This will not only be in the context of axion haloscopes, but a general characterization of cavities with a variety of geometries. This will start with a formal field formalization for cavity resonant modes, and how to label their geometries. This will then transition to defining quality factor and the techniques surrounding Q measurements. From there, a brief overview of antenna coupling. The final sections will cover considerations for multi-cavity systems, and give an overview of the ADMX multi-cavity systems in development: Run 2 and ADMX Extended Frequency Range (EFR). 
\section{Formal introduction to cavities and modes}
 A cavity is simply an empty space within a solid object; make this solid object a conductive material, and electromagnetic energy within the empty space will be trapped within as the conductor shields any external radiation and constrains the wave structure of the electromagnetic fluctuations within the cavity. The goal of this section is to solve for the structure of the preferred oscillations within a given cavity structure, which will be called the resonant modes of the cavity.
    \subsection{Solving Maxwell's equations}
    For this section I will assume a cylindrical geometry and coordinate system for most of the proofs, but this analysis can be done for a rectangular or spherical, or general cavity structure. This analysis stems mostly from Ref. \cite{jackson_classical_1999}. An easy starting point is to consider a cylindrical cavity as a cylindrical wave guide with plane end faces imposed on either side, separated by some finite distance. We will further impose that these "end-caps" will be flat and perpendicular to the cylindrical cross-section of the wave guide. Furthermore, this cavity is made entirely of a perfect electric conductor (PEC) material.
    \subsubsection{Cylindrical Wave-guide}
    \par Before assessing the boundary conditions of the end caps, let us write down Maxwell's equations for an arbitrary cylindrical wave-guide. I can first make an assumption that the propagating waves have a sinusoidal time-dependence $e^{-i\omega t}$ for fields inside the cylinder. Furthermore, I can see for a infinite cylindrical wave-guide, there is no restriction along its z-direction yet, thus it will have a spatially sinusoidal variation, $e^{\pm ikz}$, with $k$ as a free, unknown, possibly complex parameter. I can write the fields down as such:
    \begin{equation}
        \begin{array}{c}
            \vec{E}(x,y,z,t)  \\
            \vec{B}(x,y,z,t) 
        \end{array}\Biggr\}
        = \Biggl\{ \begin{array}{c}
            \vec{E}(x,y) e^{\pm ikz-i \omega t}  \\
            \vec{B}(x,y) e^{\pm ikz-i \omega t}
        \end{array}
    \label{eqn:wavefields1}
    \end{equation}
    This wave equation under these conditions reduces to the two-dimensional form:
    \begin{equation}
        [\nabla_t^2+(\mu\epsilon\omega^2-k^2)] \Biggl\{\begin{array}{c}
            \vec{E} \\
            \vec{B}
        \end{array}\Biggr\} =0
    \label{eqn:wavequation1}
    \end{equation}
    where $\mu$ and $\epsilon$ are the permeability and permittivity of the interior, non-conducting material (presumably vacuum or some dielectric material). $\nabla_t^2$ is the transverse part of the Laplacian operator $\nabla^2-\partial_z^2$. It is thus convenient to separate the fields into transverse and parallel components to the z-axis, such that $\vec{E}=E_z\hat{z}+\vec{E_t}$, where $\vec{E_t}=(\hat{z}\times \vec{E})\times\hat{z}$. We can then write out Maxwell's equations as such:
    \begin{equation}
    \frac{\partial\vec{E_t}}{\partial z}+i\omega\hat{z}\times\vec{B_t}=\nabla_t E_z \; , \; \hat{z}\cdot(\nabla_t \times \vec{E_t})=i \omega B_z
    \label{eqn:maxwell1}
    \end{equation}
    \begin{equation}
    \frac{\partial\vec{B_t}}{\partial z}+i\mu \epsilon \omega\hat{z}\times\vec{E_t}=\nabla_t B_z \; , \; \hat{z}\cdot(\nabla_t \times \vec{B_t})=i \mu \epsilon \omega E_z
    \label{eqn:maxwell2}
    \end{equation}
    \begin{equation}
    \nabla_t \cdot \vec{E_t}=-\frac{\partial E_z}{\partial z} \; , \; \\ \nabla_t \cdot \vec{B_t}=-\frac{\partial B_z}{\partial z}
    \label{eqn:maxwell3}
    \end{equation}
    Evident from the first two sets of Equations, \ref{eqn:maxwell1} and \ref{eqn:maxwell2}, is that if the z components are known, the transverse components are completely determined, with only the sign of $k$, the direction of the propagation, changing the fields' direction. These transverse fields are:
    \begin{equation}
        \vec{E_t}=\frac{i}{(\mu \epsilon \omega^2 - k^2)}[\pm k \nabla_t E_z-\omega \hat{z}\times \nabla_t B_z]
    \label{E_twaveguide}
    \end{equation}
    \begin{equation}
        \vec{B_t}=\frac{i}{(\mu \epsilon \omega^2 - k^2)}[\pm k \nabla_t B_z-\mu \epsilon \omega \hat{z}\times \nabla_t E_z]
    \label{B_twaveguide}
    \end{equation}
    This leaves us with several options for the types of waves that can propagate in this arbitrary cross-section wave-guide. The simplest case would be where both $E_z$ and $B_z$ are zero, therefore completely transverse to the propagation direction; these are called transverse electromagnetic (TEM) waves. In this case, $\vec{E_t}=\vec{E}_{TEM}$, simplifies using equations \ref{eqn:maxwell1} and \ref{eqn:maxwell3} to:
    \begin{equation}
        \nabla_t\times \vec{E}_{TEM}=0 \; ,\; \nabla_t \cdot \vec{E}_{TEM} = 0
    \label{eqn:TEM1}
    \end{equation}
    This means that $\vec{E}_{TEM}$ is a solution to an electrostatic problem in two dimensions. This imposes several restrictions on this mode. First, for Equation \ref{eqn:wavequation1} to be satisfied, $k=\sqrt{\mu\epsilon} \omega$, thus the wave number or wavelength is determined by the properties of the propagation material for an infinite variety of frequencies. Additionally, Equation \ref{eqn:maxwell2} sets $\vec{B}_{TEM}=\pm \sqrt{\mu \epsilon}\hat{z}\times \vec{E}_{TEM}$, which is just the same connection for plane waves in open infinite space. Most importantly, this means that the mode cannot exist on a single cylindrical conducting surface because that conducting surface must be an equipotential. If there are two or more separated conducting surfaces, then this TEM oscillation can propagate electrostatically by one surface supporting a positive equipotential and another operating as the negative or ground potential. Therefore, this mode is only dominant for familiar coaxial cables, or parallel-wire transmission line geometries. Cavities that contain tuning rods will also have TEM modes. Important to note, is it has no cutoff frequency in a wave-guide, and $k$ is real for all $\omega$, which won't be true for hollow cylinders of 1 surface.
    \par The next more generalized classification of modes would be if only one of the field components was transverse, meaning either $E_z=0$, Transverse Electric (TE) waves, or $B_z=0$, Transverse Magnetic (TM) waves. In this case, we still have to satisfy some surface boundary conditions at the surface of the conductor to ensure the fields vanish inside. If $B_z=0$, $E_z$ will still have to vanish at the surface of the conductor, so $E_z\vert_S=0$. If $E_z=0$, we must ensure the component of $\vec{B}$ parallel to the surface must vanish, requiring $\frac{\partial B_z}{\partial n}\vert_S=0$ where $\hat{n}$ is the normal unit vector to the cylindrical surface. One notices that when pairing these two boundary conditions with \ref{eqn:wavequation1}, it forms two distinct eigenvalue problems. For a given value of $k$, only certain $\omega$ frequencies can be supported, and vice-versa. Since TE and TM have distinct boundary conditions, their eigenvalues are not related in general, and form a natural way of dividing the type of fields propagating in the wave-guide. With these 3 categories, TEM, TE, TM waves, one can construct any arbitrary electromagnetic field disturbance within the wave-guide; they form the basis of waves within the wave-guide or cavity.
    \par In the case of TE and TM modes, we can see that the eigenvalue problem formed implies cutoff frequencies for the various waves. We can write this eigenvalue problem in terms of a scalar potential function, $\psi$, such that for TM waves:
    \begin{equation}
        \vec{E_t}=\pm \frac{ik}{\gamma^2}\nabla_t\psi
        \label{eqn:E_tPsiTM}
    \end{equation}
    or for TE waves:
    \begin{equation}
        \vec{H_t}=\pm \frac{ik}{\gamma^2}\nabla_t\psi
        \label{eqn:H_tPsiTE}
    \end{equation}
    where $\gamma^2=\mu\epsilon\omega^2-k^2$. The wave equation, Equation \ref{eqn:wavequation1}, then simplifies to the eigenequation:
    \begin{equation}
        \nabla_t^2\psi=-\gamma^2\psi
        \label{eqn:eigenpsi}
    \end{equation}
    with the boundary conditions for TM or TE modes:
    \begin{equation}
        \psi\vert_S=0 \; \rm{or} \; \frac{\partial \psi}{\partial n}\vert_S=0
        \label{eqn:eigenconditionsTMTE}
    \end{equation}
    Thus we expect a spectrum of eigenfunction solutions, $\psi_{\lambda}$, with corresponding eigenvalues $\gamma_{\lambda}^2$, where $\lambda=1,2,3,...$, which form an orthogonal set of modes (note that $\gamma^2$ must be non-negative to satisfy Equation \ref{eqn:eigenconditionsTMTE}). This means that there will be $k_{\lambda}$ that is only a real value above a cutoff frequency:
    \begin{equation}
        \omega_{\lambda}=\frac{\gamma_{\lambda}}{\sqrt{\mu \epsilon}}
        \label{eqn:cutoffomega}
    \end{equation}
    The wave number can then be written in terms of frequency:
    \begin{equation}
        k_{\lambda}=\sqrt{\mu \epsilon}\sqrt{\omega^2-\omega_{\lambda}^2}
        \label{eqn:klambda}
    \end{equation}
    This is all to show that below the cutoff frequency, certain frequencies cannot propagate within the wave-guide. The common practice is to set the dimensions of the wave-guide accordingly such that only the lowest mode can occur at the desired operating frequency.
    \subsubsection{Cylindrical Resonant Cavity}
    When we add conductive end-caps to either end of the infinite cylindrical wave-guide from the last section, a boundary condition is imposed in the z-direction such that only a standing wave structure, $A\sin kz+B\cos kz$, with value zero at either end cap, can propagate. For a cavity of length $L$, this means $k=n \pi/L$ for $n=0,1,2,...$ and the resultant fields can be written in terms of $\psi$. For 
    TM Modes these fields are:
    \begin{equation}
    \begin{aligned}
    &\vec{E_z}=\psi(x,y)\cos\left(\frac{n\pi z}{L}\right) \\
    & \vec{E_t}=-\frac{n \pi}{L \gamma^2}\sin\left(\frac{n\pi z}{L}\right)\nabla_t\psi(x,y) \\
    & \vec{H_t}=-\frac{i\epsilon \omega}{\gamma^2}\cos\left(\frac{n\pi z}{L}\right)\hat{z}\times\nabla_t\psi(x,y)
    \end{aligned}
    \label{eqn:TMFields}
    \end{equation}
    and for TE Modes the fields are:
    \begin{equation}
    \begin{aligned}
        &\vec{B_z}=\psi(x,y)\sin\left(\frac{n\pi z}{L}\right) \\
        &\vec{E_t}=-\frac{i\omega \mu}{\gamma^2}\sin\left(\frac{n\pi z}{L}\right)\hat{z}\times\nabla_t\psi(x,y) \\
        &\vec{H_t}=-\frac{n \pi}{L \gamma^2}\cos\left(\frac{n\pi z}{L}\right)\nabla_t\psi(x,y)
    \end{aligned}
    \label{eqn:TEFields}
    \end{equation}
    where $\gamma^2$ is adapted to:
    \begin{equation}
        \gamma^2=\mu\epsilon\omega^2-\left(\frac{n \pi}{L}\right)
        \label{eqn:cylindricalcav_gamma}
    \end{equation}
    and the cutoff frequency is now further adapted by $n$:
    \begin{equation}
        \omega_{\lambda n}^2=\frac{1}{\\mu \epsilon}\left[\gamma_{\lambda}^2+\left(\frac{n \pi}{L}\right)^2\right]
        \label{eqn:cylindercav_cutoff}
    \end{equation}
    Again, a cavity designer would try to adjust $L$ and $\gamma_{\lambda}$ such that the intended operating mode frequency was well-separated from other modes, to improve stability. 
    \par The important special case for ADMX, and all the test cavities used at LLNL, is the right cylindrical cavity. So far the analysis has been agnostic to this geometry to maintain a level of generality, but the only step left is the solve for $\psi(x,y)$ for this circular cross-section. It is best to work in cylindrical coordinates $\rho$ and $\phi$ in this case. If the cavity has a radius, $R$, then $E_z(\rho=R)=0$ for a TM mode, and $\psi$ has solution:
    \begin{equation}
    \psi(\rho,\,\phi)=E_0J_l(\gamma_{lm}\rho)e^{\pm i l \phi}
    \label{eqn:CavPsi}
    \end{equation}
    where:
    \begin{equation}
    \gamma_{lm}=\frac{x_{lm}}{R}
    \label{eqn:gammamnTM}
    \end{equation}
    $x_{lm}$ is the mth root of the equation, $J_l(x)=0$, where $J_l(x)$ is the lth Bessel function. $E_0$ is a normalization constant. The expression for $\psi$ works just as well for TE modes, however $\gamma_{lm}=x'_{lm}/R$ where $x'_{lm}$ is the mth zero root of the derivative Bessel function $J'_l(x)$. In this way, we now can denote cylindrical modes by 3 indices, $lmn$, with $l$ representing the number of half wavelengths in the $\phi$ direction, $m$ for the radial direction, and $n$ for the axial direction. The frequencies of this empty cylindrical cavities' modes will then be:
    \begin{equation}
    f_{lmn}^{TM(TE)}=\frac{1}{2\pi\sqrt{\mu \epsilon}}\sqrt{\frac{x_{lm}^{(')2}}{R^2}+\frac{n^2\pi^2}{L^2}}
    \label{eqn:CylinCavFreq}
    \end{equation}
    The lowest frequency mode then for TM modes is the $TM_{010}$ because $x_{01}\approx 2.405$, is the lowest value root, and less than $\pi$ for the $n=1$ case. This is referred to as the fundamental TM mode for the cavity, and, remembering Chapter \ref{chap:Haloscopes}, is the main search mode for axion searches because of its axion form factor. In the case of TE modes, notice that $n\neq0$ for this solution to work, and the lowest root, $x_{lm}'$, is $x_{11}'\approx1.841$, thus the fundamental TE mode is $TE_{111}$ with a frequency:
    \begin{equation}
        f_{111}^{TE}=\frac{1.841}{2\pi\sqrt{\mu \epsilon}R}\left(1+2.912\frac{R^2}{L^2}\right)^{1/2}
        \label{eqn:TE111freq}
    \end{equation}
    This means that the ratio of cavity length to cavity radius determines whether the $TM_{010}$ or $TE_{111}$ is the lowest mode frequency in the cavity, and taking the ratio of the two frequency expressions, one finds that if $L\gtrsim 2.03R$, then $f_{111}^{TE} < f_{010}^{TM}$. This is an important note because we will want long length cavities that maximize our detection volume for a given $R$ that sets the base $f_{010}^{TM}$ search frequency, thus there may be many TE modes at lower frequencies before we come across the first TM modes. One doesn't want to increase this ratio too much, however, because it will increase the number of $TE$ mode crossings for the $TM_{010}$ as one shifts the $TE$ spectrum lower. If one didn't have to maximize volume, then you could decrease the length of the cavity such that they are pushed far above the TM spectrum. I'll end this section with Table \ref{tab:EmptyCavities} that lists the base height and radius of the various ADMX cavities as well as their several important mode frequencies when they are empty. These frequencies are significantly lower than their loaded tuning rod configurations.
    \begin{table}
    \centering
    \renewcommand{\arraystretch}{1}
    \begin{tabular}{@{}l c c c c c r@{}}
    Cavity Name & Radius& Height (cm)& $F_{010}^{TM}$& $F_{111}^{TE}$& $F_{011}^{TM}$& $F_{020}^{TM}\;({\rm GHz})$ \\
    \hline
    ADMX Main & 21.0 & 101 & 0.547 & 0.444 & 0.567 & 1.25  \\
    \hline
    ADMX Sidecar & 4.60 & 12.1 & 2.50 & 2.28 & 2.79 & 5.73 \\
    \hline
    ADMX run 2 & 8.5 & 100 & 1.35 & 1.04 & 1.36 & 3.1 \\ 
    \hline
    ADMX EFR & 6.4 & 100 & 1.79 & 1.38 & 1.8 & 4.12 \\ 
    \end{tabular}
    \caption{A summary of the empty cavity characteristics and base frequencies for the various ADMX resonator systems. In the case of the ADMX run 2 and EFR arrays, these are the measurements of a single empty cavity within the multi-cavity system. All lengths and frequencies are listed in units of centimeters and gigahertz.}
    \label{tab:EmptyCavities}
    \end{table}
    
    \subsection{Identifying Modes}
\begin{figure*}[htb!]
    \centering
    \includegraphics[angle=0, width=.8\linewidth]{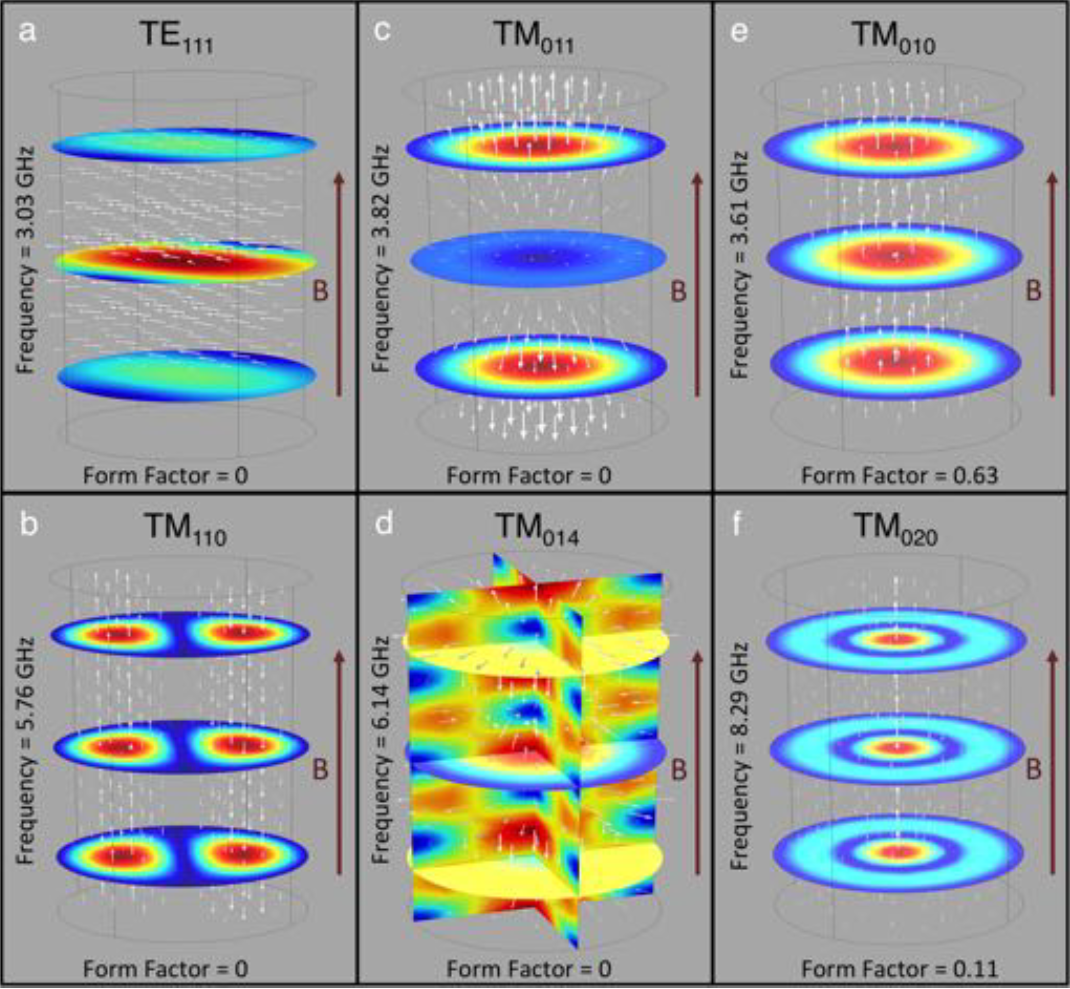}
    \caption{The electric field for different TM and TE modes in an empty Sidecar cavity \cite{Boutan}. The cross-sectional color plots show the magnitude of $\vec{E}$, while the vector plots represent $\vec{E}$, both for a single phase in the oscillation. In this way, red indicate extrema points in the electric field that can be used to label the mode.}
    \label{fig:EmptyCavModes}
    \end{figure*}

    Now that we've formally derived the fields and frequencies of an empty cavity analytically, this can inform us on how one can identify resonant cavity modes by field structure and form factors. This is typically done in RF simulation software, such as Ansys HFSS, CST microwave studio, or COMSOL, using an eigenmode solver package. A cavity geometry can be drawn or imported from a CAD file, and the resonant mode fields can be numerically solved for rather than analytically. One then typically plots several cross-sections of the field magnitude at every point in the interior cavity for a given mode, with $\vec{E}^2$ usually being the norm. This is done in Figure \ref{fig:EmptyCavModes} for several important modes.
    \par The best way visually to assess a mode from its field structure is to count the number of extrema, representing the nodes of half wave-lengths, in each of the cylindrical coordinate directions, $\phi$, $\rho$, and $z$, which would directing correspond to $l$, $m$, and $n$ respectively. Although this is easy for a human that can read plots and identify extrema by eye, its not the best for a computer. In this case, using a form factor that picks out the particular mode structure is a better route. In the case of the $TM_{010}$ mode, the axion form factor (Equation \ref{eqn:CavityFormFactor}) works perfectly because it is maximized for this mode and near-zero for most others. In fact, because TE modes have $E_z=0$, it will be exactly zero for a perfect solenoid magnet ($\vec{B}=B_0\hat{z}$). TM modes such as the $TM_{011}$ or $TM_{110}$ may have $E_z\neq0$ components, but they're opposing extrema will cancel each other once the integral over the volume is done. Thus, one can quickly see that the axion form factor is only non-zero for modes $TM_{0n0}$ with increasing $n$ corresponding to decreasing $C_{0n0}$. One can easily adapt this form factor to pick out all of the TM modes by squaring the numerator within the integral, such that:
        \begin{equation}
        C_{TM}=\frac{\int_{V}E_z^2dV}{\int_{V}\vec{E}^2dV}
        \label{eqn:TmModeFormFactor}
        \end{equation}
    This factor will be non-zero for all TM modes, and typically will decrease with increasing number of indices, $lmn$, due to more zero magnitude points in the field structure, but that is a trend rather than a hard rule. More importantly, because TE modes by definition have $E_z=0$, this form factor will also be zero. In this way, the above form factor is great for separating the TM from TE and TEM mode spectra. Similarly, you could define a TE/TEM form factor based off the tangential field:
        \begin{equation}
        C_{TE/TEM}=\frac{\int_{V}\vec{E_t}^2dV}{\int_{V}\vec{E}^2dV}
        \label{eqn:E_tFormFactor}
        \end{equation}
    However, this will simply switch for which modes the form factor is zero vs. non-zero. To distinguish TEM modes, one needs to ensure both $E_z$ and $B_z$ are zero for the mode, hence one could look at the parallel mode magnetic field form factor:
        \begin{equation}
        C_{TE}=\frac{\int_{V}B_z^2dV}{\int_{V}\vec{B}^2dV}
        \label{eqn:TEModeFormFactor}
        \end{equation}
    This form factor would be zero for TM and TEM modes, and would only be non-zero for TE modes, hence it can be used as a follow up after separating out TM modes via Equation \ref{eqn:TmModeFormFactor}. Be careful not to confuse this mode magnetic field with the external magnetic field used in the original axion form factor Equation \ref{eqn:CavityFormFactor}.
    \par Although the above derivation was for an empty, right cylindrical cavity, much of the analysis holds true for a cavity with tuning rods, other internal structures, or different geometries. The three mode indices will still be reflective of the number of half wavelengths in the base axes of the geometry. Introducing a tuning rod, there will still be $TM_{lmn}$-like and $TE_{lmn}$-like modes that are perturbations of the empty cavity fields amongst other modes that are possibly unique to the tuning rod geometry. In these cases, we are using simulation software to numerically solve and approximate the field solutions, so $E_z=0$ might not be exactly true for what is obviously a TE mode by eye, but using the TM form factor Equation \ref{eqn:TmModeFormFactor} as a follow-up check should produce a near-zero result most of the time.
\section{Quality factor}
In this section, I will cover the many definitions and ways of measuring cavity quality factor. Several definitions have already been shown in Chapter \ref{chap:Haloscopes}, such as Equations \ref{eqn:CavityQualityFactor1}, \ref{eqn:QualityfactorCircuitDef}, and \ref{eqn:QualityfactorDeltaF}, but hopefully this section will illuminate the connection between these equations and add a few more. This will start with some cavity quality factor basics for a classical copper cavity, then a derivation of the Lorentzian shape a resonant mode makes when taking a power spectrum of the cavity, which will highlight the connection between \ref{eqn:CavityQualityFactor1} as the physical definition of Q and \ref{eqn:QualityfactorDeltaF} as how we measure it. Since actually measuring Q in lab will require an understanding of antennae and coupling, I reserve a later section for this process.
    \subsection{Quality factor Basics}
    The physical definition of quality factor is Equation \ref{eqn:CavityQualityFactor1} in terms of $U$, $P_{d}$, and $\omega_0$. By conservation of energy, the time rate of change in $U$ should be equal to the negative of $P_d$, which allows us to solve for $U(t)$:
        \begin{equation}
        \frac{dU}{dt}=-\frac{\omega_0}{Q}U
        \label{eqn:cavdUdt}
        \end{equation}
    which easily solves for:
        \begin{equation}
        U(t)=U_0e^{-\omega_0t/Q}
        \label{eqn:cavU(t)}
        \end{equation}
    In this way, one can see that an initial $U_0$ just decays away with a constant inversely proportional to Q. We can write out Q using the field integral expressions for $U$ and $P_d$ as shown:
        \begin{equation}
        Q=\frac{\omega_0}{R} \frac{\mu\int\vert\vec{H}|^2dV}{\int\vert\vec{H}|^2dS}
        \label{eqn:fieldQ}
        \end{equation}
    we could have written $U$ in terms of the time-average electric field as well. This equation is often grouped in terms of a geometric factor that absorbs all the cavity mode geometric information, and separates it from the material properties of the cavity, $R$. Therefore, cavity geometric factor is written:
        \begin{equation}
        G=\frac{\mu\omega_0\int\vert\vec{H}|^2dV}{\int\vert\vec{H}|^2dS}
        \label{eqn:Gfactor}
        \end{equation}
    Where $V$ is over the total cavity interior volume, and $S$ is over the total interior surface. Quality factor in this case is simply:
            \begin{equation}
            Q=\frac{G}{R}
            \label{eqn:QGR}
        \end{equation}
    One can then write the resistance in terms of a classical surface conductivity:
            \begin{equation}
            R=\sqrt{\frac{\mu_c \pi}{\sigma}}
            \label{eqn:Rclassic}
        \end{equation}
    Where $\mu_c$ is the permeability of the cavity material, which is essentially $\mu_0$ for copper. Often the skin depth is the preferred parameter for thinking about the finite conductivity of normal metals like copper. Effectively its the length parameter measuring the penetration of the cavity field into the conductor before its fully shielded by internal currents; a perfect conductor has no field penetration and therefore zero skin depth. Skin depth of a normal conductor like copper is written:
            \begin{equation}
            \delta_{norm}=\sqrt{\frac{2}{\mu_c\omega_0 \sigma}}
            \label{eqn:dclassic}
        \end{equation}
    If we pullout a factor of $V/S$ and $\mu$ from the geometric factor expression \ref{eqn:Gfactor}, another common expression for Q in terms of skin depth can be derived:
            \begin{equation}
            Q=\frac{\mu}{\mu_c}\left(\frac{V}{S\delta}\right)\times({\rm Geometry\: Factor \approx 1})
            \label{eqn:Qclassic}
        \end{equation}  
    It is important to note that this geometry factor is not the same as the geometric factor in Equation \ref{eqn:Gfactor}; this expression is intended to illuminate the importance of volume to surface ratio and skin depth in the base cavity design, but its not great for serious calculations. Using the field results from the right cylinder, the geometric factor can be solved analytically, and the result expressed in terms of skin depth, $R$, $L$, and $lmn$ mode indices. This assumes no other external sources of power dissipation from antennae or other radiative losses.
    \begin{equation}
    Q_{lmn}=\frac{\lambda}{\delta}\frac{\sqrt{x_{lm}^2+(n\pi R/L)^2}}{2 \pi (1+2R/L)}
    \label{eqn:CavityQlmn}
    \end{equation}
    where $\lambda$ is the free space wavelength of the cavity mode.
    \par It is important to note that although the Equation \ref{eqn:dclassic} is valid for room temperature, at cryogenic temperatures the skin depth will start to shrink and conductivity will increase. When the mean free path of electrons in the metal exceeds the skin depth, surface electron currents will be more effected by the surface finish of the cavity surface itself, enough that Equation \ref{eqn:dclassic} is no longer valid \cite{Boutan,HotzThesis}. The 'anomalous' skin depth of this new regime can be written \cite{Hagmann}:
    \begin{equation}
    \delta_{anom}=\left(\frac{\sqrt{3} c^2 m_e v_F}{8 \pi^2 \omega_0 n e^2}\right)^{1/3}
    \label{eqn:danom}
    \end{equation}
    where $m_e$ is the electron mass, $e$ is the electron charge, $v_F$ is the fermi velocity of the material, $n$ is the conduction electron density. One can estimate most of these parameters for copper to write the anomalous skin depth as a function of frequency; using $v_F= 1.57 \times 10^8\,{\rm cm/s}$, $n=8.50 \times 10^{22}\,{\rm cm^{-3}}$, and the typical electron mass and charge. This expression simplifies to: 
    \begin{equation}
    \delta_{anom,Cu}=2.84\times 10^{-4}\,{\rm m} \left(\frac{1}{f}\right)^{1/3}
    \label{eqn:danomCU}
    \end{equation}
    where frequency is in ${\rm Hz}$. Often simulation software only allows for the adjustment of conductivity, assuming the classical regime for skin depth, therefore its often useful to represent this anomalous skin depth as an effective classical conductivity. This also allows us to use Equation \ref{eqn:Rclassic} to assign a conductivity to a given $R$, and subsequently a given $Q$. The effective anomalous conductivity would be:
    \begin{equation}
    \sigma_{eff,anom,Cu}=\left(\frac{2}{\mu_c 2 \pi f}\right) \left(\frac{1}{2.84\times 10^{-4}\,{\rm m} \left(\frac{1}{f}\right)^{1/3}}\right)^2= 3.14\times 10^{12}\,{\rm S/ m}\cdot f^{-1/3}
    \label{eqn:sigmaanomCU}
    \end{equation}
    This means the conductivity is from $\sigma\approx 3.95-2.41\times 10^9\,{\rm S/m}$ between $0.5-2.2\,{\rm GHz}$ at cryogenic temperatures, as opposed to a room temperature value of $\sigma=5.8\times10^7\,{\rm S/m}$. The important takeaway is copper and normal metals will start to degrade in quality factor if used for higher and higher frequency searches. Between $1-10\,{\rm GHz}$, the quality factor drops by $\approx 42\%$ for copper, slowing down the potential scan rate. By switching to a superconducting material with even marginal improvement in conductivity could avoid this degradation issue. 
    \subsection{Derivation of the Lorentzian in a RLC circuit}
    Let us return to the equivalent circuit picture. Figure \ref{fig:RLCCircuit}, that was used to calculate axion signal power in a cavity, and where we derived Equation \ref{eqn:QualityfactorCircuitDef}. We can define the impedance, $Z$, of an $RLC$ network at an arbitrary frequency, $\omega$, to be: 
    \begin{equation}
    Z= R+i \left( \omega L - \frac{1}{\omega C} \right)
    \label{eqn:RLCimpedance1}
    \end{equation}
    This means at the resonant frequency $\omega_0=1/ \sqrt{LC}$, the imaginary part vanishes. This impedance can be rewritten:
    \begin{equation}
    Z= R+iL \left(\frac{\omega^2 -\omega_0^2}{\omega}\right)
    \label{eqn:RLCimpedance2}
    \end{equation}
    We can define the de-tuning of the arbitrary frequency from the resonant frequency as $\Delta=\omega-\omega_0$. If we make the safe assumption that $\omega_0 >> \Delta$, the impedance then reduces further to: 
     \begin{equation}
    Z= R+2iL \Delta
    \label{eqn:RLCimpedance3}
    \end{equation}
    This assumption means we are looking within the vicinity of the resonant mode. The magnitude of this complex impedance will then be:
        \begin{equation}
    \vert Z \vert ^2= R^2+4\Delta^2 L^2
    \label{eqn:RLCimpedanceMag}
    \end{equation}
    Even though it will be complex off resonance, only the real part of this impedance will contribute to the power dissipated in this circuit, $V^2/R$. Playing with the complex numbers a bit, one can determine the $\cos{\theta}$ of $Z$ would be equal to $1/\sqrt{1+4 \Delta^2 (L/R)^2}$. The power dissipated can then be determined to be:
    \begin{equation}
    P(\Delta)= Re(\frac{V^2}{Z})= V_0^2 Re(\frac{1}{Z})=V_0^2 \frac{1}{\vert Z \vert} \cos{\theta}=\frac{V_0^2}{R}\frac{1}{1+4\Delta^2\left(\frac{L}{R}\right)^2}
    \label{eqn:RLCPower}
    \end{equation}
    This is the Lorentzian power spectrum we are looking for. Nonetheless, this expression can be cleaned up a bit by recognizing two things; $V_0^2/R$ is simply $P(f_0)$ the power dissipated on resonance. Secondly, the Full-Width-at-Half-Maximum (FWHM) can be recognized, $\Delta f_c=R/L$, within the Lorentzian. Using Equation \ref{eqn:QualityfactorDeltaF}, we can then re-write this expression in terms of Q and frequency:
    \begin{equation}
    P(f)=\frac{P(f_0)}{1+4Q^2\left(\frac{f-f_0}{f_0}\right)^2}
    \label{eqn:PowerLorentzian}
    \end{equation}
    Note that an assumption underlying this spectrum function is that the $RLC$ network is being driven by a voltage source, $V_0 e^{i\omega t}$; One needs an external signal source to excite the mode. By inputting a series of signals at varying $f$ about $f_0$, this peak-shape can be constructed. This will be how one ultimately measures Q in the lab; by analyzing the shape, specifically the width, of this peak.
\section{Geometric factors for multiple surfaces}
    \par Geometric Factors are critical for understanding how the different sources of power loss in the cavity breakdown amongst its design components. Equation \ref{eqn:Gfactor} describes the total surface geometric factor for a given mode. One of the novel techniques of this dissertation is extending this definition to sub-surfaces of the cavity. The motivation being for estimating and understanding hybrid, multi-material cavities' quality factor. A hybrid cavity will have different resistances for each material's surface as well as each surface having different overlap with the mode fields. This can be done by breaking up the surface integral of Equation \ref{eqn:Gfactor} into a sum of surfaces, $S=S_1+S_2+...+S_N$, we then define a sub-surface geometric factor as:
    \begin{equation}
    G_n=\frac{\mu\omega_0\int\vert\vec{H}|^2dV}{\int\vert\vec{H}|^2dS_n}
    \label{eqn:GsubFactor}
    \end{equation}
    One can then write the total geometric factor in terms of its sub-surface factors as:
    \begin{equation}
    \frac{1}{G_{total}}=\frac{1}{G_1}+\frac{1}{G_2}+...+\frac{1}{G_N}
    \label{eqn:GsubFactorSum}
    \end{equation}
    We can then write an associated sub-surface quality factor that takes into account the individual resistance of that surface:
    \begin{equation}
    Q_n=\frac{G_n}{R_n}
    \label{eqn:Qsub}
    \end{equation}
    Similarly, one combines these sub-surface quality factors in the same way as geometric factors:
    \begin{equation}
    \frac{1}{Q_{total}}=\frac{1}{Q_1}+\frac{1}{Q_2}+...+\frac{1}{Q_N}
    \label{eqn:QsubFactorSum}
    \end{equation}
    This represents a sum of resistance weighted by the geometric factors. It is not $R_{total}\neq R_1+R_2+...+R_N$, but some linear combination of them with weights $C_1,C_2,...,C_N$. If we take Equations \ref{eqn:GsubFactorSum} and \ref{eqn:QsubFactorSum} and input them into \ref{eqn:QGR}, one can solve for the weights to be:
    \begin{equation}
    C_n=\frac{\frac{1}{G_n}}{\sum_{i=1}^N \frac{1}{G_i}}
    \label{eqn:GfactorWeights}
    \end{equation}
    The total resistance for a given mode could then be written down in terms of these weights and individual resistances of the cavity surfaces:
    \begin{equation}
    R_{total}=C_1R_1+C_2R_2+...+C_N R_N
    \label{eqn:Rmode}
    \end{equation}
    In the simplest case, $S=S_1+S_2$, the weights are:
    \begin{equation}
    C_1=\frac{G_2}{G_1+G_2}\;,\;C_2=\frac{G_1}{G_1+G_2}
    \label{eqn:GfactorWeightsx2}
    \end{equation}
    This two dimensional case is important because it can be used to compare any specific feature on the cavity with the remainder of the total surface area in terms of resistance contribution. By using Equation \ref{eqn:GsubFactorSum} for different combinations of sub-surfaces, one can always reduce the problem to just two dimensions by combining surfaces back together. In fact, one can write pretty easily a python function that can combine geometric factors in iterative steps of two surfaces for a set of $N$ surfaces to result in any composite two surfaces. For instance I could combine the first half of the set together ($l=1+2+3+...+N/2$) and the second half ($k=N/2+(N/2+1)+...+N$):
    \begin{equation} 
    \begin{split} 
    \frac{1}{G_{total}}= 
    \left(\frac{1}{G_{1}}+\frac{1}{G_{2}}\right)+\left(\frac{1}{G_{3}}+\frac{1}{G_{4}}\right)+...+\left(\frac{1}{G_{N-1}}+\frac{1}{G_{N}}\right) \\
    \begin{aligned}
    =\left(\frac{1}{G_{1+2}}+\frac{1}{G_{3+4}}\right)+\left(\frac{1}{G_{5+6}}+\frac{1}{G_{7+8}}\right)+...+\left(\frac{1}{G_{(N-3)+(N-2)}}+\frac{1}{G_{(N-1)+N}}\right) \\
    ...\\
    =\left(\frac{1}{G_{1+...+(N/4)}}+\frac{1}{G_{(N/4+1)+...+(N/2)}}\right)+\left(\frac{1}{G_{(N/2+1)+...+(3N/4)}}+\frac{1}{G_{(3N/4+1)+...+N}}\right) \\
    =\frac{1}{G_{l}}+\frac{1}{G_k} \end{aligned} 
    \end{split}
    \label{eqn:combineGfactors}
    \end{equation}
    \par If starting with an odd $N$ you will instead combine the $N-2$ and $N-1$ factors and carry the $N$th factor into the next iteration, and it will eventually be combined. Since the combined expression for two G factors is a simple geometric expression, $\frac{G_1G_2}{G_1+G_2}$ this avoids using more complex solving methods. Note that I could have picked any $l$ or $k$ list of surfaces to combine, and just re-arranged terms. In fact, often $l$ will be a single surface index while $k=\sum_{j \neq l} n$, when we are comparing a single surface to compare to the rest of the cavity.
    \par These geometric factors are usually calculated using RF simulation software like Ansys HFSS or COMSOL, as mentioned earlier for identifying mode structure. The eigen-mode solver field data can be numerically integrated to calculate the field volume and surface integrals in the geometric factor as well as the frequency for that mode. One can define many sub-surfaces to calculate a $G$ value for within the model. The most typical features are each end cap and the walls/tube of both the cavity and tuning rod, totally to 6 surfaces. Often one will also define antenna port regions as well. These geometric factors are then usually exported as a data file to be manipulated with python scripts along with each mode frequency and mode identifying form factors.
\section{A simple hybrid cavity model}
\label{simplehybridcavity}
    For several reasons discussed in Chapter \ref{chap:SRF}, it is not necessarily ideal for some surfaces of a superconducting cavity to be made of superconductor when operating in high magnetic fields, and a hybrid cavity with those surfaces made of copper instead can be preferable. We can consider this 2-material hybrid cavity case without any superconducting physics to start; Just consider that each material has an associated surface resistance, $R_{s_1}$ and $R_{s_2}$, applied to the surfaces that are made of that material. To simplify this further, since we know we will be using a well-known normal metal, most likely copper, we will use this as the reference resistance value, and judge the resistance of the superconducting metal such that $R_{SC}=\kappa R_{Cu}$, where $\kappa$ is simply the ratio of the two resistances. We can write down the Q of this simply hybrid cavity, by considering the geometric factors of the surfaces made of each material, $G_{Cu}$ and $G_{SC}$; we will assume there is minimal dissipation from other surfaces such that $S_{tot}=S_{Cu}+S_{SC}$.
    \begin{equation}
    \frac{1}{Q_{hybrid}}=\frac{R_{Cu}}{G_{Cu}}+\frac{R_{SC}}{G_{SC}}
    \label{eqn:hybridkappaQ}
    \end{equation}
    If we consider the Q of this same cavity but made entirely of copper, one can find a convenient expression for the Q ratio of hybrid to a normal copper cavity in terms of the resistance ratio $\kappa=R_{SC}/R_{Cu}$ and geometric factor ratio, $g=G_{Cu}/G_{SC}$:
    \begin{equation}
    \frac{1}{Q_{Cu}}=\frac{R_{Cu}}{G_{Cu}}+\frac{R_{Cu}}{G_{SC}}
    \label{eqn:hybridkappaQCu}
    \end{equation}
    \begin{equation}
    \frac{Q_{hybrid}}{Q_{Cu}}=\frac{1+g}{1+\kappa g}
    \label{eqn:hybridkappaQratio}
    \end{equation}
    One can then also solve for the $\kappa$ from a given Q ratio, $q=Q_{Cu}/Q_{Hybrid}$, if one has a measurements from each cavity type:
    \begin{equation}
    \kappa= q+\frac{q-1}{g}
    \label{eqn:hybridkappa}
    \end{equation}
    The ideal case for the hybrid cavity is when $\kappa=0$ and therefore the superconducting surfaces have no resistance at all, giving Q its maximum boost:
    \begin{equation}
    \frac{Q_{ideal \;hybrid}}{Q_{Cu}}=1+g
    \label{eqn:idealhybridQratio}
    \end{equation}
    Therefore, the hybrid cavity improvement factor is limited by the geometric factor even in the best case scenario. Any other value of $\kappa$ from 0 to 1 will increase the Q, albeit by not as much, and predictably, any value where $\kappa>1$ will decrease the Q, because the enhancing metal is more resistive than the copper reference metal. Superconductors are often terrible normal metal conductors at room temperature, and much more resistive than copper at room temperature, so hybrid cavity will have a worse Q than an all copper version until it transitions, so this formulation is useful for tracking that as is done in Chapter \ref{chap:Sidecar1D} for the ADMX Sidecar cool-down. In this way, one expects the $\kappa$ to drop as it cools from being order $10^2$ or more at room temperature, to $\kappa=1$ as the lower bound for the cavity starting to work as intended, up to $\kappa \rightarrow 0$. Often, a $\kappa \approx 0.1$ is already well-within 90\% of the ideal hybrid Q value.
\section{Multi-mode decomposition}
\label{mulitmodedecomposition}
    The previous section on sub-surface geometric factors isn't very useful in lab for a single cavity mode, because all one can really measure is the total resistance for any given mode. Typically, we can only estimate the sub-surface resistances based off models of each cavity materials' properties. But if we have multiple modes, backed up with geometric factors for each, we can construct multiple equations of the form in \ref{eqn:Rmode}. Each equation forms a row $R_m=C_{m1}R_1+C_{m2}R_2+...+C_{mN}R_N$, for a single mode indexed $m$ instead of $lmn$ for simplicity. One can imagine constructing a matrix equation from this data:
    \begin{equation}
    \vec{R}_m=C_{ms}\vec{R}_s
    \label{eqn:RmCmsRs}
    \end{equation}
    where $\vec{R}_m$ is the list of multiple mode resistances corresponding to $\frac{G_m}{Q_m}$, the total surface geometric factor and total surface quality factor for the mode. The $\vec{R}_s$ is the list of corresponding surface resistances for each sub-surface defined; note that this relation assumes $R$ of a surface is independent of mode, while in reality this might not be true. Different modes can have very different frequencies, making the resistance of the material also very different. It was already shown for copper and other normal metals that there is some frequency dependence to account for in Equation \ref{eqn:sigmaanomCU}. Nonetheless, if they are similiar frequencies or the frequency dependence can be accounted for, this is still useful. If one can get $N$ mode measurements of $Q$, forming a list $\vec{R}_m$, with a corresponding rectangular matrix, $C_{ms}$ with $N$ sub-surfaces, this equation can be inverted, and can solve for a corresponding $N$-length list of $\vec{R}_s$:
    \begin{equation}
    \vec{R}_s=C_{ms}^{-1}\vec{R}_m
    \label{eqn:RsCms-1Rm}
    \end{equation}
    where $C_{ms}^{-1}$ is the inverse of the weight matrix. This will also only hold true if an inverse exists for $C_{ms}$, therefore $det(C_{ms})$ must not be zero. In the 2x2 case, this occurs if $G_{1i}=G_{2i}$, when the modes do not have distinct geometric factors. For higher dimensions, this pattern will hold true if all modes are identical, $G_{1i}=G_{2i}=...=G_{Ni}$, but there also may be some non-trivial solutions. Nonetheless, this suggests the best modes to use for performing this decomposition should have distinct geometric factors for the set of surfaces you are using.
    \subsection{Solving for the 2 mode decomposition case}
    The entirety of this dissertation's actual analysis will only make use of the 2-mode, 2-surface case of decomposition. This is because finding modes that have both similar frequencies and distinct geometric factors can be tough. Often modes of similar field structure will have similar frequencies but therefore not have distinct geometric factors. Also as we will see in the error analysis section, even if they are distinct, if the uncertainties in $G$ overlap, error in the inverse matrix will greatly increase.
    \par In this case, there will be four geometric factors, $G_{11}$,$G_{12}$,$G_{21}$, and $G_{22}$, and two Q measurements $Q_1$ and $Q_2$. First the Q measurements are converted to an $\vec{R}_m$ by using the total mode geometric factors:
    \begin{equation}
    G_n=\frac{G_{n1}G_{n2}}{G_{n1}+G_{n2}}
    \label{eqn:Gm2x2}
    \end{equation}
    where $n=1,2$ for each of the two modes. The $\vec{R}_m$ is then:
    \begin{equation}
    \vec{R}_m=\left[\begin{array}{c}
            G_1/Q_1 \\
            G_2/Q_2
        \end{array}\right]
    \label{eqn:Rm2x2}
    \end{equation}
    The weight matrix in this case comes out to be:
    \begin{equation}
    C_{ms}=\left[\begin{array}{cc}
            \frac{G_{12}}{G_{11}+G_{12}} & \frac{G_{11}}{G_{11}+G_{12}}  \\
            \frac{G_{22}}{G_{21}+G_{22}} & \frac{G_{21}}{G_{21}+G_{22}} 
        \end{array}\right]
    \label{eqn:C_ms2x2}
    \end{equation}
    The determinant of this weight matrix is:
    \begin{equation}
    det(C_{ms})=\frac{G_{12}G_{21}-G_{11}G_{22}}{(G_{11}+G_{12})(G_{21}+G_{22})}
    \label{eqn:DetC_ms2x2}
    \end{equation}
    Notice this is where we get the condition that geometric factors must be unique; If I use the same mode for two different surfaces or the same surface for two different modes, this numerator will be zero. The final inverted weight matrix is then:
    \begin{equation}
    C_{ms}^{-1}=\frac{1}{G_{12}G_{21}-G_{11}G_{22}} \left[\begin{array}{cc}
            G_{21}(G_{11}+G_{12}) & -G_{11}(G_{21}+G_{22})  \\
            -G_{22}(G_{11}+G_{12}) & G_{12}(G_{21}+G_{22}) 
        \end{array}\right]
    \label{eqn:Cinv_ms2x2}
    \end{equation}
    We then multiply $\vec{R}_m$ by this inverted weight matrix to get the resultant $\vec{R}_s$ of length 2 corresponding to the resistances of the two surfaces we defined. This method will be used several times Chapters \ref{chap:LLNL} and \ref{chap:Sidecar1D}.
    \subsection{Error analysis for decomposition technique}
    \label{DecompositionError}
    In addition to all of the multi-mode decomposition technique described above is the associated analysis of propagating the error into a final surface resistance. The error comes from two sources; the quality factor measurements, and geometric factor simulations. The quality factor uncertainty is going to come from laboratory measurement usually, which is fitted to a Lorentzian, of which the process is described later in this chapter. Typically a $\Delta Q / Q\approx 1-5\%$ is very achievable. The geometric factor error is a little harder to estimate; theoretically it comes from the expected error within the dimensions of your cavity. This means any vibrations of the cavity that cause the radius, length, or other dimensions to significantly change would effect the geometric factor. However, what is usually more dominant than vibrations is machining tolerances of the parts themselves, which are typically anywhere from 0.001"-0.010" depending on the quality. The $G$ factors are then calculated for variations of the geometry within this chosen length uncertainty $\Delta x$; The simulation software calculates the G factors for all combinations of each cavity dimension varied by 3 points, $x_0-\Delta x$,$x_0$, $x_0+\Delta x$. Computationally, this can be quite intensive, because the number of cavity design arrangements will be $3^D$ where $D$ is the number of dimensions in the cavity design; for the Sidecar cavity in Chapter \ref{chap:Sidecar1D}, 5 dimensions were incorporated, resulting in 243 arrangements. From there, one has a set of $G$ factors for each cavity design arrangement. What's important is that each design corresponds to a set of corresponding geometric factors; one cannot simply average each geometric factor together, but has to find the most average set of geometric factors corresponding to a real design. To do this, we must calculate the average set of geometric factors, and then find the real set that is the least different. Each term of $\vec{G}_{avg}$ can be defined by $G_{i,avg}=(1/N) \sum_{n}^N G_{i,n}$ where $N$ is the number of cavity design arrangements. The most average design arrangement minimizes the magnitude of the vector difference with $\vec{G}_{avg}$:
    \begin{equation}
    \Delta \vec{G}=\vec{G}_j-\vec{G}_{avg}
    \label{eqn:deltaG}
    \end{equation}
    The set we then choose to represent the average real set, even though not strictly a median, is:
    \begin{equation}
    \vec{G}_{median}=\vec{G}_j(min(\vert\Delta\vec{G}\vert^2))
    \label{eqn:G_med}
    \end{equation}
    On the other end, the set with the most error would be the design arrangement that maximized $\vert\Delta \vec{G}\vert^2$, $\Delta \vec{G}_{max}$. The final estimated uncertainty in the geometric factors is determined by taking the difference in these two vectors that represent real cavity designs and are all correlated with each other. 
    \begin{equation}
    \Delta\vec{G}_{final}=\vert \vec{G}_{max}-\vec{G}_{median} \vert
    \label{eqn:deltaGfinal}
    \end{equation}
    This produces typically a fractional error, $\Delta G/G\approx 5\%$. This list of $\Delta G$'s can then be combined with errors in quality factors, propagating through the analysis steps of a multi-mode decomposition. The total mode resistance error is simply the fractional errors of $G$ and $Q$ combined in quadrature:
    \begin{equation}
    \vert\frac{\Delta R_m}{R_m}\vert=\sqrt{\left(\frac{\Delta G_m}{G_m}\right)^2+\left(\frac{\Delta Q_m}{Q_m}\right)^2}
    \label{eqn:RmError}
    \end{equation}
    where $G_m=\sum_s G_{ms}$, the total mode geometric factor, and its corresponding error is $\Delta G_m=\sqrt(\sum_s \Delta G_{ms}^2$. Up until now, this error analysis hasn't assumed a dimension. But going forward one has to solve for the error in each weight index, $C_{ms}$, which depends on the matrix dimension. For simplicity, I will assume the 2x2 case to calculate $\Delta C_{ms}$. To make things easier to read, I will denote each index in the matrix, $C_{ix}=\frac{G_{iy}}{G_{ix}+G_{iy}}$, where if $x=1$, then $y=2$ and vice-versa:
    \begin{equation}
    \vert\frac{\Delta C_{ix}}{C_{ix}}\vert=\sqrt{\left(\frac{\Delta G_{iy}}{G_{iy}}\right)^2+\frac{\Delta G_{ix}^2+\Delta G_{iy}^2}{(\Delta G_{ix}+\Delta G_{iy})^2}}
    \label{eqn:CmsError}
    \end{equation}
    Taking the inverse of the 2x2 matrix gives us an expression for $\Delta C_{ms}^{-1}$ in terms of error in $C_{ms}$ components as well as error in the determinant of the matrix, $\Delta (det(C))$, which can be written:
    \begin{equation}
    \vert\frac{\Delta (det(C))}{det(C)}\vert=\sqrt{C_{11}^2\Delta C_{22}^2+C_{22}^2\Delta C_{11}^2+C_{12}^2\Delta C_{21}^2+C_{21}^2\Delta C_{12}^2}
    \label{eqn:DetCmsError}
    \end{equation}
    \begin{equation}
    \vert\frac{\Delta C_{ix}^{-1}}{C_{ix}^{-1}}\vert=\sqrt{\left(\frac{\Delta C_{xy}}{C_{xy}}\right)^2+\left(\frac{\Delta (det(C))}{det(C)}\right)^2}
    \label{eqn:CmsinvError}
    \end{equation}
    I use the indexing $xy$ here, because these are no longer specifically mode and surface indexes after the inversion. Note that if the determinant is near zero, it will inflate this inversion error significantly. This can make or break your ability to perform a decomposition because error bars can quickly grow to multiples of the resistance values you are trying to solve for. Later measurements will show that even for good mode combinations, the fractional error can commonly be about $10 \%$ for the inverse matrix. Since this expression is entirely dependent on the simulated $G$ factors, one can check the fractional error in the determinant for a variety of mode combinations to find pairs that work well together. The final step is to perform the matrix multiplication with $\vec{R}_m$ to make $\vec{R}_s$, so the error will incorporate $\Delta C^{-1}_{xy}$ as well as $\Delta R_m$
    \begin{equation}
    \vert\frac{\Delta R_{s}}{R_{s}}\vert=\sqrt{\sum_{x=1}^2\left(R_x^2\left(\Delta C^{-1}_{sx}\right)^2+\left(C^{-1}_{sx}\right)^2\Delta R_x^2\right)}
    \label{eqn:RsError}
    \end{equation}
    This covers all the equations needed to propagate the uncertainties for a 2-mode decomposition. See chapters \ref{chap:LLNL} and \ref{chap:Sidecar1D} for examples of this method in action.
\section{Antenna coupling}
Up until now, this analysis has assumed a closed cavity with no antennae or losses to the outside. In reality, one needs at least a single antenna to pull some power out of the cavity to detect a resonant mode in the lab. An antenna simply acts as another power loss source for the cavity. The complication is we are actually measuring the power deposited in this loss source from the cavity, not the intrinsic power of the cavity. The more coupled an antenna is to the cavity, the more power will be drawn from the cavity into the antenna, which in turn will spoil the cavity quality factor with a high $P_d$. This is referred to as the "loaded" cavity quality factor $Q_L$ vs. the "unloaded" or intrinsic Q of the cavity, $Q_0$, that is independent of antenna coupling.
    \subsection{Calculating coupling coefficient}
    Similarly to how we defined sub-surface quality factors for dealing with multiple surfaces, we can define a $Q_{e}$ that is the associated Q of the antenna within the cavity, which implies the loaded quality factor would be:
    \begin{equation}
    \frac{1}{Q_L}= \frac{1}{Q_0}+\frac{1}{Q_c}
    \label{eqn:Qloaded}
    \end{equation}
    This expression implies that if $Q_c>>Q_0$, it will have a small effect on the $Q_L$, whereas if $Q_c\leq Q_0$, it will be a greater contribution. To characterize this more precisely, one defines the coupling coefficient:
    \begin{equation}
    \beta=\frac{Q_0}{Q_c}
    \label{eqn:Beta1}
    \end{equation}
    This creates a convenient expression for $Q_0$, the intrinsic cavity value, in terms of $Q_L$ and $\beta$, the measurable values:
    \begin{equation}
    Q_0=Q_L(1+\beta)
    \label{eqn:Beta2}
    \end{equation}
    This means if $\beta \approx 0$, that antenna coupling doesn't have a significant effect on $Q$ and the measured $Q_L \approx Q_0$.
    \subsection{Antenna Design}
    \begin{figure*}[htb!]
    \centering
    \includegraphics[angle=0, width=.4\linewidth]{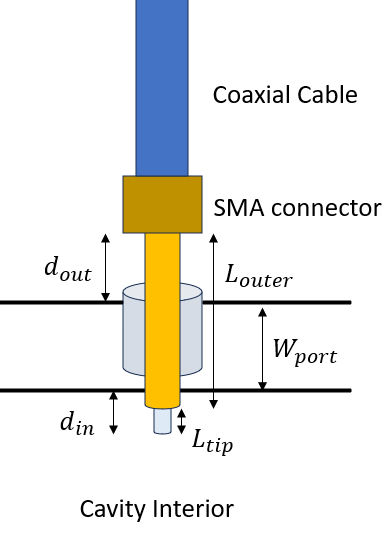}
    \caption{A schematic of a coaxial antenna inserted into a cavity antenna port. The antenna itself only has two major parameters, $L_{tip}$, the length of exposed inner conductor and $L_{outer}$, the length of outer conductor before the coaxial connector. Based on the width of the cavity port, $W_{port}$, and the length of exposed outer conductor outside the cavity, $d_{out}$, the insertion depth, $d_{in}$, can be calculated.}
    \label{fig:antennaschematic}
    \end{figure*}
    ADMX typically uses a simple coaxial antenna design for its searches because it couples well to the electric field of the $TM_{010}$ mode. Since the $TM_{010}$ has an axial electric field, an axially oriented segment of the conductor will have surface currents moving along its length, which are then transferred into a coaxial cable and subsequent transmission line. This obviously means putting the antenna port on an end cap of the cavity for the right orientation. Since this field is maximal towards the cavities' radial center, there is somewhat of a preference to put the antenna port towards the center, but it tends not to be that important. Figure  \ref{fig:antennaschematic}
    outlines what this antenna setup looks like: A semi-rigid coaxial cable is cut such that a small length of the inner conductor is exposed, $L_{tip}$, leaving a length, $L_{outer}$, of the outer conductor before connectorized, usually with an SMA-type connection. This antenna is inserted into the port with a threaded, silver-plated feed-through to minimize any radiation losses through gaps in the port. This also grounds the outer conductor to the outer cavity itself. Based on the width of the end plate, $W_{port}$, the port is drilled through and the antenna is inserted with an exposed length of outer conductor above, $d_{out}$, and the insertion depth of the antenna can be estimated, $d_{in}$, which is useful for quantifying geometrically how far the antenna is inside of the cavity. This parameter can also be used for simulating antenna coupling. 
    \subsection{1-port Measurements}
\begin{figure*}[htb!]
    \centering
    \includegraphics[angle=0, width=0.25\linewidth]{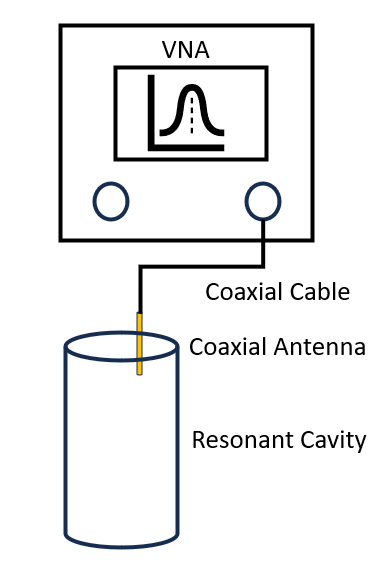}
    \caption{A schematic of a typical 1-port cavity measurement with a vector network analyzer (VNA).}
    \label{fig:VNA1portsetup}
    \end{figure*}

    Measurements with a single antenna into the resonant cavity can be classified as a 1-port measurement. Up until now, I have not formally introduced the primary instrument for making cavity measurements in the lab: The network analyzer. A network analyzer is essentially two instruments working in conjunction with one another: A signal generator makes known RF signals to input into the "RF network", our antenna and cavity setup, via one of its ports, while a spectrum analyzer measures the power spectrum coming from that network into a network analyzer port. In the case of 1-port measurements, the network analyzer will only have 1 port in use, but typically they come with 2 ports. Fig \ref{fig:VNA1portsetup} is a simple schematic of this setup. SMA connections are typical, and connect the transmission line, coaxial cable, to the network analyzer and antenna. 
    \par The only measurement one can make in this set up is known as reflection measurement. This is a swept measurement, where a fixed power signal is generated and sent down the transmission line at various frequencies (this is often referred to as a "sweep"), and the resultant amount of power that is reflected back down the input line is measured with the spectrum analyzer. The input signal will not only be attenuated in magnitude, but there could be an associated phase shift in the signal as well, therefore the response will be a complex value $\Gamma$ at each frequency. The ability to measure this phase information as well as magnitude makes this network analyzer a vector network analyzer (VNA). One could do this measurement without the input signal, but it won't necessarily have the same power being inputted at each frequency hence $\Gamma(f)$ will not be accurate. Nonetheless, an axion search digitization simply removes the input signal and looks for ambient axion power exciting the modes (after one has characterized the cavity behavior). The magnitude represents the ratio of reflected to input power, $\vert \Gamma \vert$, a reflection coefficient. In this way, $\Gamma(f)=\vert \Gamma(f) \vert e^{i \angle \Gamma(f)}$ where $\angle \Gamma(f)$ is the associated phase shift for that frequency. Unless the input power is near a resonant frequency, most of the power will be reflected back up the transmission line resulting in a reflection ratio $\vert \Gamma \vert \approx1$ or $S\approx 0\;{\rm dB}$. However, if the antenna is coupled to the cavity, some portion of the power will be absorbed by the cavity near the resonant mode frequency, producing a Lorentzian dip! This will make $0< \vert \Gamma \vert < 1$ and therefore $S$ will be a negative dB value. This Lorentzian dip can be related to $\beta$, the coupling coefficient, by the relation:
    \begin{equation}
    \beta= \frac{1+sign(\angle \Gamma (f_0) -\pi) \vert \Gamma (f_0) \vert}{1-sign(\angle \Gamma (f_0) -\pi) \vert \Gamma (f_0) \vert}
    \label{eqn:BetaReflection}
    \end{equation}
    Where $f_0$ is the mode resonant frequency and therefore $\Gamma(f_0)$ refers to the values of $\Gamma$ at the bottom of the Lorentzian dip. The $sign(x)$ function simply outputs the sign of the argument: $sign(x)=+1$ for $x>0$, $sign(x)=-1$ for $x<0$, and $sign(0)=0$. This is how one measures the $\beta$ and couples antennae in lab; look at the mode reflection dip using a network analyzer, write down the $\vert \Gamma(f_0) \vert$ and $\angle \Gamma (f_0)$, and calculate $\beta$. One can see that from Equation \ref{eqn:BetaReflection}, $\beta$ is to some degree proportional to the dip depth, but if  $\vert \Gamma(f_0) \vert \approx 1$, no power is coupled into the cavity, could respond to either $\beta=0$ or $\beta \rightarrow \infty$, depending on the phase shift. 
    \subsection{Coupling regimes} 
\begin{figure*}[htb!]
    \centering
    \includegraphics[angle=0, width=1\linewidth]{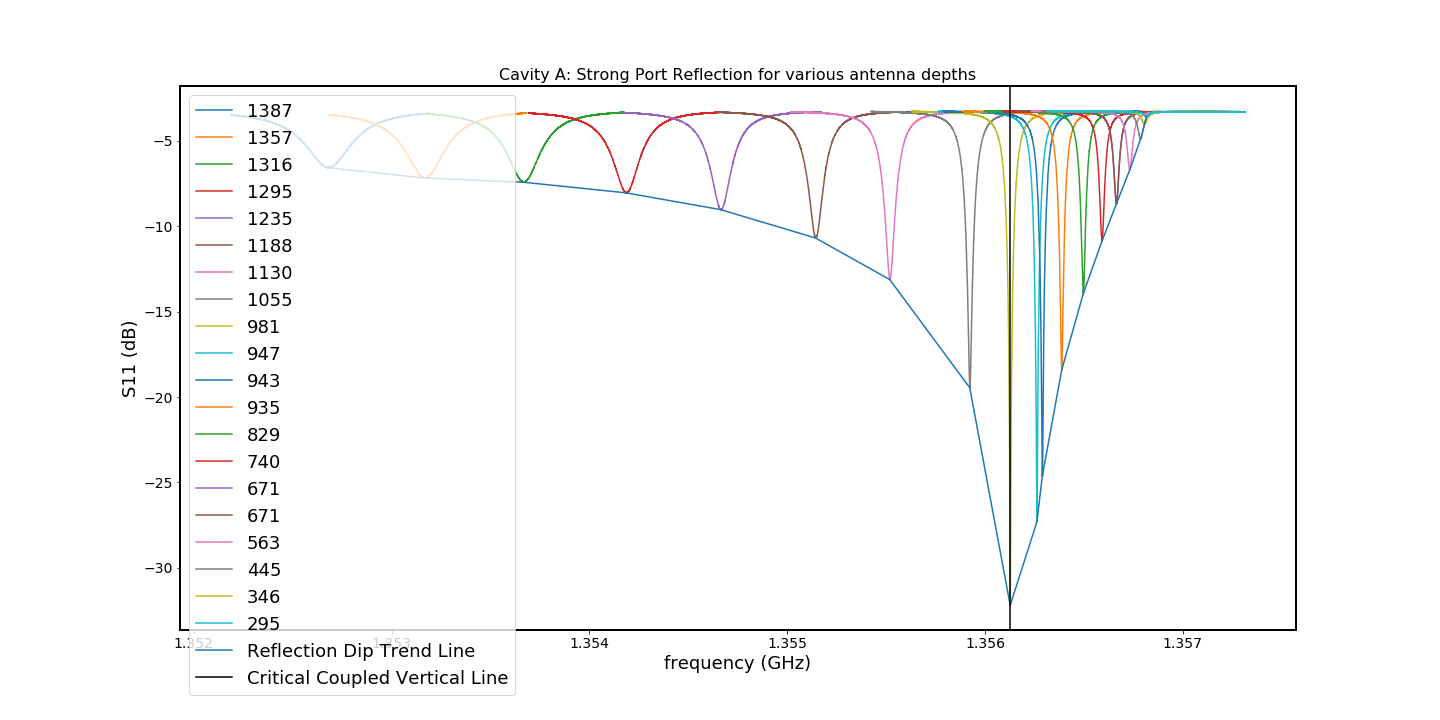}
    \caption{A series of network analyzer reflection measurements on a single ADMX run 2A cavity $TM_{010}$ for varying antenna depths. They are indexed in the legend by how many thousandths of an inch the antenna is inserted into its port from the antenna tip. Additionally, a peak trend-line is interpolated across the series of Lorentzian dips, and a vertical line marks the critically coupled point.}
    \label{fig:antennasweepR}
    \end{figure*}

    The coupling coefficient $\beta$ can be used to characterize 3 regimes of antenna coupling. In Figure \ref{fig:antennasweepR}, a successive number of reflection measurements from a network analyzer are made on a cavity as an antenna is inserted at fixed length intervals. 
    One starts on the right side of the plot where insertion depths are lowest and the antenna isn't very coupled to the cavity. If the antenna isn't inserted at all, one would expect a flat spectrum, and as it is inserted slowly, small dips would appear around the resonant frequencies of the modes. The coupling coefficient would be $\beta\approx 0$ for these first couple sweeps, when the dip is $>-1 dB$, but will start to increase towards $\beta=1$ as the dip gets deeper. This regime where $\beta< 1$, is known as being weakly coupled; the antenna is only a perturbation on the cavity. Notice that there is some downward shift in the resonant frequency as $\beta \rightarrow 1$, but it is less than 1 MHz over the range for a 1.35 GHz cavity. 
    \par When $\beta=1$, the antenna will have roughly the same Q as the cavity itself, and theoretically will be taking half the power out of the cavity compared to the surface resistance losses. In terms of the reflection dip, this should be the maximum dip depth one sees; for $\beta > 1$, the antenna Q will dominate over the cavity Q, meaning a smaller fraction of inputted power will transmit into the cavity, lowering the dip depth. In Figure \ref{fig:antennasweepR},  you can see this maximum depth right around 1.356 GHz, marking the $\beta=1$ insertion depth. This regime, where $\beta\approx 1$, is called critically coupled; typically $\beta$ values near 1 are still considered critically coupled even out to $\beta=2$ (more on this in the next section).
    \par The third regime is known as being over-coupled, and is when $\beta>1$. This is represented by the entire left side of Figure \ref{fig:antennasweepR} past the critically coupled minimum. Not only does the dip depth degrade, but one can see that the width of the Lorentzian dip starts to widen significantly, indicative of the $Q_L$ degrading from the antenna coupling. The antenna starts to be a considerable object within the internal cavity geometry in this regime, causing the resonant frequency to shift on the order of several MHz for this cavity. 
    \subsection{Axion search antenna coupling}
     Equations \ref{eqn:axionpower4} and \ref{eqn:ScanRate2}, both include the effect of antenna coupling on axion power and scan rate, but we did not cover where these factors come from or its optimization. Clearly the cavity quality factor at a given frequency will degrade inversely, according to Equation \ref{eqn:Qloaded}, introducing the first factor of $\frac{1}{1+\beta}$, as $\beta$ is increased. Increasing the coupling will also increase the fraction of power being pulled out of the cavity and into the digitizer; this introduces a factor $\beta$. The overall signal power is therefore enhanced by a factor of $\frac{\beta}{\beta+1}$, implying that an over-coupled antenna will draw the most power out of the cavity. What this doesn't include though is, by increasing $\beta$ towards the critically coupled regime, one also effectively spreads out their cavity search bandwidth, $\delta f_{c}$ by a factor of $1+\beta$, which will allow one to take bigger frequency steps during an axion search. These two effects are combined in Equation \ref{eqn:ScanRate2}, enhancing the scan rate by a factor of $\frac{\beta^2}{(\beta+1)^3}$. This factor can be differentiated to find critical points, one of which is a non-negative maximum at $\beta=2$. For this reason, the optimal coupling for an axion search is $\beta=2$, slightly over-coupled. When calibrating an antenna, its easiest to first critically couple it, and then incrementally increase it a tiny bit more to a $\beta=2$; therefore its often said that an axion search is best performed with a critically coupled antenna, rather than the more nuanced $\beta=2$ answer.
\section{Measuring Q in the lab}
This section will cover how one typically measures cavity Q in a lab using a network analyzer. In the antenna coupling section we covered 1-port measurements for determining a coupling coefficient of a single antenna, but typically one uses a 2-port setup for cavity characterization.
    \subsection{2-port Measurements}
    Adding a second port increases the number of possible measurements from 1 to 4; The signal generator can input power on either port 1 or port 2, and the spectrum analyzer can measure the resultant signal at either port. This forms a $2x2$ matrix known as the scattering parameters, or S-matrix, for the system, and they are indexed by the port measured and power inputted respectively; $S_{ij}$ is produced from measuring at port $i$ and inputting a signal at port $j$.
    \par The two possible reflection measurements are $S_{11}$ and $S_{22}$, indexed by each antenna port; They will each have an associated $\beta_{1}$ and $\beta_{2}$. This doesn't change much to the analysis of the 1-port measurement except that Q will degrade more based on both antennae. A simple proof, following the analysis of how we defined $\beta$ in terms of $Q_0/Q_c$, will show that $\beta_{total}=\sum_{i=1}^N \beta_i$ when combining $N$ antenna ports, therefore in the 2 port case, $\beta_{1+2}=\beta_1+\beta_2$. As outlined in Chapter \ref{chap:Haloscopes}, haloscopes typically run with a strong, variably coupled port antenna ($\beta_{strong}\approx 1-2$) that takes the axion search data, and a fixed, weakly coupled port antenna where $\beta_{weak} \approx 0$ for periodic cavity characterization. This means the strong antenna is the majority contributor to the total coupling most of the time.
    \par In the 2-port scenario, transmission measurements are now possible by looking at the $S_{21}$ and $S_{12}$ parameters. In this case, the cavity will act like a filter for many frequencies; unless an input frequency is near a resonant mode, no power will be transmitted through the cavity from the input port to output port. This makes the transmission coefficient, sometimes called gain $\vert G \vert$, with linear units, near-zero off resonance. When the input signal frequency is in the vicinity of a cavity mode, the cavity will resonate and transmit the signal through, creating a non-zero $+\vert G \vert$; this generates a Lorentzian peak on the transmission S-parameter plots, unlike the dip in reflection measurements. Since both of the antennae aren't necessarily coupled in the same way, $S_{21}$ and $S_{12}$ might not be reciprocal to each other. In the case of a strong vs weak port, a signal inputted from the weak port will not transmit as much of the signal into the cavity because $\beta$ is much lower than the strong port. A strong port that is measuring the transmitted signal from the weak port will be more sensitive because of its higher $\beta$ than if measuring with the weak port. In the case of a low power axion search, it is then best to input power from the weak port, minimizing the external RF power put into the cavity system, and measure with the strong port that is more coupled to the cavity. If one is just characterizing a cavity, and doesn't need a critically coupled antenna, for simplicity, one often just sets both $\beta_1$ and $\beta_2$ to be as weakly coupled as possible, to minimize the antennae effecting the cavity system; when $\beta_1=\beta_2$, assuming the antenna ports are also placed in symmetric positions to one another, the network should then be reciprocal and $\vert S_{21} \vert = \vert S_{12} \vert$ . 
    \par The reason one wants to take transmission measurements is that they generally produce much more consistent Lorentzian shape to measure and calculate quality factor from. In principle, one could use the following methods on reflection data alone to calculate Q, but typically reflection measurements will contain more noise, standing waves, etc. from the transmission line and upstream RF components of the measuring port. Additionally, the Lorentzian dip changes drastically depending on the associated coupling coefficient, whereas a transmission measurement, by depending on two $\beta$ values, can be less sensitive.
    \begin{figure*}[htb!]
    \centering
    \includegraphics[angle=0, width=1\linewidth]{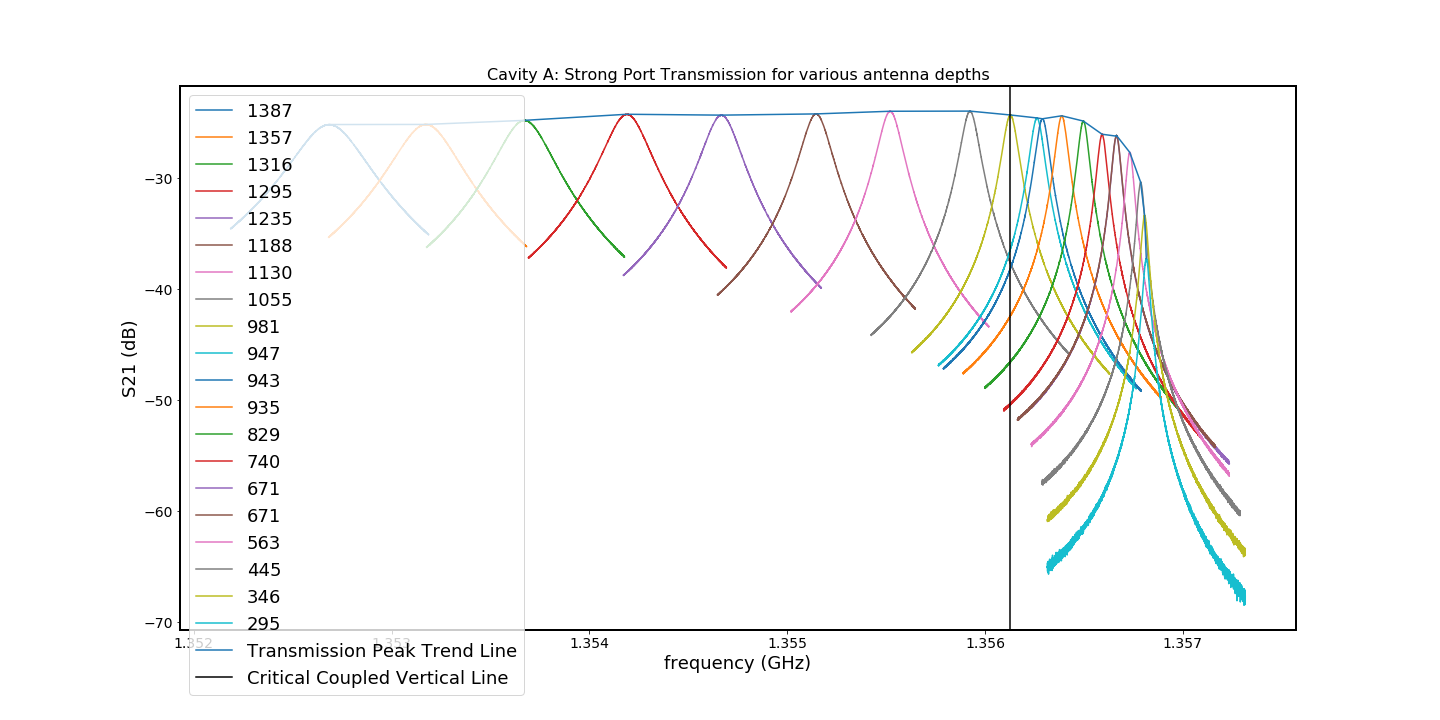}
    \caption{A series of network analyzer transmission measurements on a single ADMX run 2A cavity $TM_{010}$ for varying antenna depths. In this case, port 1 is being inserted into the cavity, while port 2 is fixed in a weakly coupled position. They are indexed in the legend by how many thousandths of an inch the antenna is inserted from its tip. Additionally, a peak trend-line is interpolated across the series of Lorentzian peaks, and a vertical line marks the critically coupled point.}
    \label{fig:antennasweepT}
    \end{figure*}
    This can be seen in Figure \ref{fig:antennasweepT}, which is the partner plot to Figure \ref{fig:antennasweepR}. The peak height in this case is much stabler than the dip depth in reflection measurements, and the widening of $\Delta f_c$ is much easier to see in the weakly to critically coupled regime. However, one can see that picking out the critically coupled scan is much less intuitive in this plot compared to reflection measurements. For this reason, transmission measurements are used for getting a $Q_L$, and reflection measurements are used to get $\beta$ as well as a secondary, less reliable $Q_{L, \, reflection}$. We are now in a position to talk about various methods of measuring Q.
    \subsection{3-dB method}
    This method is by far the simplest method, which often can be more stable than the more complex methods, but less robust. First, one wants to window their VNA scan such that the resonant peak or dip is the only feature. In this way, the maximum $S_{21}$ or $S_{12}$ for the scan corresponds to $G(f_0)$, the peak value at the resonant frequency. In terms of reflection measurement, the minimum $S_{11}$ or $S_{22}$ corresponds to $\Gamma (f_0)$. One can then simply find the frequency that has the maximum S value in the scan, and $f_0$ is determined. This can be tricky for reflection measurements that can more often have large standing waves that put the minimum S-value within the noise region, and not on the resonant dip.
    \par The Lorentzian shape defines $\Delta f_c$ to be the Full-Width-at-Half Maximum (FWHM) of the resonant peak or dip of interest. Therefore the goal is to find the two points on either side of the peak (or dip) that correspond to half the peak value, $G(f_0)/2$, (or twice the dip value, $2\times \Gamma (f_0)$). Because the S-parameters are usually expressed in dB units, rather than linear units, this means looking for the two points $\approx - 3dB$ from the peak (or +3 dB from the dip). This can be done by splitting the original scan at the measured resonant frequency into left and right arrays, and looking for the frequency value that corresponds closest to $S_{21}(f_0)-3dB$ (or $S_{12}(f_0)-3dB$, $S_{11}(f_0)+3dB$, $S_{22}(f_0)+3dB$) in each array. From there, $\Delta f_{3dB}=\vert f_R- f_L \vert$ where $f_R$ is the right side 3 dB frequency and $f_L$ is the left side 3dB frequency. From there the $Q_{3dB}$ is simply:
    \begin{equation}
    Q_{3dB}=\frac{f_0}{\Delta f_{3dB}}
    \label{eqn:f3dB}
    \end{equation}
    The main issue with this method is it doesn't take into account other factors within the Lorentzian equation, like the noise level, phase, etc. Additionally, there is some tiny error introduced by 3 dB not exactly corresponding to $\frac{1}{2}$ in linear units, but this can be corrected for by simply converting the set into linear units through out this whole process. If all 4 S-parameters are taken, theoretically you have 4 equivalent values of Q assuming a reciprocal network. The subsequent fitting methods often take the $Q_{3dB}$ as the initial guess for the quality factor, and re-fit the value accordingly.
    \subsection{Lorentzian Fit}
    The next step up in complexity would be to perform a least-squares fit according to the power spectrum function in Equation \ref{eqn:RLCPower}. However, this function is adapted to have an arbitrary constant noise level, $P_{noise}$, usually. This fit then has 4 parameters to fit to; $P_{noise}$, $P_{peak}$, $f_0$, and $Q$. If one has done the 3-dB method before this, they will already have on hand values for $P(f_0)$, $f_0$, and $Q$ which are convenient for giving a fitting routine software an initial guess. In this setup, one must be careful to recognize $P(f_0)=P_{peak}+P_{noise}$; this can even be reduced to a single parameter, $\Delta P$, the prominence of the peak, but it still involves making some separate guess or fit to the noise level beforehand. The best initial guess for the noise level is usually done by taking the average of either end of the scan, which should be the most off-resonant values in the set. This can be finicky at times, because it is correlated with the width of the VNA scan. If the VNA scan is a bit narrow, it will over-estimate the noise level by incorporating part of the Lorentzian into its estimate. If the scan is too wide, one might not have enough resolution for fitting the Lorentzian itself. This can be corrected for by taking VNA scans at multiple window span sizes; one wider scan that captures the noise level estimate, and a narrow scan that captures the Lorentzian parameters. Because $Q$ depends on $f_0$, $\Delta f$ is sometimes used instead of $Q$ directly for the fit, and $Q$ is calculated subsequently from the fit results; this is just personal preference, and shouldn't make a difference.  Typically, the python package we use is \texttt{Scipy.optimize.curve\_fit}; this will not only output the fit parameters but a covariance matrix from the fit, that will give uncertainties for all the fit parameters, including $Q$. Although the 3 dB method may get you a value for Q very quickly, and even if this Lorentzian fit doesn't change the Q value by that much, this fit method produces a smooth and refined function for the cavity Lorentzian that can be used in subsequent analysis. 
    \subsection{Complex Q-circle Fit}
\begin{figure*}[htb!]
    \centering
        \includegraphics[angle=0, width=0.6\linewidth ]{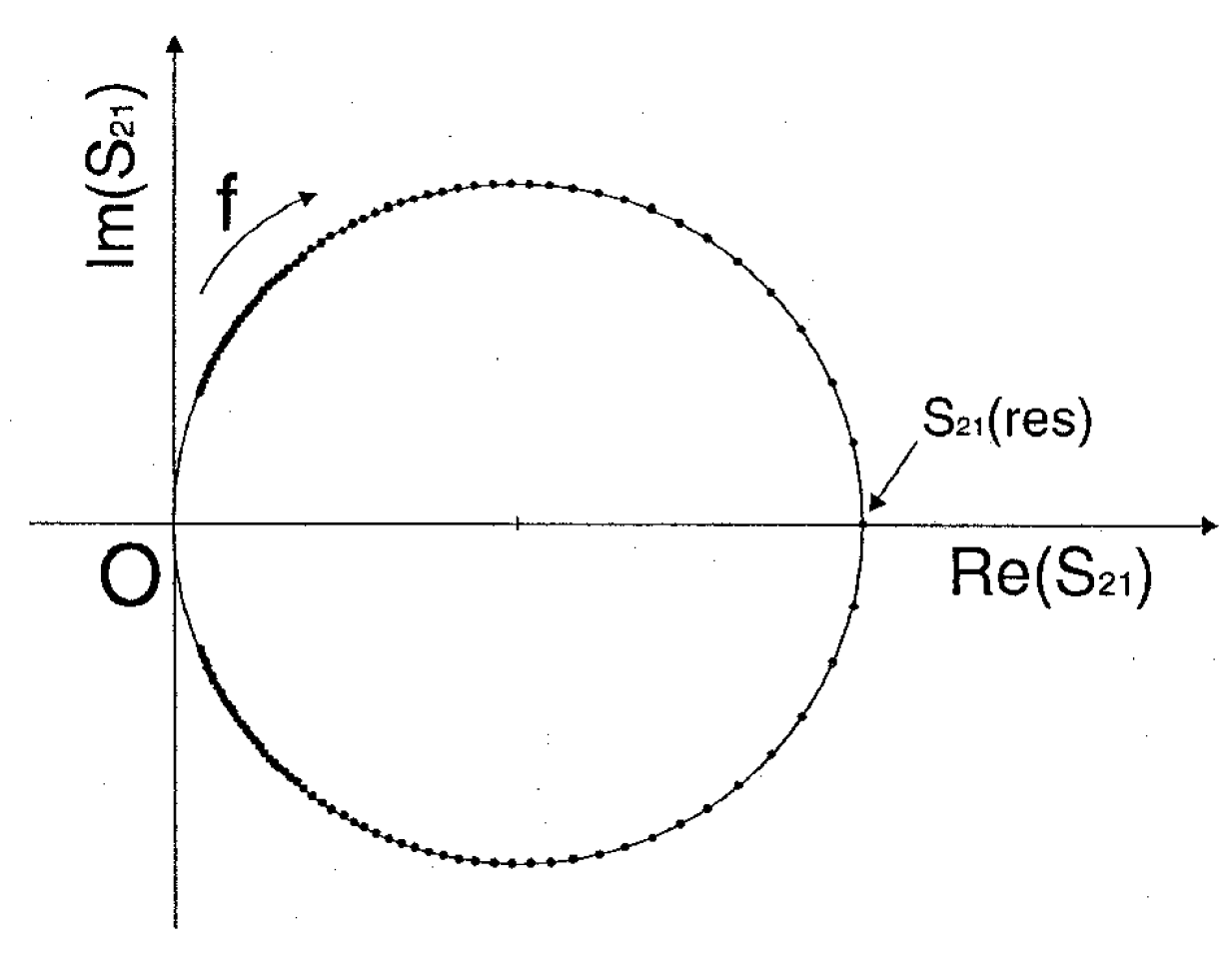}
        \caption{An ideal $S_{21}$ Q-circle. In principle, an ideal reflection Q-circle is the same, however in practice it is a much noisier fit \cite{Qcircle1}.}    
    \label{fig:Qcircleideal}
    \end{figure*}

    The final Q-fitting method that is used  by the ADMX collaboration is a complex fit, and the most rigorous assessment of the cavity's performance. This method makes use of the shape of the complex plot of the S-parameters for the resonant cavity: a circle. This complex data is often plotted on a smith chart, which is a convenient tool for visualizing and measuring the related parameters to this measurement: impedance, admittance, reflection coefficients, etc. How this circle looks plotted in the complex plane also captures many of the exterior phenomenon to the cavity: phase delay and losses from transmission lines, connectors and antennae, impedance mismatch, and cross-talk \cite{Qcircle2,Qcircle1}. Each of these has an effect on shifting the Q-circle from the ideal case. Figure \ref{fig:Qcircleideal} represents the ideal case where none of these effects take place: The real and imaginary parts of the S-parameter near a resonance form a circle that passes through the origin at the edges of the scan (the detuned point), whereas the S-parameter at the resonant frequency passes through the real-axis at a point $(Re(S_{21}(f_{res})) \neq 0, Im(S_{21}(f_{res}))=0)$; notice the center of the Q-circle will also be along the real axis as well. 
    \par
\begin{figure*}[htb!]
    \centering
        \includegraphics[angle=0, width=0.5\linewidth ]{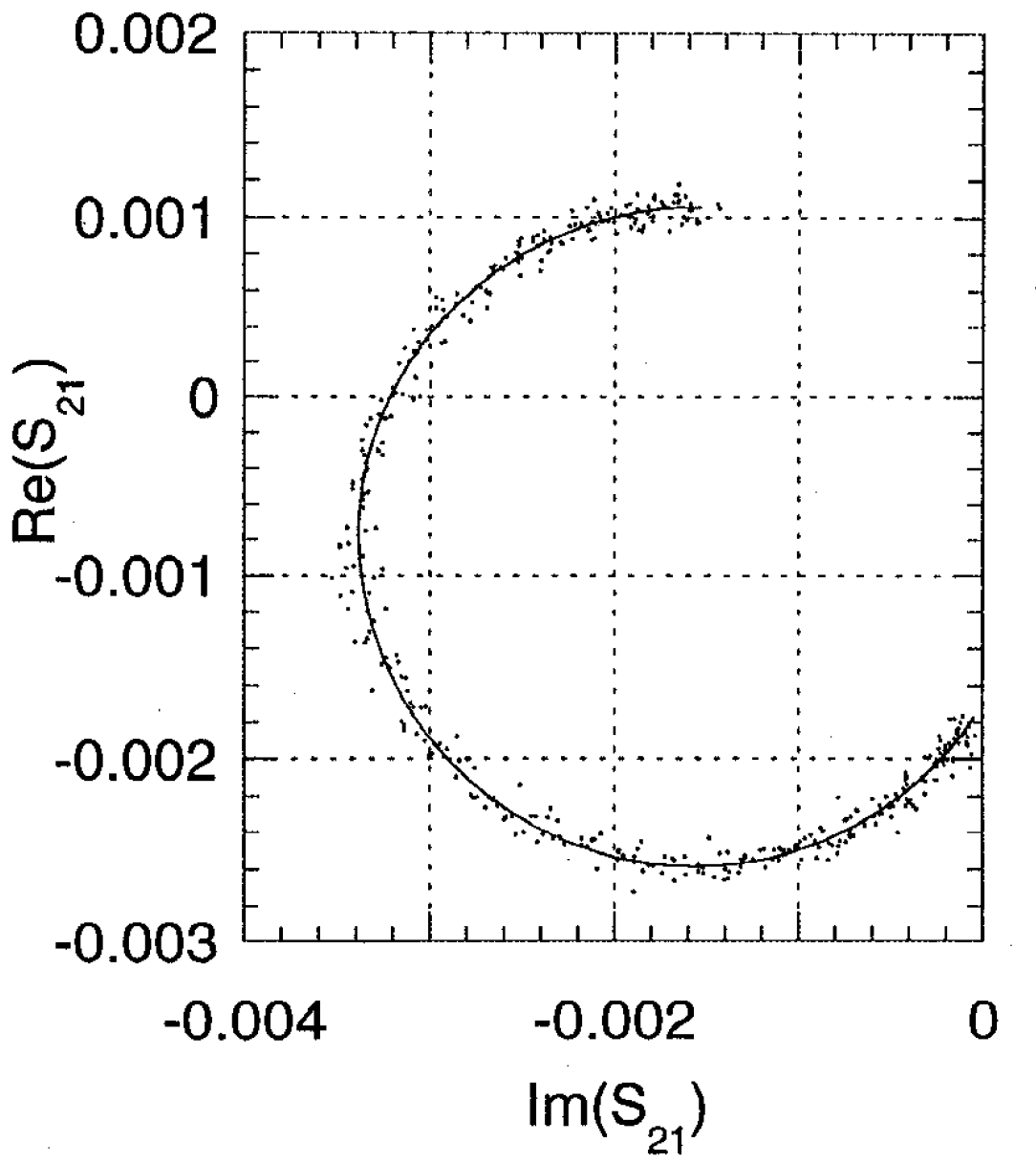}
        \caption{A practical $S_{21}$ Q-circle fit that reflects the effects of the transmission line and antenna setup \cite{Qcircle1}.}    
    \label{fig:QcirclePractical}
    \end{figure*}

 In practice, a Q-circle more often looks like Figure \ref{fig:QcirclePractical}. The cross-talk between the two antennae (coupling loops) offsets the circle away from the origin. A phase shift due to the transmission lines between the VNA and resonator causes the Q-circle to be rotated about the origin. Losses in the connectors, cables, and antennae have the effect of reducing the diameter of the circle. Any noise in the data translates into the roughness of the circle; In general, reflection Q-circles are much rougher than transmission measurements, hence why we use transmission measurements to measure quality factor.
    \par
\begin{figure*}[htb!]
    \centering
        \includegraphics[angle=0, width=0.9\linewidth ]{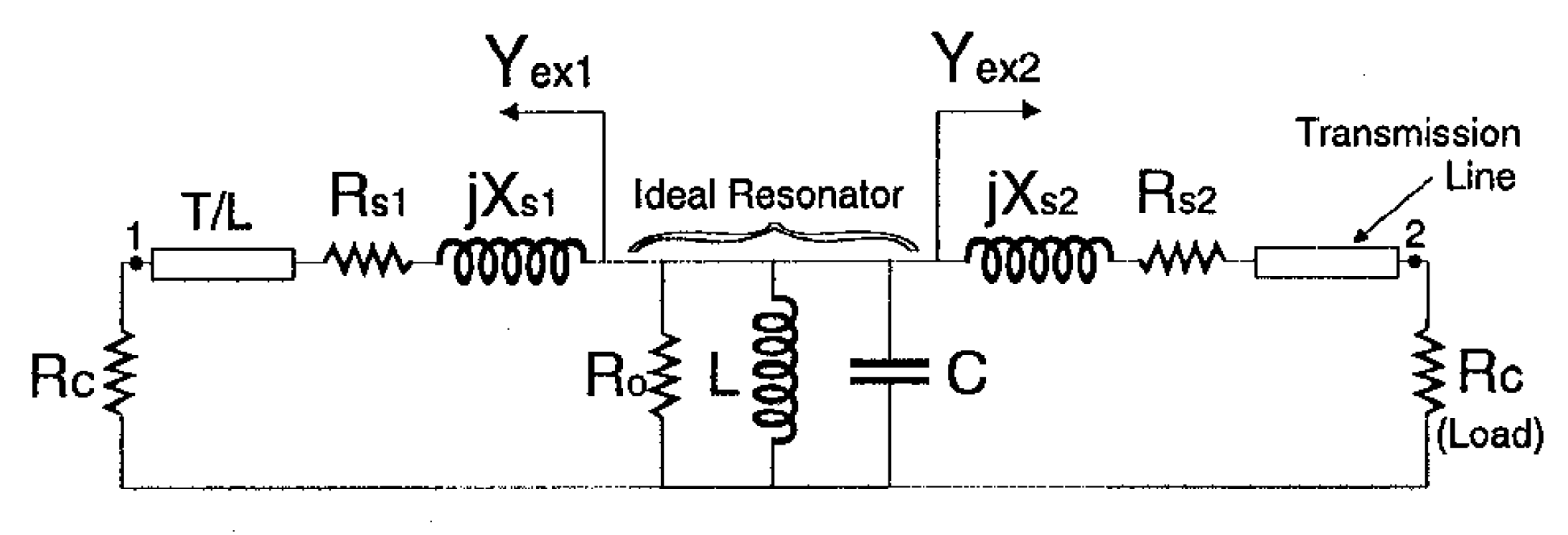}
        \caption{An equivalent circuit model of a two-port cavity measurement with antennae and coaxial cables connected to a VNA: an ideal resonator in parallel with two transmission lines with complex impedance \cite{Qcircle1}.}    
    \label{fig:RLCw/Lines}
    \end{figure*}

 These effects can be better quantified by expanding the equivalent circuit model for the resonator (Figure \ref{fig:RLCCircuit}) to include the impedance of the two transmission lines, $Z_{si}=R_{si}+iX_{si}$, and a load representing the VNA itself, $R_c$, as pictured in Figure \ref{fig:RLCw/Lines}.
    The admittance, $Y_{exp}$, the reciprocal of impedance, as well as conductance, $G_{exp}$, the reciprocal of resistance, are useful to define here for ports $p=1,2$:
    \begin{equation}
        Y_{exp}=\frac{(R_c+R_{sp})-iX_{sp}}{(R_c+R_{sp})^2+X_{sp}^2}=G_{exp}+iB_{exp}
        \label{eqn:RLCTLAdmittance}
    \end{equation}
    \begin{equation}
        G_{exp}=Re(Y_{exp})
        \label{RLCTLconductance}
    \end{equation}
    \begin{equation}
        B_{exp}=Im(Y_{exp})
    \label{eqn:RLCTLconductance}
    \end{equation}
    Additionally, the conductance of the cavity, $G_0=1/R_0$, can be used to define the coupling coefficients of each port:
    \begin{equation}
        \beta_p=\frac{G_{exp}}{G_0}
        \label{eqn:RLCTLbeta}
    \end{equation}
    It is worth noting that although $G_0 \neq Q_0$ exactly, $Q_0 \propto G_0$ and will only differ by a constant that is completely independent of the transmission lines. The cavity itself also will have an admittance:
    \begin{equation}
        Y_0=G_0\left[1+i2Q_0\frac{\omega-\omega_0}{\omega_0}\right]
        \label{eqn:RLCY_0}
    \end{equation}
    The $ABCD$ matrix formalism described in most microwave engineering textbooks, such as Ref. \cite{pozar2011microwave}, is useful for describing this circuit as 3 cascaded networks:
    \begin{equation}
        \left[\begin{array}{cc}
            A & B \\
            C & D
        \end{array} \right]
        = \left[\begin{array}{cc}
            1 & R_{s1}+iX_{s1} \\
            0 & 1
        \end{array}] \right]
            \left[\begin{array}{cc}
            1 & 0 \\
            Y_0 & 1
        \end{array} \right]
        \left[\begin{array}{cc}
            1 & R_{s2}+iX_{s2} \\
            0 & 1
        \end{array} \right]
        \label{eqn:RLCABCD}
    \end{equation}
    The S-parameters can be derived from the ABCD parameters using conversion tables from Ref. \cite{pozar2011microwave}. In the case of $S_{21}$:
    \begin{equation}
        S_{21}=\frac{2R_c}{AR_c+B+CR_c^2+DR_c}
        \label{eqn:RLCS21ABCD}
    \end{equation}
    Which follows that:
    \begin{equation}
        S_{21}=\frac{2R_cY_{ex1}Y_{ex2}}{Y_{ex1}+Y_{ex2}+Y_0}
        \label{eqn:RLCS21YR}
    \end{equation}
    We can define the total admittance of the loaded resonator, $Y_L$:
    \begin{equation}
        Y_L=Y_0+Y_{ex1}+Y_{ex2}=G_0+G_{ex1}+G_{ex2}+i(2Q_0G_0\frac{\omega-\omega_0}{\omega_0}+B_{ex1}+B_{ex2})
        \label{eqn:RLCYL}
    \end{equation}
    Comparing equation form of Equation \ref{eqn:RLCY_0} to Equation \ref{eqn:RLCYL}, one can see that the loaded resonator will have an effective shift in the resonant frequency based on the susceptance of the two transmission lines. This can be solved for by remembering that the resonance point will be when the impedance is entirely real, and therefore the susceptance must be zero:
    \begin{equation}
        \omega_L=\omega_0 \left[ 1-\frac{B_{ex1}B_{ex2}}{2Q_0G_0}\right]
        \label{eqn:RLCWL}
    \end{equation}
    What is important to note here is the denominator of the shift term containing $Q_0 G_0$; this can effectively be thought of as $\propto Q_0^2$, therefore, this shift term is negligible for $Q_0>>1$. This means for high-Q cavities, usually above 1000, $\omega_L \approx \omega_0$. Further examination of Equation \ref{eqn:RLCYL} shows that we can also define the total conductance of the loaded circuit, $G_L+G_0+G_{ex1}+G_{ex2}$, which will, similarly to $G_0$, be proportional to the loaded quality factor, $Q_L$ such that:
    \begin{equation}
        Q_L=Q_0\frac{G_0}{G_L}
        \label{eqn:RLCQL}
    \end{equation}
    The complex transmission coefficient for the mode of the cavity, $S_{21r}$, can then be derived from \ref{eqn:RLCS21YR},\ref{eqn:RLCTLbeta},\ref{eqn:RLCYL}, \ref{eqn:RLCWL}, and \ref{eqn:RLCQL} as:
    \begin{equation}
        S_{21r}=\frac{2R_cY_{ex1}Y_{ex2}}{G_0(1+\beta_1+\beta_2)\left[1+i2Q_L\frac{(\omega-\omega_L)}{\omega_0}\right]}
        \label{eqn:RLCTLS21r}
    \end{equation}
    If we use the high-Q approximation, $\omega_L \approx \omega_0$, Equation \ref{eqn:RLCTLS21r} can be parameterized into the form,
    \begin{equation}
        S_{21}=\frac{a_1t+a_2}{a_3t+1}
    \label{eqn:RLCTLS21rat}
    \end{equation}
    Where:
    \begin{equation}
        t=2\left(\frac{\omega-\omega_0}{\omega_0}\right)
    \label{eqn:RLCTLS21t}
    \end{equation}
    and in this case:
    \begin{equation}
        a_1=0, \; a_2=\frac{2R_cY_{ex1}Y_{ex2}}{G_0(1+\beta_1+\beta_2)}, \; a_3=iQ_L
    \label{eqn:RLCTLS21ra}
    \end{equation}
    Equation \ref{eqn:RLCTLS21rat} represents a fractional linear transformation in the complex plane. This is a class of transformations that maps equations of lines to circles and vice-versa. In this case, the parameter, $t$, is the horizontal line in the complex plane, $y=0$, the x-axis, being that $\omega$ and $\omega_0$ are fully real. This special case of coefficients will always map this line to a circle centered on a point on the x-axis, with $S_{21r} \rightarrow 0$ for $f\rightarrow\pm\infty$, and $Im(S_{21r}(\omega_0))=0$, just as Figure \ref{fig:Qcircleideal} shows. The radius of this circle is the measure of the system loss, because $Re(S_{21r}(\omega_0))=2R_{Q-circle}$ will be entirely resistive. The point that $S_{21}$ approaches as $f\rightarrow\pm\infty$, in this case the origin, is generally called the detuned point, $S_{ppd}$, where $p$ is the port numbers and $d$ just represents "detuned."
    \par This analysis can be repeated for the reflection S-parameters as well, and will result in the same fractional linear form as Equation \ref{eqn:RLCTLS21rat}, but with different coefficients:
    \begin{equation}
        a_1=iQ_LS_{ppd}, \; a_2=S_{ppd}+\frac{2R_cY_{exp}^2}{G_0(1+\beta_1+\beta_2)}, \; a_3=iQ_L
    \label{eqn:RLCTLSppra}
    \end{equation}
    The general geometric role of these three coefficients on the Q-circle become a little more clear here. $S_{ppd}$ can be solved to be $a_1/a_3$. $a_2$ is the reflection coefficient at the resonant frequency, $S_{pp}(\omega_0)$. Most importantly, $Q_L=Im(a_3)$ in both transmission and reflection cases. 
    \par Two effects that aren't included in the transmission analysis are an explicit phase-delay in transmission lines and any cross-talk that may exist between the coupling structures. Any non-zero length cable will introduce some phase delay. This is a frequency dependent phase shift of the form, $\phi(f)=\phi_{ref}f/f_{ref}$, where $\phi_{ref}$ is the phase introduced by the delay at an arbitrary $f_{ref}$. This will have the effect of offsetting the phase angle as the scan approaches the resonance, resulting in a rotation of the Q-circle by an angle, $\phi \propto L_{line}$,  where $L_{line}$ is the length of the transmission lines. Because of the frequency dependence of this shift, a higher frequency regime will result in a larger angle rotation. One also expects this to distort the shape of the Q-circle slightly to a balloon-shape, with the strength of the effect proportional to the frequency regime of the resonance; this is harder to quantify and will be ignored for now. Cross-talk simply has the effect of a complex translation, re-centering the Q-circle to some point $X=(x_0,y_0)$. These two effects can be combined into a corrected $S_{21}$ parameter:
    \begin{equation}
        S_{21}=(S_{21r}+X)e^{i\phi}
    \label{eqn:S21wcrosstalk}
    \end{equation}
    This doesn't change much other than changing the location of the Q circle. However, the order of this translation is unique; cross-talk shifts the circle first, and then it is rotated about the origin. The general geometric form of Equation \ref{eqn:RLCTLS21rat} still holds, but the explicit coefficients in Equation \ref{eqn:RLCTLS21ra} do not. The geometric interpretation of $a_1,a_2,a_3$ does hold; $S_{21d}$ won't be at the origin, but will be at $a_1/a_3$ and $S_{21}(\omega_0)=a_2$. Since these two points will always be on opposites sides of the circle, despite its center in the complex plane, one can derive a useful expression for the radius of the Q-circle:
    \begin{equation}
        R_{Q-circle}=\frac{1}{2} \vert a_2 - \frac{a_1}{a_3} \vert 
    \label{eqn:RQcircle}
    \end{equation}
    \par At this point one has everything needed to give a baseline process for estimating the Q-circle parameters from a VNA scan. One can determine the point closest to $S_{ij}(\omega_0)$ relatively easily from just finding the point with maximum or minimum $\vert S_{ij} \vert$ depending on if it is a reflection or transmission parameter; $a_2$ is simply this complex point. The detuned point can be estimated by taking the two edge frequency points from the scan (the ones closest to $\omega \rightarrow \pm \infty$) and averaging them together:
    \begin{equation}
        S_{ij}(\omega \rightarrow \pm \infty)\approx \frac{1}{2}(S_{ij}(\omega_{max})+S_{ij}(\omega_{min}))
    \label{eqn:SdetunedScan}
    \end{equation}
     This gives you $a_1/a_3$. From there, the center of the Q-circle should be the average of these two points coordinates in the complex plane:
    \begin{equation}
        X_{center}\approx \frac{1}{2}(S_{ij}(\omega_0)+S_{ij}(\omega \rightarrow \pm \infty))
    \label{eqn:CircleCenterScan}
    \end{equation}
    The radius can then be calculated from Equation \ref{eqn:RQcircle}. There isn't an intuitive geometric way of getting $a_3$ alone from the circle, but if one takes a Lorentzian fit of the magnitude data, they now have an estimate. This forms a nice set of initial guesses to feed into a complex curve fitting process.
    \par In terms of using this for a fitting process, there are many routes one can take to implement this geometric interpretation. If the goal is simply to measure $Q_0$, there's a relatively simple way: perform a complex fit of an $S_{21}$ scan to Equation \ref{eqn:RLCTLS21rat}, solving for the 3 coefficients (using the estimates as initial parameter values). Once done, simply take the imaginary part of $a_3$ to get a revised $Q_L$. Next, you can either determine $\beta_1$ and $\beta_2$ from a reflection Lorentzian measurement as described earlier, or you can also perform a complex fit to Equation \ref{eqn:RLCTLS21rat} on the reflection Q-circles through the same process. In this case, one uses their $a_{2i}$ from each reflection fit on ports $i=1,2$ to get  $\beta_i$:
    \begin{equation}
        \beta_i=\frac{1-\vert a_{2_{ii}}\vert}{\vert a_{2_{11}}\vert+\vert a_{2_{22}}\vert}
    \label{eqn:betacomplex}
    \end{equation}
    and from there Equation \ref{eqn:Qloaded} still applies for finding $Q_0$. Combining all of this together in terms of fit parameters:
    \begin{equation}
        Q_0=Im(a_{3_{21}})\left(1+\frac{2-\vert a_{2_{11}}\vert-\vert a_{2_{22}}\vert}{\vert a_{2_{11}}\vert+\vert a_{2_{22}}\vert}\right)
    \label{eqn:Q0complex}
    \end{equation}
    \par One can take this much further if they wish to quantify more specifics about the system. Further analysis in ref. \cite{Qcircle1} gives you exact more exact expressions for the diameters of each S-parameter's Q-circle; this is useful for quantifying the losses in the measurement setup. Inverse mapping techniques outlined in Ref. \cite{QmethodsReview} can give you good values for $\phi$ and $X$ to quantify phase-delay and cross-talk; this reference also contains several other fitting techniques of note.
    \subsubsection{Short note on Impedance Mismatches}
    \par Impedance mismatches are also often common in measurement setups between cables, connectors, and antennae. This can usually be diagnosed by just looking at the magnitude of a reflection measurement, $\vert S_{11} \vert$ or $\vert S_{22} \vert$: A well-matched setup will have a constant response off-resonance, whereas the presence of standing waves indicates an impedance mismatch. In the complex plane, this will manifest itself in a second Q-circle in the off-resonance region, near the detuned point, whose center does not coincide with the center of the Smith circle; this is often called a coupling loss circle. See Ref. \cite{Qcircle2,Qcircle1} for more details. 
    \begin{figure*}[htb!]
    \centering
    \includegraphics[angle=0, width=0.4\linewidth ]{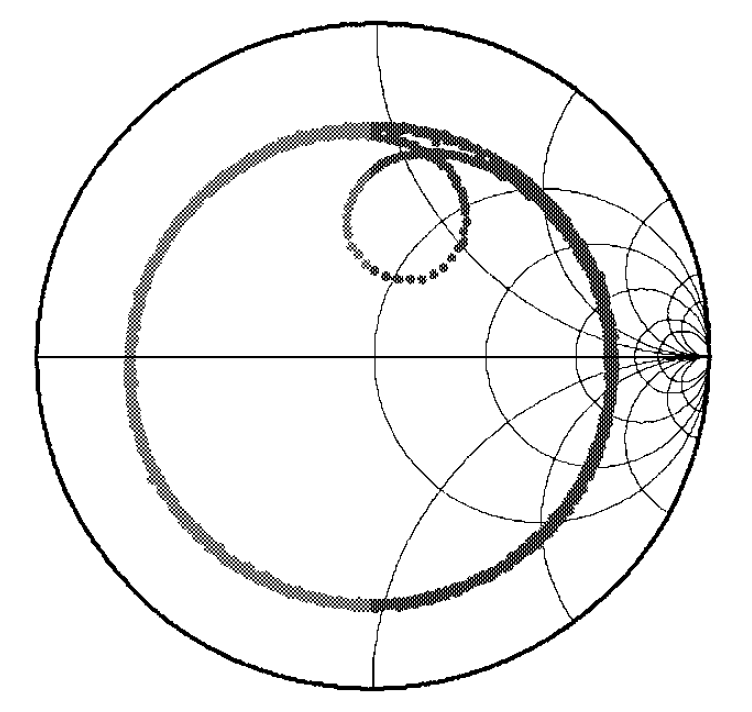}
    \caption{A Q-circle with an impedance mismatch \cite{Qcircle2}.}    
    \label{fig:QcircleImpedanceMismatch}
    \end{figure*}
\section{High frequency axion Searches}
    \subsection{Motivation: the problems at high frequencies}
    Upon further examination of Equation \ref{eqn:ScanRate2}, it is apparent that many of the experimental variables depend on the search frequency. Most notably, since $f_{TM_{010}} \propto 1/R$, one will have to create a smaller cavity for higher frequency searches. Since $\frac{df}{dt} \propto V^2$, this would imply that volume contributes a factor $f^{N_V}=f^{-4}$ slow down in scan rate! But this isn't the only factor that will degrade at higher frequencies; Equation \ref{eqn:sigmaanomCU} implies that the conductivity of copper cavities will degrade the $Q$ as the anomalous skin depth grows at higher frequencies. One can show that $Q \propto f^{-2/3}$ for a classic copper cavity, and since Q is linear with the scan rate, this means a power of $N_Q=-2/3$ added to the scan rate slow-down. Finally, although currently ADMX operates above the standard quantum limit for system noise temperature, this noise level grows with frequency such that $T_{sys} \propto f^1$. This results in a $N_{T_{sys}}=-2$, because $\frac{df}{dt} \propto T_{sys}^{-2}$. One is saved a tiny bit by the fact that scan rate explicitly depends on frequency, $\frac{df}{dt} \propto f^{2}$, but all in all this isn't enough and the total frequency dependence can be written:
    \begin{equation}
        \frac{df}{dt} \propto f^{N_f+N_{T{sys}}+N_Q+N_V}= f^{2-2-(2/3)-4}=f^{-14/3}
        \label{eqn:dfdtslowdown}
    \end{equation}
    Clearly, the biggest problem here is the degradation from the volume drop off, and any high frequency search needs to resolve or mitigate the $N_V=-4$ factor. Future ADMX systems, such as run 2 and ADMX EFR, do this by having multiple smaller cavities whose total volume is comparable to the current ADMX main volume. A naive estimate for the number of cavities needed can be made with the expression, $N(f)=(f/f_0)^2$, where $f_0$ is the frequency of the original single cavity system. However this assumes cavities of the same length and ignores the wasted space in the magnet bore due to packing multiple cavities together. Using Table \ref{tab:EmptyCavities}, one can see that going from the ADMX-main single cavity to the run 2A frequency range implies $N \approx 6$ is necessary to keep the same volume; however this doesn't account for the tuning rod perturbation. Run 2 will end up being a 4 cavity system.
    \par Beyond the geometry, the trick with these systems is how and when one combines these separate cavity signals. If one does not care about matching these cavities signals coherently, they then constitute $N$ independent haloscope experiments, and the total signal power of each cavity must be combined in quadrature such that $P_{sig,N}=\sqrt{N}P_{sig,0}$. However, if the cavity signals can be combined in phase, one can simply sum the powers $P_{phase-matched}= N P_{sig,0}$, offering the potential of a $\sqrt{N}$ improvement. Original prototype multi-cavity systems were to power combine cavity signals before the primary amplifier to minimize the number of RF receivers for the system. This obviously also requires resonant frequencies of these cavities to be matched when combining. A mismatch in summing each cavity's Lorentzian curve will have the effect of spreading it out, lowering the net quality factor of the system. Tuning each cavity adds to the complexity of this matching process, requiring some sort of cavity frequency "locking" protocol to make sure everything is stable before a net digitization would take place.
    \par For these reasons and more, it was settled on having each cavity have its own RF receiver. Data for each cavity is stored separately, and they are combined digitally in the analysis. This gives each cavity the flexibility of tuning through its range at its own pace, not being slowed down by the other cavities, or additional calibration processes. There's still the possibility of phase-combining their separate signals digitally, trying to get at the $\sqrt{N}$ improvement for at least some of the run. The only down-side to this route is the number of amplifiers, cables, and hardware also increases with $N$, requiring more fridge space and generally increasing the number of things to be maintained on site.
    \subsection{ADMX Run 2}
\begin{figure*}[htb!]
    \centering
    \begin{subfigure}{}
        \includegraphics[angle=0, width=0.6 \linewidth]{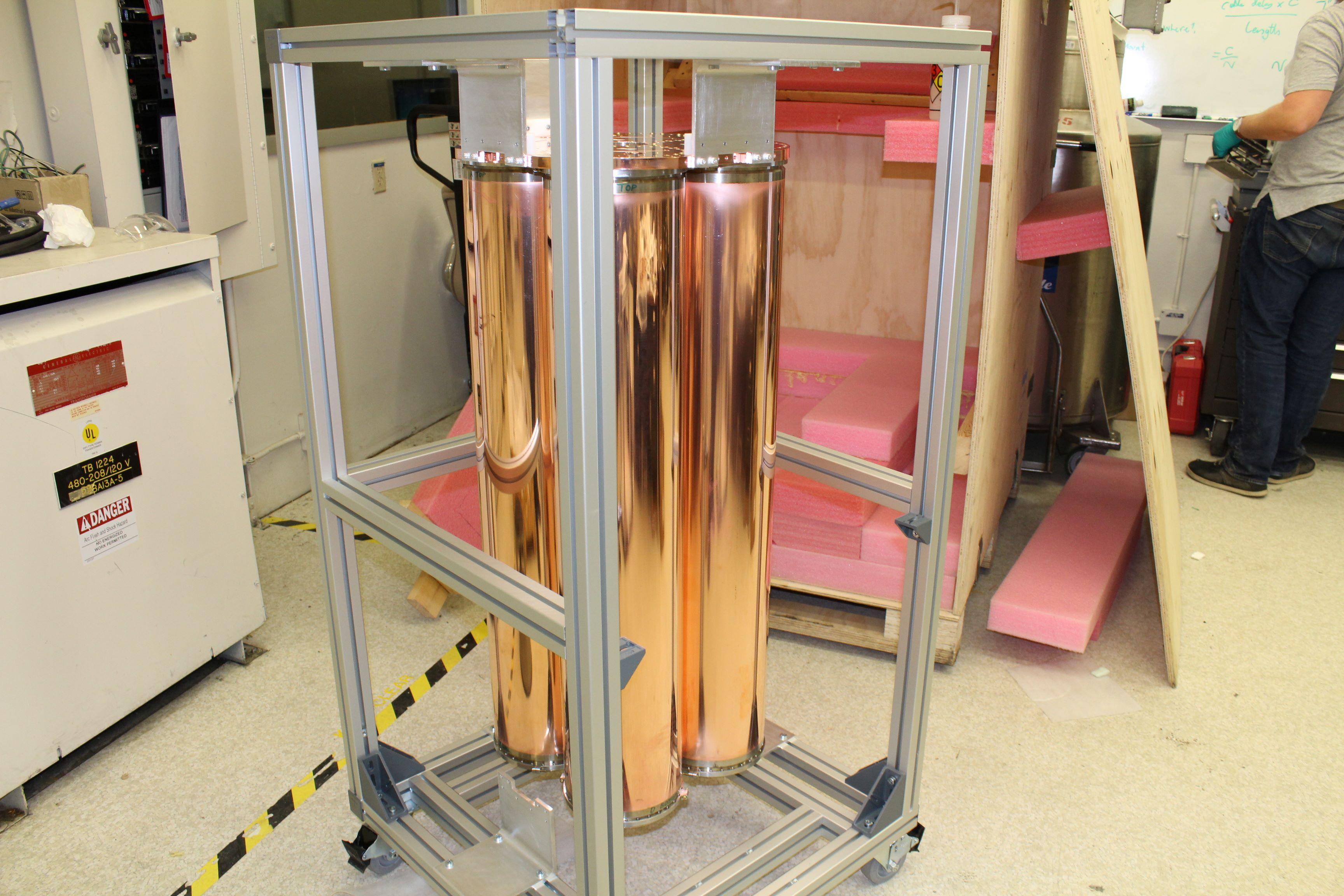}
    \end{subfigure}
    \begin{subfigure}{}
        \includegraphics[angle=0, width=0.6\linewidth]{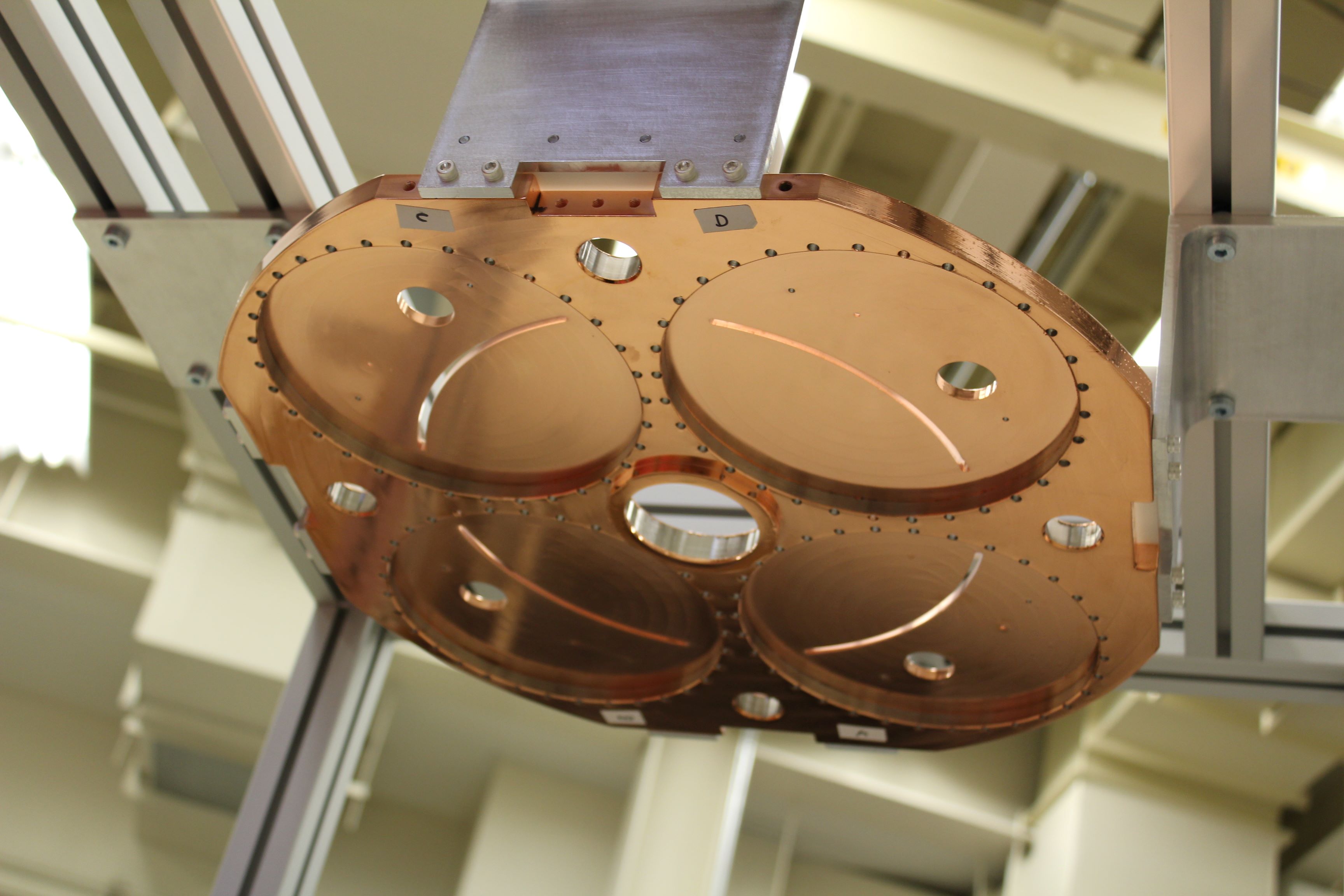}
    \end{subfigure}
    \caption{The run 2 cavity system assembled at LLNL.}
    \label{fig:ADMXRun2A}
    \end{figure*}

    ADMX run 2 is a 4-cavity array designed to cover the $1.3-2 \; {\rm GHz}$ range. Similar to the ADMX-main, each tube has a knife-edge on its end that is clamped to a recessed area on the end cap plate via a stainless steel collar as pictured in Fig \ref{fig:ADMXRun2A}.
    Each is copper-plated in the same manner as ADMX-main. As shown in Figure \ref{fig:ADMXRun2A}, the original prototype had all four cavities attached to the same end cap plate, with a joint tuning rod system. This system would have one coarse tuner that turned 4 tuning rods all at once, as well as 4 individual fine tuners that would allow for adjustment of the cavity frequencies within $1 \,{\rm kHz}$ for locking them together. However, upon switching from an analog to digital combining system, this was changed to 4 different cavity end plates and tuners that join together. This cuts down on the tolerance requirements for the end plate, and allows for each cavity to be assembled and tuned individually.
    \begin{figure*}[htb!]
    \centering
    \begin{subfigure}{}
        \includegraphics[angle=0, width=0.5\linewidth ]{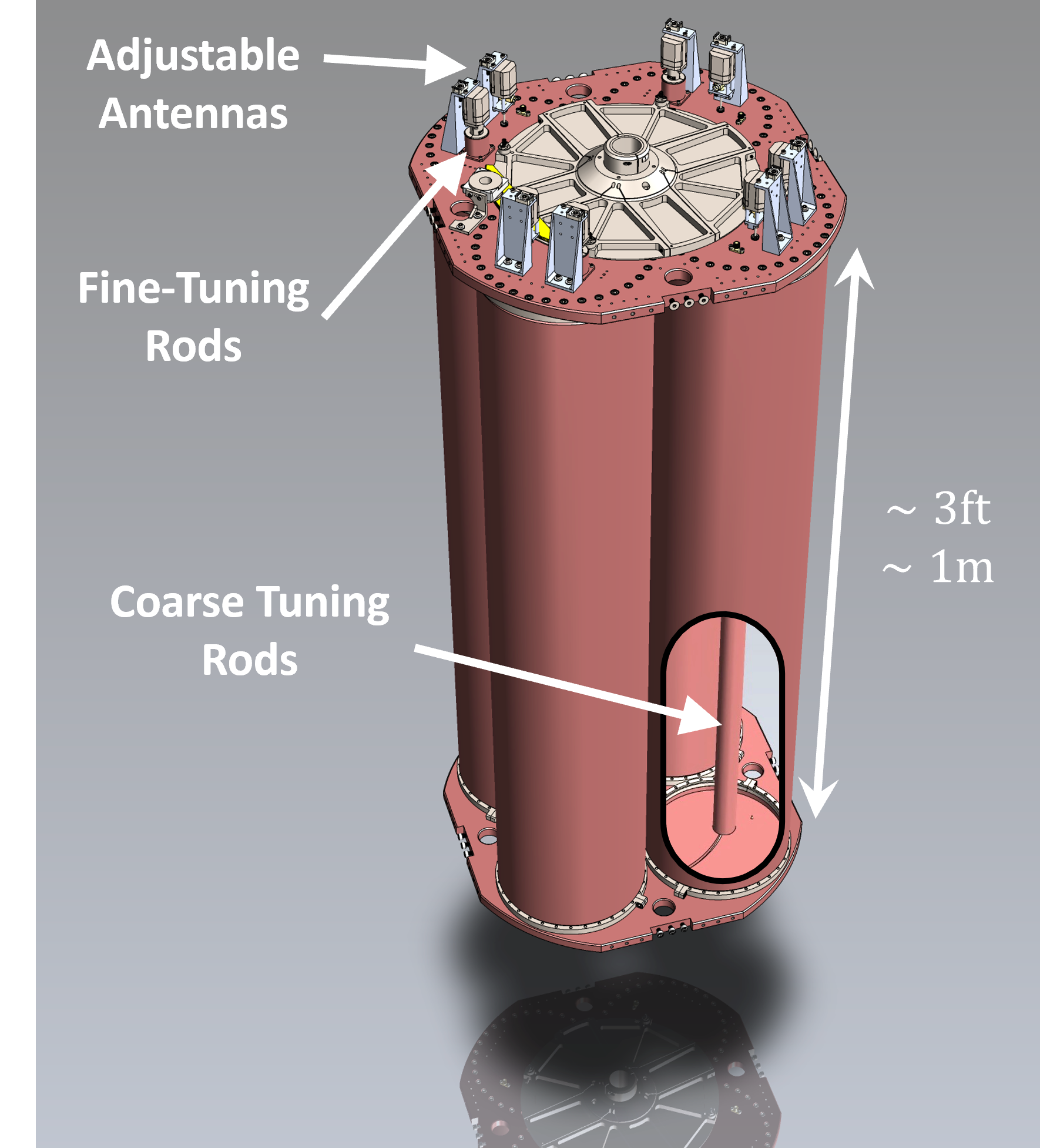}
        \caption{The run 2A cavity array original coarse tuner design.}    
    \end{subfigure}
    \label{fig:ADMXRun2CADassembly}
    \end{figure*}
    It has an overall packing footprint such that it can operate in the current ADMX magnet bore at UW, although it will still require many changes to the site and insert to accommodate the necessary electronics. This includes converting the current stepper motors and gearboxes into an 8 piezoelectric motor system, covering each tuning rod and strong port antenna. Additionally, 4 separate RF cryogenic receiver chains, instead of 2, need to be able to fit within the system. Because of these space requirements, ADMX Sidecar will also be retired before this system is commissioned. The overall system has a volume of $85 \, {\rm L}$, which is a slight reduction (62.5\%) from ADMX run 1 volume of $136 \, {\rm L}$. Table \ref{tab:EmptyCavities} lists the empty cavity frequency characteristics.
    \par
\begin{figure*}[htb!]
    \centering
    \includegraphics[angle=0, width=1\linewidth]{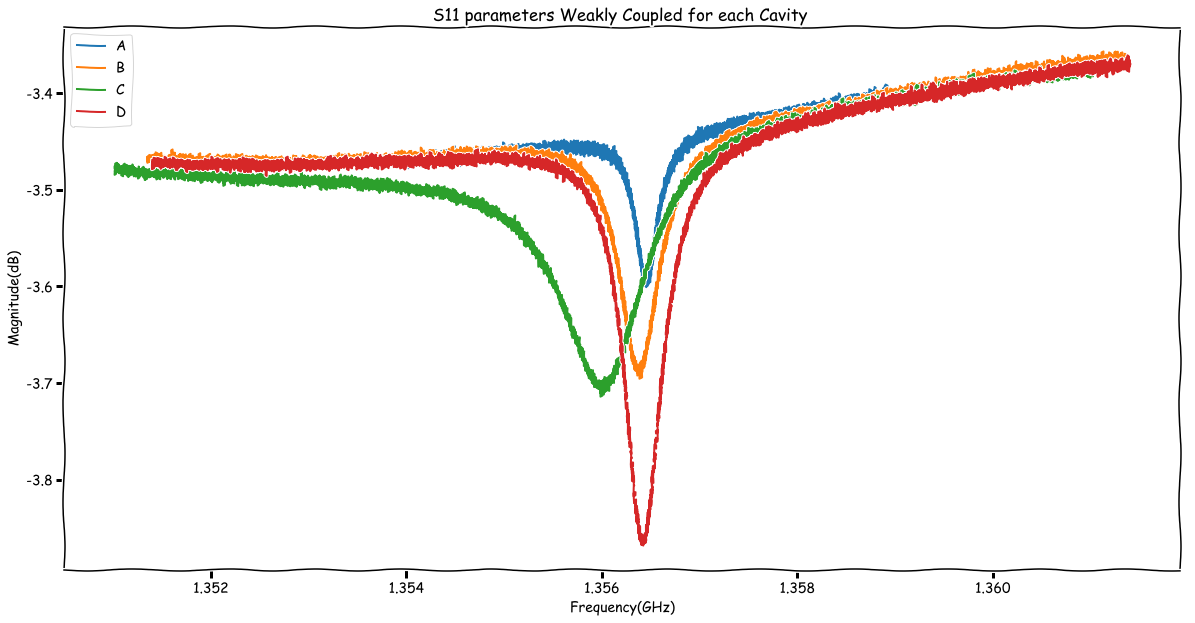}
    \caption{Comparison of the $TM_{010}$ mode weakly coupled reflection measurement for each cavity of the run 2 system. Drifts in the dip are due slightly different couplings. Only cavity C had an appreciably different frequency $>1 \, {\rm kHz}$.}
    \label{fig:Run2Aemptysweeps}
    \end{figure*}

 The system was first assembled at LLNL by myself and former LLNL staff scientist Nathan Woollett, where we also measured the empty cavities. The previous plots, Figure \ref{fig:antennasweepR} and \ref{fig:antennasweepT}, were generated from a single cavity of this system. Figure \ref{fig:Run2Aemptysweeps} shows the empty reflection measurement of each cavity, with the frequency drift due to the difference in machining between them. The appreciable difference in Cavity C's frequency partially motivated the switch to separate tuning systems, because it was a larger frequency drift than the original fine tuning system could handle. It has since been shipped to Fermilab where it resides currently for cryogenic testing. It will be shipped to UW for commissioning in the insert after the conclusion of run 1D. 
    \subsection{ADMX EFR}
\begin{figure*}[htb!]
    \centering
    \includegraphics[angle=0, width=1\linewidth]{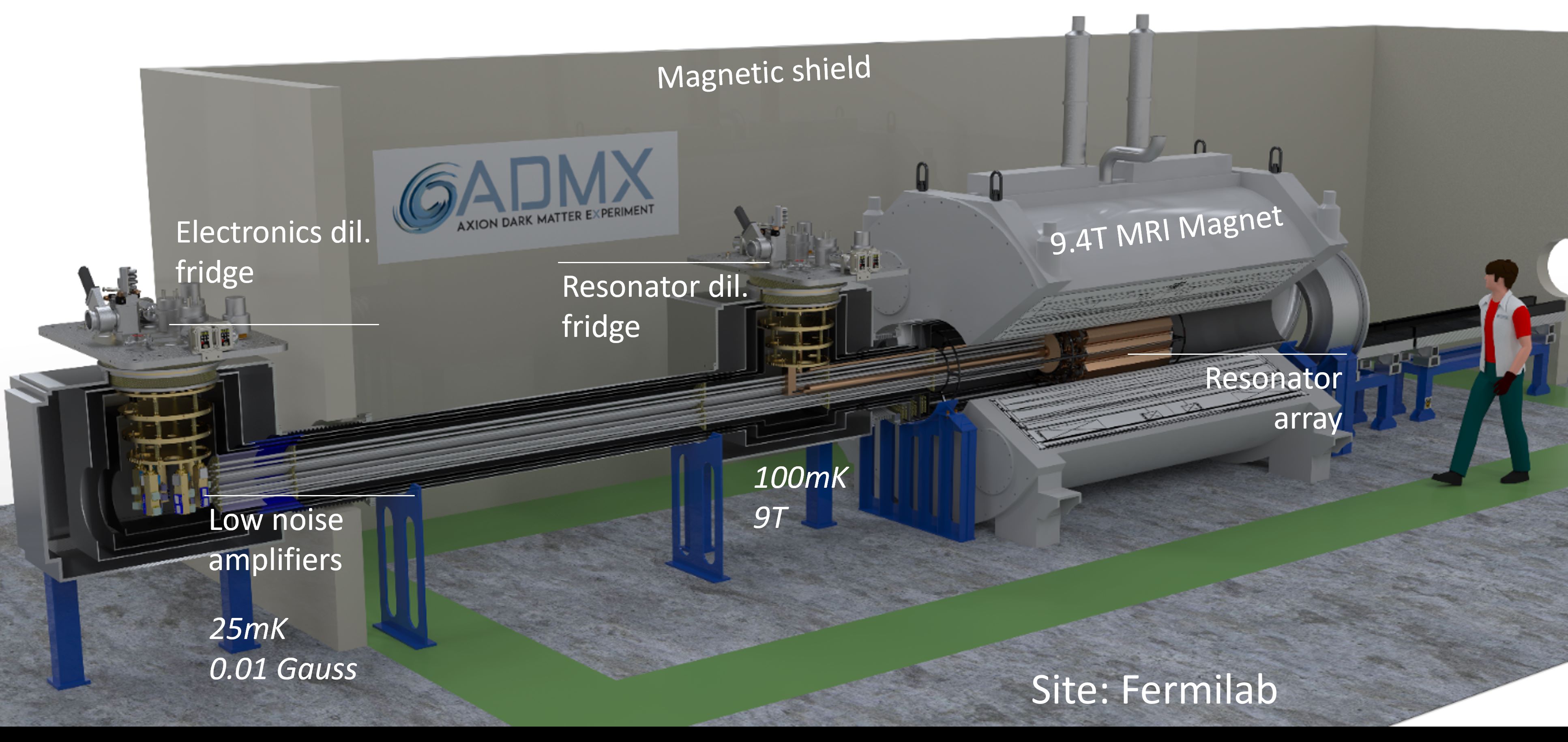}
    \caption{Diagram of the new EFR site at Fermilab. The resonator system and amplifier systems are stored and cooled separately. This allows the cold electronics to be cooled further and minimize the magnetic shielding requirements from the $9.4 \, {\rm T}$ magnet. The system is horizontal to accommodate the MRI magnet design.}
    \label{fig:EFRsitediagram}
    \end{figure*}

    ADMX Extended Frequency Range is an 18-cavity array that will cover the $2-4 \,{\rm GHz}$ range to DFSZ sensitivity. The goal is to have a scan speed that is 5x greater than the current ADMX, which makes siting it at UW in the current magnet bore not possible. Instead, ADMX-EFR will be located in its own building at Fermilab with a completely new site design, outlined in Figure \ref{fig:EFRsitediagram}. 
    \par A $9.4 \, {\rm T}$ research MRI magnet manufactured by GE Medical Systems has been requisitioned from University of Illinois Chicago for the project. With a 80 cm bore allowing for more volume, this means a 47\% increase in $B_0V_{bore}$ over the current ADMX magnet. If this volume increase is fully utilized for cavities, one expects an increase in scan rate by $\approx 2.65-4.1 \times$ depending on what operating field value one chooses for the new magnet, $\approx 8.4-9.4 \, {\rm T}$. Because this an MRI magnet, the entire system must be horizontal as pictured in Figure \ref{fig:EFRsitediagram}.
    \par The other major change to the site construction is the separation of the cold electronics and resonator cooling system. One dilution refrigerator will cool the cavity system to $100 \, {\rm mK}$, while the electronics package will be cooled separately to $25 \, {\rm mK}$. This allows the 18 JPA SQUID amplifiers to be stored out of the magnetic field, negating the need for an active bucking coil system; a passive magnetic shield will be used instead. Quantum squeezing techniques will hopefully allow this electronics package to operate below the standard quantum limit. Low-loss coaxial cabling will be critical for connecting this system to the cavities.
    \begin{figure*}[htb!]
    \centering
    \includegraphics[angle=90, width=0.7\linewidth]{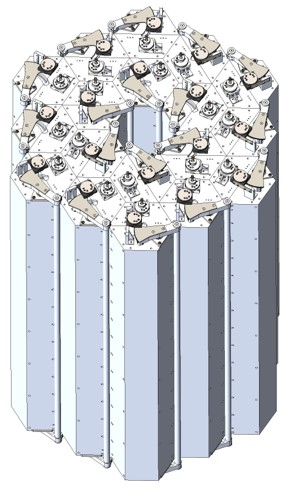}
    \caption{The proposed ADMX-EFR 18 cavity array. Each cavity is a hexagonal clam-shell to maximize packing efficiency and allow for superconducting coatings. Every cavity is topped with a rotor piezoelectric motor and geared tuning arm, as well as a linear piezoelectric motor for individual adjustment of the antenna coupling.}
    \label{fig:EFRcavityarray}
    \end{figure*}
    \par
\begin{figure*}[htb!]
    \centering
    \includegraphics[angle=0, width=0.6\linewidth]{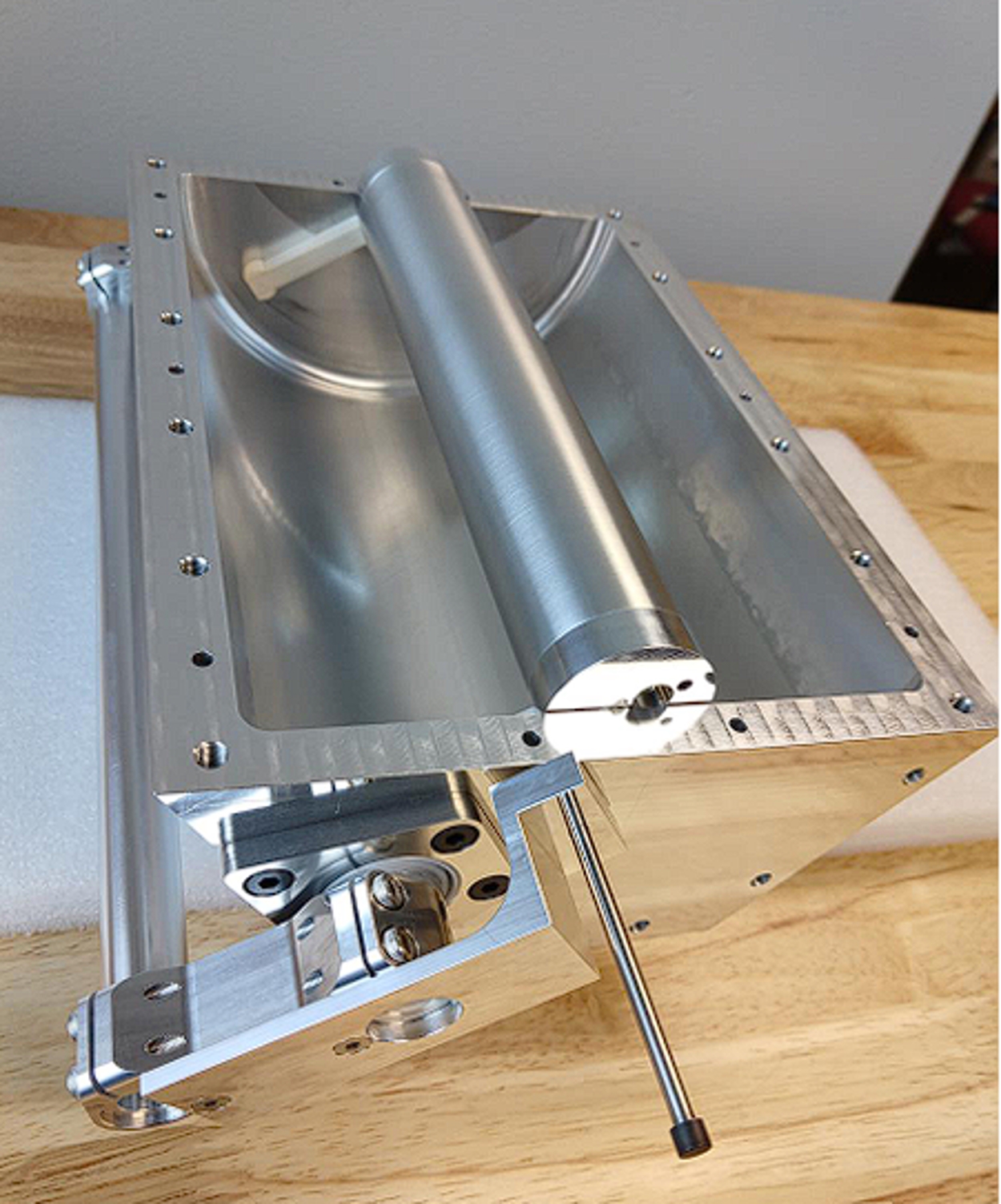}
    \caption{An open aluminum reduced length EFR prototype cavity being tested at LLNL.}
    \label{fig:EFRprototype}
    \end{figure*}

\begin{figure*}[htb!]
    \centering
    \includegraphics[angle=0, width=0.6\linewidth]{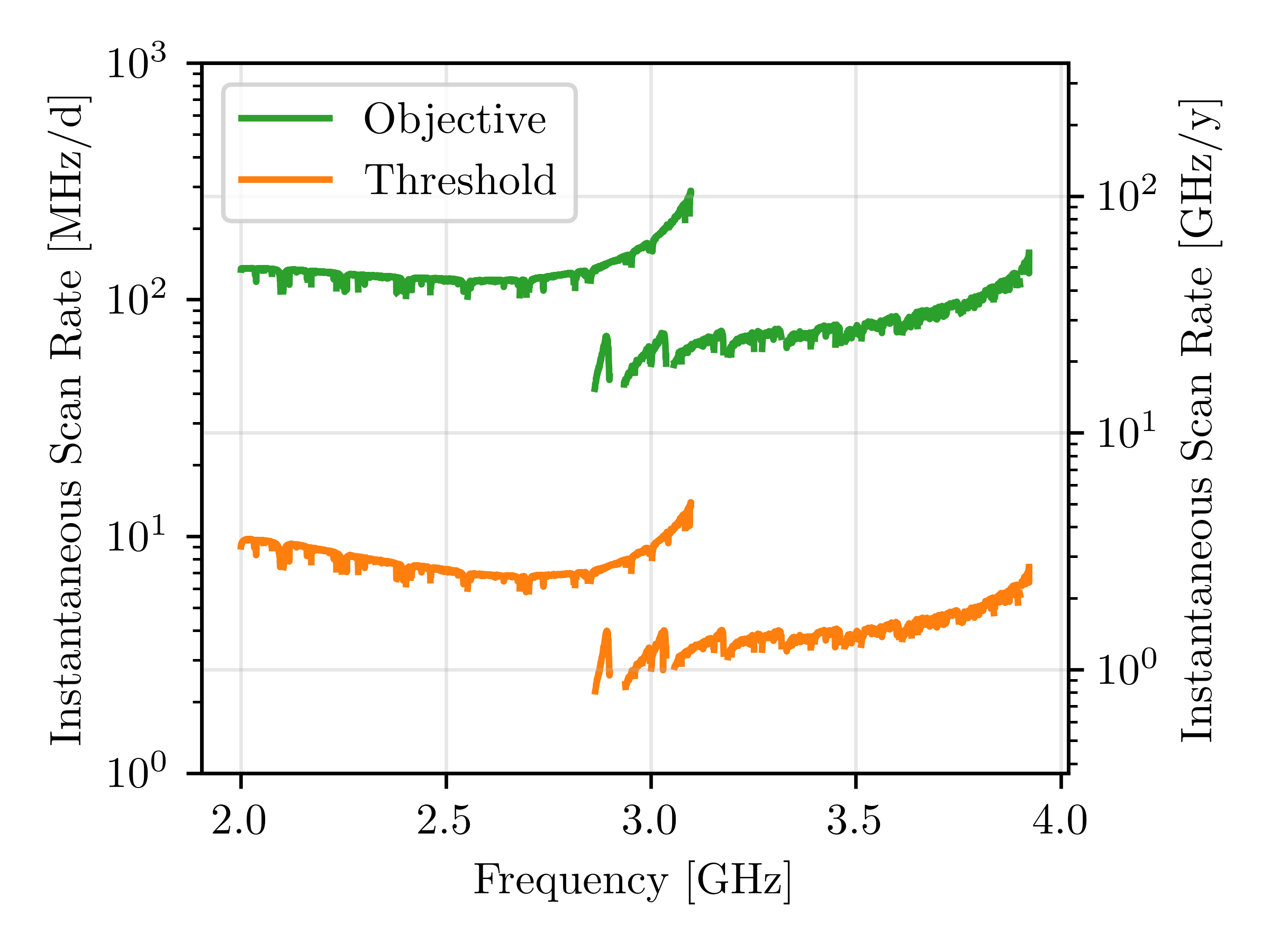}
    \caption{The instantaneous scan rate estimated for full ADMX-EFR run. The orange line represents our conservative estimate \emph{(threshold)}. The green line represents our most optimistic estimate \emph{(objective)} assuming optimal performance of all components, superconducting cavities and squeezing.}
    \label{fig:EFRscanspeed}
    \end{figure*}

 The actual cavity system itself will be 18 "clam-shell" cavities with hexagonal exteriors to maximize the packing volume, pictured in Figure \ref{fig:EFRcavityarray}. The nominal cavity diameter and height are 128 mm and 1 m respectively, with empty cavity characteristics listed in Table \ref{tab:EmptyCavities}. This gives each cavity a loaded volume of $12.1 \, {\rm L}$ for the 2-3 GHz range, and $10.1 \, {\rm L}$ for the 3-4 GHz range; larger tuning rods will be implemented half-way through. This improves the total volume to $182-218 \, {\rm L}$, a 34-60\% improvement over the current ADMX cavity. Currently several reduced length aluminum prototypes, pictured in Figure \ref{fig:EFRprototype} are being tested at LLNL before the design is finalized.
    Considerations are being made for this system to be superconducting; the clam-shell design lends itself well to the coating processes. Additionally, cost-benefit analyses are being done on the efficacy of niobium substrate; the results of this dissertation are informing many EFR decisions! (see Chapter \ref{chap:Sidecar1D}). A projected scan speed of the EFR system is shown in Figure \ref{fig:EFRscanspeed}. Stay tuned for a publication soon on the design and projection of this future ADMX system.
    \subsection{ADMX Orpheus}
\begin{figure*}[htb!]
    \centering
    \includegraphics[angle=0, width=0.7\linewidth]{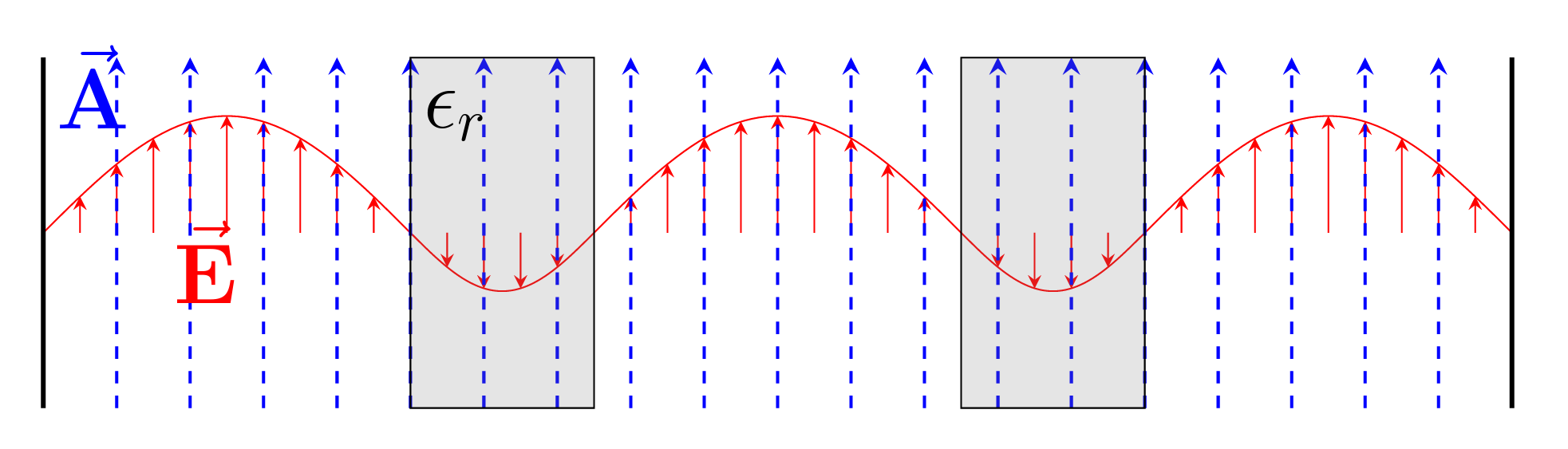}
    \caption{Diagram of a dielectrically-loaded cavity operating at a multi-wavelength mode. The dielectrics (grey boxes) are placed at every anti-node such that they suppress parts of the mode wavelength ($\vec{E}$) so that the axion form factor is non-zero ($\propto \vec{E}\cdot \vec{A}$) \cite{ADMXOrpheus}.}
    \label{fig:OrpheusFormfactor}
    \end{figure*}

    ADMX Orpheus is a path-finder experiment sited at the UW CENPA site that implements an alternative approach to the volume degradation problem; it is a tunable, dielectrically-loaded cavity operating at higher order mode, allowing the detection volume to remain the same. Although I did not work on Orpheus during my time at UW, I believe its important to include in order to emphasize that multi-cavity systems are not the only approach ADMX is taking to tackle high frequency searches. Dielectrics are strategically placed at every 1/4th of a half wavelength of the operating mode to suppress the electric field in those regions as shown in Figure \ref{fig:OrpheusFormfactor}.
    This results in a $\vec{E}\cdot \vec{B} \neq 0$ for the higher order mode. One can think of this as producing a non-zero form factor. However, because Orpheus is a Fabry-Perot cavity, with an open-walled design, typically the form factor is combined with cavity volume and referenced as an effective volume, $V_{eff}$:
    \begin{equation}
    V_{eff}=\frac{\vert \int dV \vec{B_0} \cdot \vec{E} \vert ^2}{B_0^2 \int dV \epsilon_r \vert E \vert ^2}
        \label{eqn:OrpheusVeff}
    \end{equation}
    \par
\begin{figure*}[htb!]
    \centering
    \includegraphics[angle=0, width=0.5\linewidth]{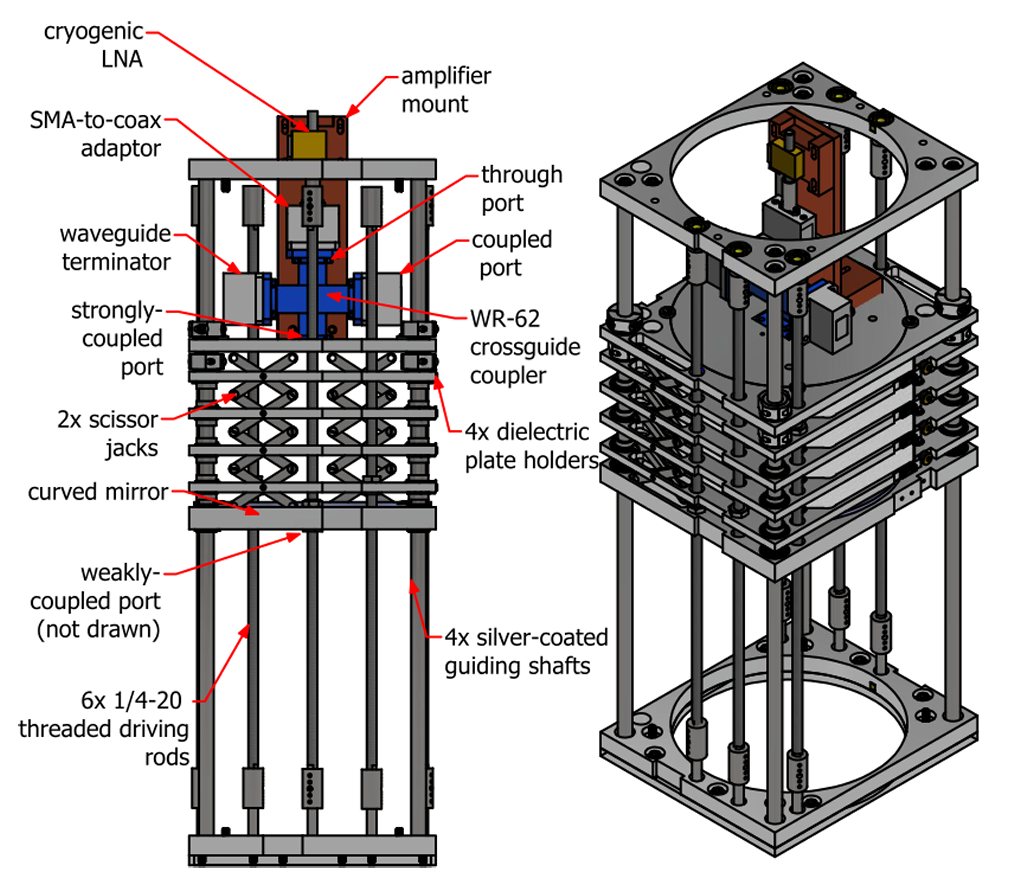}
    \caption{CAD model diagram for Orpheus Cavity \cite{ADMXOrpheus}.}
    \label{fig:OrpheusCAD}
    \end{figure*}

 The model of this Fabry-Perot cavity is pictured in Figure \ref{fig:OrpheusCAD}. Mirrors on either end form the resonator, reflecting the electromagnetic waves back and forth between the two. A small aperture on the top mirror acts as the strong port, and the bottom aperture is the weak port. The operating mode is the $TEM_{00-18}$. The cavity is tuned by adjusting the position of the bottom mirror, changing the length between the mirrors. Scissor jacks evenly adjust the spacing of the dielectric plates as it tunes so that they remain in the right relative locations to create the non-zero $V_{eff}$.
    \par
\begin{figure*}[htb!]
    \centering
    \includegraphics[angle=0, width=0.9\linewidth]{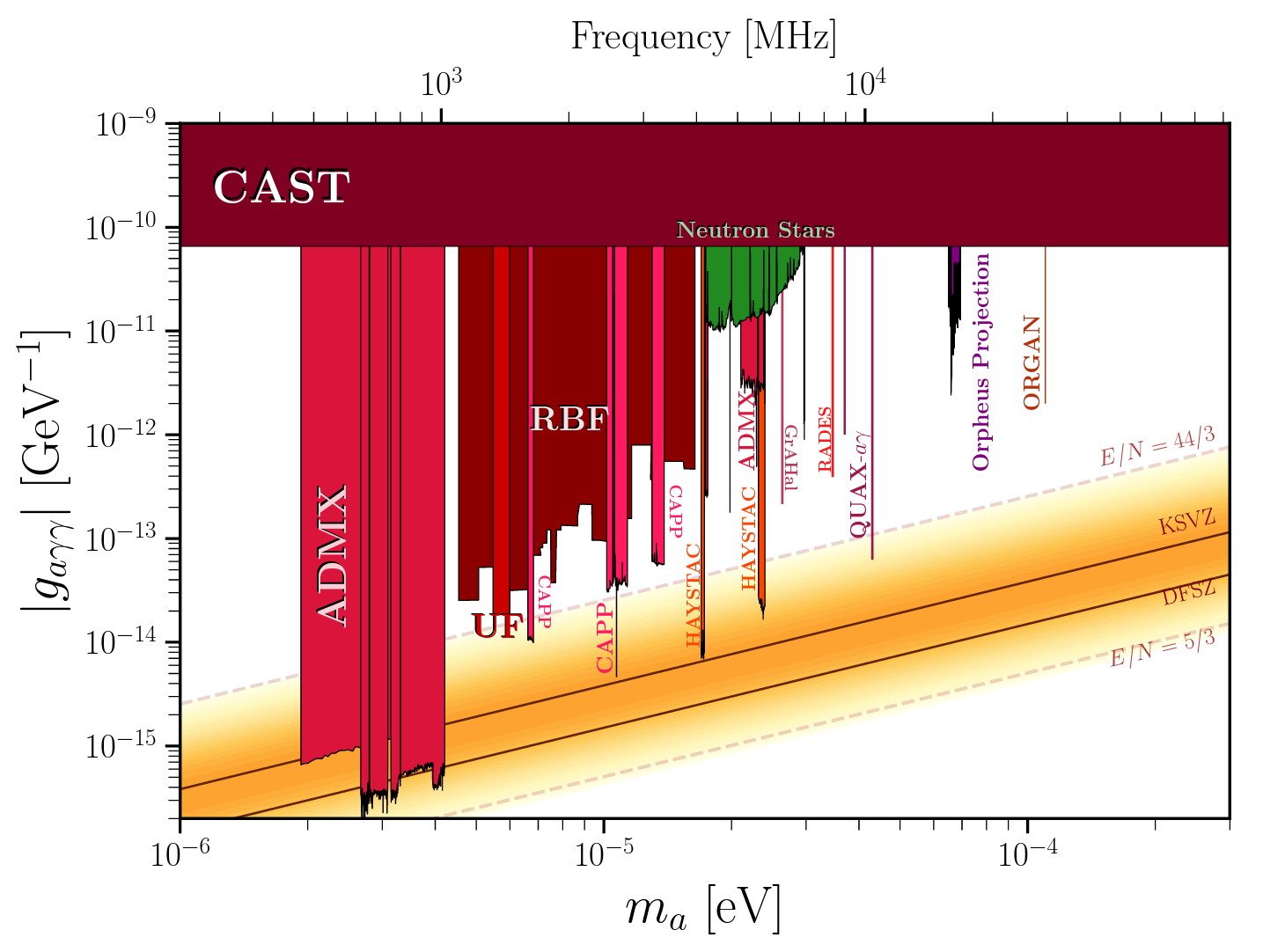}
    \caption{The project Orpheus axion limit assuming a $1.5 \, {\rm T}$ dipole magnet\cite{ADMXOrpheus}. Figure adapted from \cite{o_hare_axionlimitplot}.}
    \label{fig:OrpheusProposedAxionLimits}
    \end{figure*}

 Orpheus has a tuning range from 15.8 GHz to 16.8 GHz. In the summer of 2021, ADMX Orpheus performed a search for dark photons in this range, the results of which can be found in Ref. \cite{ADMXOrpheus}. A $1.5 \, {\rm T}$ dipole magnet is being constructed at UW for an axion search in this same range to a nominal $g_{a \gamma \gamma} \approx 3 \times 10^{-12} GeV^{-1}$, twice the sensitivity as the CAST limit. The magnet design can be found in ref. \cite{RaphaelThesis}. The proposed limits are shown in Figure \ref{fig:OrpheusProposedAxionLimits}. Stay tuned for results!

\chapter{SRF Cavity Theory}
\label{chap:SRF}
Superconducting Radio Frequency Cavities, although challenging to implement on axion haloscopes, are an established tool in the physics community, and the standard for achieving the highest Q cavities in the laboratory. They are essential in the accelerator community for efficiently transferring the large RF power along particle beams \cite{CAS02}. In recent years, they have become a crucial tool for housing and high fidelity read out of qubits for the Quantum Information Science (QIS) community \cite{SQMS}. Within the dark matter community, superconducting cavities have been used to look for dark photons at an incredibly high sensitivity \cite{RaphaelSRF}. The axion community has done several feasibility studies of SRF cavities for a variety of materials over the past few years, particularly NbTi \cite{BraineMultiMode,QUAX1,QUAX2}, $\mathrm{Nb}_3\mathrm{Sn}$ \cite{PosenNbSN}, and YBCO \cite{CAPPYBCO1,CAPPYBCO2,CAPPYBCO3}. The CAPP axion institute will soon be publishing their first axion search results using a YBCO cavity and tuning rod \cite{CAPPPatras}.
\par
\begin{figure*}[htb!]
    \centering
    \includegraphics[angle=0, width=1\linewidth]{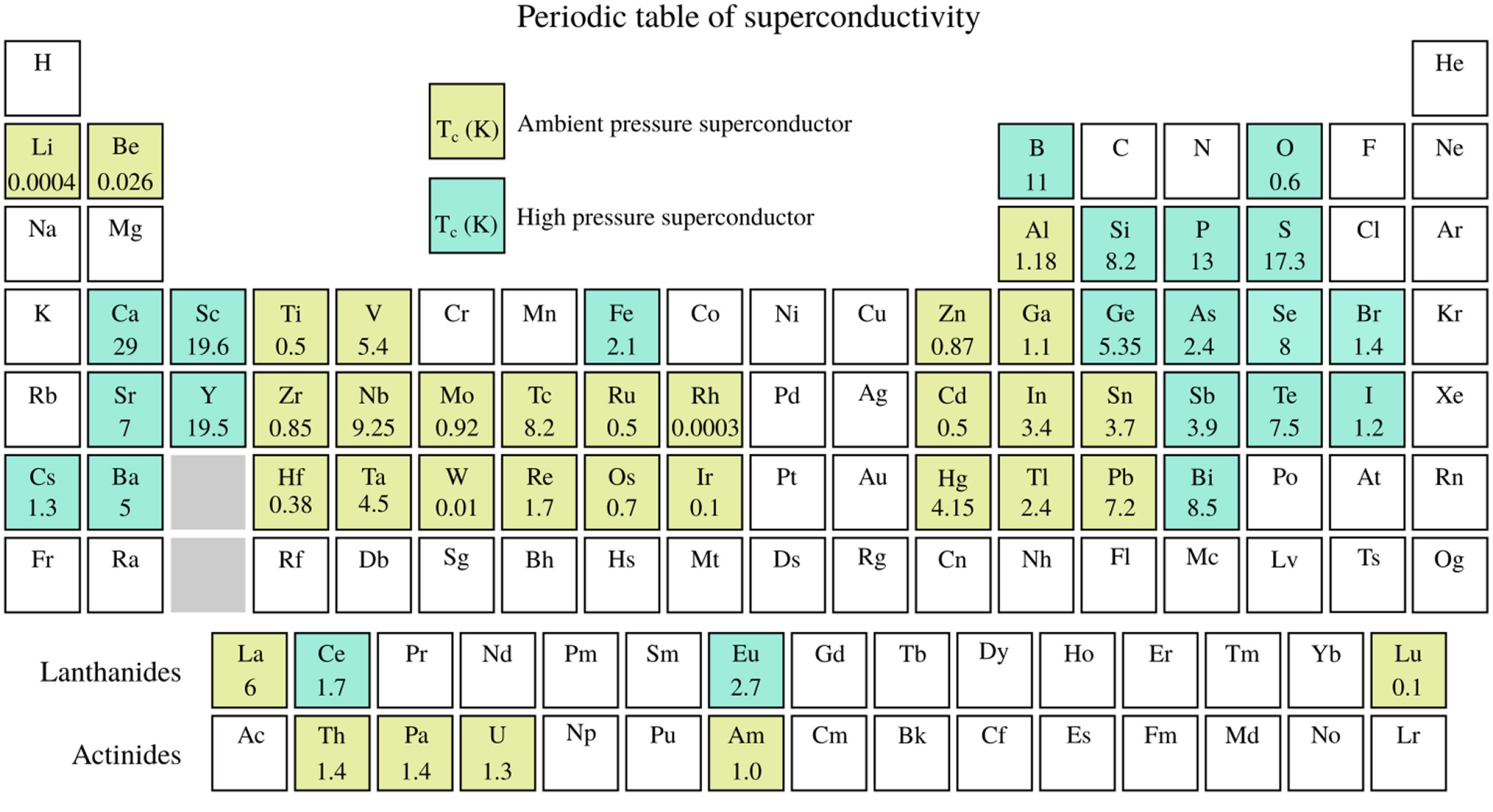}
    \caption{The periodic table of elements with the critical temperature of known superconductors listed. Some elements only exhibit superconductivity above atmospheric pressure, known as high-pressure superconductors \cite{SCperiodicTable}.}
    \label{fig:SCperiodictable}
\end{figure*}

 Superconductivity refers to the property certain solid materials exhibit where internal electrical resistance vanishes. Normal conductors, like copper, exhibit an internal resistance that decreases with temperature gradually until absolute zero. However, superconductors each have a characteristic critical temperature, $T_c$, below which their resistance drops abruptly to zero. This phenomenon was first observed in 1911 by Heike Kamerlingh Onnes, who immersed a solid mercury wire in liquid helium that went superconducting at $4.2 \, {\rm K}$ (he was also the first physicist to liquefy helium in 1908, for which he won the 1913 Nobel Prize in Physics). Since then, a myriad of materials have been shown to be superconductors with critical temperatures ranging from just thousandths of a degree above $0 \, {\rm K}$ to record holding yttrium and lanthanum hydride compounds that range from $T_c\approx 224-288 \,{\rm K}$ \cite{YH6,LH10}. However, this upper limit is routinely pushed to higher levels with the hope of one day having a room temperature superconductor.  Figure \ref{fig:SCperiodictable} shows the critical temperature for elements of the periodic table, although this does not do justice to the vast variety of compounds that exhibit the property \cite{SCperiodicTable}. Throughout this section I will be drawing from two introductory textbooks on superconductivity, Ref. \cite{Parks, Tinkham}.
    \section{The Meissner effect}
    An additional effect of superconductivity is the expulsion of magnetic fields from a superconductor during its superconducting transition. This was first noticed in 1933 by German physicists Walther Meissner and Robert Ochsenfeld when measuring the magnetic field distribution outside samples of superconducting lead and tin, and is now referred to as the Meissner effect \cite{Meissner}. An external magnetic field was applied to samples that were subsequently cooled below their critical temperature, causing the samples' internal magnetic field to drop to zero. They detected this effect through the conservation of magnetic flux; when the internal field decreased, the exterior field around the superconductor increased. This gave a defining feature to superconductors versus a classical perfect conductor for the first time. What is clear, even without a theory of superconductivity, is that the superconductor itself must be generating a magnetization, $\vec{M}$, in response to the external field, $\vec{H}$. As $\vec{H}$ is increased, eventually the superconductor reaches a limit where $\vec{M}$ can no longer cancel out the external field, known as the critical field, $\vec{H}_c$, forcing the material back into a normal state. Being that currents generate magnetic fields, often this is also expressed as a critical current in the case of superconducting wires. This critical field was found empirically to be well approximated by the parabolic relation:
    \begin{equation}
        H_c(T) \approx H_c(0 \,K) \left(1-\left(\frac{T}{T_c}\right)^2\right)
        \label{eqn:HcVsT}
    \end{equation}
    where $H_c(0\, K)$ is the projected critical field at 0 K.
    \par This effect is the heart of the problem for using superconducting cavities in axion haloscopes. Axion haloscopes can only operate effectively in high fields, in excess of several Tesla, as discussed in Chapter \ref{chap:Haloscopes}. For reasons that will be clearer by the end of this chapter, there are no known materials that exhibit perfect superconductivity in excess of around $1 \,{\rm T}$. This means most materials will be forced back into a normal state within a haloscope, where their resistance is higher than copper. This is why ADMX has so far settled on using a high conductivity normal conductor, copper.
    \section{London theory}
    Although superconductivity can only be explained in full with quantum mechanics, the first theory of superconductivity was purely classical in nature. Only two years after the discovery of the Meissner effect, in 1935, Fritz and Heinz London worked out a phenomenological theory of superconductors now known as London theory \cite{London}. The crux of superconductors is that Ohm's law, $\vec{j}=\sigma\vec{E}$, clearly cannot apply because $\sigma=0$. The brothers imagined the superconducting current as free electrons within an external uniform electric field, which would follow the Lorentz force law, $\vec{F}= m \dot{v}=-e\vec{E}+e\vec{v}_s \times \vec{B}$, with $\vec{j}_s=-n_se\vec{v}_s$, a superconducting current density. Assuming the electrons are just being driven by the electric field, one can derive the first London equation:
    \begin{equation}
        \frac{\partial \vec{j}_s}{\partial t}=\frac{n_se^2}{m_e}\vec{E}
        \label{eqn:LondonEqn1}
    \end{equation}
    where $e$ and $m_e$ are the electron charge and mass, and $n_s$ is a number density of charges. The 2nd equation can be derived by applying Faraday's law to the first to obtain:
    \begin{equation}
        \frac{\partial}{\partial t}\left(\vec{\nabla} \times \vec{j}_s + \frac{n_s e^2}{m_e} \vec{B} \right) = 0
        \label{eqn:LondonEqnInt}
    \end{equation}
    They recognized any solutions that were non-zero constants would be non-physical due to the recent discovery of the Meissner effect, and required the time derivative be exactly zero, resulting in the 2nd London equation:
    \begin{equation}
        \vec{\nabla} \times \vec{j}_s = -\frac{n_se^2}{m_e} \vec{B}
        \label{eqn:LondonEqn2}
    \end{equation}
    Notice that this is the key step that differentiates it from a perfect conductor; it can only be said that a perfect conductor has $\dot{B}=0$ within its bulk, whereas the field itself is zero, $\vec{B}=0$, within a superconductor. This means a perfect conductor can hold trapped magnetic flux whereas a superconductor cannot. Often, these two equations can be expressed as one equation in terms of a superconducting vector potential, $\vec{A}_s$:
    \begin{equation}
        \vec{j}_s = -\frac{n_se^2}{m_e} \vec{A}_s
        \label{eqn:LondonEqn3}
    \end{equation}
    of which they defined the 'London gauge' for $\vec{A}_s$:
    \begin{equation}
        \vec{\nabla}\cdot\vec{A}_s=0 ,\; \vec{A}_s \cdot \hat{n}=0 
        \label{eqn:LondonGauge}
    \end{equation}
    where $\hat{n}$ is the surface normal vector of the superconductor. Additionally, they required $\vec{A}_s=0$ within the superconducting bulk. This becomes clearer by taking the curl of Equation \ref{eqn:LondonEqn2} resulting in a Helmholtz equation:
    \begin{equation}
        \nabla^2\vec{B}=\frac{1}{\lambda_s^2} \vec{B}
        \label{eqn:LondonHelmholtz}
    \end{equation}
    where $\lambda_s$ is the London penetration depth,
    \begin{equation}
        \lambda_L=\sqrt{\frac{m_e}{\mu_0 n_s e^2}}=\sqrt{\frac{m_ec^2}{4 \pi n_s e^2}}
        \label{eqn:LondonPenetrationdepth}
    \end{equation}
    where the second definition in terms of $c$ is more common. This name makes more sense when one solves Equation \ref{eqn:LondonHelmholtz} for the case of an infinite plane superconducting bulk sitting in a constant external magnetic field, $B_0$, oriented parallel to the surface of the superconductor along the $\hat{z}$. Defining $\hat{x}$ to be the perpendicular direction where $x<0$ is free space and $x>0$ is within the superconductor, one can solve for $B_z$ within the superconductor to be:
    \begin{equation}
        B_z(x)=B_0e^{-x/\lambda_L}
        \label{eqn:LondonPenetrationdepthBz}
    \end{equation}
    Hence, this is the length scale by which an external field penetrates within a superconductor before it is exponentially suppressed. Typical values are 50 to 500 nm with $\mathrm{Nb}_3\mathrm{Sn}$ having a value of about 160 nm. One can reason from this that the super-currents must also flow near the surface of the superconductor in order for this magnetic field distribution to hold. This theory triumphs at giving a clear classical picture of the Meissner effect and surface super-currents, yet it still fails at a microscopic explanation of superconductivity. Even more so, the charge carriers in this theory, electrons, being fermions and bound to the Pauli exclusion principle, pose problems when developing a quantum theory of superconductivity.
    \section{Ginzberg-Landau theory}
    The 2nd phenomenological theory of superconductivity was developed by Soviet physicists Lev Landau and Vitaly Ginzburg, first published in 1950 \cite{Landau1}, who approached the problem with quantum mechanics in mind. Lev Landau made use of his theory of 2nd order phase transitions, published in 1937 \cite{Landau2}, describing the free energy density of the superconductor, $f_s$, in terms of a complex order parameter field, $\vec{\psi}$, analogous to a wave function; the quantity $\vert \psi \vert ^2$ is a measure of the local density of superconducting electrons, $n_s$. As Landau's explanation of super-fluidity in liquid helium was only years earlier, in 1941, their thought was the electrons in a superconductor form a super-fluid, and this would represent the fraction of electrons that had condensed into this state. They assumed $\vert \psi \vert$ would be non-zero below the transition temperature, $T_c$, but also small with small fluctuations ($\vert \nabla \psi \vert$ is small), which actually holds well for most low-temperature, conventional superconductors we deal with in this dissertation, but not the more exotic modern high-temperature (HTC) superconductors. This assumption allows one to assume the form of the free energy density to have the form of a field theory exhibiting a U(1) gauge symmetry:
    \begin{equation}
        f_s=f_n+\alpha \vert \psi \vert ^2 + \frac{1}{2} \beta \vert \psi \vert ^4 + \frac{1}{2m^*} \vert \left(-i\hbar \vec{\nabla} -\frac{e^*}{c}\vec{A}\right) \psi \vert ^2 +\frac{\vec{B}^2}{8\pi}
        \label{eqn:GLFreeEnergy}
    \end{equation}
    where $\alpha$ and $\beta$ are the phenomenological parameters of the theory, $f_n$ is the free energy density of the normal conductor state, $m^*$ and $e^*$ are the effective charge and mass of the superconducting current carrier, and $\vec{B}$ and $\vec{A}$ correspond to the external magnetic field and its vector potential. One wants to minimize the total free energy, $F=\int f_s d^3r$, with respect to variations in the order parameter, $\psi$, which gives the first Ginzburg-Landau (GL) equation:
    \begin{equation}
        \alpha\psi+ \beta \vert \psi \vert ^2 \psi + \frac{1}{2m^*}\left(-i\hbar \vec{\nabla} -\frac{e^*}{c}\vec{A}\right)^2\psi=0
        \label{eqn:GL1}
    \end{equation}
    One can then use Ampere's law to derive the superconducting current in the presence of the external field to be:
    \begin{equation}
        \vec{\nabla}\times \vec{B} = \frac{4 \pi}{c} \vec{J} \; ; \; \vec{J}= \frac{e^*}{m^*} Re \left\{ \psi^* \left(-i\hbar \vec{\nabla} -\frac{e^*}{c}\vec{A}\right) \psi \right\}
        \label{eqn:GL2}
    \end{equation}
    This is referred to as the 2nd GL equation. A student of quantum mechanics will notice how this is remarkably similar to the time-independent Schrodinger equation with the exception of the non-linear term, $\beta \vert \psi \vert ^2 \psi$. In the absence of an external field this simplifies greatly:
    \begin{equation}
        \alpha\psi+ \beta \vert \psi \vert ^2 \psi=0
        \label{eqn:GLNofield}
    \end{equation}
    This equation has the trivial solution $\psi =0$, which would correspond to the normal conducting state, $T>T_c$, but below the transition temperature one expects a non-trival solution of the form $ \vert \psi \vert ^2= - \alpha/\beta$. This means in order for a real non-trivial solution to exist below $T_c$, but not above, one can assume the functional form of $\alpha$ to be $\alpha(T)=\alpha_0(T-T_c)$ where $\alpha_0$ must be positive. The linear dependence can be made here under the assumption that one is near the transition temperature. Note this also requires $\beta$ to be positive and temperature independent. Hence the order parameter is zero close to the transition, which is ideal for second order phase transitions. Therefore in the superconducting state absent of magnetic fields, $ \vert \psi \vert ^2$ is :
    \begin{equation}
        \vert \psi \vert ^2 = -\frac{\alpha_0 (T-T_c)}{\beta}
        \label{eqn:GLNofieldpsi}
    \end{equation}
    Considering the 1-dimensional case, where we use a normalized wave function, $s(x)=\psi(x)/\vert \psi \vert ^2 $, on the first GL equation, \ref{eqn:GL1}, for the case of zero field, $\vec{A}=0$, one gets a differential equation of the form:
    \begin{equation}
        \xi^2 \frac{d^2s(x)}{dx^2}+s(x)-s(x)^3=0
        \label{eqn:GL1Dnofield}
    \end{equation}
    where,
    \begin{equation}
        \xi = \sqrt{\frac{\hbar^2}{2m^*|\alpha|}}
        \label{eqn:GLcoherencelength}
    \end{equation}
    This equation shows how $\xi$ is a length scale over which the superconducting order parameter will fluctuate; this is referred to as the coherence length of the superconductor. Note that it will have a temperature dependence of ~$(1-T/T_c)^{-1/2}$. 
    \par If we return to the 2nd GL equation, Equation \ref{eqn:GL2}, considering the non-zero field case, one can perform the same analysis as in the London theory: apply the curl to the equation to get the form $\nabla^2\vec{B}=\lambda^{-2}\vec{B}$. This gives us the penetration depth for GL theory to be:
    \begin{equation}
        \lambda_{GL}=\sqrt{\frac{m^*c^2}{4\pi e^{*2}\vert\psi\vert^2}}=\sqrt{\frac{m^*c^2\beta}{4 \pi e^{*2}|\alpha|}}
        \label{eqn:GLpenetrationdepth}
    \end{equation}
    This similarly has the ~$(1-T/T_c)^{-1/2}$ temperature dependence. In principle, this should be equal to the London penetration depth, Equation \ref{eqn:LondonPenetrationdepth}, and is commonly just referred to as $\lambda$. The ratio of these two length scales, $\lambda$ and $\xi$ ,are critical for understanding the behavior of the superconductor in fields, and one can see that it must be temperature independent:
    \begin{equation}
        \kappa=\frac{\lambda}{\xi}=\frac{m^*c}{e^* \hbar}\sqrt{\frac{\beta}{2 \pi}}
        \label{eqn:GLparameter}
    \end{equation}
    This is called the Ginzburg-Landau parameter, or GL parameter. Ginzburg and Landau were able to show that for $\kappa \lessapprox 1$ the superconductor had a positive free energy at the superconducting boundary; this was the case for pretty much all superconductors known at the time. Intuitively, one can see that if $\kappa>1$, meaning the penetration depth is greater than the coherence length of the superconductor, the superconducting state will run into instability problems at its boundaries as an external field is raised (more on this later).
    \par If one considers again the zero field transition of superconductor, the energy density gap between the normal and superconducting states, $f_s-f_N$, can be found from Equation \ref{eqn:GLFreeEnergy} to be $-\alpha^2/2\beta$. This can be thought of as the transition energy between the two states, which would be indicative of a drastic change in electron behavior since such a large spike in conductivity occurs below the phase transition. One can imagine the hypothetical scenario where a superconductor is cooled through its superconducting transition all the way to absolute zero. Once at absolute zero, an external field is applied to the superconductor, and slowly raised. The superconductor would respond according to the Meissner effect, repelling the field according to its field strength, until the magnetic field exceeds a critical value, $H_{c_0}$, denoting the $0 \; {\rm K}$ value. One might then think that this field value corresponds with a magnetic energy, $H_{c_0}^2/8\pi$, equal to that energy gap, $\alpha^2/2\beta$. However, this drastically underestimates the critical fields that are observed in the laboratory for superconductors, only giving values on the order of $10^{-2}$ T.
    \par Finally, notice that this energy gap, $-\alpha^2/2\beta$, has a temperature dependence ~$(1-T/T_c)^{-1}$. This seems to suggest that if the superconducting state has some sort of binding energy, it is gradually weakened as its approaches the critical temperature; perhaps electrons are entering a bound state that allows them to form a superfluid. Remember electrons are spin-1/2 fermions, and cannot form a Bose condensate.
    \section{BCS theory}
    The first microscopic theory of superconductivity was published in 1957 by John Bardeen, Leon Cooper, and Robert Schrieffer, now called by the last initials of the three, BCS theory \cite{BCS1957}. They would win the Nobel prize for this work in 1972. The fundamental problem with electrons condensing into a charged super-fluid is that they do not obey the Bose-Einstein statistics necessary for its formation. BCS theory solves this by suggesting the formation of composite bosons made from two electrons called Cooper pairs. Cooper showed that even an arbitrarily small attraction between electrons in the crystalline lattice of a metal can cause a paired state with an energy lower than the Fermi energy. This is conventionally explained as a electron-phonon interaction; in the classical picture of a metal, an electron distorts the local lattice of positive ions towards itself due to its electrical attraction, increasing the local positive charge density, which can attract electrons (with opposite spin) that would otherwise be repelled by the electron. The collective motion of this positive lattice acts as the phonon. 
    \par This pairing distance must be long enough to shield the electron-electron repulsion, to cause the paired state. This is a fundamentally quantum mechanical phenomenon, highlighted by the fact that this minimum shielding distance between these bound electrons is usually 100-1000 nm, which is much larger than the spacing of conduction electrons themselves; this means cooper pairs overlap significantly, enough so that there may be up to $10^7$ electrons between the two electrons in a cooper pair! This is also why some may frown at my description of a composite boson being formed from two individual electrons; the paired state is energetically preferential to each electron and electrons will go in and out of it accordingly with many different partners. 
    \par This binding energy is very weak, of the order $10^{-3}$ eV, thus thermal energy easily breaks these pairs up unless significantly cooled. This means upon cooling, electrons start to form cooper pairs within the metal, which then form a composite charged bosonic gas, allowing the condensation of a super-fluid below its $T_c$. Once pairs are condensing together, the energy required to break any given pair depends on all the other pairs breaking as well, making the energy barrier to break the state macroscopically high. Any collisions/friction of the positive ions with the cooper pairs has insignificant energy in comparison, thus all the pair states are preserved and electron flow is unabated with zero resistance! 
    \par Another cool piece of evidence for this theory is that it supports the isotope effect. The phonon frequency is inversely proportional to square root of the lattice ions mass; the critical temperature of many conventional superconductors follow the same inverse square root relation. In the classical picture, heavier lattice ions are displaced less by electrons, lowering the binding energy, and requiring more cooling before a condensate can form.
    \par Another important prediction of BCS theory is an energy gap, $\Delta$, half the energy to break a cooper pair, being dependent on the critical temperature. The value was predicted to increase from zero at $T_c$ to a limiting value,
    \begin{equation}
        \Delta(0)=3.528k_BT_c
        \label{eqn:BCSenergygap}
    \end{equation}
    This result agreed very well with experiments and gave the theory credence. This energy gap can also be used to approximate a surface resistance of superconductor:
    \begin{equation}
        R_{BCS}=A\frac{f^2}{T}e^{-\frac{\Delta}{k_bT}}
        \label{eqn:BCSResistance}
    \end{equation}
    which will be useful for predicting surface resistances of SRF cavities. All in all the BCS theory was capable of explaining all the phenomenon of superconductivity predicted by London theory and GL theory. Where it starts to fail is in the explanation of HTC superconductors, such as YBCO, but that is not relevant to the conventional superconductors discussed in this dissertation.
    \section{Type-I and type-II superconductors}
\begin{figure*}[htb!]
        \centering
        \includegraphics[angle=0, width=0.5\linewidth]{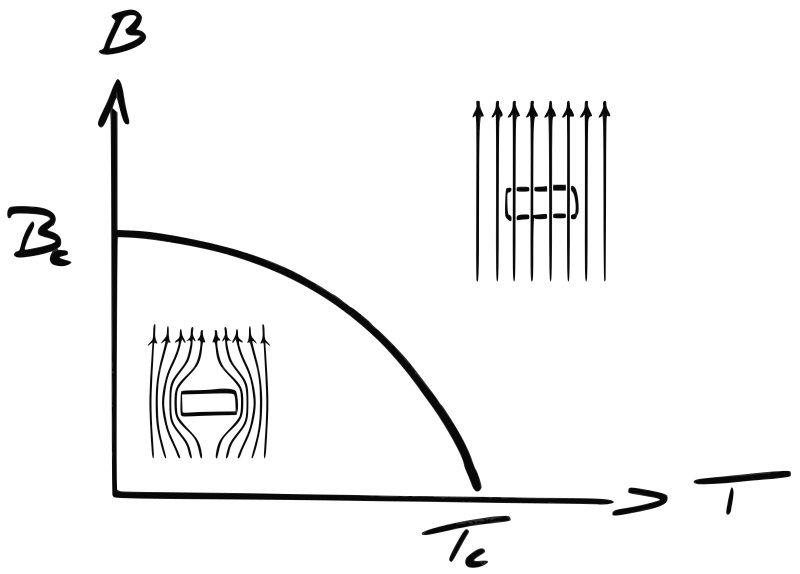}
        \caption{The phase diagram of a type-I superconductor. The transition line represents the parabolic relation of $B_c$, the critical field, with temperature. If $B<B_c$ and $T<T_c$, the critical temperature, the object is superconducting. Note that $B_c$ and $H_c$ are often used interchangeably \cite{TypIphasediagram}.}
        \label{fig:TypeIPhaseDiagram}
    \end{figure*}

    In the same year BCS theory was published, 1957, Alexei Abrikosov published work on superconductors with high GL parameters, $\kappa$ \cite{ABRIKOSOV1957}. The experimental discovery of high $\kappa$ superconductors in Lead-Bismuth alloys was during 1935-36 \cite{RJABININ1935}, but hardly noticed at the time. Much of this work was inspired by the theory work of Richard Feynman and Lars Onsager on quantum vortices in superfluids \cite{FEYNMAN1955,Onsager1949}. Fritz London also had showed the quantization of magnetic flux in superconducting solenoids, which was also very relevant to the work \cite{London1950}. Ginzburg and Landau had shown in their work that above some value of $\kappa$, superconductors would exhibit a negative surface energy, as mentioned earlier. Nonetheless, Abrikosov was able to take GL-theory to precisely show that for $\kappa> 1/\sqrt{2}$, superconductors would exhibit this negative surface energy and generalize the ideas around this distinction between superconductors. What are now called Type-I superconductors, $\kappa<1/\sqrt{2}$, behave as described so far; upon super-conduction, they repel any external magnetic fields below their critical field completely, $H_c$. However, above this singular critical field, the Meissner effect breaks down entirely, and the material returns to the normal state. In accordance with Equation \ref{eqn:HcVsT}, the phase diagram for Type-I superconductors is shown in Figure \ref{fig:TypeIPhaseDiagram}.
    \begin{figure*}[htb!]
        \centering
        \includegraphics[angle=0, width=0.4\linewidth]{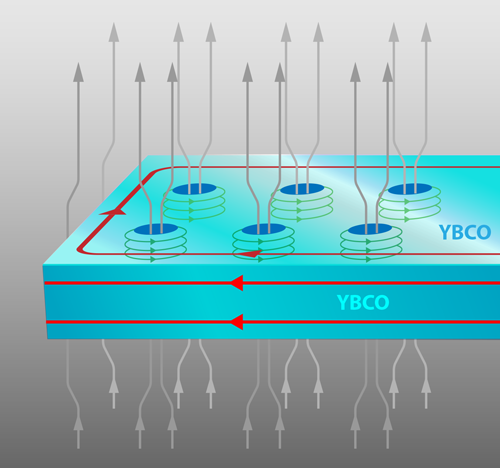}
        \caption{Type-II superconducting film in the intermediate vortex phase, gray arrows represent the magnetic field penetrating through the quantized flux vortices, which generate their own circular surface currents (green arrows). The red arrow represents the Meissner surface current \cite{YBCOdiagram}.}
        \label{fig:YBCOComic}
    \end{figure*}
    \par
\begin{figure*}[htb!]
        \centering
        \includegraphics[angle=0, width=0.4\linewidth]{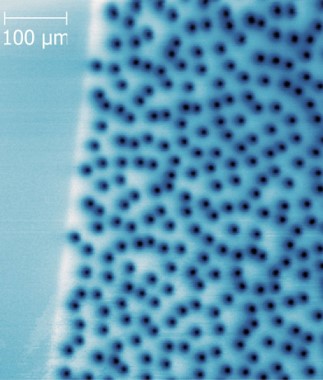}
        \caption{A scanning SQUID microscopy image of a 200 nm thick YBCO film (Type-II superconductor) after field cooling at $6.93 \,{\rm \mu T}$ to 4 K. The black dots are flux vortices or fluxons. \cite{YBCOFLuxonMicroscopy}.}
        \label{fig:SQUIDFLUXONS}
    \end{figure*}

\begin{figure*}[htb!]
        \centering
        \includegraphics[angle=0, width=0.7\linewidth]{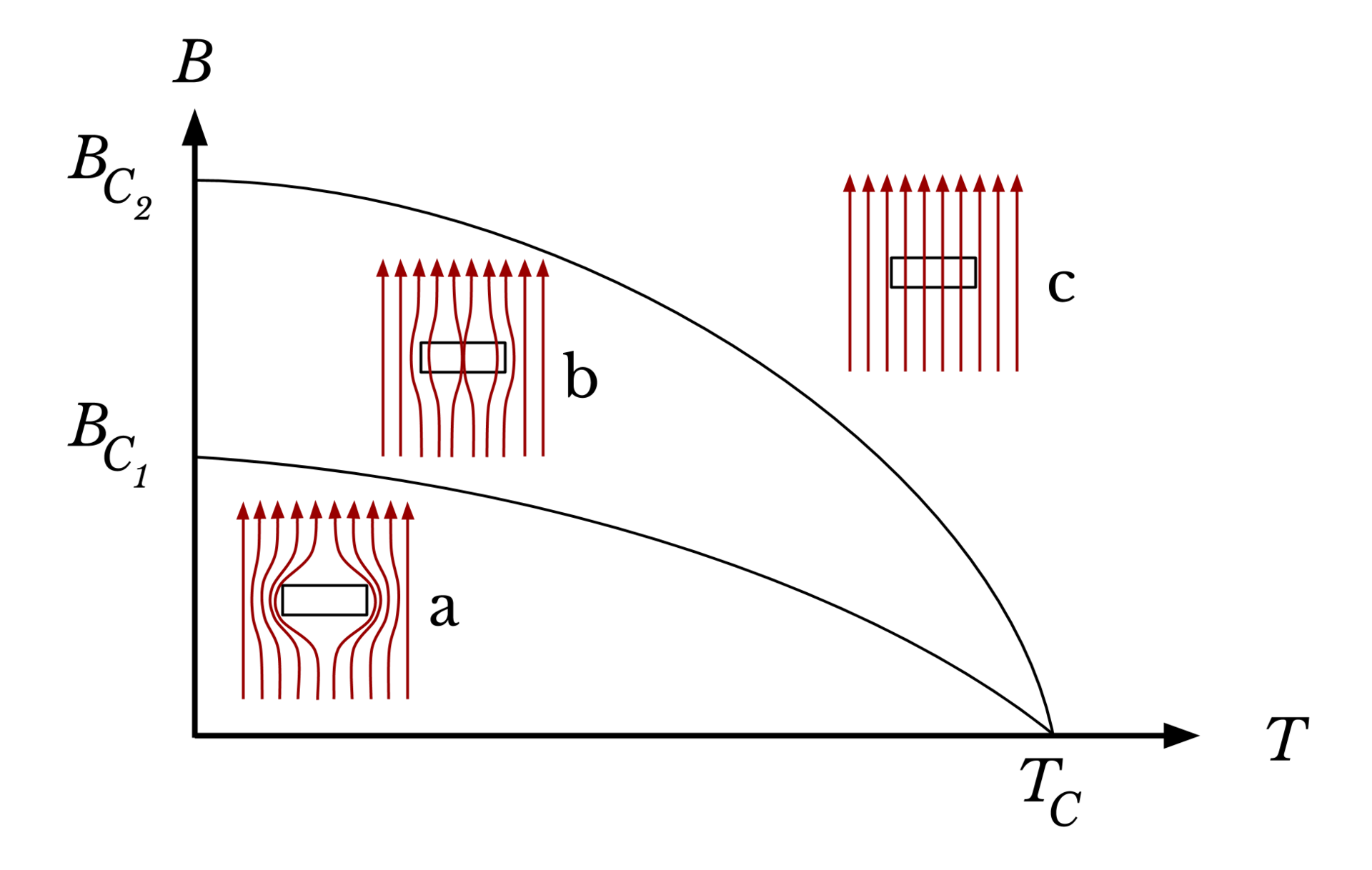}
        \caption{The phase diagram of a type-II superconductor. Both critical fields follow a parabolic relation with temperature. In zone a, $B<B_{c1}$, the object is completely superconducting. In zone b, $B_{c1}<B<B_{c2}$, the object is the mixed superconducting state, where the external field penetrates through quantized flux vortices, called fluxons. In zone c, $B>B_{c2}$, the field penetrates fully and the object is in the normal state \cite{TypIIphasediagram}.}
        \label{fig:TypeIIPhaseDiagram}
    \end{figure*}

 For $\kappa> 1/\sqrt{2}$, the superconductor would exhibit a transitional phase where the external magnetic field would partially penetrate the superconductor via vortices in the normal state, while the remainder would remain superconducting (see Figure \ref{fig:YBCOComic}). The magnetic flux passing through each vortex is quantized by the magnetic flux quantum $\Phi_0=h/2e$, giving them the name fluxons. Because of this, the number of vortices is directly proportional to the flux passing through the superconductor, $N_{vortex}\approx \Phi_{ext}/\Phi_0$. Furthermore, the relative size of each vortex is proportional to the penetration depth, $\approx \pi \lambda^2$, meaning there is a sharp cutoff between the superconducting and normal state about the vortex core. Abrikosov found that these vortices try to arrange themselves in a lattice structure, now known as an Abrikosov Lattice. This lattice of vortices can be imaged with SQUID scanning microscopy, as pictured in Figure \ref{fig:SQUIDFLUXONS}.  Being an intermediate phase, this implies there are two critical fields: one field below which the superconductor is fully in the Meissner state, $H_{c1}$, and one field above which the material is fully in the normal state, $H_{c2}$. The phase diagram for the type-II superconductor is pictured in Figure \ref{fig:TypeIIPhaseDiagram}.
    \par An important practical feature of type-II superconductors is the typical difference between $H_{c1}$ and $H_{c2}$ values. Both type-I superconductors' $H_c$ and type-II superconductors' $H_{c1}$ are pretty much never in excess of a single Tesla, and are more commonly on the order of $100$ mT \cite{TypeIRohlf}. However, common type-II superconductors' $H_{c2}$ are often multiple Teslas and can be much higher than the highest laboratory magnetic fields generated; this isn't surprising because the magnets' themselves are made out of these type-II superconductors! This why type-II superconductors are most commonly used in applications: Magnet wire, Josephson devices, and SRF cavities. In the case of SRF cavities for axion haloscopes, with $B>1$ T at a minimum, type-II superconductors operating in this mixed vortex state are the only option; it is therefore important to understand what drives the resistance in this mixed state.
    \section{Fluxons in SRF cavities}
    At the high frequencies of RF cavities, similar to Chapter \ref{chap:Cavities}, the response of the superconductor must be modelled with a complex surface impedance $Z_s= R_s+i X_s$, where $R_s$ and $X_s$ are the surface resistance and reactance respectively. Often it is the case to consider this the result of a complex resistivity or complex conductivity, depending on your preference. In the case of resistivity, $Z_s=\sqrt{i\omega\mu_0\Tilde{\rho}}=i\omega\mu_0 \Tilde{\lambda}$, where $\Tilde{\lambda}$ refers to a complex shielding length, this length is a function of $H$ and $T$, and should capture the quasi-particle scattering and vortex motion that governs the impedance of a type-II superconductor. This means being a function of the complex conductivity of the superconducting area, $\sigma_f=\sigma_1+i\sigma_2$, as well as the resistivity of the vortices and their motion, $\rho_{vm}$ \cite{Tinkham}. If one is not close to a transition, near $T_c$ or $H_{c2}$, one can assume that the imaginary part of the complex conductivity dominates, $\sigma_1<<\sigma_2$. This simplifies the impedance to:
    \begin{equation}
        Z_s=i\omega\mu_0\sqrt{\lambda^2-\frac{i\rho_{vm}}{\omega \mu_0}}
        \label{eqn:SRFZ_s}
    \end{equation}
    where $\lambda$ is the penetration depth. In the zero-field case, there will be no fluxons in the superconductor and therefore $\rho_{vm}=0$, which simplifies this equation to $R_{s,0}\approx0$ and $X_{s,0}=\omega\mu_0 \lambda$. 
    \par The fluxon motion is what is driving the resistivity; the cavity mode fields displace the fluxons on the cavity surface, which dissipates power in the system. In order to minimize this motion, superconductors are often made intentionally with defects causing smaller grain boundaries; these "pin" down the fluxons to a smaller area, therefore limiting the power absorption from their motion. This process is called flux pinning. One can imagine at a high frequency mode, the induced surface currents of the mode will displace the fluxons from their equilibrium position in the lattice by just a small amount. Therefore one can ignore any mutual interactions between fluxons during their displacement. Therefore we will consider the gross vortex motion an average effect of single fluxon model; the Gittleman-Rosemblum (GR) model in this case  \cite{GRfluxonmodel} can be used to write out the vortex motion resistivity \cite{PompeoSilva}:
    \begin{equation}
        \rho_{vm,GR}=\rho_{vm}'+i\rho_{vm}''=\rho_{ff}\frac{1}{1-i\frac{\omega_p}{\omega}}
        \label{eqn:GRrhovm}
    \end{equation}
    This introduces two new parameters: The flux-flow resistivity, $\rho_{ff}$, and the de-pinning frequency, $\omega_p$. The flux-flow resistivity is associated with the flux-flow viscosity, $\eta$, by the relation:
    \begin{equation}
        \rho_{ff}=\frac{\Phi_0 B}{\eta}
        \label{eqn:GRrhoff}
    \end{equation}
    where $\Phi_0$ is the magnetic flux quantum. This highlights the fact that these fluxons behave very much like a fluid within the crystal lattice. In principle, $\eta$, should be independent of field, therefore it is commonly set by its value near the 2nd critical field:
    \begin{equation}
        \eta=\frac{\Phi_0 B_{c2}}{\rho_0}
        \label{eqn:GRviscosity}
    \end{equation}
    The value of $\rho_0$ is typically a constant set by the manufacturing; in the case of $\mathrm{Nb}_3\mathrm{Sn}$, $\rho_0=0.71-9.7\times 10^{-7} \,{\rm \Omega\, m}$ is used depending on the observed grain size; a smaller grain size means a higher resistivity \cite{Posen_2017}. 
    \par The pinning frequency, $\omega_p$, is very important to our purposes because it marks the boundary between fluxons behaving elastically, $\omega<\omega_p$, and vortex oscillations being purely dissipative, $\omega>\omega_p$; the operating mode frequency must be less than the pinning frequency of the superconductor. It is sometimes also called the de-pinning frequency. It is related to the flux-flow viscosity by the equation:
    \begin{equation}
        \omega_p=\frac{k_p}{\eta}
        \label{eqn:PinningFreq}
    \end{equation}
    where $k_p$ is called the pinning constant. One can see that from Equation \ref{eqn:GRrhovm}, that $\omega_p$ can be obtained directly from the real and imaginary parts of the resistivity:
    \begin{equation}
        \omega_p=\omega \frac{\rho_{vm}''}{\rho_{vm}'}
        \label{eqn:GRPinningFreq}
    \end{equation}
    This is useful for measuring the pinning frequency experimentally.
    \par A more comprehensive solution to this problem would be to consider a full equation of motion for the fluxons that not only takes into account their harmonic motion from the resonance, but also the Lorentz force acting on them and a pinning force from the grain boundaries. The Lorentz force arises because each vortex is a loop of current, and the magnetic field of the mode as well as any external field will push them around. One can write out the magnetic field of each vortex outside its core as:
    \begin{equation}
        B(r)=\frac{\Phi_0}{2\pi\lambda^2}K_0\left(\frac{r}{\lambda}\right)
        \label{eqn:FluxonB-field1}
    \end{equation}
    where $K_0(z)$ is the first Bessel function. The function diverges for $r<\lambda$, when in reality the entire inside of the fluxon can be approximated by:
    \begin{equation}
        B(0) \approx \frac{\Phi_0}{2\pi\lambda^2}ln(\kappa)
        \label{eqn:FluxonB-fieldcore}
    \end{equation}
    This field will be normal to the surface of fluxon, whereas the current will subsequently be circling within the surface plane. This current can be approximated as constant from the core field:
    \begin{equation}
        I_{fluxon} \approx \frac{\Phi_0}{\mu_0\lambda}ln(\kappa)
        \label{eqn:FluxonI-fieldcore}
    \end{equation}
    The actual Lorentz force will then depend on the angle between the normal vector of the fluxon and the direction of the net external field as it overlaps with the fluxon. For our purposes, the external field of the magnet will be significantly greater in magnitude than the resonant mode field, and therefore $H_{mode}$ can be ignored when considering the strength of the force. In this case, because the external field is axial and static, the Lorentz force motion arises from the harmonic motion of the fluxons induced by the magnetic field of the resonant mode.
    \par If we consider a right cylindrical cavity, fluxons on the end caps will have a normal vector parallel to the external magnetic field; a small displacement of the current loop in any direction will have maximal Lorentz force in the direction of the displacement. These fluxons will be in an unstable equilibrium, where their motion will be increased, subsequently increasing the power dissipated in the cavity. Therefore, surface perpendicular to the external field will exhibit maximal losses.
    \par On the other hand, a fluxon on the wall of the cavity, when displaced a small amount, will actually be in a semi-stable arrangement; displacements along the axis of the field will exhibit no Lorentz force, and displacements along the circumference of the cavity would result in procession about its original center if undisturbed. This makes superconducting surfaces parallel to the magnetic field have minimal losses! This is the motivation for using hybrid SRF cavities in very high magnetic fields for axion detection. By keeping surfaces perpendicular to the field made of copper, one avoids the maximal losses of the superconductor, while parallel surfaces should hypothetically still be superconducting with a lower resistance than copper.
    \par Putting this together, one can write the 1-dimensional Lorentz force acting on the fluxon for parallel surfaces, with strength $\gamma$:
    \begin{equation}
        F_{Lorentz}(t,z) \approx \frac{\Phi_0 B_0}{\mu_0\lambda}ln(\kappa) e^{-\frac{z}{\lambda}} \cos{\omega t}=\gamma e^{-\frac{z}{\lambda}} \cos{\omega t}
        \label{eqn:Florentz}
    \end{equation}
    where $B_0$ is the peak field of the magnet. The factor of $e^{-\frac{z}{\lambda}} \cos{\omega t}$ takes into account the field penetration into the surface, and harmonic motion of the fluxons from the mode.
    \par The pinning force is a bit of a mess to approximate, as it depends on the position function of the fluxon from its origin in the surface plane, $\vec{x}(t,z)$; $z$ represents the depth of the fluxon into the surface in this case. This force can be written:
    \begin{equation}
        F_{pinning}(t,z)=-\frac{2e\kappa^2 U_0}{\lambda} \vec{x}(t,z)
        \label{eqn:Fpinning}
    \end{equation}
    where $e$ is the electron charge and $U_0$ is called the fundamental pinning energy. This energy can be found by finding $\vert\vec{x}_{max} \vert$, the distance from the fluxon origin that maximizes the pinning force, which should be correlated to the grain size of crystals in the superconducting material. This can be done by finding the zero of the first derivative of force, which can actually be written out for the 1-dimensional case:
    \begin{equation}
        F_{pinning}'(x)=\frac{8 x^2 \xi^2}{(x^2+\xi^2)^3}-\frac{2\xi^2}{(x^2+\xi^2)^2}
        \label{eqn:Fpinprime}
    \end{equation}
    The fundamental pinning energy can then be calculated:
    \begin{equation}
        U_0=\frac{f_p \xi^2}{2ex_{max}} \left(\left(\frac{x_{max}}{\xi}\right)^2+1\right)^2
        \label{eqn:U0pinning}
    \end{equation}
    $f_p$ is a fundamental pinning force constant roughly equal to $8 \times 10^{-6} {\rm N/m}$ in the case of $\mathrm{Nb}_3\mathrm{Sn}$.
    \par The vortex equation of motion can then be written out in full: 
    \begin{equation}
        \eta \lambda^2 \frac{dx(t,z)}{dt}=\epsilon\frac{d^2x(t,z)}{dz^2} +F_{Lorentz}(t,z)+ F_{pinning}(t,z)
        \label{eqn:FluxonMotion}
    \end{equation}
    where all the variables have been defined except for $\epsilon$; this is called the vortex line elasticity. It can be calculated from the expression:
    \begin{equation}
        \epsilon \approx \frac{\Phi_0^2}{4 \pi \mu_0 \lambda^2} \left(ln(\kappa) + \frac{1}{2}+e^{-0.4-0.8ln(\kappa)-0.1(ln(\kappa))^2}\right)
        \label{eqn:VortexLineElasticity}
    \end{equation}
    This equation can then be solved under three boundary conditions:
    \begin{enumerate}
        \item The displacement is zero at $t=0$, $x(0,z)=0$
        \item There exists a maximum depth where the displacement drops to zero, $x(t,Z_{max})=0$. This will be $Z_{max}\approx 10 \lambda$ where the field is roughly zero, or even better, $Z_{max}=t_{coating}$ the thickness of the superconducting coating.
        \item The derivative of displacement with respect to depth must be zero at the surface of the superconductor, $\frac{dx(t,0)}{dz}=0$.
    \end{enumerate}
    If the solution to this equation converges, one expects the plot of $x$ vs $\frac{dx}{dt}$ to be an ellipse, representative of the harmonic motion. The sensitivity of the surface resistance to magnetic fields ($\Omega/{\rm T}$) can then be calculated:
    \begin{equation}
        S= \frac{\mu_0 \omega^2}{2 \pi^2} \int_0^{2 \pi/\omega}\int_0^{Z_{max}} \frac{dx(t,z)}{dt} e^{-z/\lambda} \cos{(\omega t)} dz dt
    \label{eqn:VortexSensitivity}
    \end{equation}
    This sensitivity is irrespective of the cavity wall geometry, which if we are to calculate a total surface resistance, must be integrated over. In this case, we will assume an axial symmetry along the cavity, such that the angle of the cavity surface with respect to the axial direction is just a function of axial direction, $\theta(z)$ (this is not the surface thickness direction, which we already integrated over). Additionally, we will ignore the end caps and assume an empty cavity, such that the radius of the surface from the center-line can be defined, $r(z)$; this will allow us to integrate over circumferences of the wall in the azimuthal direction of the cavity $\phi$. We only consider the axial component of the magnetic field, $B_z(z)$, along the length of the solenoid, however, one can define a cavity tilt parameter, $\Delta\theta_{cav}$, which is the angle of the cavity center-line with respect to the magnetic field center-line. In this way one can integrate over the surface of the cavity to get the power dissipated:
    \begin{equation}
        P_d= \frac{1}{2 \mu_0^2 B_0} \int_0^{L_{cav}}\int_0^{2 \pi} S B_z(z)^2 \vert \sin{\Delta\theta_{cav}}\cos{\phi} \cos{\theta(z)}+ \cos{\Delta\theta_{cav}} \sin{\theta(z)} \vert r(z) d\phi dz
        \label{eqn:VortexPower}
    \end{equation}
    This can then be paired with the energy stored in cavity, $U_{cav}$, which is independent of any superconducting physics to get the quality factor according to equation \ref{eqn:CavityQualityFactor1}.
    \par Typically, the best way to perform these calculations is numerically using Python or Mathematica and importing a data file that contains a table of the cavity geometry values of interest; $\theta(z)$, $r(z)$, and $B(z)$.


\chapter{Experimental SRF cavity work at LLNL}
\label{chap:LLNL}
This chapter covers the experimental work on superconductors and SRF cavities I performed at LLNL. This includes DC resistivity measurements of superconductors, as well as some prototype cavity measurements. The highlight of this work is my first published paper that I was first and corresponding author for; it makes use of the multi-mode decomposition technique, outlined in Chapter \ref{chap:Cavities}, to characterize a NbTi SRF cavity. 
\par The work done on superconducting cavities at Lawrence Livermore National Laboratory (LLNL) had two goals: Qualify potential superconducting materials for use in ADMX operating conditions, and demonstrate the efficacy of future cavity geometries such as the hybrid SRF cavity and clam-shell cavity design. This meant being able to test materials under cryogenic and high magnetic fields in order to measure materials' critical temperature and fields. On the RF side, being able to measure the RF impedance properties, such as the pinning frequency would also be useful. Finally, an actual direct measurement of the increased losses from a cavity end cap perpendicular to field, versus a parallel wall, would give credence to the hybrid cavity concept.
    \section{PPMS system}
    \label{PPMSsystem}
\begin{figure*}[htb!]
        \centering
        \includegraphics[angle=0, width=0.4\linewidth]{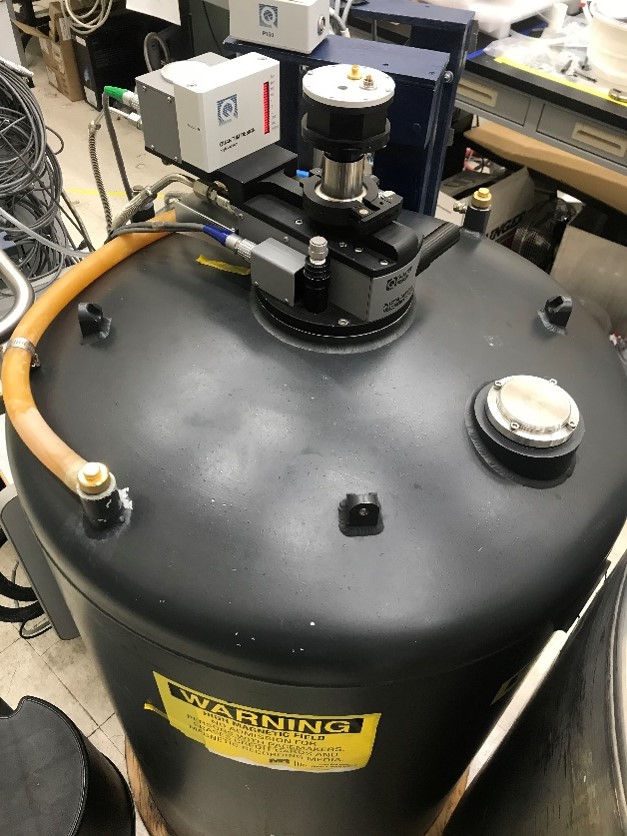}
        \caption{Quantum Designs Physical Properties Measurement System (PPMS) from the outside.}
        \label{fig:PPMS}
    \end{figure*}

\begin{figure}
            \centering
            \includegraphics[angle=0, width=0.4\linewidth]{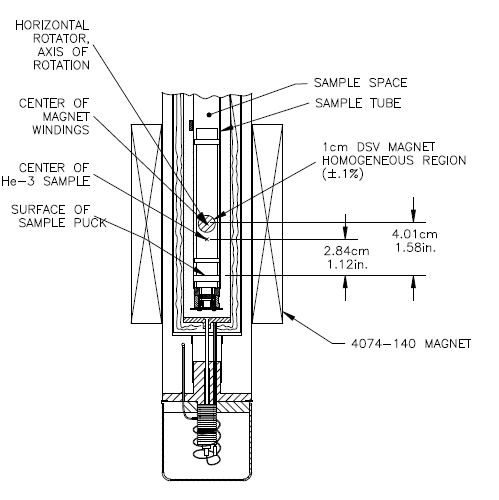}
            \caption{Diagram of the PPMS bore.}
            \label{fig:PPMSbore}
    \end{figure}

    The bulk of this work was done in a Quantum Designs Physical Properties Measurement System (PPMS). Pictured in Figure \ref{fig:PPMS} and Figure \ref{fig:PPMSbore}, the PPMS is an exchange gas cryostat with a $26.5\,{\rm mm}$ bore, equipped with a superconducting magnet capable of fields up to $16\,{\rm T}$ applied axially along the bore; the minimum temperature is $1.8\,{\rm K}$. Because all of the potential superconductors had critical temperatures well above $1.8\,{\rm K}$, this cryostat was fine for our purposes; field performance at milli-Kelvin temperatures could be extrapolated using Equation \ref{eqn:HcVsT}. 
    \par
\begin{figure*}[htb!]
        \centering
        \includegraphics[angle=0, width=0.55\linewidth]{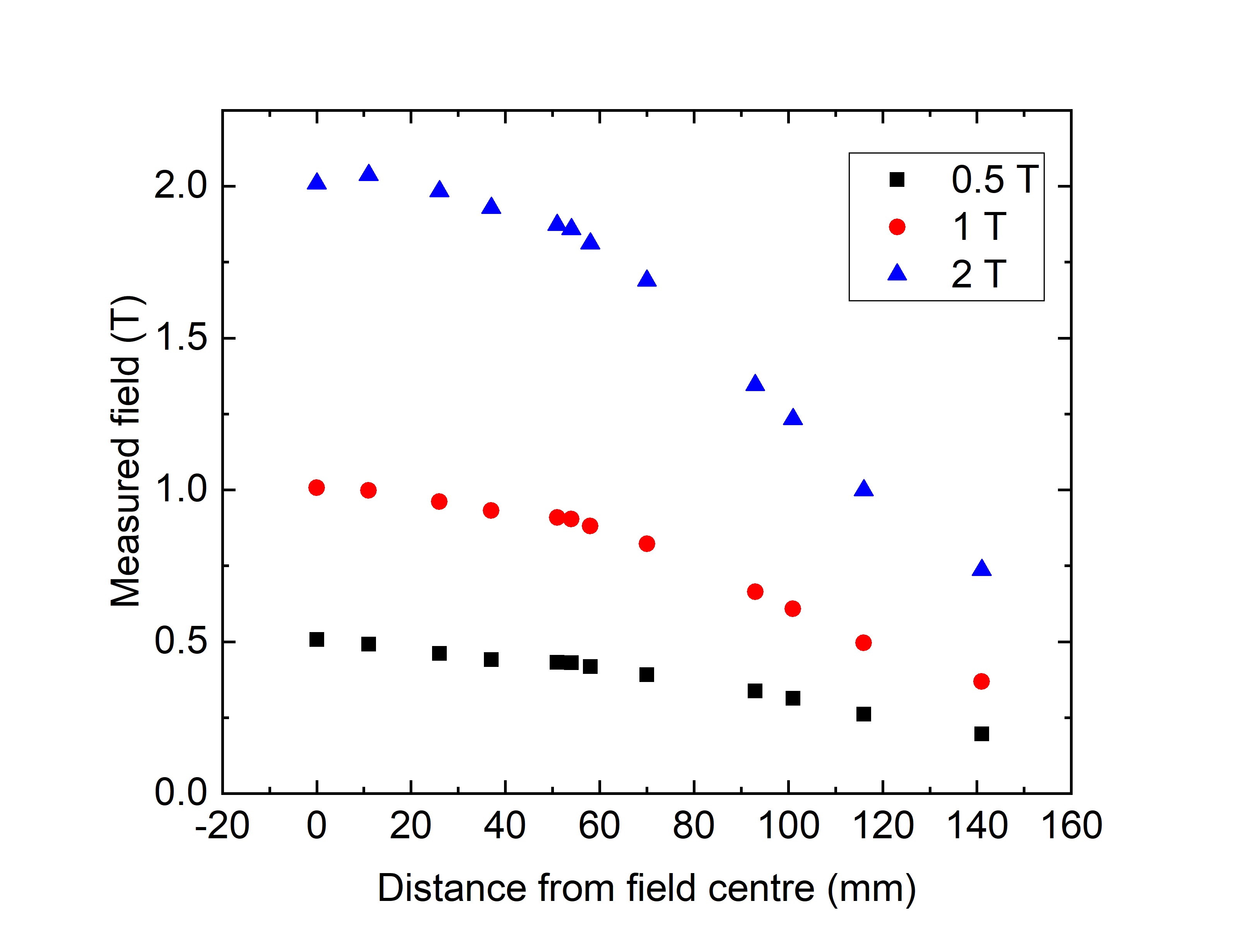}
        \caption{Absolute field deviation of the 16 T PPMS magnet as function of distance from field center. Data was taken at three different magnet field strengths: 0.5, 1, and 2 T.}
        \label{fig:PPMSfielddeviation1}
    \end{figure*}

\begin{figure*}[htb!]
        \centering
        \includegraphics[angle=0, width=0.55\linewidth]{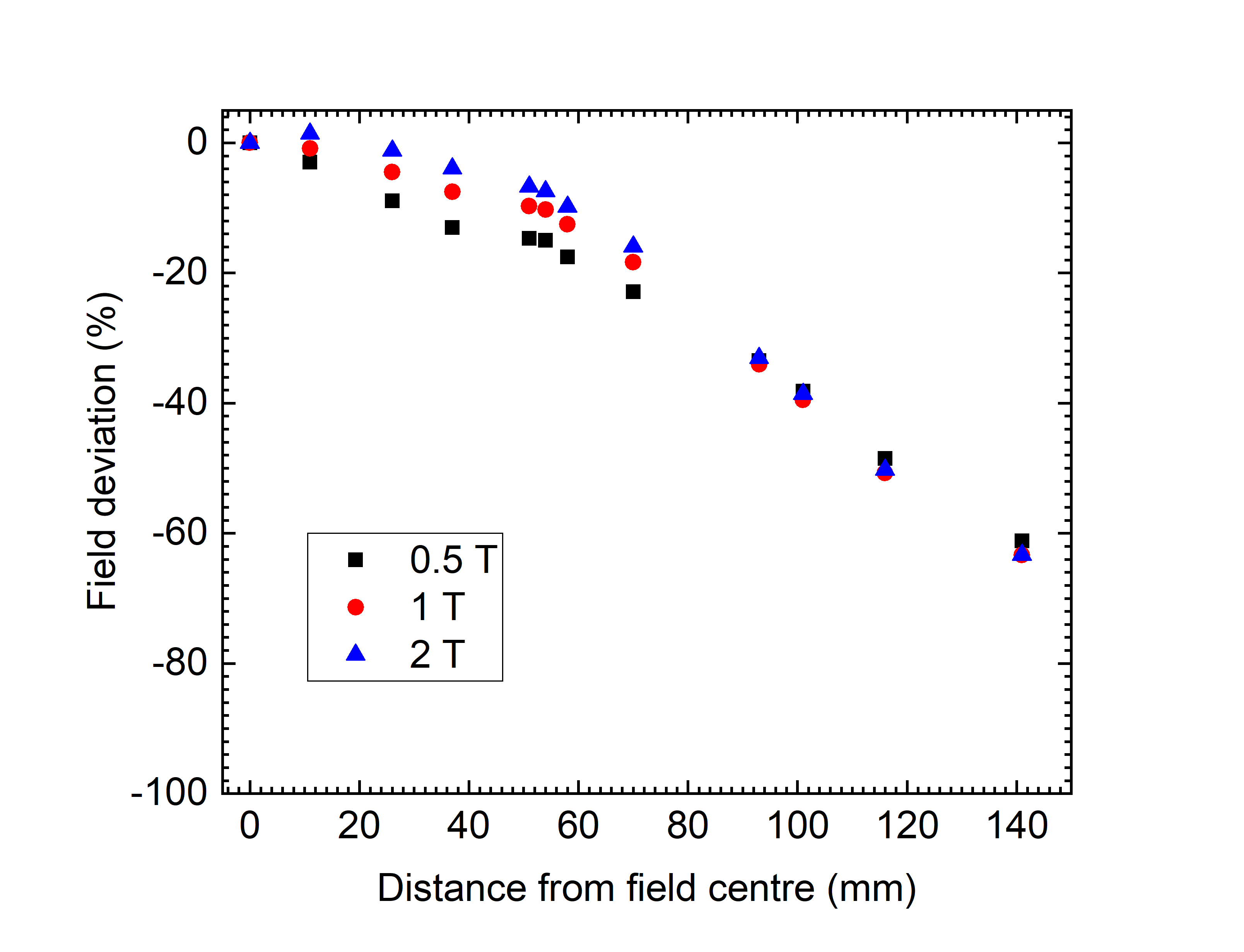}
        \caption{Percent field deviation of the PPMS magnet as function of distance from field center.}
        \label{fig:PPMSfielddeviation2}
    \end{figure*}

 The uniform field region height of the magnet was approximately $40\,{\rm mm}$. The field deviation of magnet was directly measured with a hall probe as a function of distance from the field center, as shown in Figure \ref{fig:PPMSfielddeviation1} and Figure \ref{fig:PPMSfielddeviation2}; past 60 mm, the field deviates beyond 20\% of its center value. This deviation tends to go down for higher magnet strength settings, at least for distances less than 60 mm from the field center.
    \par The proprietary use of the PPMS system is material characterization under various field and temperature conditions via DC resistance measurements. Samples are mounted on "pucks" that are inserted into the PPMS. As shown in Figure \ref{fig:PPMSbore}, the sample of the puck's surface is actually located about 40 mm below the field center, deviating the field value by roughly 5-15\% depending on the set strength. Depending on the measurement setup, care was taken to account for these deviations accordingly. 
    \section{DC measurements}
\begin{figure*}[htb!]
        \centering
        \includegraphics[angle=0, width=0.65\linewidth]{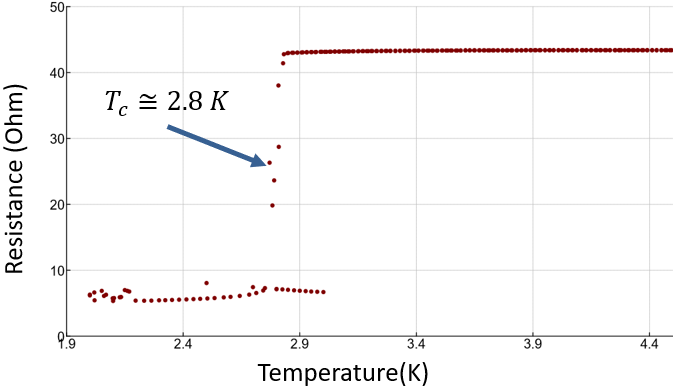}
        \caption{A typical DC resistance versus temperature graph for determining $T_c$. Resistance measurements are periodically taken as the sample is cooled through its critical temperature, At the critical temperature, the superconductor's resistance abruptly drops, making it easy to read off $T_c$. The residual resistance is from the leads connecting to the sample, in this case, being only a two-wire measurement. This particular sample is a coating of NbTiN made at LLNL that had an abnormally low $T_c$ of $\approx$ 2.8 K; typically it is between 10-12 K. This transition temperature was also not particularly stable; upon cycling the sample back through critical temperature, several resistance measurements imply it was superconducting up to 3.1 K. All of this indicated that the coating was compromised by impurities in the deposition process.}
        \label{fig:NbTiNT_c}
    \end{figure*}

    Proprietary to the PPMS's uses, DC resistance measurements of superconductor samples provided a fast and effective way of measuring their critical field and temperature. A typical superconductor exhibits a finite DC resistance above its critical temperature, which abruptly transitions to zero at the critical temperature, $T_c$, so by cooling the sample and continuously taking resistance measurements, this transition can be observed, and $T_c$ measured, as pictured in Fig \ref{fig:NbTiNT_c}.
    \par
\begin{figure*}[htb!]
        \centering
        \includegraphics[angle=0, width=0.65\linewidth]{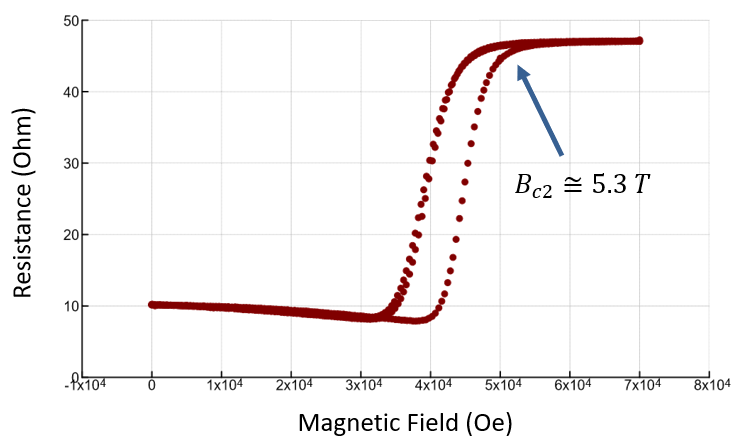}
        \caption{A typical DC resistance versus magnetic field strength graph for determining $B_{c2}$. Resistance measurements are periodically taken as the magnet is ramped through the samples' critical fields. For type II superconductors, such as NbTiN in this case, the resistance will start to increase and plateau into a fully normal state above $B_{c2}$. The residual resistance is from this being only a two-wire measurement. This particular sample is a coating of NbTiN, the same used in Figure \ref{fig:NbTiNT_c}, made at LLNL. It had an abnormally low $B_{c2}$ of $\approx$ 5.3 T; typically it is in excess of 15 T. The sample was cycled back through its critical field when the magnet was ramped down, generating the hysteresis loop structure; this is from trapped flux from the initial ramp shifting and increasing the measured resistance on the ramp back down. Both the low $B_c$ and large hysteresis loop, indicate that the coating was compromised by impurities in the deposition process.}
        \label{fig:NbTiNB_c}
    \end{figure*}

 By ramping the PPMS magnet, one can force the sample, if a Type II superconductor, into the intermediate vortex state, above the first critical field, $B_c1$. Because one is measuring a DC resistance, one doesn't actually expect to see a change at this first transition, because the frequency of the applied current source is so low, it won't move the fluxons very much, and the "swiss-cheese" structure of the superconductor will let the current through unimpeded. It is only when the density of fluxons is great enough that large portions of the superconductor are normal and very few superconducting DC "paths" remain that the DC resistance will start to increase. Eventually, when the sample has gone fully normal, at $B_{c2}$, this DC resistance plateaus; this resistance should also be relatively equal to the resistance of the sample measured just above $T_c$. This process is outlined in Fig \ref{fig:NbTiNB_c}. Since we are qualifying materials that will be operating in the intermediate phase no matter what, measurement of this second critical field is sufficient.
        \subsection{4-wire Measurements}
\begin{figure*}[htb!]
            \centering
            \includegraphics[angle=0, width=0.4\linewidth]{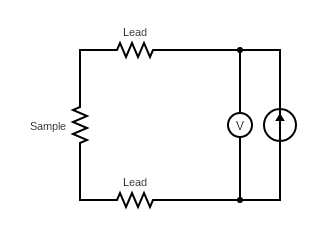}
            \caption{The circuit diagram for a 2-wire resistance measurement. A fixed current source and voltmeter are connected in parallel internally within the DMM. Two wire leads are then connected in series with the sample load of interest.}
            \label{fig:2wirecircuit}
        \end{figure*}

        A 4-wire measurement is a much more precise way of measuring resistance across a particular load, such as a superconducting sample, rather than a classical, 2-wire digital multi-meter (DMM) measurement. In the two-wire measurement, the DMM supplies a known current to two wire leads that the sample is connected in series to, and then measures the voltage drop between these two leads with a voltmeter in parallel; the resistance, by Ohm's law, is then the ratio of this voltage to the supplied current (the equivalent circuit is shown in Figure \ref{fig:2wirecircuit}). However, being connected in series, this includes the resistance of the leads themselves, rather than just the sample. This can be fine for the purpose of determining critical temperatures and fields, as long as the lead resistance is relatively small compared to the change in resistance of the sample upon superconduction; this is the case for Figure \ref{fig:NbTiNT_c} and \ref{fig:NbTiNB_c}. But if one's goal is precisely measuring the sample resistivity as a function of temperature and field, the lead resistance must be calibrated out. This is also useful for measuring near-zero resistances, like superconductors, where the leads resistance will dominate.
        \par
\begin{figure*}[htb!]
            \centering
            \includegraphics[angle=0, width=0.4\linewidth]{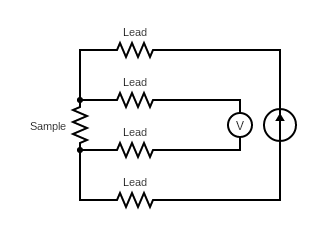}
            \caption{The circuit diagram for a 4-wire resistance measurement. One set of leads connects the sample to a fixed current source, while another connects to a voltmeter reading the voltage drop solely across the sample.}
            \label{fig:4wirecircuit}
        \end{figure*}

 The solution to this is relatively simple: Separate the fixed current source leads from the voltmeter measurement leads that connect to the sample, giving it 4 connected wires. This circuit is pictured in Figure \ref{fig:4wirecircuit}. Because a voltmeter is designed to have an exceedingly high impedance, very little current passes through the leads connected to the voltmeter, hence their individual voltage drop is negligible. This means the voltage measured across the voltmeter is practically the same as voltage drop across the sample, whereas it wasn't in Figure \ref{fig:2wirecircuit}. 
        \subsection{Mounting Samples}
\begin{figure*}[htb!]
            \centering
            \includegraphics[angle=0, width=0.3\linewidth]{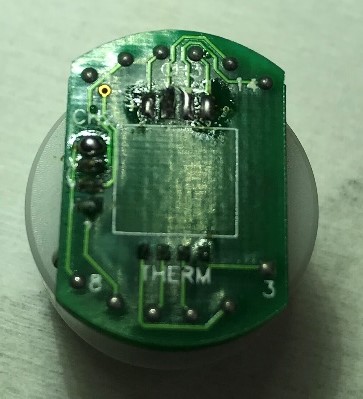}
            \caption{A PPMS puck for 4-wire resistance measurements of samples.This particular puck has 3 sets of 4 solder contacts ($V+$,$V-$,$I+$, and $I-$) to handle 3 simultaneous sample characterizations. The central square where samples would be placed is about $1 \,{\rm cm^2}$.}
            \label{fig:PPMSpuck}
        \end{figure*}

        Samples were mounted to proprietary PPMS pucks that had 2-4 sets of 4-wire solder contacts (see Figure \ref{fig:PPMSpuck}); This meant a single puck could hypothetically take 2-4 samples at a time if space permitted. These pucks had a sample space just over $1 \,{\rm cm^2}$, so samples had to be chosen and sometimes cut to size to accommodate. It was best for a sample, whether bulk or a coating, to have an elongated rectangular shape, creating a wire-like current path through it. Four very thin (20 micron diameter) gold wires would be epoxied to the sample with two leads on either short-end of the sample, one short-end being the positive voltage and current side, and the other end the negative lead side. These wires would be placed perpendicular to the long side of the sample, with the inner two leads being the voltage connection. This was a silver conductive epoxy, called H20E EPO-TEK from Ted Pella Inc, that cured at roughly 150 degrees Celsius for 5 minutes. The use of silver and gold is to still minimize the lead resistance as much as possible. Once cured, the gold wire leads would be soldered to the 4 puck contacts appropriately. A small amount of N-grease was used to hold the sample to the puck surface. Throughout this process a microscope was used to see the very thin wires and where their tips were actually being placed. Because they are very delicate, it was common for a gold wire to break at either the epoxy or solder stage of mounting, or even thermally upon cooling in the PPMS. If this happened late in the mounting or cooling process, we could still run a two-wire measurement; Often the 4 wires were just redundant connections rather than critical to a 4-wire measurement. Once a sample was properly mounted and the puck connectivity was confirmed, it would be inserted into the PPMS for cooling and subsequent measurement. 
        \subsection{Operating Procedure}
\begin{figure*}[htb!]
            \centering
            \includegraphics[angle=0, width=0.8\linewidth]{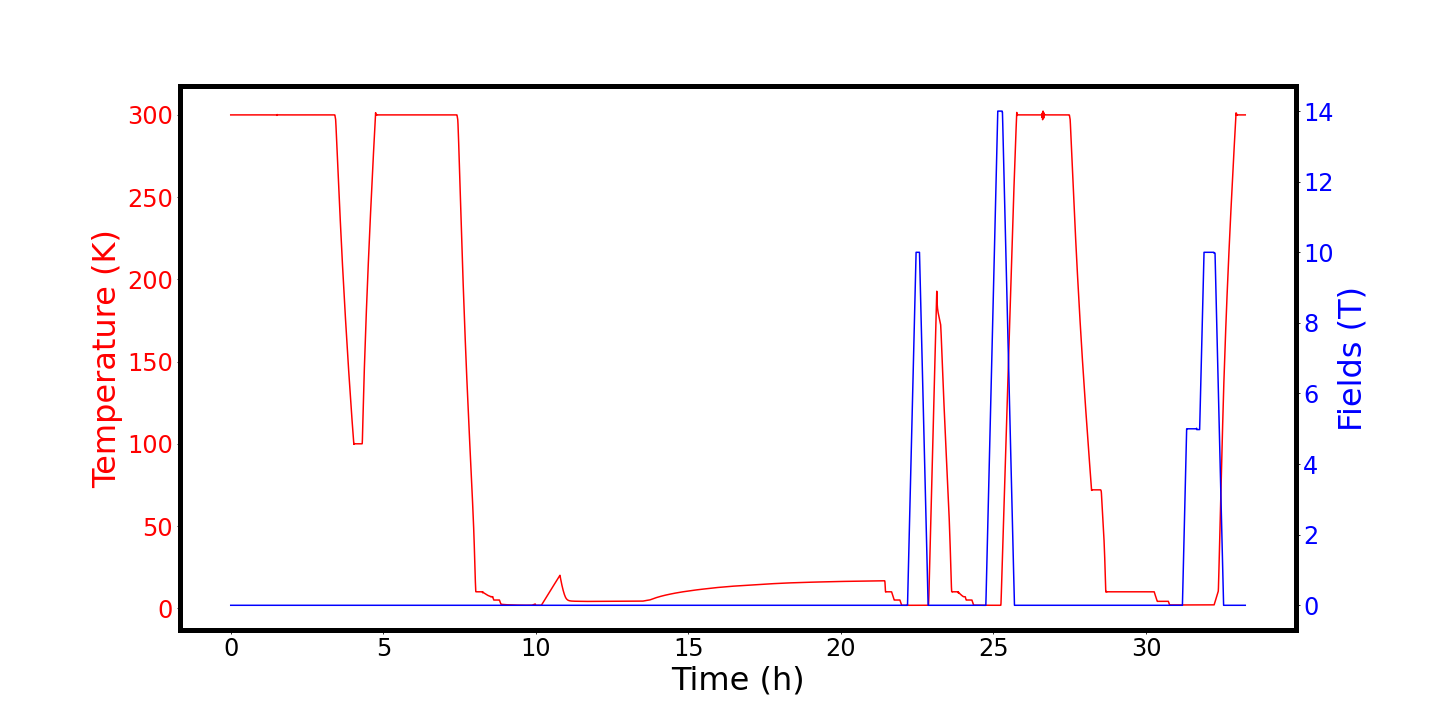}
            \caption{A plot of the PPMS operating temperature and magnetic field versus time during a 33 hour measurement of a bulk NbTi sample. Samples are typically cooled overnight, and magnet operations begin the following day. Sample may lose connectivity due to thermal or magnetic strains, at which point the system is warmed and sample is removed for re-mounting (This was the case after the 2nd magnet ramp in the plot). Warming above the critical temperature in between magnet ramps can also be useful for expelling any trapped flux in the sample.}
            \label{fig:PPMSProcedure}
        \end{figure*}

        Once samples are mounted and loaded into the PPMS, the PPMS is cooled overnight. This is done in two stages typically; a rapid cooling, 10 K/min, to just above the critical temperature, typically around 10-20 K, and a slow cool, 1 K/min, from there to base fridge temperature, 2 K. The sample would then be given some extra time to ensure full thermalization. Usually a procedure starts at the end of a work day, is set to cool overnight, and by morning it has been sitting for several hours at the base fridge temperature (see Figure \ref{fig:PPMSProcedure}). Depending on the smoothness of the cool down through the critical temperature and the data quality, a subsequent warming might be done to get a clearer transition temperature measurement. Then magnet ramps are performed accordingly to measure the critical field. In the case of Figure \ref{fig:PPMSProcedure}, the bulk NbTi sample went through 3 ramps; the first two of these we failed to measure the critical field because a wire broke in situ, giving intermittent readings of the resistance. After warming, and re-mounting the sample, the third ramp produced the needed data. 
        \subsection{Results}
        \begin{table}
            \centering
            \begin{tabular}{||c|| c| c||} 
            \hline
            Material (Source) & $T_c$ (K) & $B_{c2}$ (T) \\ [0.5ex] 
            \hline\hline
            NbTiN coating on silicon (LLNL) & 2.8 & 5.3  \\ 
            \hline
            Bulk NbTi (American Elements Alloy Nb-55\% Ti-45\%) & 8.5 & $>\,$14 T  \\
            \hline
            $\mathrm{Nb}_3\mathrm{Sn}$ on Bronze (FSU) & 10.3 & $>\,$14 T \\
            \hline
            \end{tabular}
            \caption{Critical temperature and fields of various materials measured at LLNL. The NbTiN coating was produced at LLNL. The NbTi was acquired from American Elements. The $\mathrm{Nb}_3\mathrm{Sn}$ was supplied by Lance Cooley's group at Florida State University (FSU).}
            \label{tab:DCTCHCResults}
        \end{table}
        In Table \ref{tab:DCTCHCResults}, I highlight three main DC measurement results: NbTiN coatings made at LLNL, a bulk chip of NbTi metal used to machine the test cavity in following sections, and $\mathrm{Nb}_3\mathrm{Sn}$ coatings on bronze substrate given to us by FSU. Unfortunately, we were unable to make measurements of Fermilab's $\mathrm{Nb}_3\mathrm{Sn}$ coating on pure Niobium substrate, because we couldn't get any samples of the right size made; this was the coating process used to make the Sidecar SRF tuning rod described in Chapter \ref{chap:Sidecar1D}. Similarly, the YBCO samples acquired were too large to fit within the PPMS, and remain untested for the next LLNL student. 
        \par At the beginning of my stay at LLNL, a local deposition shop on site made several NbTiN samples in their system. Unfortunately, we discovered that the base pressure of their system was too high $10^{-5}$ Torr than what is needed to produce clean coatings ($\approx \, 10^{-9}$ Torr). This resulted in very low $T_c$ and $B_{c2}$ measurements, as shown in Figure \ref{fig:NbTiNT_c} and \ref{fig:NbTiNB_c}. Without the proper deposition system and cheaper alternatives available, these nitride coatings were ultimately abandoned for our purposes.
        \par
\begin{figure*}[htb!]
            \centering
            \includegraphics[angle=0, width=0.7\linewidth]{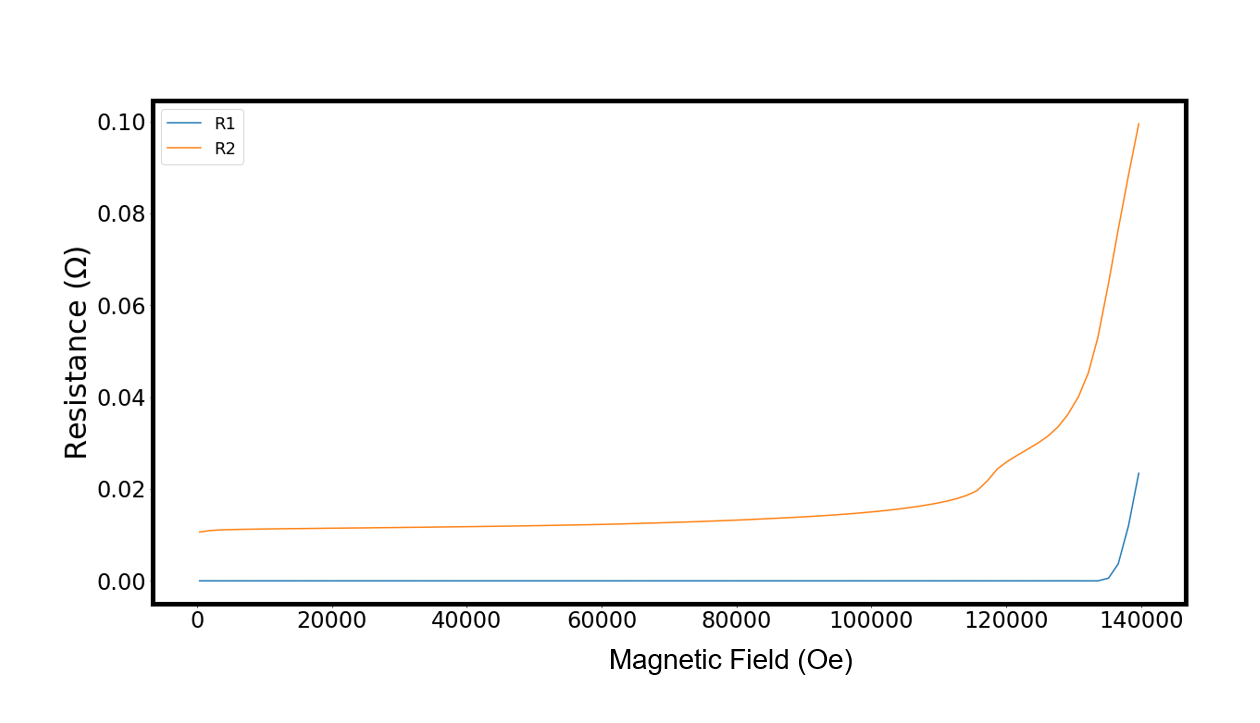}
            \caption{Measuring the $B_{c2}$ of a bulk NbTi sample. $R1$ and $R2$ denote two different resistance bridges in the PPMS, that each were connected to a NbTi sample. Because the magnet was limited to 14 T, it can only be said that $H_{c2} >14$ T. This is in agreement with similar alloys that exhibit an $H_{c2}\approx \, 14-15$ T.}
            \label{fig:BulkNbTiHc}
        \end{figure*}

 During the manufacturing of a bulk NbTi test cavity, we were able to ask the machinist to chip off a small portion of the square rod stock for DC testing. This cavity will be discussed in the following two sections in detail, but the DC results of this chip are included here: $T_C\approx 8.5$ K and $H_{c2} >14$ T. We were unable to determine the full critical field value because it was not permitted to ramp the magnet above 14 T to its maximum at 16 T. As shown in Figure \ref{fig:BulkNbTiHc}, the NbTi sample had begun transitioning to a normal state, but had yet to plateau; typically Nb-45\% Ti-55\% alloys such as this have critical fields between 14-15 T, so this agrees well with that figure.
        \par
\begin{figure*}[htb!]
            \centering
            \includegraphics[angle=0, width=0.6\linewidth]{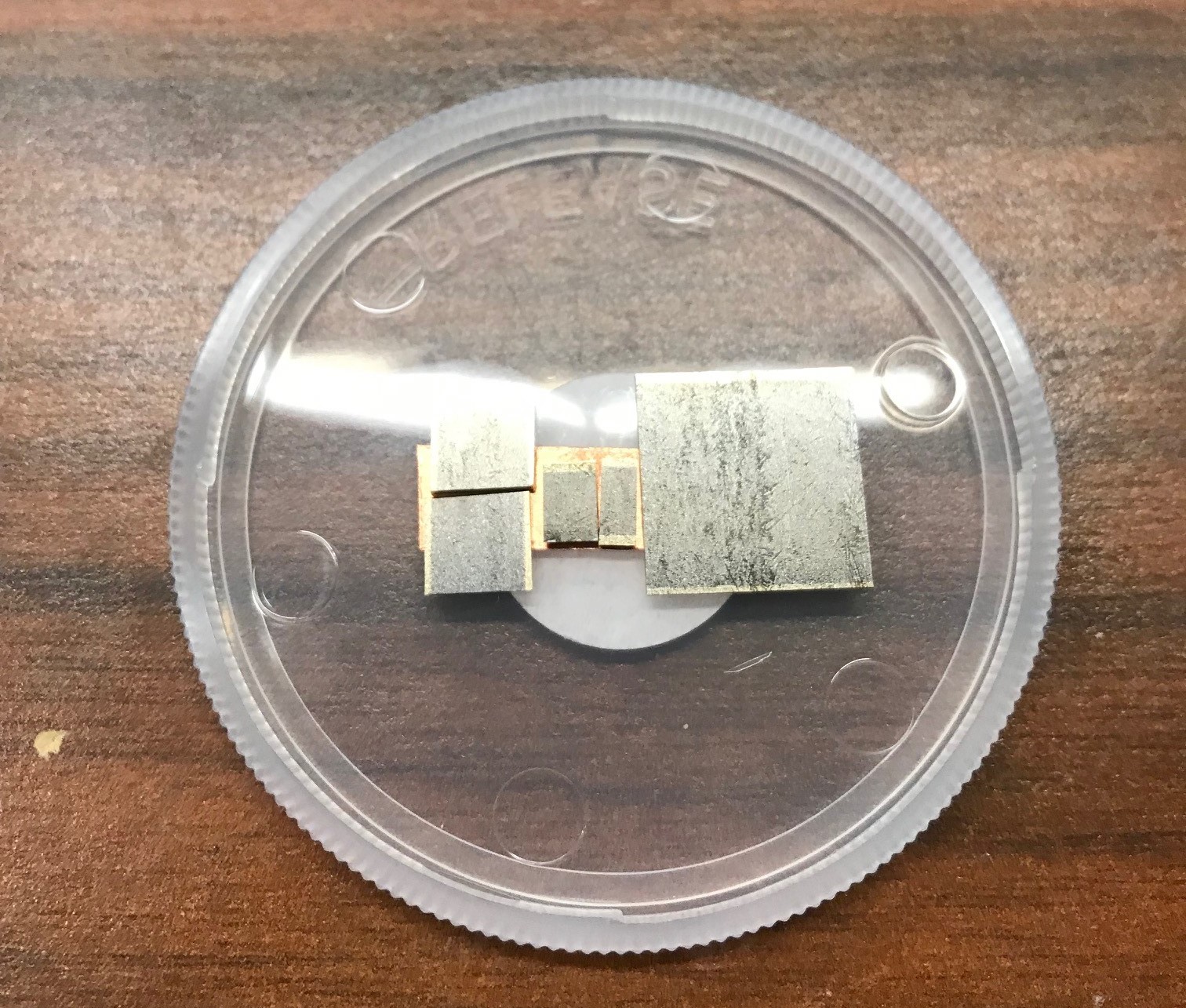}
            \caption{$\mathrm{Nb}_3\mathrm{Sn}$ coating deposited on bronze substrate. This deposition work was done at Florida State University by Lance Cooley's group, see Ref. \cite{Withanage_2021} for details.}
            \label{fig:Nb3SnBronzeSample}
        \end{figure*}

 The $\mathrm{Nb}_3\mathrm{Sn}$ sample highlighted in Table \ref{tab:DCTCHCResults} was made by Dr. Wenura Withanage, working under Dr. Lance Cooley at the time, and uses an innovative coating process allowing for a bronze substrate (see Figure \ref{fig:Nb3SnBronzeSample}). This was of interest to us at ADMX, because a bronze substrate would have the heat capacity properties of copper, as well as a the lower price tag when compared to a pure Niobium substrate. As expected for a $\mathrm{Nb}_3\mathrm{Sn}$ coating, the critical field was above 14 T, and therefore could not be precisely measured; Typically it is around 18 T. 
        \par
\begin{figure*}[htb!]
            \centering
            \includegraphics[angle=0, width=0.5\linewidth]{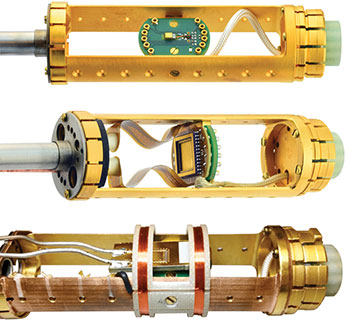}
            \caption{The Quantum Designs PPMS rotating puck stage. This allows for samples to be rotated with respect to the magnetic field axis from its typical perpendicular arrangement.}
            \label{fig:PPMSrotator}
        \end{figure*}

\begin{figure*}[htb!]
            \centering
            \includegraphics[angle=0, width=0.9\linewidth]{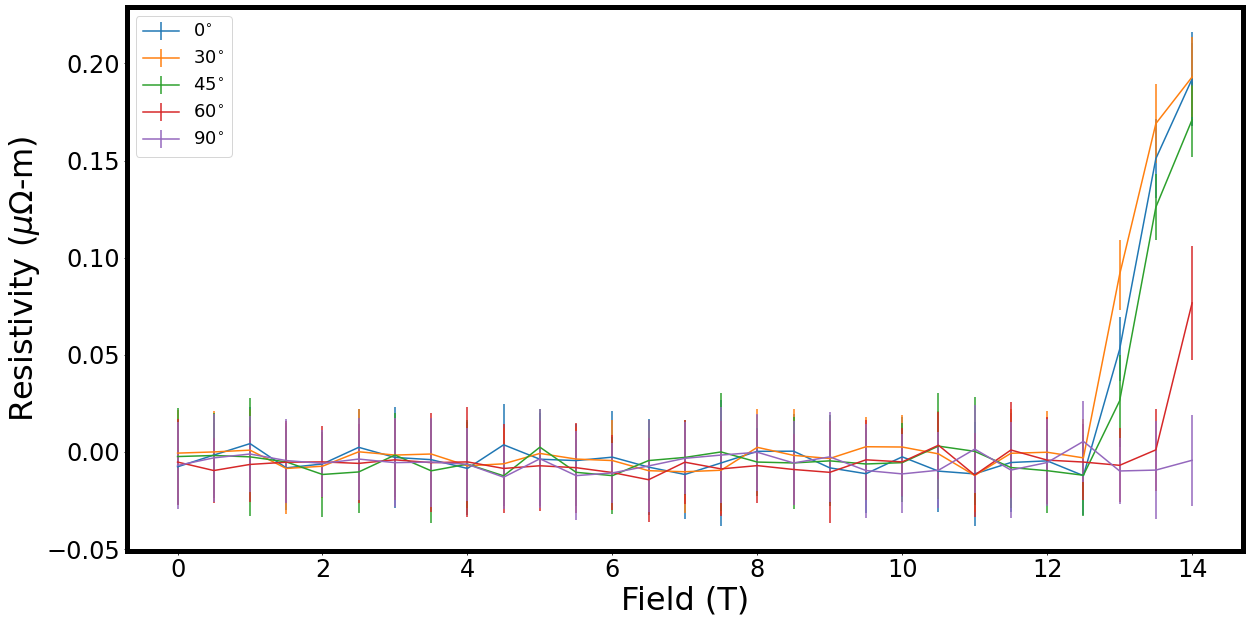}
            \caption{The resistance versus magnetic field of $\mathrm{Nb}_3\mathrm{Sn}$ on bronze substrate for multiple angles with respect to the field axis; 0 degrees corresponds to the sample being perpendicular to the field axis, while at 90 degrees the sample surface is parallel. Successive Magnet ramps at 5 different angles were made. Note that the parallel arrangement does not begin its superconducting transition, while the other angles with perpendicular components begin to transition above $\approx 12.5$ T.}
            \label{fig:Nb3SnRotated}
        \end{figure*}

 This sample was also tested with a variable angle rotating sample stage made by Quantum Designs (see Figure \ref{fig:PPMSrotator}). This allowed an operator to rotate the sample surface with respect to the magnet field, rather than just a perpendicular arrangement. The operating procedure for this becomes a bit longer; one has to perform a full ramp up and down, ensure there's no trapped flux, then rotate, and then repeat. For this reason, we only tested at 5 angles: 0, 30, 45, 60, and 90 degrees. As shown in Figure \ref{fig:Nb3SnRotated}, the sample actually remained superconducting in the parallel arrangement (90 degrees in the figure) all the way to the 14 T limit. The other angles accordingly start to transition at lower field values with the perpendicular arrangement having the lowest transition field strength at $\approx 12.5$ T; this is exactly the angular dependence one hopes to see.
    \section{PPMS cavity design}
\begin{figure*}[htb!]
         \centering
        \includegraphics[width=1\linewidth]{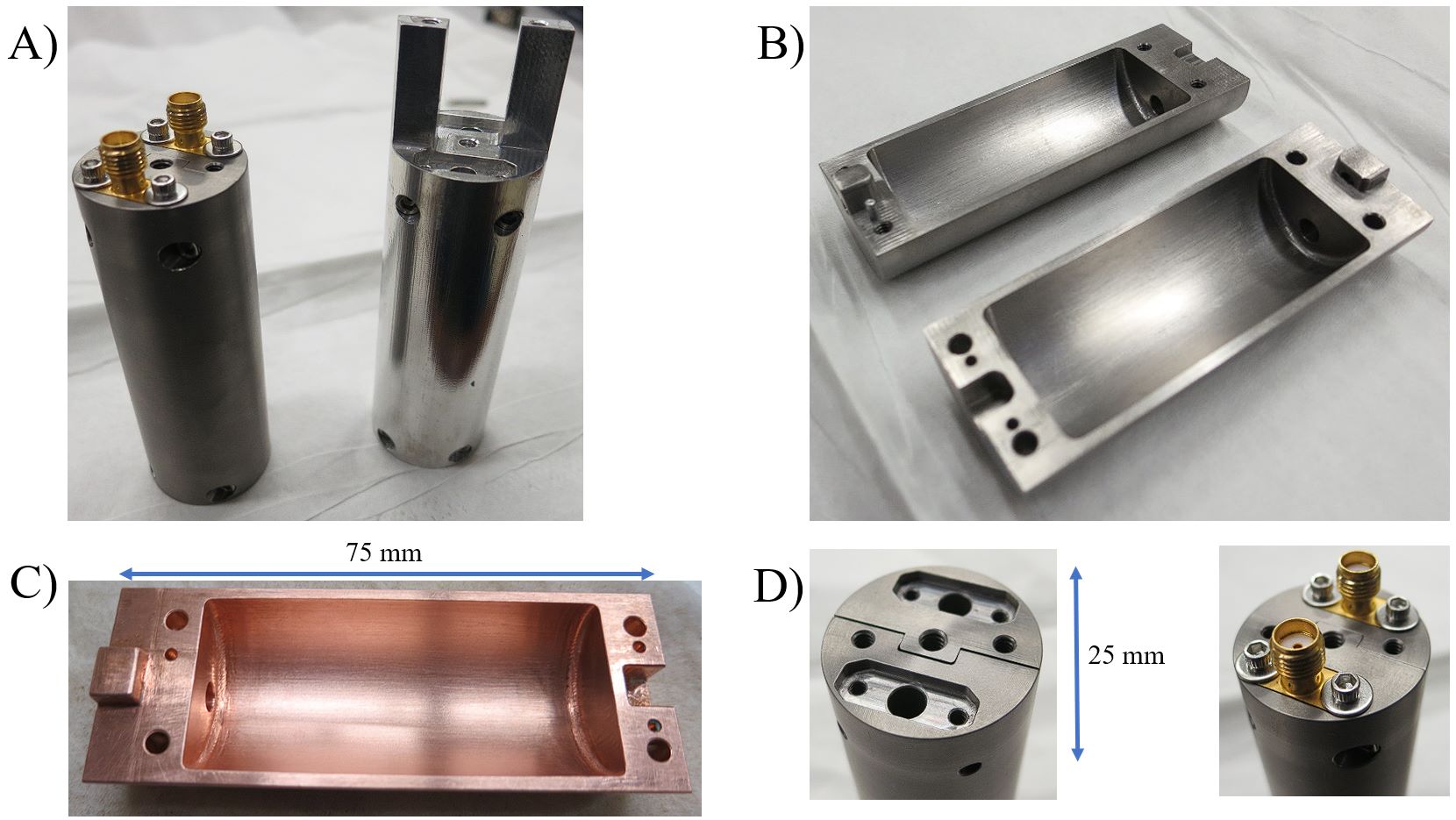}
        \caption{PPMS clamshell cavity machined from a bulk NbTi square stock. A) The NbTi cavity assembled next to aluminum prototype. B) A profile of both NbTi halves side by side. C) One half of the copper clamshell close-up. D) A close-up of the cavity top without and with fixed pin antennae attached; additional holes are for mounting to probe.}
        \label{fig:NbTiCavity}
    \end{figure*}

    In order to perform RF measurements within the system, a cavity and readout system had to be designed to fit within the PPMS bore. This design was used to machine cavities out of a variety of materials; cavities made out of copper, aluminum, and bulk NbTi have been made, as pictured in Figure \ref{fig:NbTiCavity}A-C. This enables a direct comparison of material performance. The clam-shell design lends itself well to being coated, although no cavity has been coated yet as of the time of writing. The clam-shell halves are joined by 4 4-40 screws, and are aligned by 3 alignment pins pictured in Figure \ref{fig:NbTiCavity}B-C.
    \par The bore, being very narrow, limited the internal diameter of any cavity that could fit to 23 mm, which put the frequency of the $TM_{010}$ mode at $\approx 10 \; {\rm GHz}$ . Additionally, the internal height was limited to be $53 \; {\rm mm}$ because of the magnet coil height; this would put the bottom end cap within $\approx 25 $ mm below the field center, whereas the top end cap would be $\approx 28 $ mm above. Since most measurements would be taken at high field strengths, this was deemed within the uniform field range, and close to symmetric about the field center such that field deviations could be ignored. As shown in Table \ref{tab:cavgeofactor}, limiting the cavity length had the added benefit of keeping the end cap geometric factor within an order of magnitude of the wall geometric factor; any longer and the walls would dominate over any end cap contributions (this was important for making multi-mode decomposition measurements).
    \begin{table}
        \centering
        \begin{tabular}{||c|| c| c |c||} 
        \hline
        & $TM_{010}$ & $TM_{011}$ & $TM_{012}$ \\ [0.5ex] 
        \hline\hline
        Walls & $457.0\pm2.2$ & $494.6\pm3.4$ & $549.9\pm5.5$ \\ 
        \hline
        Top End Cap & $3846\pm41$ & $2045\pm15$ & $2214\pm6$ \\
        \hline
        Bottom End Cap & $4127\pm55$ & $2029\pm26$ & $2192\pm21$ \\
        \hline
        Total Endcaps & $1991\pm38$ & $1019\pm17$ & $1101\pm12$ \\
        \hline
        Total & $371.7\pm9.3$ & $332.9\pm7.0$ & $366.8\pm6.2$ \\ [1ex] 
        \hline
        \end{tabular}
        \caption{Geometric factors for clamshell cavity design for first three $TM$ modes by interior surfaces. These were calculated from simulation in the HFSS software package and are expressed in Ohms.}
        \label{tab:cavgeofactor}
    \end{table}
    \par
\begin{figure*}[htb!]
        \centering
        \includegraphics[width=0.8\linewidth]{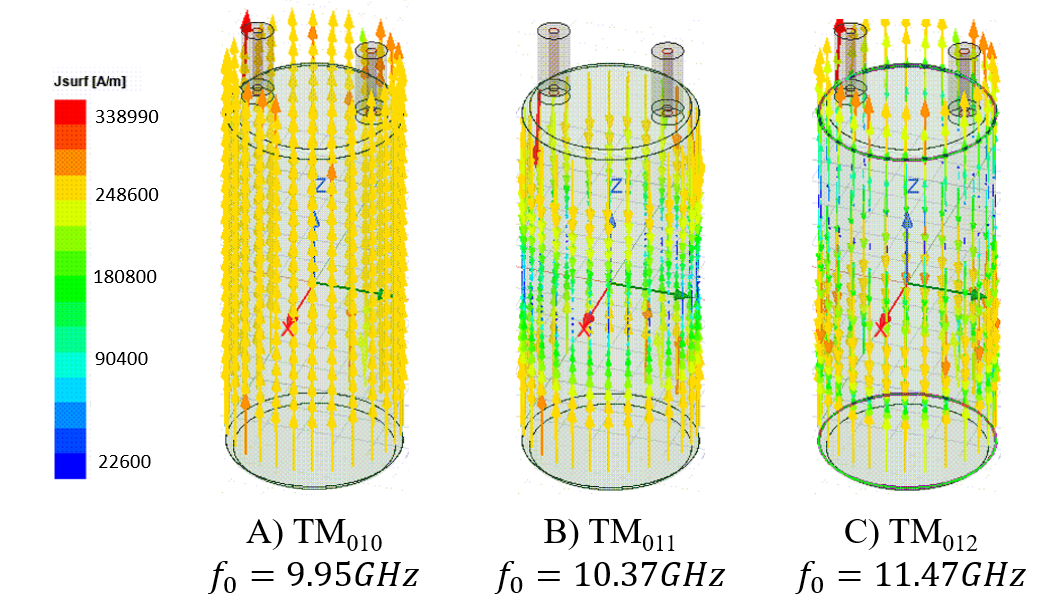}
        \caption{A field overlay of the cavity modes' surface currents in Ansys\texttrademark  ~ HFSS software. A) The $TM_{010}$ mode. B) The $TM_{011}$ mode. C) The $TM_{012}$ mode. Scale bar for the magnitude of the surface currents picture left in $\mathrm{A}/\mathrm{m}$.}
        \label{fig:CavityCurrents}
    \end{figure*}

 The geometric factors shown in Table \ref{tab:cavgeofactor} were simulated using the Ansys HFSS software package. The simulated cavity modes' field structure and surface currents (see Figure \ref{fig:CavityCurrents}) were used to calculate the surface and volume integrals for geometric factor in Equation \ref{eqn:Gfactor}. The set of nine geometric factors correspond to the three modes of interest, $TM_{010}$ , $TM_{011}$, and $TM_{012}$, as well as the three sub-surfaces of interest, top end cap, bottom end cap, and walls. One may notice that the top and bottom end caps do not differ very much in values; nonetheless it was important to quantify this difference as the top end cap contained the antenna port holes. Python software enabled one to combine these geometric factors in various combinations, most commonly the total end cap geometric factor, as well as their associated error. 
    \par
\begin{figure*}[htb!]
         \centering
        \includegraphics[width=0.9\textwidth]{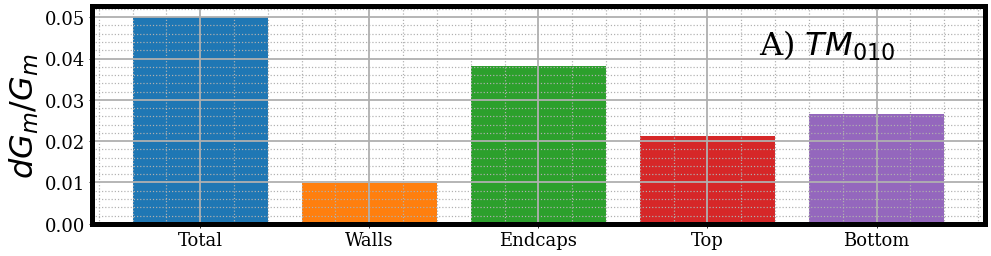}
        \includegraphics[width=0.9\textwidth]{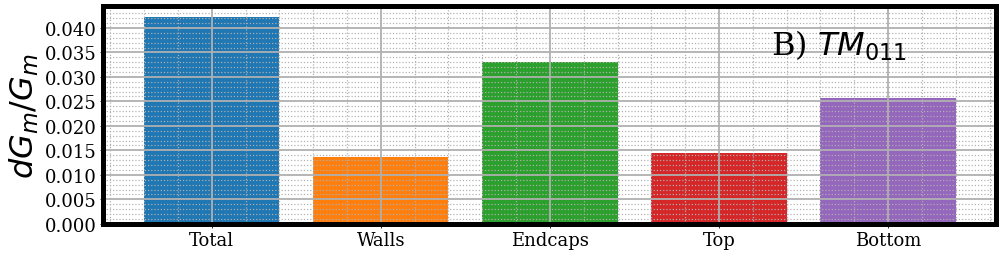}
        \includegraphics[width=0.9\textwidth]{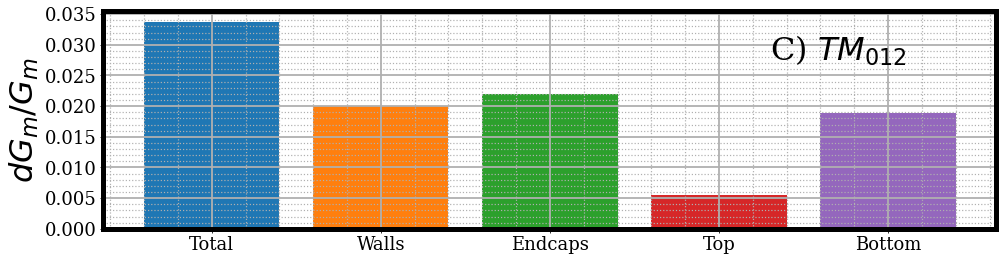}
        \caption{The estimated fractional error in geometric factor by mode and subsurface. Note that the end caps factor is the sum of the top and bottom end cap contributions, and the total geometric factor is the sum of the end caps and walls. A) The error in the $TM_{010}$ mode geometric factors. B) The error in the $TM_{011}$ mode geometric factors. C) The error in the $TM_{012}$ mode geometric factors.}
        \label{fig:dGBargraph}
    \end{figure*}

 The error in these G-factors were calculated through a parametric analysis according to the cited machine tolerance, $\Delta X= \pm 0.005"$. The height and radius were varied by three points, $X+\Delta X$, $X$, and $X-\Delta X$, generating 9 configurations of the cavity design; the nine geometric factors were subsequently calculating for each configuration. The final uncertainty was calculated according to Equations \ref{eqn:deltaG}. \ref{eqn:G_med}, and \ref{eqn:deltaGfinal}. The fractional uncertainty was kept to less than 5\% as shown in Figure \ref{fig:dGBargraph}. Unfortunately, subsequent calculation of the weight matrix and its determinant for the three possible 2-mode decompositions showed that the $TM_{011}/TM_{012}$ mode combination had a statistically zero determinant and couldn't be used.
    \par
\begin{figure*}[htb!]
        \centering
        \includegraphics[width=1 \linewidth]{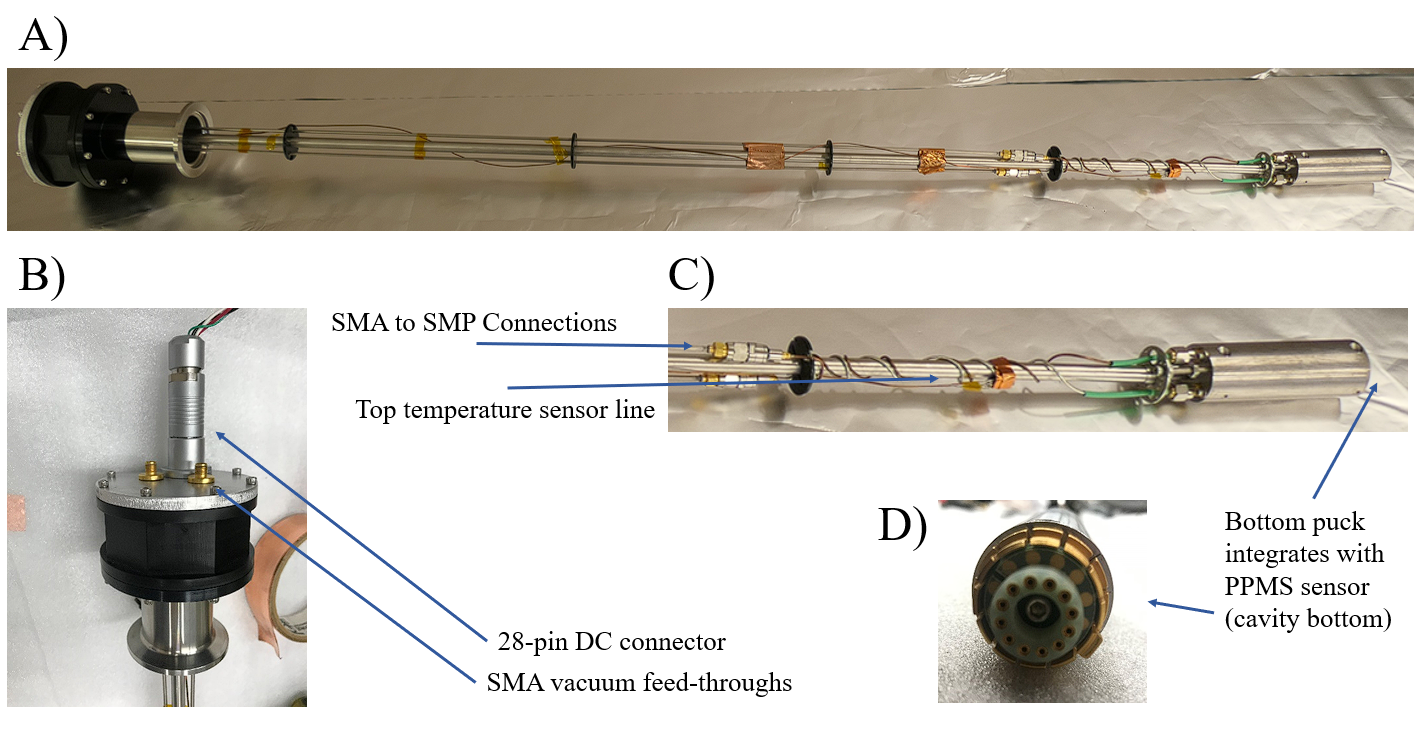}
        \caption{The PPMS readout probe with an aluminum clamshell cavity mounted after a test. A) The entire probe with NbTi cavity mounted on bottom. B) The probe top with DC and SMA vacuum feed-through connections. C) The probe bottom with cavity, SMA to SMP connections, and temperature sensor line. D) Cavity bottom puck that ensures thermal connection to PPMS internal temperature sensor.}
        \label{fig:Probe}
    \end{figure*}

 Two antenna ports on the top of the cavity enabled one to perform two-port measurements and characterization (see Figure \ref{fig:NbTiCavity}D). Fixed pin antennae attached with low-profile 4-40 screws; these antennae were cut such that $\beta<0.01$, making the cavity weakly coupled and direct measurement of $Q_0$ could be made. The cavity was mounted to a custom multi-use probe, shown in Figure \ref{fig:Probe}C. The fixed pin antennae were attached by SMA connections with an SMA to SMP junctions several inches above, with stainless steel semi-rigid coaxial cable up to the top of the probe where SMA vacuum feedthroughs enabled connection to the warm space. A 28-pin DC connector next to the SMA feed-throughs (Figure \ref{fig:Probe}B) connected to a temperature sensor located on the top of cavity. A puck screwed into a center-line hole on the bottom of the cavity, which thermally sunk the cavity to the PPMS cryostat as well as connecting it to the system's internal temperature sensor. A vacuum flange at the top of the probe sealed it to the PPMS bore with an ISO-K clamp. 
    \section{Results of a bulk NbTi clamshell surface decomposition}
    Using the PPMS clamshell design outlined in the last section, as well as the novel multi-mode decomposition technique outlined in Chapter \ref{chap:Cavities}, we were able to directly show the losses in an SRF cavity were dominated by the perpendicular field surfaces, the end-caps, while the surfaces parallel to the field remained superconducting, with minimal loss. The superconducting NbTi cavity was machined from a Nb-45\% Ti-55\%  square stock with 99\% purity purchased from American Elements by LLNL. An annealed copper clamshell was used as the control in this measurement.
    \par Each clamshell cavity was cooled in the PPMS to $2\,{\rm K}$ in zero field by cooling to $10\,{\rm K}$ at $10\,{\rm K/min}$, then to $2\,{\rm K}$ at $1\,{\rm K/min}$ and then held for $\approx\,10$ minutes to allow the cavity to fully thermalize. This cool down shifts the resonant frequencies from room temperature by about $25\,{\rm kHz}$ upward. The magnet was then ramped in steps ($0.1\,{\rm T}$ steps from $0$ to $1\,{\rm T}$ and $1\,{\rm T}$ steps from 1 to $10\,{\rm T}$), and held in between until the mode frequencies and Q factors settled to constant values ($\approx\,1$ minute), at which point the resonant peaks were recorded from the VNA for the three $TM$ modes of interest ($TM_{010}$, $TM_{011}$, and $TM_{012}$). 
    \begin{figure*}[htb!]
        \centering
        \includegraphics[width=0.8\linewidth]{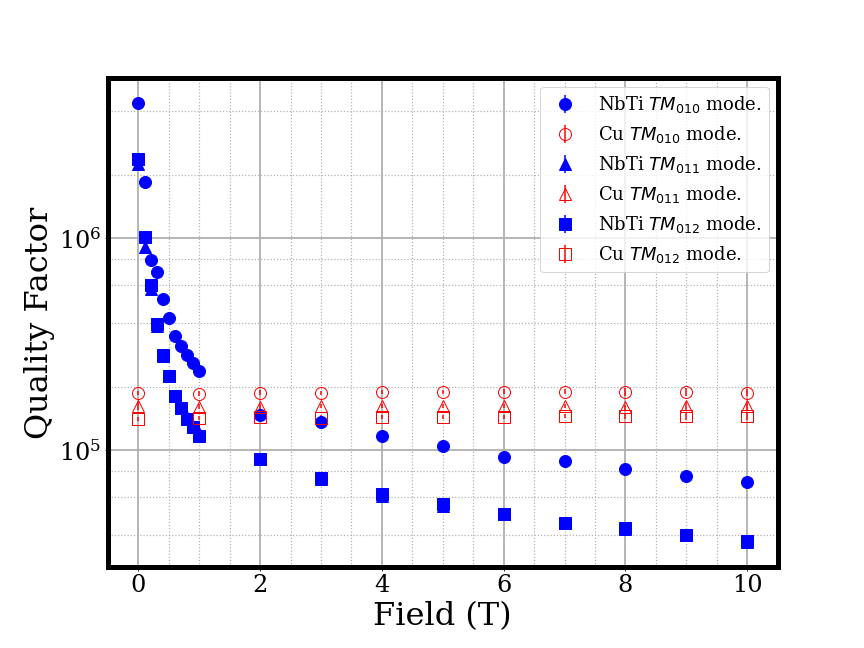}
        \caption{NbTi versus copper clamshell in PPMS system: The cavity quality factor for the first three $TM$ modes as function of applied magnetic field. Data was taken at a temperature of 2\,K.}
        \label{fig:ModeQvsB}
    \end{figure*}
    \par
\begin{figure*}[htb!]
        \centering
        \includegraphics[width=0.8\linewidth]{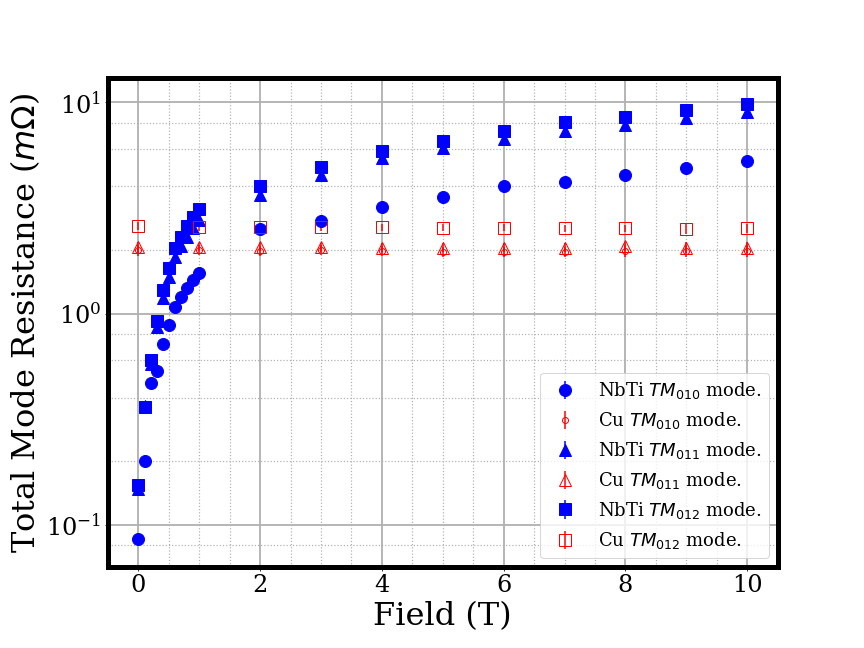}
        \caption{NbTi versus copper clamshell in the PPMS system: The total mode surface resistance for the first three $TM$ modes as function of applied magnetic field.}
        \label{fig:ModeRvsB}
    \end{figure*}

 The zero field Q factor for the NbTi cavity was measured to be $\approx\,70,000$ at $2\,{\rm K}$ directly after the cavity was machined. The cavity was then sent to Jefferson Laboratory for a cleaning and annealing process. Post-annealing, the cavity had a zero field quality factor at $2\,{\rm K}$ of nearly 1 million; annealing and cleaning has a significant effect on the zero field Q. For comparison, the copper cavity had a Q of about 180,000 at $2\,{\rm K}$. The critical temperature of the NbTi was observed to be approximately $\approx\,8\,{\rm K}$ based on the beginning of the superconducting transition for both the pre-annealed and post-annealed cavity; this is actually slightly under the typical  $\,10\,{\rm K}$ usually cited for NbTi \cite{TypeIIBlatt}. The mode quality factor as a function of applied magnetic field is shown in Figure\,\ref{fig:ModeQvsB}. As expected, the NbTi cavity Q degrades exponentially as a function of field. Eventually these quality factors were then converted to total mode resistances via Equation \ref{eqn:Rmode} shown in Figure \ref{fig:ModeRvsB}.
    \par
\begin{figure*}[htb!]
        \centering
        \includegraphics[width=0.6\linewidth]{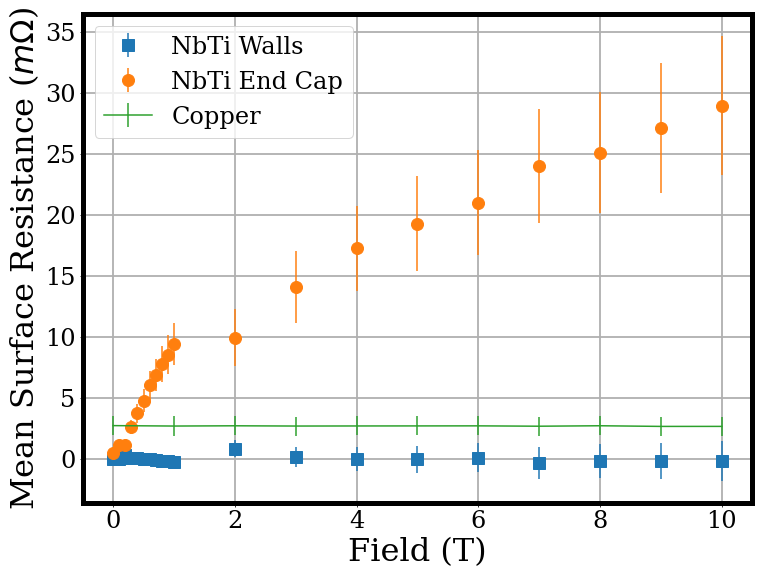}
        \caption{The mean surface resistance contributions of the walls and end caps of the NbTi cavity based on two different two-mode decompositions: $TM_{010}$/$TM_{011}$ and $TM_{010}$/$TM_{012}$. This is compared to the mean copper cavity resistance, averaged over surface type for the same mode combinations.}
        \label{fig:NbTiTM010RSvsB}
    \end{figure*}

 In order to test the consistency of the multi-mode decomposition, a two-mode decomposition to the walls and total end caps contributions was performed with combinations of the three modes measured. The $TM_{010}$/$TM_{011}$ and $TM_{010}$/$TM_{012}$ decomposition surface resistances agreed very well and confirmed the hypothesis that the end caps were the primary contribution to the surface resistance in field  for the NbTi cavity (Figure\,\ref{fig:NbTiTM010RSvsB}). 
    \par The $TM_{011}$/$TM_{012}$ combination unfortunately did not produce a physical result; this is because the geometric factors of the two modes were nearly identical, making the determinant of the weights matrix very nearly zero as discussed in Chapter \ref{chap:Cavities} (almost two orders of magnitude smaller than the other two combinations), inflating errors when taking the inverse (see Table \ref{tab:PPMSdeterminants}). Having determined what caused this mode-combination to fail, it was excluded from the final data set.
    \begin{table}
        \centering
        \begin{tabular}{||c|| c| c||} 
        \hline
         Mode Combination & $\mathrm{Det(C_{ms})})$ & $\mathrm{\Delta(Det(C_{ms}))}$ \\ [0.5ex] 
        \hline\hline
        $TM_{010}/TM_{011}$ & 0.140 & 0.016  \\ 
         \hline
        $TM_{010}/TM_{012}$ & 0.146 & 0.016  \\
         \hline
        $TM_{011}/TM_{012}$ & 0.006 & 0.013  \\
        \hline
        \end{tabular}
        \caption{PPMS clamshell cavity multi-mode decomposition: determinant of weight matrix and associated error for the three possible mode combinations.}
        \label{tab:PPMSdeterminants}
    \end{table}
    \par These results can be compared to the results of the QUAX hybrid NbTi cavity in Ref. \cite{QUAX1} and Ref. \cite{QUAX2}; the NbTi film is thick enough (3-4 Microns) to be in the bulk limit and was tested in a similar temperature and field regime. The first analysis in Ref. \cite{QUAX1} estimates a similar resistance for their copper cone end-caps at $4\,{\rm K}$ of $5\,{\rm m\Omega}$. However they must calculate their NbTi resistance using this copper resistance value with their methodology; they get a value of $20\pm20\,{\rm m\Omega}$ for NbTi in zero field, much higher than our measurement at zero field, and closer to the resistance we saw at $5\,{\rm T}$ (Figure \ref{fig:NbTiTM010RSvsB}). As for higher field values, the quality factor drops off at a similar rate, starting at 1.2 million at zero field to around 35,000 at $5\,{\rm T}$. It is a bit curious and discouraging considering our cavity is entirely NbTi while the QUAX is a hybrid design. This can be explained partially by operating at $4\,{\rm K}$ versus $2\,{\rm K}$, a significantly different cavity geometry, and lower frequency regime. The results of the analysis found in \cite{QUAX2} are in better agreement with ours; they cite a Q of 1.2 million at $4\,{\rm K}$ and $0\,{\rm T}$ versus our 1.4 million at $2\,{\rm K}$, and this dropped to 200,000 at $5\,{\rm T}$ versus our 100,000. This implies their cavity must have had a lower total resistance at $5\,{\rm T}$ because of the hybrid construction.
    \par As discussed in Chapter \ref{chap:Cavities}, the two sources of error in this measurement are the quality factor and geometric factor. The geometric factor errors in the PPMS cavity design were discussed earlier and reported in Table \ref{tab:cavgeofactor}. The quality factor uncertainty was determined from the estimated covariance matrix of the Lorentzian magnitude fit; this is an output of the \verb|Scipy.optimize.curve_fit| package. Typically this error was constant with $\Delta Q$ on the order of $10^2$. Any measurements with $\frac{\delta Q}{Q} > 0.10$ were considered outliers, and the fit was subsequently reviewed by plotting the data; this occurred only seven times in the data set, all for the superconducting NbTi cavity at low field. Upon further inspection, it was clear that the fit function failed because of the large window size compared to the very narrow resonant peak width; this could be corrected for by reducing the window size and by removing some of the off-resonant data from the fit in post-processing.
    \par The maximum estimated uncertainty for each parameter along the analysis chain for the decomposition is then reported in Table \ref{tab:PPMSerrorbudget}. As you can see, the inverse weight matrix compounds a very high increase in fractional uncertainty from the amount of multiplications needed in its construction. Since the surface resistance was statistically zero for the NbTi wall measurements, it was better to report the maximum absolute uncertainty, which was in the NbTi endcaps, $ \pm \, 5.7 \, {\rm m \Omega }$, rather than a misleading high fractional uncertainty in the NbTi walls.
    \begin{table}
        \centering
        \renewcommand{\arraystretch}{1.3}
        \begin{tabular}{@{}lcr@{}}
        Parameter & Variable & Maximum Uncertainty \\
        \hline
        Quality Factor & $\Delta Q$ & $4.2\%$ \\
        \hline
        Geometric Factor & $\Delta G_{ms}$ & $5.0\%$ \\
        \hline
        Total Mode Resistance& $\Delta R_{m}$ & $6.6\%$ \\  
        \hline
        Weight matrix & $\Delta C_{ms}$ & $4.9\%$ \\  
        \hline
        Inverse weight matrix & $\Delta C_{ms}^{-1}$ & $12.6\%$ \\
        \hline
        Surface Resistance & $\Delta R_{s}$ & $5.7  m\Omega$ \\ 
        \hline
        \end{tabular}
        \caption{A summary of the maximum estimated uncertainties for the surface decomposition measurement setup. They are expressed as percentages of their nominal values with the exception of surface resistance, since many values measured were effectively zero while the cavity was superconducting. Note that $\Delta C_{ms}^{-1}$ and $\Delta R_{s}$ does not include the $TM_{011}$/$TM_{012}$ decomposition because of near-zero determinant.}
        \label{tab:PPMSerrorbudget}
    \end{table}
    \par
\begin{figure*}[htb!]
        \includegraphics[width=0.48\textwidth]{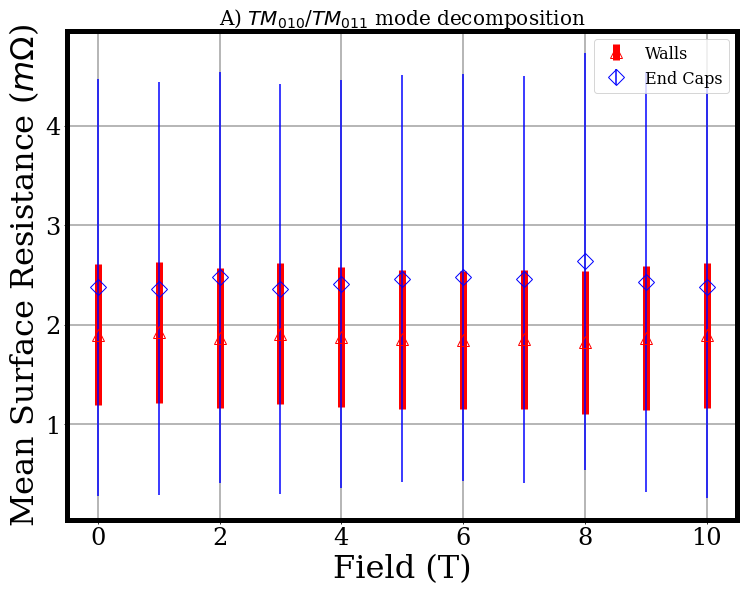}
        \includegraphics[width=0.48\textwidth]{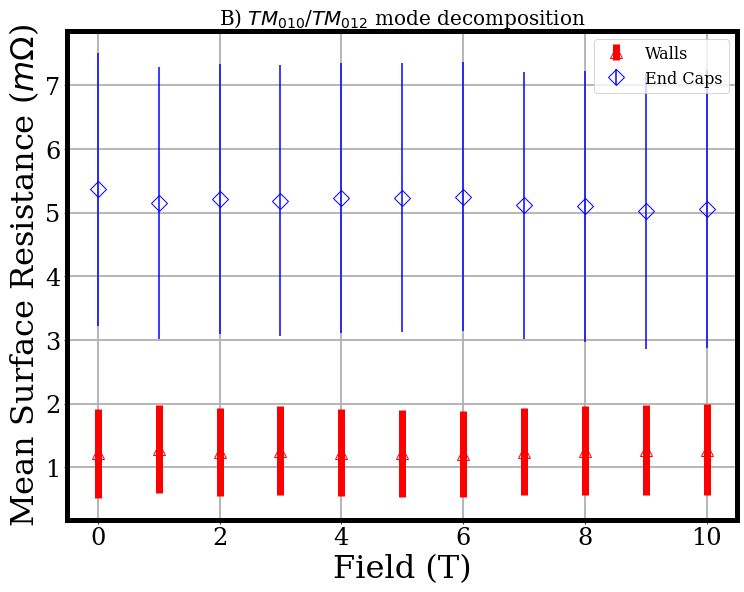}
        \includegraphics[width=0.48\textwidth]{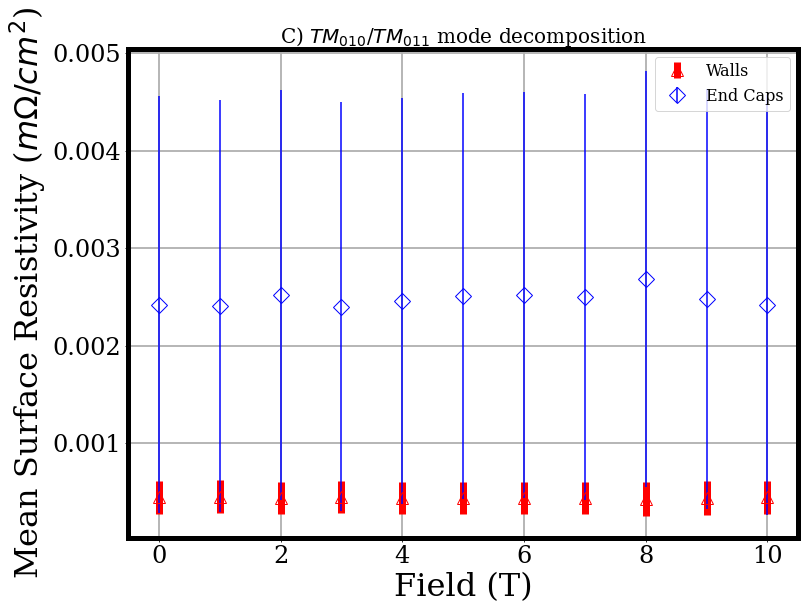}
        \includegraphics[width=0.48\textwidth]{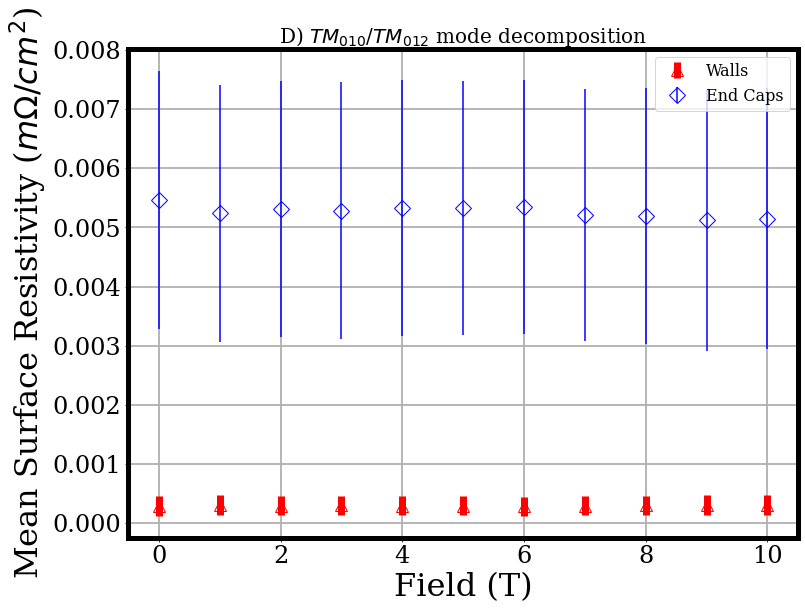}
        \caption{The two two-mode surface resistance decompositions for the pure copper clamshell cavity vs field. A) Surface resistance of the cavity sub-surfaces for the $TM_{010}/TM_{011}$ combination. B) The same plot for the $TM_{010}/TM_{012}$ combination. C) The corresponding resistance per unit surface area for the $TM_{010}/TM_{011}$ combination. D) The same plot for the $TM_{010}/TM_{012}$ combination. These bottom plots account for the differing surface areas of the walls and end caps. Note the differing error bars are due to the higher uncertainty in the end cap geometric factors.}
        \label{fig:CuDecompgraph}
    \end{figure*}

 A final check on the efficacy of this method was to look closely at the control copper cavity, and its variance for each mode combination decomposition. Being a non-magnetic material, one expects to not see a difference between the end-cap and wall resistivity (resistivity referring to resistance per unit surface area). The resistance may be slightly different based on the differing surface areas, but overall this should be captured by the geometric factor. As you can see in Figure \ref{fig:CuDecompgraph}, the $TM_{010}/TM_{011}$ combination does follow this pattern; however, the $TM_{010}/TM_{012}$ does not, indicating it might not be a stable or effective mode pairing. Nonetheless, this observable difference between the two surfaces in this mode pairing is on the order of the error bars themselves, and is far smaller than the difference observed for the NbTi cavity surfaces in Figure \ref{fig:NbTiTM010RSvsB}. Overall, this gives confidence in the efficacy of the decomposition method for its future use.
    \par The final take-away is that this result supports the construction of hybrid SRF cavities. The NbTi walls had a lower resistance than copper up 10 T, and its overall degradation seems to only be from the increased resistance in the end caps. If one takes the wall resistance with maximum positive error, giving the highest possible resistance, one sees that it is still lower than copper by a factor of $\approx 2.3$. Using this minimum improvement factor to project a Q for this cavity if it were to be a hybrid, with copper end caps and NbTi walls, one expects a minimum Q of 315,000 up to 10 T, which is a 67\% improvement over the all copper PPMS clamshell. The details of this measurement were reported in Review of Scientific Instruments in March 2023 \cite{BraineMultiMode}.
    \section{Design and status of a Hakki-Coleman PPMS Cavity}
\begin{figure*}[htb!]
        \centering
        \includegraphics[width=1.0\linewidth]{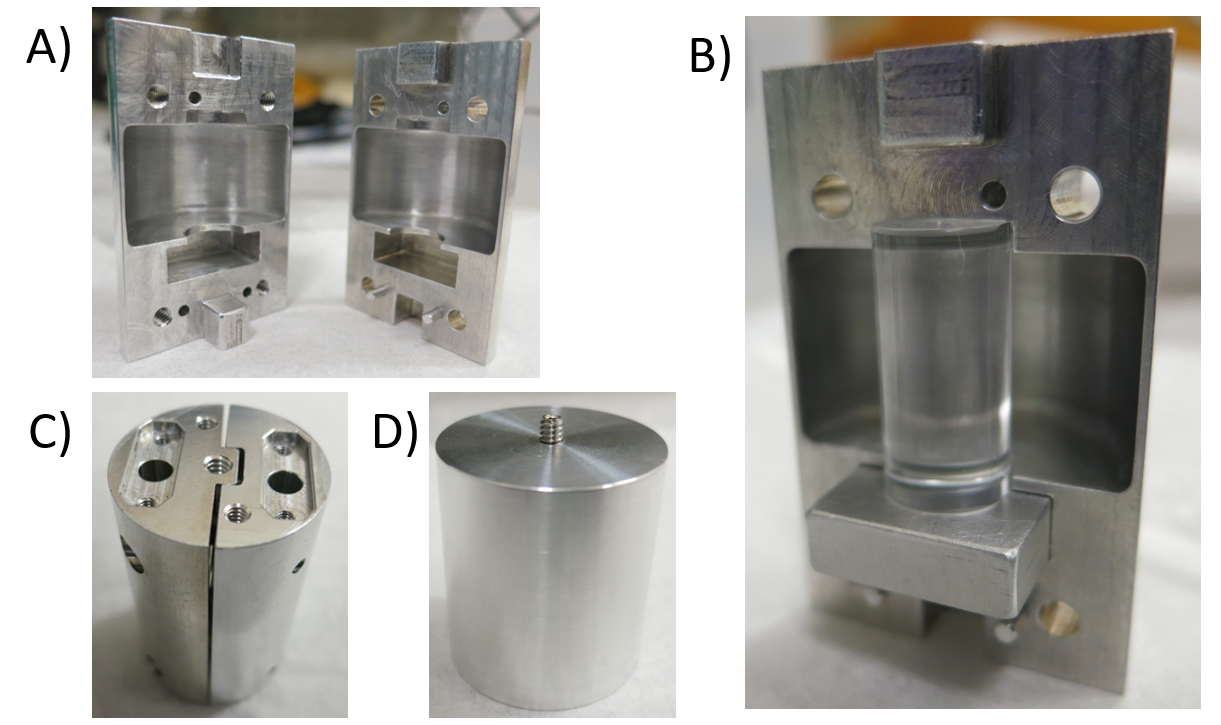}
        \caption{An aluminum clamshell dielectric loaded cavity for precise surface impedance measurements of samples. A) The two halves have circular insets for the sapphire cylinder and rectangular inset below for inserting sample blocks. The bottom circular inset acts as a mask to remove edge defects from the exposed surface. B) The cavity loaded with sapphire rod and aluminum sample block. Superconducting coatings would be sputtered to a substrate block of similar size and can be exchanged easily. C) The cavity contains the same exterior features for compatibility with the PPMS. D) A rising cylinder lifts the shorter cavity to center the sample surface in the PPMS magnet.}
        \label{fig:PPMSRodCavity}
    \end{figure*}

    Another cavity design compatible with the PPMS was pursued for precise RF surface impedance measurements of planar superconducting samples. Pictured in Figure \ref{fig:PPMSRodCavity}, this cavity is loaded with a dielectric sapphire crystal cylinder, which focuses the electromagnetic field of the mode away from the wall region and onto the end cap regions above and below the cylinder. Additionally, the increased permittivity of the dielectric ($\epsilon \approx 10$) shifts the mode frequency downward, allowing one to probe lower frequency regimes with smaller physical cavity dimensions. This is not a new technique, and is used for high sensitivity impedance measurements \cite{Alimenti2019,ChenMicrowave}. This is similar to a Hakki-Coleman type resonator \cite{HakkiColeman}; however, it was made into a clam-shell to accommodate the small bore size of the PPMS and did not contain any springs or vernier drive that typically clamp the cylinder to the sample surface (more on this later). 
    \par The $TE_{011}$ mode is the principle mode of operation because the circular currents are easily focused away from the walls, and they also will generate maximal vortex motion on the sample surface. Since this sample surface is also oriented perpendicular to the external field of the PPMS, this gives the maximum expected surface resistance for a sample. A sapphire crystal of 4.5 mm diameter was chosen to give an expected mode frequency of ~8.5 GHz. This was chosen over an 8 mm cylinder design, similar to the resonator Ref. \cite{Alimenti2019}, which had an operating frequency of ~14.5 GHz. The height of the cavity was reduced from 53 mm to 13 mm, with two 1 mm circular insets (Figure \ref{fig:PPMSRodCavity}A-B) that cradled the 15 mm length rod; this was intended to reduce the wall surface area, and subsequently increase the geometric factor of the walls, making it a negligible contribution to the quality factor. The bottom inset additionally exposed a uniform circular region from the center an inserted sample block; this masked any edge defects on the sample coating, and controlled the geometry of the exposed sample region.
    \par To make this compatible with the PPMS system, all of the exterior features were kept the same as the original clam-shell design (Figure \ref{fig:PPMSRodCavity}C). Since this cavity had the reduced length, a spacing cylinder was attached below the cavity, which lifted the sample surface to the exact center of the magnet by 40 mm (Figure \ref{fig:PPMSRodCavity}D). This also kept the overall length of the assembled system the same, so that the PPMS puck and probe seated correctly within the bore. 
    \par The surface impedance measurement worked as follows. The Quality factor and resonant frequency of the mode are monitored by measuring and fitting to the S parameters periodically, with $T$ and $H_{ext}$ being varied. The coupling of the antennae again are kept weak such that $Q_L\approx Q_0$. Additionally, running the network analyzer with low power input signals minimized the RF amplitude of the mode magnetic field, so $Z_s$ was unaffected by the measurement itself, which Niobium and $\mathrm{Nb}_3\mathrm{Sn}$ can be very sensitive to \cite{Hall2017HighPN,NbZ_sField1}. The total geometric factor of this cavity, in principle, is dominated by the sample surface, such that all other sub-surfaces are negligible. This then enables us to write down the surface impedance of the sample by the relation:
    \begin{equation}
            Z_s=R_s+i\Delta X_s=\frac{G_s}{Q}-i2G_s\frac{\Delta f_0}{f_0}-C
            \label{eqn:RodCavityImpedance}
    \end{equation}
    where $\Delta$ represents the variation with respect to the reference zero-field reactance, $X_{s,0}=\omega\mu_0 \lambda$, and frequency, and C represents a background constant. The background constant is determined by calibration with a matched metal sample with the cavity in field. After this, a superconducting sample can be loaded, and zero-field calibration can be made: $R_s(H=0, T\rightarrow 0) \approx 0$ as well as $R_s=X_s$ for $T>T_c$ are used as fixed point relations to center the absolute impedance values (this is a real quasi-particle conductivity, also called the Hagens-Rubens limit). The data is then taken by fixing the magnet strength at several values and then varying the temperature for each field strength at a fixed rate from the fridge base temperature to slightly above zero field critical temperature. Once the absolute complex impedance data is in hand, Equation \ref{eqn:SRFZ_s} can be used to find the complex resistivity values, $\rho_{vm}'$ and $\rho_{vm}"$, the ratio of which, $r=\rho_{vm}"/\rho_{vm}'$, is directly related to the pinning frequency, $r=f_p/f=\omega_p$. As stated in Chapter \ref{chap:SRF}, this pinning frequency provides the upper frequency limit for elastic vortex motion in the superconductor.
    \par This project unfortunately hit a standstill during my PhD tenure due to several design flaws. The principle design flaw was that the rod was too narrow and the walls were too long to achieve the desired geometric factor arrangement. As you can see from Table \ref{tab:PPMSRodcavityGfactors}, the cavity is dominated by the mask region of the bottom end cap, rather than the sample surface as intended. Furthermore, the top end cap and walls themselves, having a smaller geometric factor, contribute more loss to the Q than the sample. The cavity used in Ref. \cite{Alimenti2019}, which was 8 mm in diameter and 5 mm in height, had an incredibly high wall geometric factor of 420,00 $\Omega$; this is due to the shorter wall height as well as the increased focusing power from a larger diameter crystal. This design also had a top end cap with a lower geometric factor than the sample surface, but the ratio between the two in our design is much closer to equality. The solution to this is relatively simple: Build a new shorter cavity with a wider crystal cylinder. However, this does drastically change the frequency regime of the measurement; our original research goal was to measure the impedance in the frequency regime below 10 GHz.
    \begin{table}
        \centering
        \begin{tabular}{||c|| c| c||} 
        \hline
         $TE_{011}$ Geometric Factor ($\Omega$) & Our design (4.5x15 mm Rod) & 8x5 mm Rod from Ref. \cite{Alimenti2019} \\ [0.5ex] 
        \hline\hline
        Walls & 1160 & 420,000  \\ 
        \hline
        Mask & 265 & 64  \\
        \hline
        Top End Cap & 1260 & 42\\
        \hline
        Total Non-sample surface & 166 & 25  \\
        \hline
        Sample Surface & 2940 & 58  \\
        \hline
        Total & 173 & 18  \\
        \hline
        \end{tabular}
        \caption{A comparison of geometric factors expressed in Ohms between the LLNL PPMS-compatible clamshell Hakki-Coleman Resonator versus a previous, more traditionally designed Hakki-Coleman resonator used in Ref. \cite{Alimenti2019}. This highlights a design flaw in our resonator; the sample surface is much less significant than the other surfaces in the resonator, therefore making accurate surface impedance measurements of the sample very difficult.}
        \label{tab:PPMSRodcavityGfactors}
    \end{table}
    \par In practice, this clam-shell cavity design had several other flaws. As stated earlier, because the design lacked a spring/compression system to ensure the crystal was pressed against the sample surface, thermal contractions likely caused a gap to develop upon cooling. This manifested in unpredictable mode structures due to the gap. It also would further increase the sample geometric factor, further decreasing its contribution to the quality factor. Additionally, the sample block had clear issues with thermalizing at cryogenic temperatures because it lacked a tight connection with the cavity, causing it to be much hotter than the PPMS temperature. Attempts were made to mitigate this by pressing small amounts of indium metal into the gaps between the block and rectangular insert, but were unreliable. Furthermore, because the surface current structure of the $TE_{011}$ mode crosses the seam of the clam-shell, the surface resistance of the cavity surfaces are increased, decreasing their sub-surface Q and drowning out the sample surface Q signal in measurement. For all these reasons, a real surface impedance measurement of a superconductor was never made with this cavity during my time at LLNL.
    \par I bring this project up because it is a great example of the lessons learned through practical failures in grad school. This was a great idea in principle, and had been done before, but it was only by doing it and dissecting the design details that we were able to see the deeper requirements that make this cavity design work. It is my hope that someone in the future can take note of this if pursuing the same measurement scheme.
    
\chapter{ADMX Sidecar run 1D preliminary results}
\label{chap:Sidecar1D}
This chapter will go over the ADMX Sidecar preparations, operations, and results so far in run 1D. The highlight of this Sidecar run is the first implementation of a $\mathrm{Nb}_3\mathrm{Sn}$ superconducting tuning rod. It is late April 2024 at the time of writing. ADMX run 1D began data-taking operations in earnest in mid-December 2023 after multiple delays.  The main experiment is expected to run for about a year for its first half of data-taking with the next expected insert extraction taking place around December 2024. The ADMX Sidecar is as always "along for the ride" so to speak and is unable to be extracted for repairs or testing. Everything must be done "in-situ" and must also be done in accordance with the main experimental operations. Unlike the LLNL PPMS system discussed in section \ref{PPMSsystem}, the Sidecar is in a much less controllable environment in terms of adjusting temperatures and magnetic fields; this makes characterizing the cavity challenging and sometimes impossible. The goal of this chapter is to not only highlight the work that has been done so far in characterizing the superconducting rod, but the axion search analysis that could be done given less than ideal operating conditions.
\section{Previous Sidecar runs overview}
\begin{figure*}[htb!]
\centering
\includegraphics[width=0.52\textwidth]{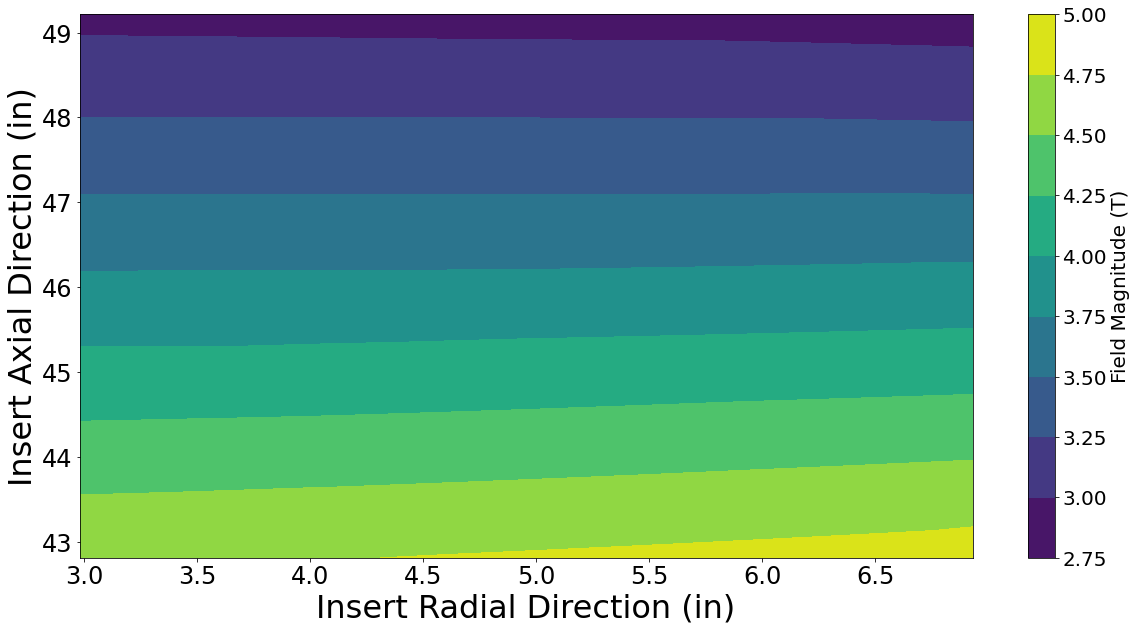}
\includegraphics[width=0.52\textwidth]{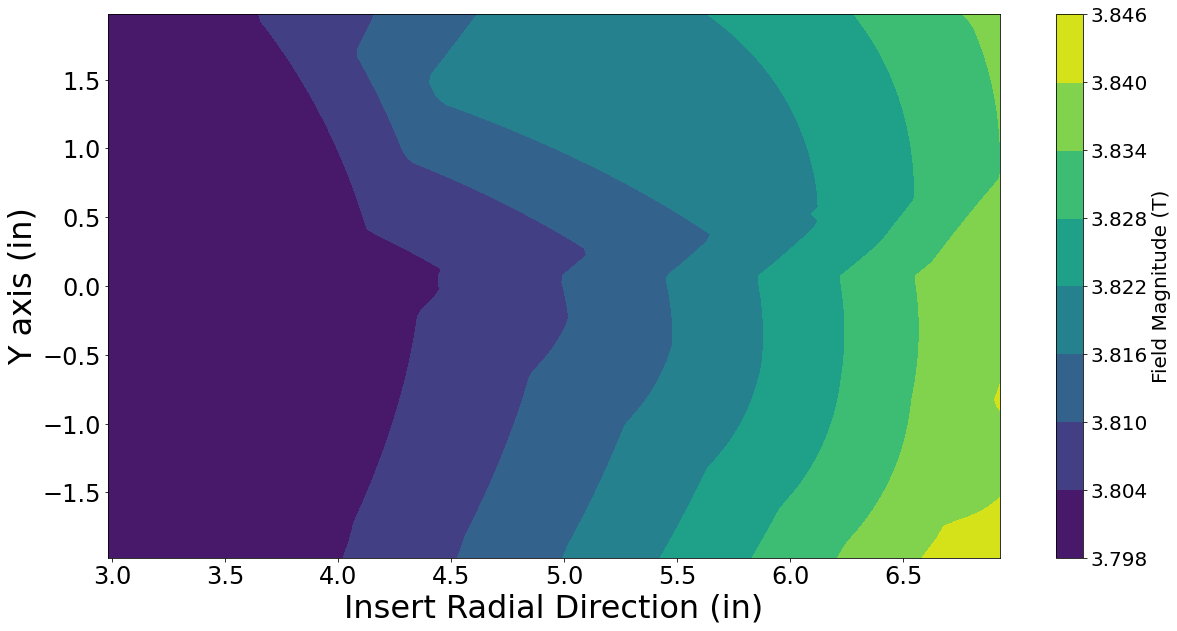}
\includegraphics[width=0.52\textwidth]{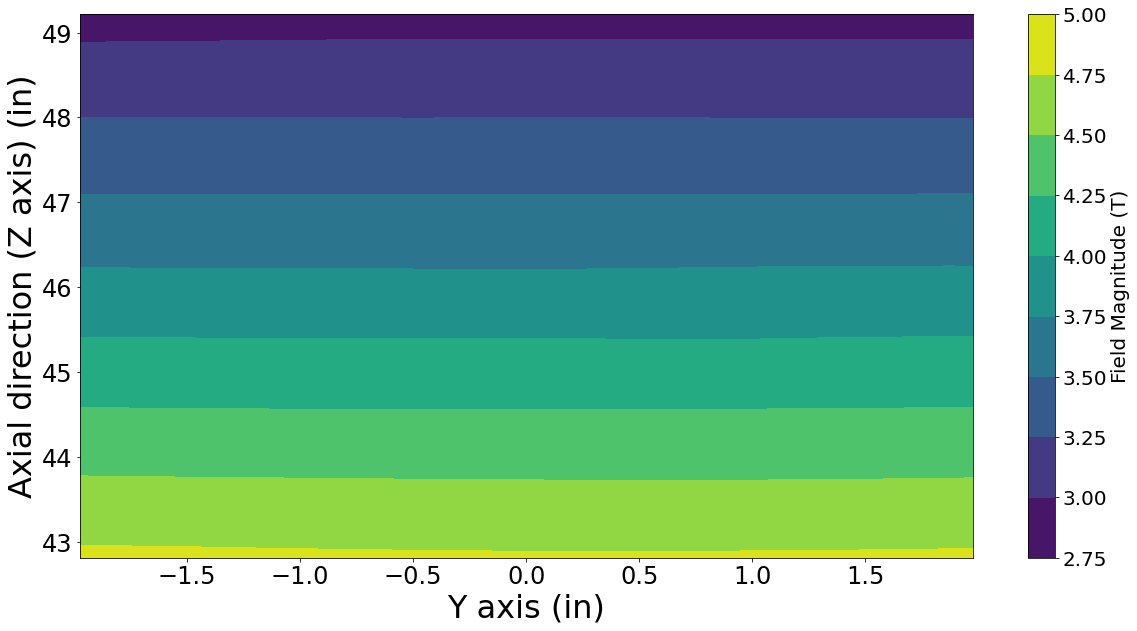}
\caption{The ADMX Sidecar Magnetic Field Region.The magnitude of the magnetic field is plotted for each cross-section passing through the center of the Sidecar cavity. The radial coordinate using the ADMX insert center as zero. The axial 'Z' coordinate uses the center of the magnet coils as its zero. The 'Y' coordinate is the only coordinate that uses the Sidecar center as its zero. Note that this plot doesn't display the direction of the field vector, which is predominantly in the axial direction; off-axis components are quite small with the average $B_z/|B|$ over the region being $\approx 0.981$.}
\label{fig:SidecarField}
\end{figure*}

The ADMX Sidecar was introduced by former ADMX UW graduate student Christian Boutan, as a high-frequency (4-7 GHZ) side axion experiment for not only testing emerging technologies that could help solve the problems with high frequency axion searches, but for taking real axion search data at these frequencies, albeit at a much lower sensitivity than the main experiment. This was only possible because of a bit of free space on the top of the main cavity and below the 1 Kelvin plate above. This region is in the fringe field of the main magnet, giving the Sidecar only an average magnetic field of 3.58 T at full current, 220 A, significantly reducing its sensitivity to axions; this is the main reason why it does not achieve of KSVZ or DFSZ sensitivity. This fringe field region has been mapped, with the magnitude of the field plotted through each center cross section in Figure \ref{fig:SidecarField}. Although the field is slightly curved in this region, the off-axis components are still relatively small, with an average $\frac{B_z}{|B|}\approx 0.981$ over the region. In addition to the cavity sitting in this reduced field region, the Sidecar must have all the other necessary systems for a haloscope: Tunable rod and antennae, a cold, low-noise RF receiver chain, and the warm electronics to support these systems. 
\par Sidecar's premier run was during run 1A, where it's emerging technology was the use of the piezoelectric motors for tuning the rod and antennae. This was one of the first successful demonstrations of cavity resonant tuning with such a motor. The stick-slip motors were manufactured by Attocube \cite{Attocube} to operate at 10 mK, 31 T, and high-vacuum. The Attocube ANR240/RES rotory motor and ANPz101eXT12/RES linear motor were mounted to the cavity's top end cap and were used for tuning rod control and antenna coupling adjustment respectively. These motors can produce a significant heat load, so extra care had to be taken not to heat the Sidecar cavity or overall insert too much. The linear piezoelectric motor also did not function for a majority of this run, rendering the antenna coupling stationary. This run only implemented a Low Noise Factory HFET amplifier, LNF-LNC1\_12A, thermally sunk to the top of the helium reservoir, as its primary amplifier. They measured a noise temperature of $4.0 \pm K$ using the Y-factor method 1 highlighted in section \ref{hotload1}. The cavity itself was a knife-edge construction similar to the main cavity, with two end caps bolted to a 0.121 m tall, 0.064 m inner diameter cylindrical barrel with very sharp edges that 'bite' into the end caps; the tuning rod was a 0.013 m diameter, hollow copper tube with fused end caps. This cavity had an unloaded Q of 22,000 empty, and 11,000 with the tuning rod. The system was able to exclude axions in 3 regions, 4202-4249 MHz, 5086-5799 MHz, and 7173-7203 MHz, to varying sensitivities beyond the CAST limit; the last region was done using the higher $TM_{020}$ mode, which was a novel technique at the time. For more details of this run, see ref. \cite{Sidecar1A}.
\par
\begin{figure*}[htb!]
\centering
\includegraphics[width=1\textwidth]{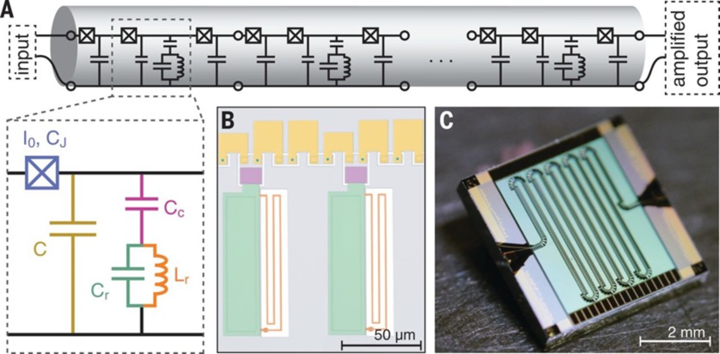}
\caption{The Sidecar JTWPA. A) The circuit diagram of the amplifier. B) A diagram of the amplifier micro structure. C) A picture of the entire amplifier chip.}
\label{fig:TWPAschematic}
\end{figure*}

\begin{figure*}[htb!]
\centering
\includegraphics[width=0.7\textwidth]{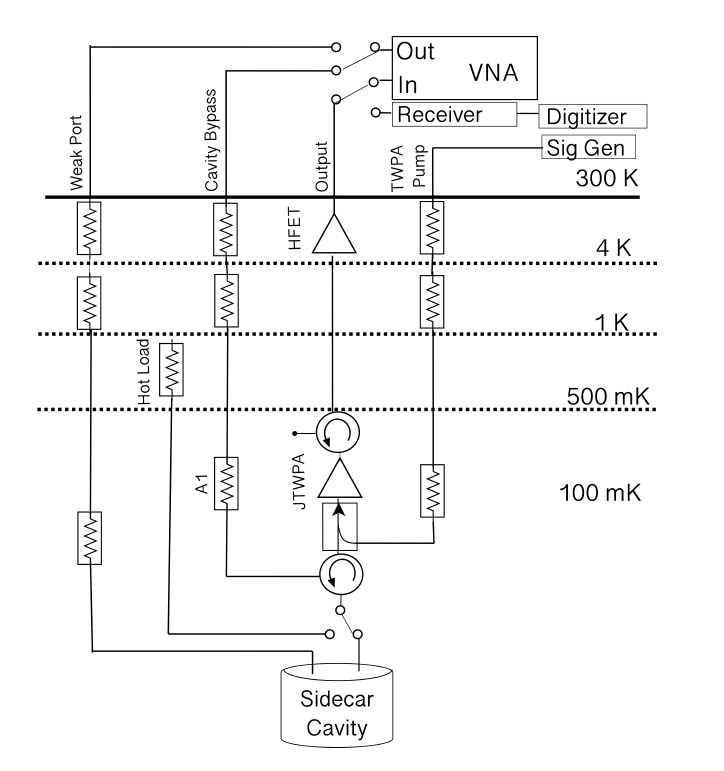}
\caption{An RF schematic of the run 1C and 1D Sidecar receiver. The JTWPA is placed between two circulators to prevent power being reflected back into the cavity and to allow for periodic RF calibrations. Although a completely separate chain, these magnetic-sensitive cold electronics resided in the SQUIDadel along side the main cavities' sensitive electronics protected by the bucking coil magnet.}
\label{fig:Run1CSidecarRFchain}
\end{figure*}

 The first major set of upgrades to Sidecar came together in time for ADMX run 1C. The premier technology for this run was the introduction of a quantum noise-limited amplifier, a Josephson Traveling Wave Parametric Amplifier (JTWPA), as the primary amplifier. JTWPAs are known for delivering high gain over a wide (3 GHz) bandwidth, which is ideal for more broadband axion searches. This amplifier, pictured in Figure \ref{fig:TWPAschematic} was fabricated at and acquired from MIT Lincoln Laboratory \cite{LincolnLabTWPA}. It is made of niobium with a critical temperature of roughly 9.3 K. Being a field-sensitive component, this amplifier was mounted within the SQUIDadel, so that it would be protected by the bucking coil magnet. This amplifier, because of its broadband nature, has an operational gain typically limited to around 10 dB. When the JTWPA was paired with the Sidecar HFET as a secondary amplifier, which had a noise temperature of $3.7 \pm 0.2 K$ using Y factor method 2 (outlined in Section \ref{hotload2}) the system had an overall system noise of $925 \pm 80 mK$. That HFET amplifier towards the end of this run failed, and was sent for repair after the insert was extracted. The RF schematic of this amplifier chain is pictured in Figure \ref{fig:Run1CSidecarRFchain}.
\par
\begin{figure*}[htb!]
\centering
\includegraphics[width=0.8\textwidth]{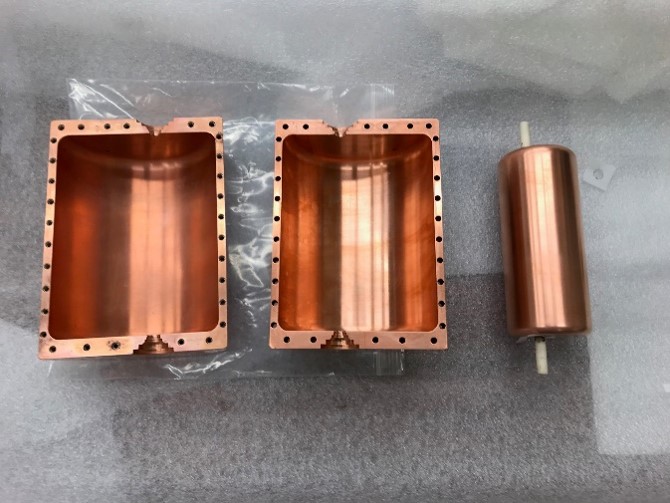}
\caption{The run 1C sidecar clam-shell cavity disassembled with 2" copper tuning rod and alumina axles.}
\label{fig:Run1CSidecarCavity}
\end{figure*}

 Concurrent with the upgrade to the RF chain, a whole new Sidecar cavity was fabricated at LLNL, and pictured in Figure \ref{fig:Run1CSidecarCavity} disassembled. This cavity was a clam-shell construction, with two cylindrical halves machined from a solid copper block and bolted together with machine screws around the flat face of either half. The clam-shell seam, in theory, being along the axial direction, should not disturb the currents of the $TM_{010}$ mode or any higher axion-search modes, therefore maintaining Q. The 2" diameter copper tuning rod was hollowed out and the end caps welded back on to minimize the mass, and therefore the torque required by the piezoelectric motor to rotate. The alumina axles were epoxied to copper tabs that bolted into an inset on the top and bottom of the rod. These axles would pass through the apertures that are seen slightly off center axis of either half of the cavity in Figure \ref{fig:Run1CSidecarCavity}. These tuning port holes were made diagonal with varying steps to accommodate different size ball-bearings to optimize the smoothness of the rod turning; however, it was later suspected that this may have been a source of RF leakage, and would inform the cavity design in run 1D. During run 1C operations, the cavity unfortunately had a very low Q of about 700. This was thought to be because of RF leakage, most likely through the tuning rod port, but also possibly through the clam-shell seam if it had loosened up significantly under cryogenic conditions.
\par Finally, the piezoelectric motors were moved from being mounted directly to the top of the Sidecar cavity to the 1 Kelvin plate above. The exposed top alumina axle was clamped by a flexible bellow stainless steal rod coupler, which then clamped to a stainless steel rod that was directly attached to the rotor piezoelectric motor above through a port hole in the 1 Kelvin plate. Because the alumina was an insulator, there shouldn't be a significant thermal connection between the 1 kelvin plate and the cavity. The linear piezoelectric motor was mounted to the underside of the 1 Kelvin plate, and a NbTi superconducting coaxial cable antenna was clamped to it; the antenna would be superconducting during operations, and would not have the thermal conductivity to create a link between the cavity and 1 kelvin plate. The whole idea behind this change was to minimize the heating of the cavity and mixing chamber plate during tuning operations; tuning rates in run 1A were limited by the heat generated by the motors. The piezoelectric motors had mixed success during this run with both motors failing at certain times. The rotor motor, although working initially, failed before the JTWPA could be run for axion search data, and as a result the axion limits reported in Ref. \cite{Sidecar1C} only covered 2 MHz of frequency space, roughly the bandwidth of the digitization scans taken.
\par The successes and failures of these past two runs would inform the preparations and operations of run 1D Sidecar. As one will see, run 1D Sidecar seems to still suffer from some of the problems of its past: A low Q possibly from RF leakage, and failing piezoelectric motors.
\section{Preparations and design changes for run 1D}
This section will go over the design changes to the Sidecar system, warm testing, and installation into the ADMX insert.
    \subsection{Superconducting Tuning Rod}
\begin{figure*}[htb!]
    \centering
    \includegraphics[width=0.7\textwidth]{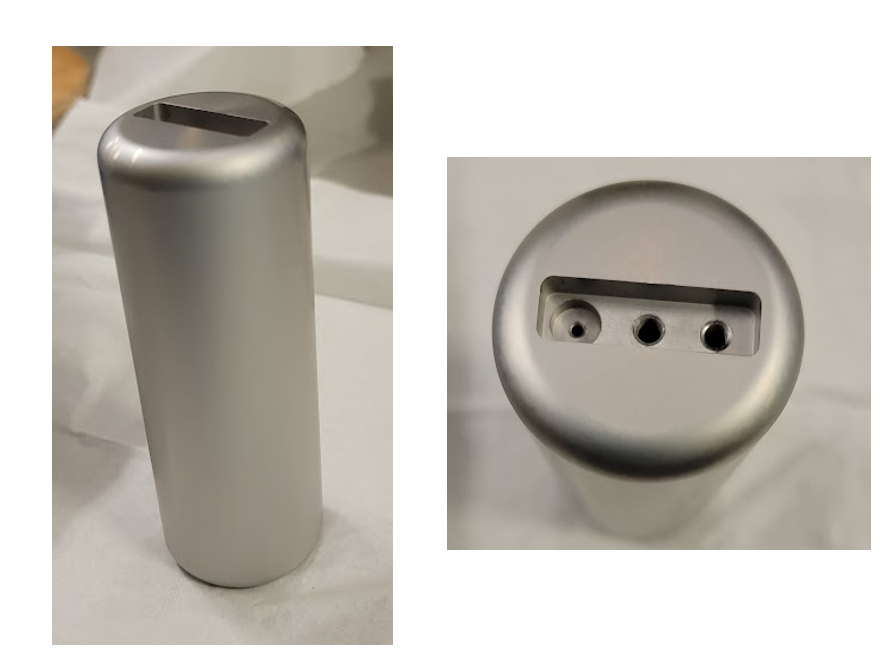}
    \caption{The $\mathrm{Nb}_3\mathrm{Sn}$ coated tuning rod from a side and top viewpoint. The rod is 1.75" in diameter, and features an inset for mounting the copper sapphire axle tabs. The center and right holes are tapped with 8-32 threads for mounting these tabs. The center hole goes through the length of the rod so that it could be hung by a wire in the coating furnace.}
    \label{fig:Nb3SnRod}
    \end{figure*}

    The highlighted novel technology of this run is a $\mathrm{Nb}_3\mathrm{Sn}$ coated tuning rod manufactured at Fermilab and overseen by Fermilab senior scientist, Dr. Sam Posen. The coating process comes from the accelerator cavity world, using pure niobium metal as the substrate for the coating, and is outlined in Ref. \cite{Posen_2017}. It is meant to be a thick film, on the order of about 3 microns, thick enough to prevent the field from penetrating into the bulk niobium. At their request, the rod diameter was reduced from 2" to 1.75", and the axle tab inset was redesigned to accommodate a center line hole; this would allow for hanging the rod by a wire strung through the center line hole when it was placed in the furnace for coating. The completed rod is pictured in Figure \ref{fig:Nb3SnRod}. An inadvertent benefit of this was that the detection volume of the cavity increased slightly, from 0.58 L to 0.61 L. This also shifted the tuning frequency range downward to 3.6-6.2 GHz. Two versions were made: One solid Nb Rod, and one that was hollowed out and the end-caps welded back in place. The hollow version was preferred by us because it would have a reduced torque for the piezoelectric motor. The solid version, however, was much easier for the Fermilab engineers to manufacture initially. In the end, the hollow rod was the rod installed in the cavity from which data was taken. This additional solid rod, however, is planned to be installed in a cavity at LLNL for further testing soon. 
    \subsection{Copper Clamshell Cavity}
\begin{figure*}[htb!]
    \centering
    \includegraphics[width=0.7\textwidth]{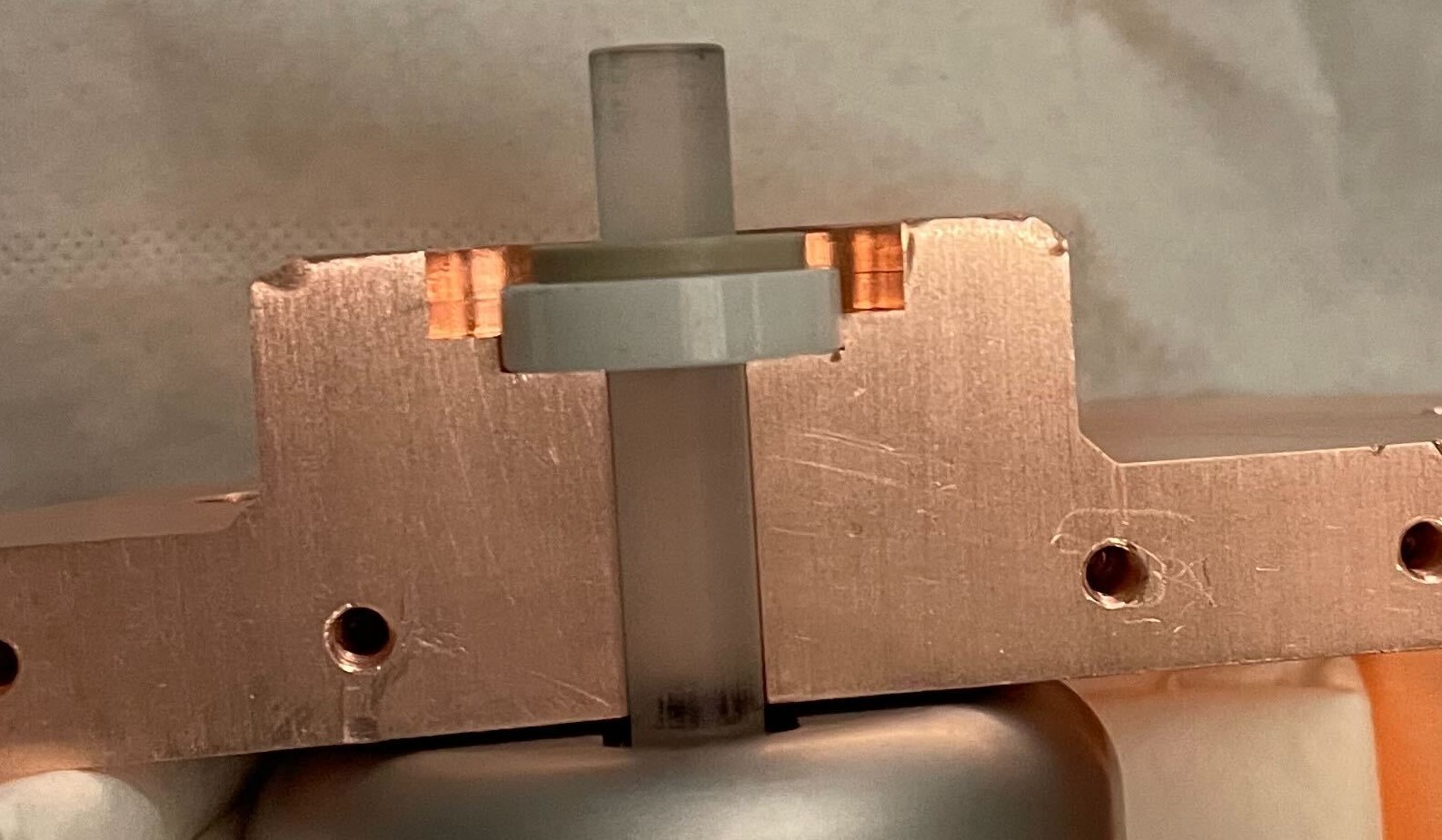}
    \caption{A close up of the run 1D Sidecar cavity tuning rod port profile as seen from one clam-shell half. The extension increases the length of port, minimizing RF power leakage out.}
    \label{fig:SidecarNewPort}
    \end{figure*}

    The copper clamshell cavity was re-made for this run by collaboration colleagues at the University of Sheffield. The design remained almost entirely the same with the exception of the tuning rod axle port; the port was extended out from the top of the cavity exterior and the number of ball-bearing inset steps was reduced from 4 to 2, as pictured in Figure \ref{fig:SidecarNewPort}. The idea here was to reduce any RF leakage through the port coming from the free space around unused steps, by creating a much longer port. Additionally, the tuning rod axles were changed from alumina to sapphire, in order to increase the thermal conductivity and dielectric constant of the axles. This was all motivated by the low Q seen in run 1C Sidecar operations.
    \subsection{Simulation Work}
    The cavity assembly was simulated using the COMSOL emw software package. The eigenmode solver was used to extract simulated values for the mode frequency, axion form factors, as well as an array of geometric factors for performing multi-mode measurements later.
    \par
\begin{figure*}[htb!]
    \centering
    \includegraphics[width=0.8\textwidth]{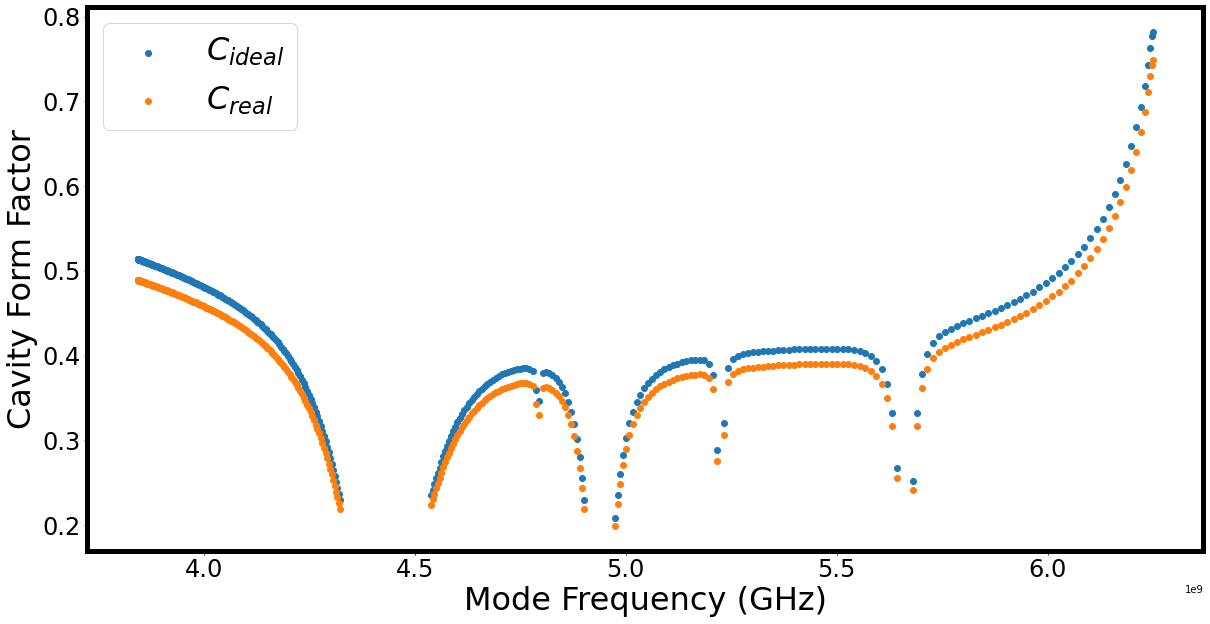}
    \caption{The simulated real and ideal axion form factor for the $TM_{010}$ mode as a function of the mode resonant frequency from COMSOL simulations. The ideal form factor assumes a perfectly straight magnetic field, whereas the real form factor data interpolates the magnetic field map data shown in Figure \ref{fig:SidecarField}. Because the real field isn't straight and constant, it is less than the ideal form factor. Both form factors drop to zero in mode crossing regions, of which there are 5 crossings visible in this plot. The real form factor data is used later in the axion search analysis for calculating expected axion signal power for each digitization scan.}
    \label{fig:sidecarformfactor}
    \end{figure*}

 The real axion form factor was calculated using the magnetic field map shown in Figure \ref{fig:SidecarField}, as well as an ideal form factor that assumed the magnetic field had perfect axial structure ($\vec{B}=B_0\hat{z}$). This was necessary for the $TM_{010}$ mode, because these values would be required for feeding into the axion search analysis using that mode later. For the other modes, it proved useful for identifying $TM$ modes of interest, The $TM_{010}$ form factors, real and ideal, over the entire tuning range are plotted in Figure \ref{fig:sidecarformfactor}.
    \par
\begin{figure*}[htb!]
    \centering
    \includegraphics[width=0.8\textwidth]{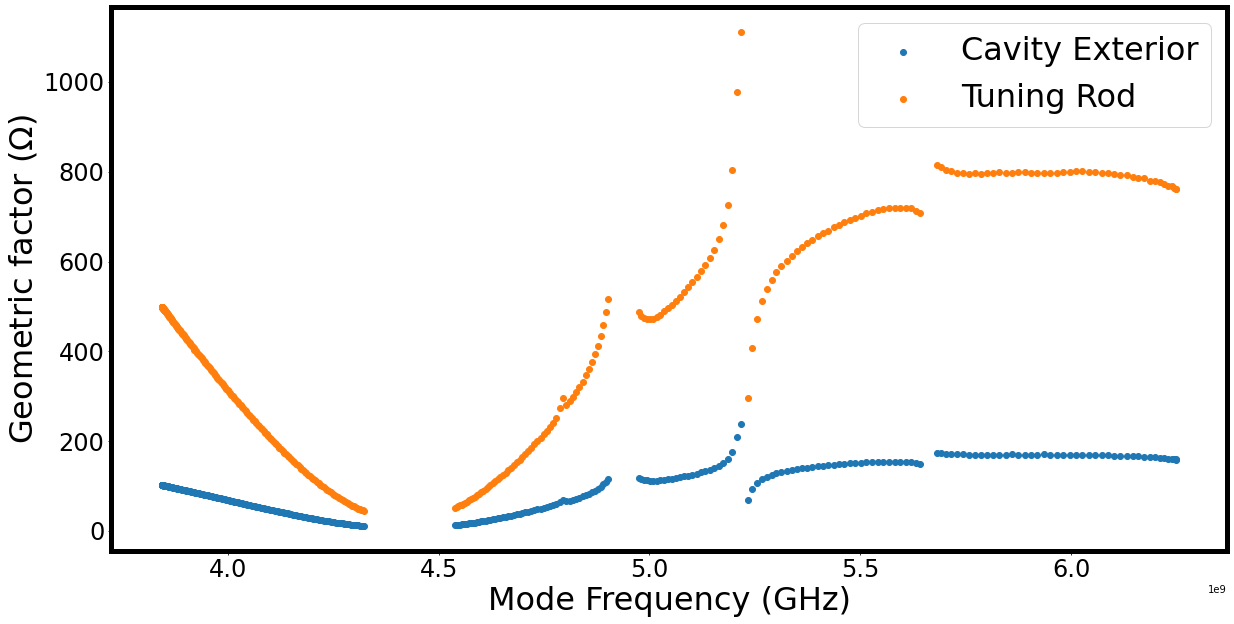}
    \caption{The simulated $TM_{010}$ mode geometric factors for the rod and exterior cavity as a function of mode resonant frequency over the entire tuning range of the cavity. The geometric factor tends to drop to zero near mode crossings, which makes it indistinguishable with TE modes, and therefore these regions are cut from the data set, producing the gaps in the plot data. When making a run plan, it was best to choose a region between mode crossings with relatively constant, well-separated geometric factors, such as the region between 5.2-5.6 GHz.}
    \label{fig:SidecarTM010GfactorCR}
    \end{figure*}

 A total of 12 sub-surface geometric factors were tracked in simulation: The exterior copper cavity surface, the tuning rod axle assembly, the entire tuning rod, the tube wall of the rod, and the end caps including the curved fillet surfaces that curved towards the wall. The rod wall region was further subdivided into 5 sub-surfaces along its axis based on height within the magnetic field: rod bottom, rod middle/bottom, rod middle, rod top/middle, and rod top. This was in anticipation of the magnetic field changing from the bottom to the top of the cavity interior, and the hope that one might be able to see a change in surface resistance along the length of the rod. The tuning rod angle was the primary geometric parameter varied, where 0 degrees represented a centered rod and 180 degrees represented a rod at the wall of the cavity, which as it was varied would give us a simulated mode map for the modes of interest. The first 5 TM modes were identified at first using axion form factor, but were eventually followed up by plotting the simulated electric fields and identified by eye especially near mode crossings. This initial simulation pass used the base design dimensions and took half degree steps in the tuning rod angle. Because of the large amount of data in this set, it has been chosen just to show the $TM_{010}$ mode geometric factor for the tuning rod and exterior cavity over the whole tuning range, pictured in Figure \ref{fig:SidecarTM010GfactorCR}. The gaps in this plot are due to the geometric factor dropping to near-zero values near mode crossings, and are cut from the data set. One would not want to operate the cavity in these frequency regions anyway. This initial pass didn't take into account error in the geometric factor because of the computational time.
    \par
\begin{figure*}[htb!]
    \centering
    \includegraphics[width=0.99\textwidth]{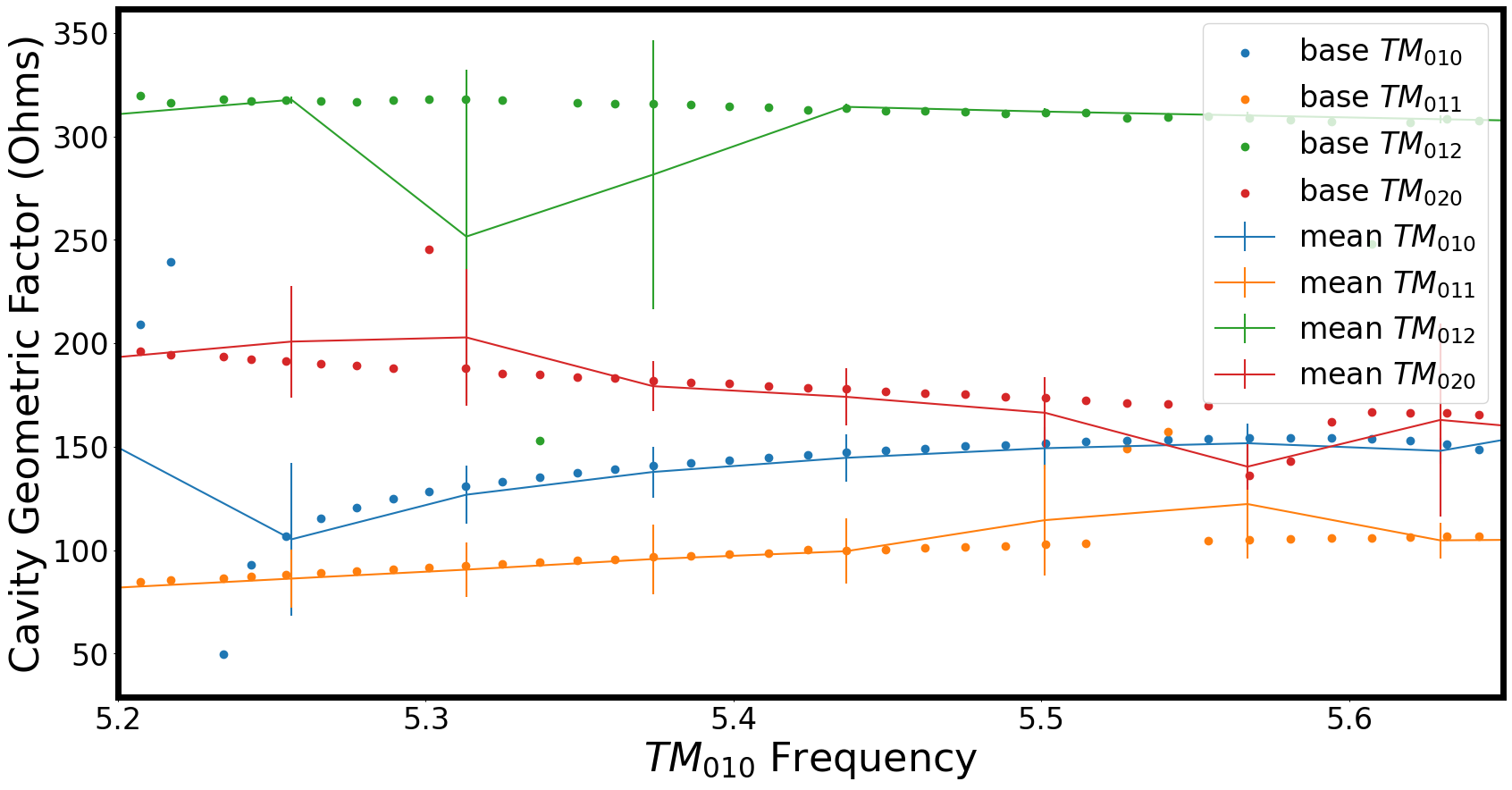}
    \caption{The simulated Sidecar exterior cavity geometric factors with error bars for each mode of interest in the target run range. The error was calculated by simulating $3^5=243$ possible geometries based on the machine tolerances of the design at each tun rod angle; because of the large computational time necessary, this was only done by moving the tuning rod in steps of 2.5 degrees from 10 to 50 degrees. This median value set of G factors is then plotted against the initial base geometry that used half degree steps. It agrees well in between mode crossings, but the uncertainty increases greatly near mode crossings.}
    \label{fig:sidecargfactorsw/error}
    \end{figure*}

 Once the target frequency range for the run was chosen to be 5.2-5.6 GHz follow up simulations were made just in that run range to lower computational time. This allowed us to get error estimates of the geometric factors by adding a nested parametric loop within the rod angle loop that varied the major dimensions of the cavity according to its machine tolerance for each value of tuning rod angle. Five major dimensions were chosen for this analysis: The diameter of the cavity, the height of the cavity, the diameter of the tuning rod, the height of the tuning rod, and the radius of the fillet on the rod and cavity end cap regions. Each variable received 3 possible values based on the base design value, $x_0$, and the machine tolerance, $\Delta x$:  $x_0+\Delta x/ 2$,\;$x_0$,\;$x_0-\Delta x /2$. This created a set of $3^5=243$ possible geometries of the cavity at each possible tuning angle of which standard deviations in the geometric factor could be taken, as outlined in section \ref{DecompositionError}. The tuning rod steps had to be increased for this simulation to 2.5 degree steps because computational time was significantly increased otherwise. These were then plotted over the data from the initial pass with error bars, as pictured in Figure \ref{fig:sidecargfactorsw/error}, specifically showing the exterior cavity geometric factor for each mode. The uncertainty in geometric factor, unsurprisingly, got quite large near mode crossing regions, and even if one was not in a mode crossing, but close to it, the error in the geometric factor would be higher. This work is still on-going as how to best optimize where in the tuning rod angle space is it best to perform a decomposition. The cavity characterization section of this chapter will go into the challenges faced performing the decomposition on the Sidecar system. Table \ref{tab:sidecarGfactors} will show the values of all the geometric factors and their uncertainties for the starting $TM_{010}$ frequency of Run 1D, 5521 MHz, which corresponds to a tuning rod angle of roughly 26 degrees.
    \begin{table}
    \centering
    \begin{tabular}{||c|| c| c |c | c||} 
    \hline
    & $TM_{010}$ & $TM_{011}$ & $TM_{012}$ & $TM_{020}$  \\ [0.5ex] 
        \hline\hline
        Frequency (MHz) & $5521.0\pm33.6$ & $5655.8\pm20.0$ & $5842.5\pm26.3$ & $6060.7\pm26.0$\\ 
        \hline
        Exterior cavity & $150.05\pm10.05$ & $116.89\pm26.67$ & $311.44\pm1.70$ & $158.60\pm15.41$\\ 
        \hline
        Tuning rod (whole) & $694.62\pm66.67$ & $568.02\pm132.28$ & $1299.9\pm7.94$ & $786.53\pm78.16$\\ 
        \hline
        Rod axle assembly & $54160\pm9966$ & $38018\pm21386$ & $3971305\pm670343$ & $67609\pm20900$\\ 
        \hline
        Rod walls & $1386\pm49$ & $2032\pm387$ & $1322\pm 8$ & $1691\pm72$\\ 
        \hline
        Rod end caps & $1405\pm222$ & $793\pm199$ & $78789\pm 10196$ & $3047\pm576$\\ 
        \hline
        Top end Cap & $2814\pm471$ & $1585\pm398$ & $157656\pm 20445$ & $3047\pm576$\\ 
        \hline
        Bottom end Cap & $2808\pm427$ & $1585\pm398$ & $157508\pm20366$ & $3047\pm576$\\
        \hline
        Rod wall top & $9628\pm338$ & $10640\pm66697$ & $26992\pm232$ & $11555\pm252$\\
        \hline
        Rod wall top/middle & $5497\pm265$ & $9419\pm904$ & $5102\pm44$ & $10329\pm4604$\\
        \hline
        Rod wall middle & $6649\pm315$ & $38101\pm17901$ & $3435\pm32$ & $6912\pm2525$\\
        \hline
        Rod wall middle/bottom & $5514\pm265$ & $9412\pm892$ & $5122\pm43$ & $10365\pm4623$\\
        \hline
        Rod wall bottom & $9631\pm355$ & $10634\pm6701$ & $27049\pm267$ & $11534\pm261$\\
        \hline
        Total & $123.1\pm17.4$ & $96.7\pm36.9$ & $251.2\pm2.4$ & $131.73\pm21.4$\\
        \hline
        \end{tabular}
        \caption{The geometric factors in Ohms of the Sidecar cavity for a tuning rod angle of 26.74 degrees. This angle corresponds to the $TM_{010}$ mode frequency that the Sidecar cavity was left at for the start of run 1D, and therefore they are the relevant values used for characterizing the cavity in cool down and magnet ramp. Unfortunately, this was near a mode-crossing for the $TM_{020}$ mode so some errors are quite large for that mode. There is anomalous behavior in the values for the $TM_{012}$ mode as well, which has much larger values and smaller errors than the other modes. $TM_{012}$ mode also had a measured mode frequency that did not correspond to this rod angle.}
        \label{tab:sidecarGfactors}
        \end{table}
    \subsection{Warm cavity assembly and testing at LLNL}
\begin{figure*}[htb!]
    \centering
    \includegraphics[width=0.3\textwidth]{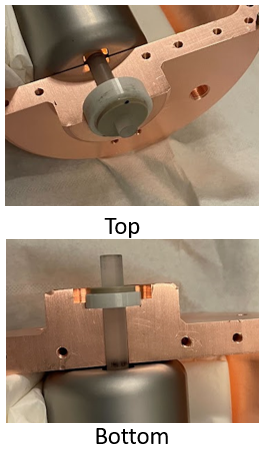}
    \caption{A close up of the run 1D Sidecar cavity tuning port. The cavity is opened with tuning rod and axle resting inside and ball bearing inserted in place above. Special care had to be taken to ensure there was no rubbing of the rod with the cavity interior, and no rubbing of the axle assembly within the port.}
    \label{fig:sidecarRodGap}
    \end{figure*}

\begin{figure*}[htb!]
    \centering
    \includegraphics[width=0.3\textwidth]{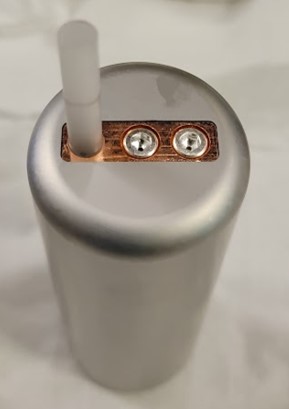}
    \caption{The Sidecar tuning rod with copper axle tab installed. The sapphire axle was epoxied to the copper tab with stycast. Vented, silver-plated 8-32 screws attached the tab to the Nb-substrate tuning rod.}
    \label{fig:sidecarrodwaxle}
    \end{figure*}

\begin{figure*}[htb!]
    \centering
    \includegraphics[angle=90, width=0.8\textwidth]{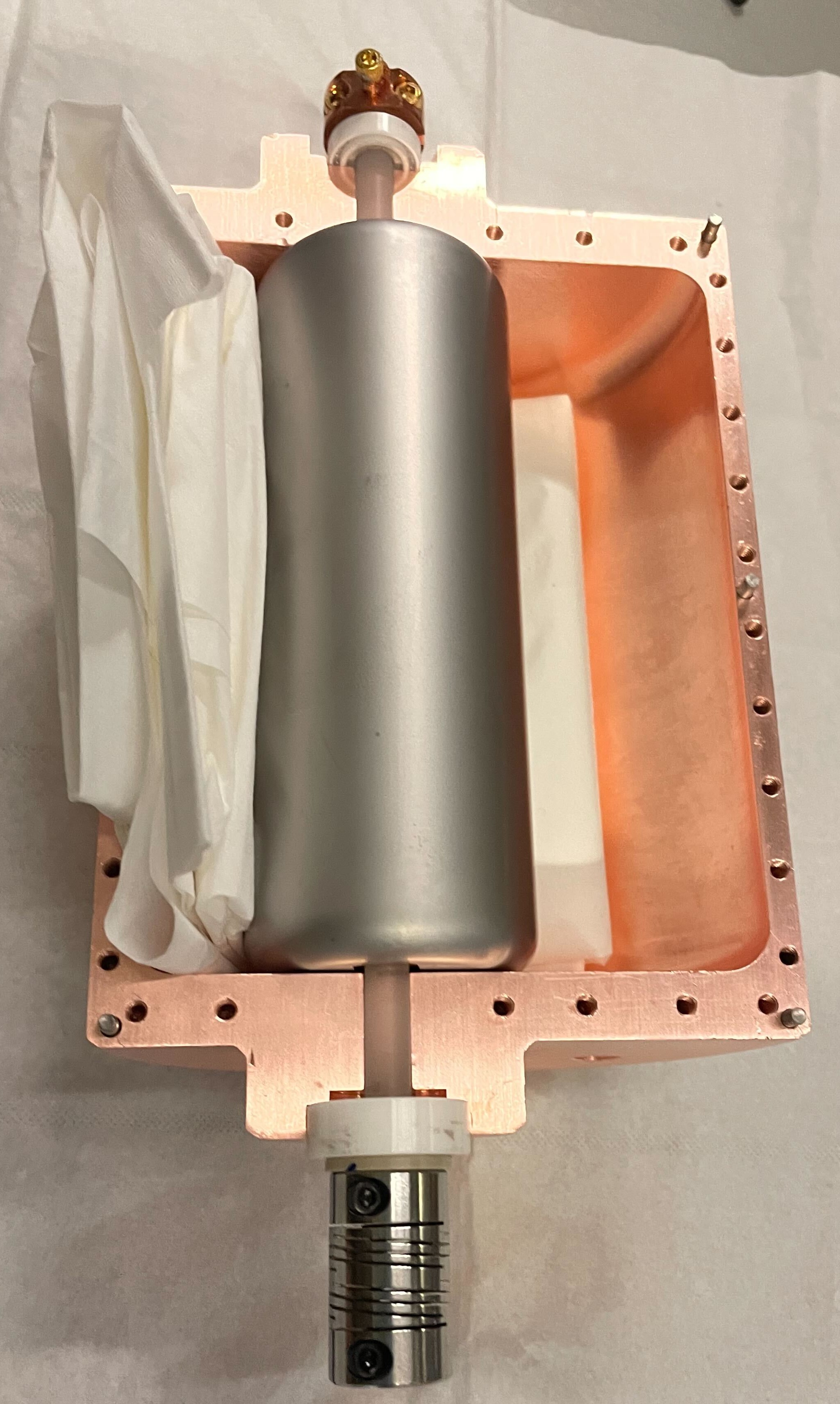}
    \caption{The Sidecar cavity with rod, axles, ball-bearings and stainless steel axle coupler all installed.The copper clamp on the cavity bottom was used for rod thermalization, whereas the axle coupler on top connected to the rotor motor. Both of these locked the gaps on either side of the rod in place, ensuring no rubbing, before the other cavity half was added.}
    \label{fig:Sidecarw/Rod}
    \end{figure*}

    The cavity and tuning rod were received at LLNL in the summer 2023. Minor alterations to the cavity and tabs were made to ensure proper fitting and rotation of the tuning rod at room temperature. Very special care had to be taken to prevent rubbing of the tuning rod against the interior end-cap surfaces; the design accommodates for 0.005" gaps on either side between the rod and interior end cap surface, as pictured in Figure \ref{fig:sidecarRodGap}. The sapphire axles were epoxied to their copper tabs with stycast; several sets had to be made to ensure the tab was flush with the rod surface in its inset, and the axle was perpendicular with the tab (pictured in Figure \ref{fig:sidecarrodwaxle}). Any deviations from this produced friction in the rod rotation. There were several issues with the silver-plated screws cold-welding themselves to the Nb substrate, so moly-powder dry lubricant was used, as well as much shorter screws than initially designed for. Additionally, the axle ball bearings had to sit in their insets on the cavity exterior with very tight tolerances; when the clam-shell halves were fully tightened together, any misalignment could cause the bearing to clamp on the axle and cause friction. After the rod with axles had been placed and aligned within one cavity half, and ball-bearings placed in their insets, a stainless steel coupler and copper clamp were placed on the cavity top and bottom respectively (pictured in Figure \ref{fig:Sidecarw/Rod}) and tightened flush to the ball-bearing bushing. This locked in the position of the rod and its gap. The other cavity half would then be installed on top and tightened in a star-pattern formation.
    \par A test installation of the cavity was done on a mock-up test stand of the ADMX insert T-plate and 1 kelvin plate, including the piezoelectric motor assembly. At the time of testing, only the solid Nb rod had arrived at Livermore, which was used for these assemblies and tests; the hollow rod would only arrive in time to be immediately sent to the University of Washington for installation in the insert. Nonetheless, the warm temperature tests with the piezoelectric motors informed us on how inconsistent torque required to turn the rod was in certain angles of its rotation path. Initial tests required using 60 V from the power supply to get consistent stepping. After re-assembling the cavity many times, this number was brought down to about 30 V to reliably tune the cavity through its range; however, certain smoother turning regions could be operated with a voltage as low as 18 V. For context, the copper rod used in run 1C that had gotten stuck at cryogenic temperatures, had operated in warm testing at 28 V; the copper rod being 36\% heavier than the solid niobium rod indicated that there were frictional issues in the new assembly and that it wasn't just the weight of the solid rod that was causing them. 
    \par
\begin{figure*}[htb!]
    \centering
    \includegraphics[width=0.99\textwidth]{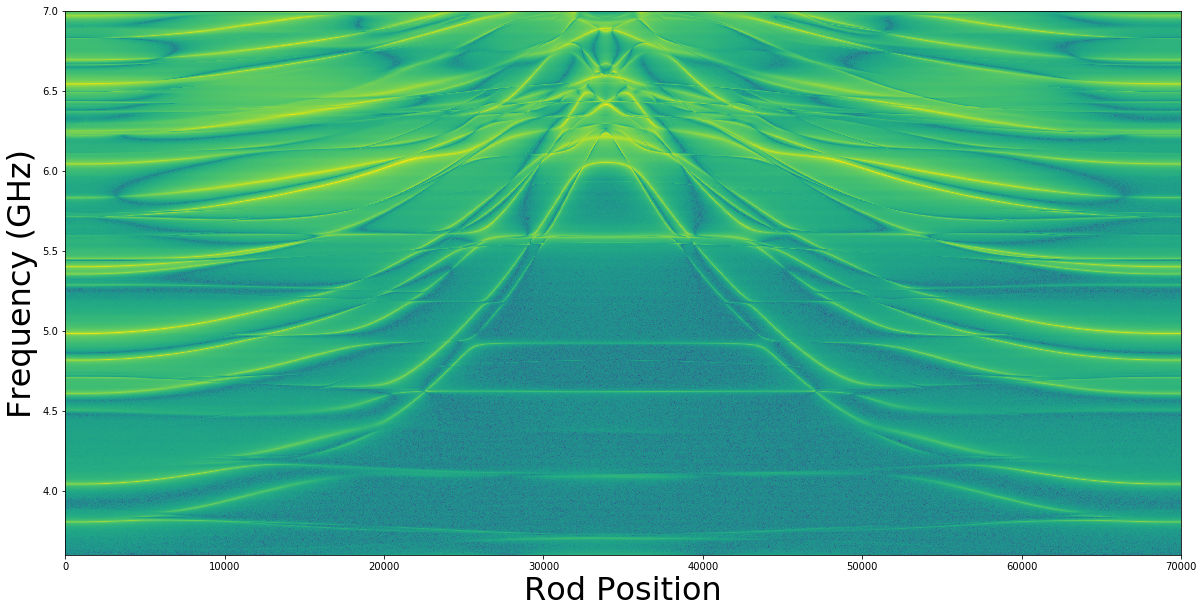}
    \caption{A room temperature mode map of the run 1D Sidecar taken at LLNL. Because of the high resistance of room temperature $\mathrm{Nb}_3\mathrm{Sn}$, $Q_0$ was only around 500-2500 throughout the tuning range.}
    \label{fig:sidecarModeMap}
    \end{figure*}

 Nonetheless, warm RF testing proceeded with taking a mode map of the cavity pictured in Figure \ref{fig:sidecarModeMap}. Because of the high resistance of room temperature $\mathrm{Nb}_3\mathrm{Sn}$, $Q_0$ was only around 500-2500 throughout the tuning range for the $TM_{010}$ mode. The mode frequencies lined up with simulation values, as well as the expected mode crossings. It was at this point that a target frequency range of 5.2 to 5.6 GHz for the run was settled on, because it offered the longest track of frequency space without any mode crossings, and had only been probed by Sidecar run 1A in 2016 to a low sensitivity value. With hints at the time that the CAPP institute was going to release axion limits up to 5.3 GHz, the cavity would be set to start at the upper end of this range, at about 5.5 GHz when installed at UW.
    \subsection{HFET repairs}
    The Sidecar HFET amplifier was repaired by Low Noise Factory in the interim, after failing towards the end of run 1C. It was sent back to UW in August 2023, and passed all room temperature and liquid nitrogen dunk testing, following in line with the factory specs given by LNF. It was installed in the insert without helium testing or a full noise temperature measurement.
    \subsection{Installation of Cavity on ADMX insert}
\begin{figure*}[htb!]
    \centering
    \includegraphics[width=0.4\textwidth]{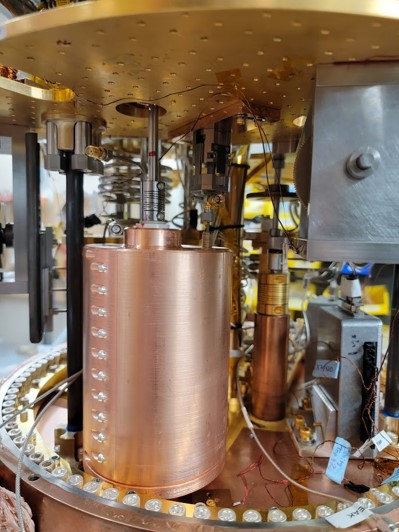}
    \caption{The run 1D Sidecar cavity installed in the ADMX insert with piezoelectric motors mounted.}
    \label{fig:sidecarinstalled}
    \end{figure*}

\begin{figure*}[htb!]
    \centering
    \includegraphics[width=0.5\textwidth]{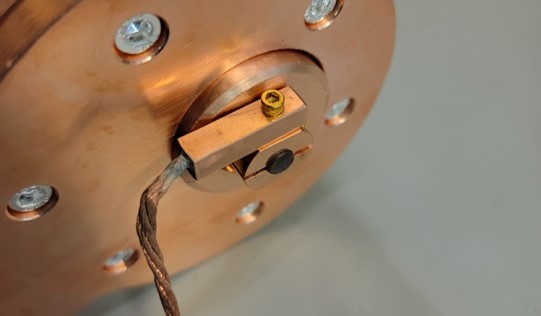}
    \caption{The copper tuning rod cooling strap attached to the bottom axle. This strap is then directly bolted to the mixing chamber plate. The sapphire rod keeps the rod electrically isolated, while still providing thermal conductivity for cooling.}
    \label{fig:sidecarcoolingstrap}
    \end{figure*}

    The cavity was finally assembled and installed at UW in September 2023 with the freshly received hollow tuning rod, as pictured in Figure \ref{fig:sidecarinstalled}. The rod assembly turned very smoothly in comparison with previous assemblies done at LLNL. A copper strap was fabricated on site to give a thermal link to the tuning rod and cool it through its bottom axle, pictured in Figure \ref{fig:sidecarcoolingstrap}. The strap has an estimated cooling power of $77 \mu W$, and the cooling would be limited by the single sapphire axle, which could only transfer heat at $0.52 \mu W$. Following the specific heat value models given in Ref. \cite{NiobiumSpecificHeat}, a cooling time constant of 26.3 hours was roughly estimated for the rod. 
    \par Because of the high resistance of the insert wiring at room temperature used for the piezoelectric motors, motor testing was limited and unreliable. Voltages from the power supply had to be set very high as a result, always in the 55-70 V range to actuate the motors. The linear motor successfully actuated the antenna at room temperature, but the rotor showed clear signs of missed actuation or "skipping". An oscilloscope was used to check the wave form of the saw-tooth wave originating from the power supply, and it was discovered that the wave-form was being deformed in the warm space wiring. The wiring started at the power supply as a BNC cable that lead to a breakout board where the leads were split into 5 parallel paths to reduce the overall resistance; these 10 leads then went into a dB-50 cable, which stepped down to a micro-dB cable that attached to a vacuum feed-through on the insert. The 5 parallel lines would then be recombined on the 1-K plate, right before the connection to the motor itself. Oscilloscope measurements showed that this degradation was occurring at the point where the signal was split into 5 parallel paths. We were able to bypass this problem temporarily by wiring with other cables at the time, and test the tuning rod motor successfully at 70 V. The decision was made that the wiring within the insert was operating correctly, and that the warm wiring would be replaced after the insert went into the magnet, when on-site personnel were not preoccupied with other insert-specific work. It was still not clear if the tuning rod would move once things got cold. The strong port antenna was left in a weakly coupled state with a coupling constant of 0.1, with the expectation that upon cool down this would drop to roughly 0.01. This was to keep the cavity in a state for characterizing it during the cool down and magnet ramp. After the Q characterization data was taken, it would be adjusted to critical coupling for an axion search starting at 5.5 GHz.
\section{Hybrid cavity characterization}
    This section will cover the quality factor measurements taken on the hybrid Sidecar cavity during the cool down and magnet ramp. As stated earlier, this was a challenging task because the insert is not designed for controlled cooling and warming. Additionally, there was no way of measuring the superconducting transition of the tuning rod, because its critical temperature should be higher than the NbTi SQUID loops within the JTWPA; the JTWPA would be opaque until the system had cooled below at least 9 Kelvin. Simulations of the geometric factors were done beforehand with the hope of using the multi-mode decomposition technique outlined in chapters \ref{chap:Cavities} and demonstrated in \ref{chap:LLNL} with the NbTi PPMS cavity. The plan was to measure the first 5 TM modes: $TM_{010}$, $TM_{011}$, $TM_{012}$, $TM_{013}$, and $TM_{020}$. It was then noticed that the $TM_{013}$ was too close to a mode crossing with the $TM_{020}$ in the current tuning configuration, and could not be observed well. What further complicated this was the rotor piezoelectric motor failed to move the tuning rod, even after the warm wiring had an attempted repair.
    \par Each multi-mode measurement was taken using the same process on the Sidecar DAQ. Because there is only one vector network analyzer for Sidecar, modes couldn't be tracked simultaneously. Furthermore, the DAQ software couldn't be easily adapted to automate this measurement process, and the center frequency of the VNA had to be adjusted via a remote operator on our DAQ website. The $TM_{010}$ was the default mode that the system would be left in after a measurement, so it was only necessary to periodically cycle through the higher modes. The process went as follows:
    \begin{enumerate}
        \item Turn off axion search data cycles such as digitizations and TWPA re-biasing, but keep the network analyzer scans running each data cycle: transmission, reflection, and transmission wide scan.
        \item Set the network analyzer to the mode target frequency and ensure good windowing parameters for Q fitting. Note the starting time stamp for the first scan.
        \item Wait at least 15 minutes while the network analyzer does many Q fittings. These values of frequency, coupling, and loaded Q are stored in the database to be pulled later. By waiting 15 minutes and getting many fits, a standard deviation and uncertainty in these values can be set.
        \item Set the network analyzer to the next mode frequency and record the time. 
        \item Repeat until all modes of interest have associated start and stop time stamps. Set the frequency back to the $TM_{010}$ mode, and record its start time stamp to be at least 15 minutes before the first time stamp recorded.
    \end{enumerate}
    A script then took the recorded time stamps from each measurement, and pulled the relevant sensor values during those time periods from the database, including values for relevant temperature sensors, and the magnet current. A weakness in this method was that, especially during the cool down, temperatures were not guaranteed to be the same for each mode in a given measurement; if the temperature changed significantly over the course of the hour it took to measure, then the last mode measured would not be equivalent to the first. This meant that performing a surface resistance decomposition between those two modes wouldn't produce necessarily an accurate result. This was less relevant when ramping the magnet because it is done in steps, and so the magnet current was pretty much constant over the measurement; however, the magnet ramp does heat the insert by eddy currents, meaning differences in temperatures arose based on how long it had been since the field last changed. Additionally, these had to be made with a static antenna, so the coupling could not be changed mode to mode; in principle this shouldn't be an issue as long as one measures beta well. This all accumulated into a data set that has too many variables to really know what is going on with the SRF tuning rod. Nonetheless, I will now present the results of it so far.
    \subsection{Cool Down Studies}
\begin{figure*}[htb!]
    \centering
    \includegraphics[width=0.99\textwidth]{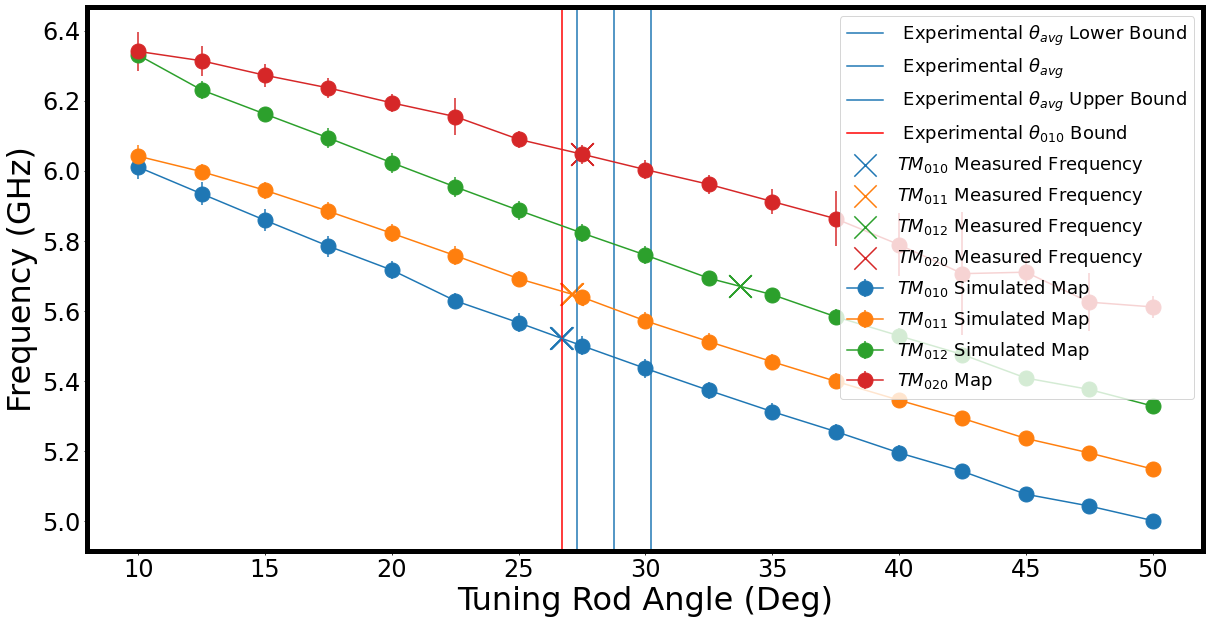}
    \caption{The measured mode frequencies during cool down plotted onto the simulated mode map of which simulated geometric factors are derived. One hopes that all mode frequencies would align closely with a single tuning rod angle, but that is not the case here; the $TM_{012}$ is at an anomalously lower frequency than the other 3, giving it a corresponding tuning angle of about 33.7 degrees, whereas the other lay much closer to the $TM_{010}$ average angle of 26.7 degrees. This $\theta_{010}$ angle is plotted as a vertical line, whereas the $\theta_{avg}$ for all four modes is plotted with the upper and lower 1-sigma bounds plotted as well.}
    \label{fig:cooldowntheta}
    \end{figure*}

    The first cool down measurement was taken until the first signs of cavity structure appeared in the VNA, which occurred when the Sidecar cavity temperature sensor was at about 5.02 K. Three more multi-mode measurements were made over the course of the cool down roughly at 3.5 K, 2.5 K, and 0.67 K. One strange aspect was that although an appreciable frequency shift had occurred as expected upon cool down for all the modes, the coupling constants for each mode were close to their room temperature values: The $TM_{010}$ was about 0.15, $TM_{011}$ about 0.9, $TM_{012}$ was 0.12, and $TM_{020}$ was 0.7. This may suggest that the antenna didn't fully thermalize yet when the measurements were taken. Another important check on the system was to see if the measured mode frequencies all aligned on the simulated mode map in correspondence with a single tuning rod angle as show in figure \ref{fig:cooldowntheta}; this would be necessary for the multi-mode decomposition because one needs to pick one set of G factors corresponding to one tuning rod angle. All but the $TM_{012}$ mode seem to correspond well to a single angle, with an average of about $27.1 \pm 0.4$ degrees. The $TM_{012}$ is anomalously lower, and when included in the average, shifts things to $28.7 \pm 2.9$ degrees. This was not noticed in room temperature testing. It is therefore hard to say that the system is well aligned with simulations, and a multi-mode decomposition result shouldn't be taken with confidence.
    \par
\begin{figure*}[htb!]
    \centering
    \includegraphics[width=0.92\textwidth]{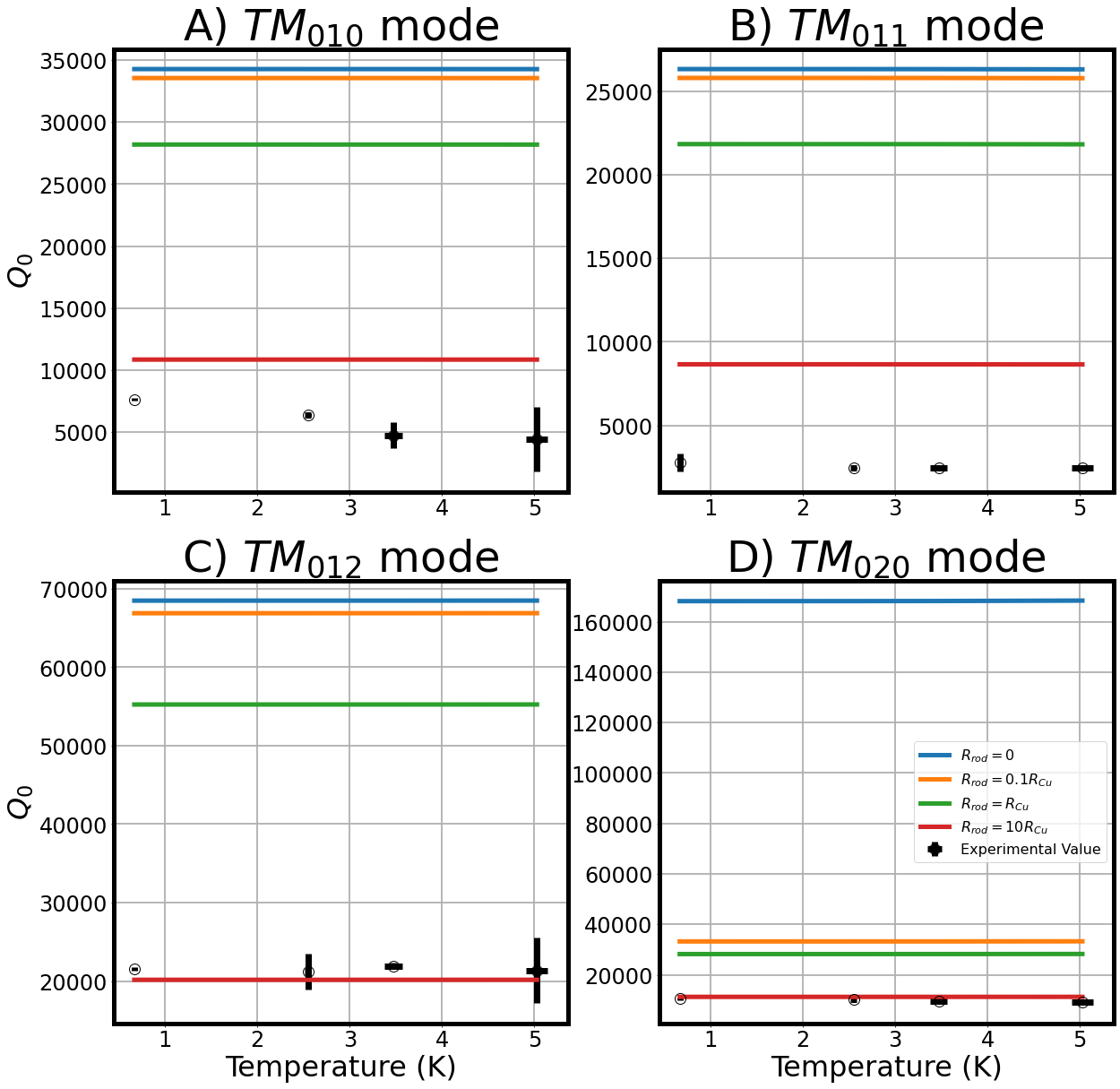}
    \caption{The unloaded quality factors versus the Sidecar temperature during the insert cool down for each of the relevant modes: A) $TM_{010}$, B) $TM_{011}$, C) $TM_{012}$, and D) $TM_{020}$. In addition, horizontal lines of the expected cryogenic quality factor for each mode with varying tuning rod surface resistance is shown. The $R_{rod}=R_{Cu}$ line corresponds to the expected cryogenic Q for an all copper cavity, whereas the $R_{rod}=0$ would be the upper bound where the rod contributes zero surface resistance to the Q.}
    \label{fig:sidecar_cooldown_Qmodes}
    \end{figure*}

 Most importantly, the unloaded quality factor measurements were much lower than expected, even much lower than what is expected for an all copper cavity. Figure \ref{fig:sidecar_cooldown_Qmodes} shows the experimental unloaded quality factors versus temperature for each of the four TM modes. The horizontal lines in this plot show the expected cryogenic quality factor for various surface resistances of the tuning rod; the $R_{rod}=R_{Cu}$ corresponds to an all copper cavity, whereas the $R_{rod}=0$ corresponds to the best possible hybrid Q where the superconducting rod doesn't contribute any surface resistance. In this way, one expects as the system cools, the experimental data should cross these lines, and ideally settle somewhere above the $R_{rod}=R_{Cu}$, but below the $R_{rod}=0$ line. As one can see, this did not happen, and the quality factor, although increasing as the system cooled, was lower than the expected value for a rod exhibiting 10x the surface resistance of cryogenic copper. 
    \par
\begin{figure*}[htb!]
    \centering
    \includegraphics[width=0.82\textwidth]{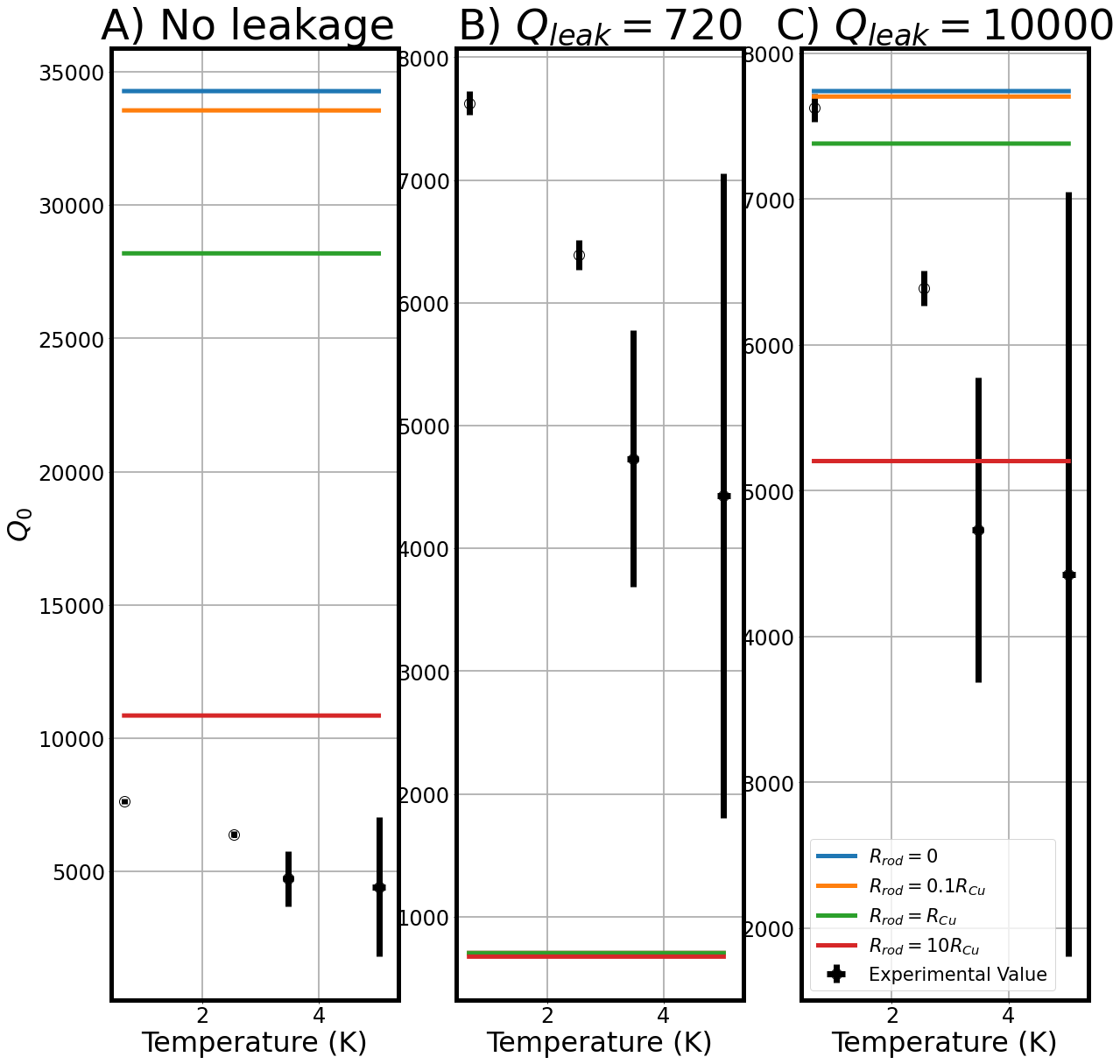}
    \caption{The unloaded quality factors of the $TM_{010}$ mode versus Sidecar cavity temperature. The horizontal lines are the same as Figure \ref{fig:sidecar_cooldown_Qmodes}, however they introduce an associated $Q_{leak}$ value. A) shows the values with no associated leakage, $Q_{leak}\rightarrow \infty$. B) $Q_{leak}=720$, which was the expected leakage Q for Sidecar during its previous run. One can see that a leakage Q this low is not possible because the experimental data greatly exceeds the expectation for an ideal hybrid cavity. C) $Q_{leak}=10000$ represents a prescribed lower bound on the leakage Q where the experimental Q settles into a good hybrid cavity value between an all copper cavity and ideal superconducting hybrid. However, nothing prevents the actual $Q_{leak}$ from being much higher than this, and the cavity surface resistances actually behaving less than ideally.}
    \label{fig:sidecar_cooldown_Qleak}
    \end{figure*}

 The expected quality factors in Figure \ref{fig:sidecar_cooldown_Qmodes}, however, do not account for any possible RF leakage in the cavity. As mentioned when describing Sidecar run 1C, it was presumed that there must have been RF leakage causing the very low Q of 700 during that run. One can calculate based off the expected quality factor for that cavity that it would've had a $Q_{leak}$ of about 720 to have such a low result. There were no tests between run 1C and 1D that confirmed that this leakage issue had been resolved by the efforts made adapting the cavity tuning rod ports to their new configuration in run 1D. Furthermore, it was noticed in room temperature testing that the torque applied to the screws holding the clam-shells together did have an effect on Q if they were too loose, although this was never quantified. As the copper contracts in cool down, it is possible that gaps may have opened up, increasing the $Q_{leak}$ and spoiling the unloaded Q, however, this is very hard to quantify or prove. Figure \ref{fig:sidecar_cooldown_Qleak} explores this idea by introducing an associated $Q_{leak}$ to the expected Q value lines for the $TM_{010}$ mode: A) shows the same data from Figure \ref{fig:sidecar_cooldown_Qmodes} for $TM_{010}$, B) shows the values assuming a $Q_{leak}=720$ in accordance with Sidecar in run 1C, and C) shows a $Q_{leak}=10000$. One can see that $Q_{leak}$ can't be as bad as Sidecar run 1C, because the implied intrinsic quality factors would be far higher than even the ideal hybrid cavity, which isn't physically possible. However, if one sets a higher value, $Q_{leak}=10000$, the experimental data does start to follow the expected values; in this way, $Q_{leak}=10000$ represents the rough lower bound on the $Q_{leak}$, because any lower value would imply intrinsic Q values that are in excess of what is physically possible. However, this is a totally prescribed value; one still doesn't have any upper bound on the $Q_{leak}$, which is the critical value to determining the actual surface resistance of the tuning rod. Without an upper bound, one could just pick a $Q_{leak}$ that makes the behavior of the rod look agreeable, as is shown in \ref{fig:sidecar_cooldown_Qleak}C.
    \par
\begin{figure*}[htb!]
    \centering
    \includegraphics[width=0.92\textwidth]{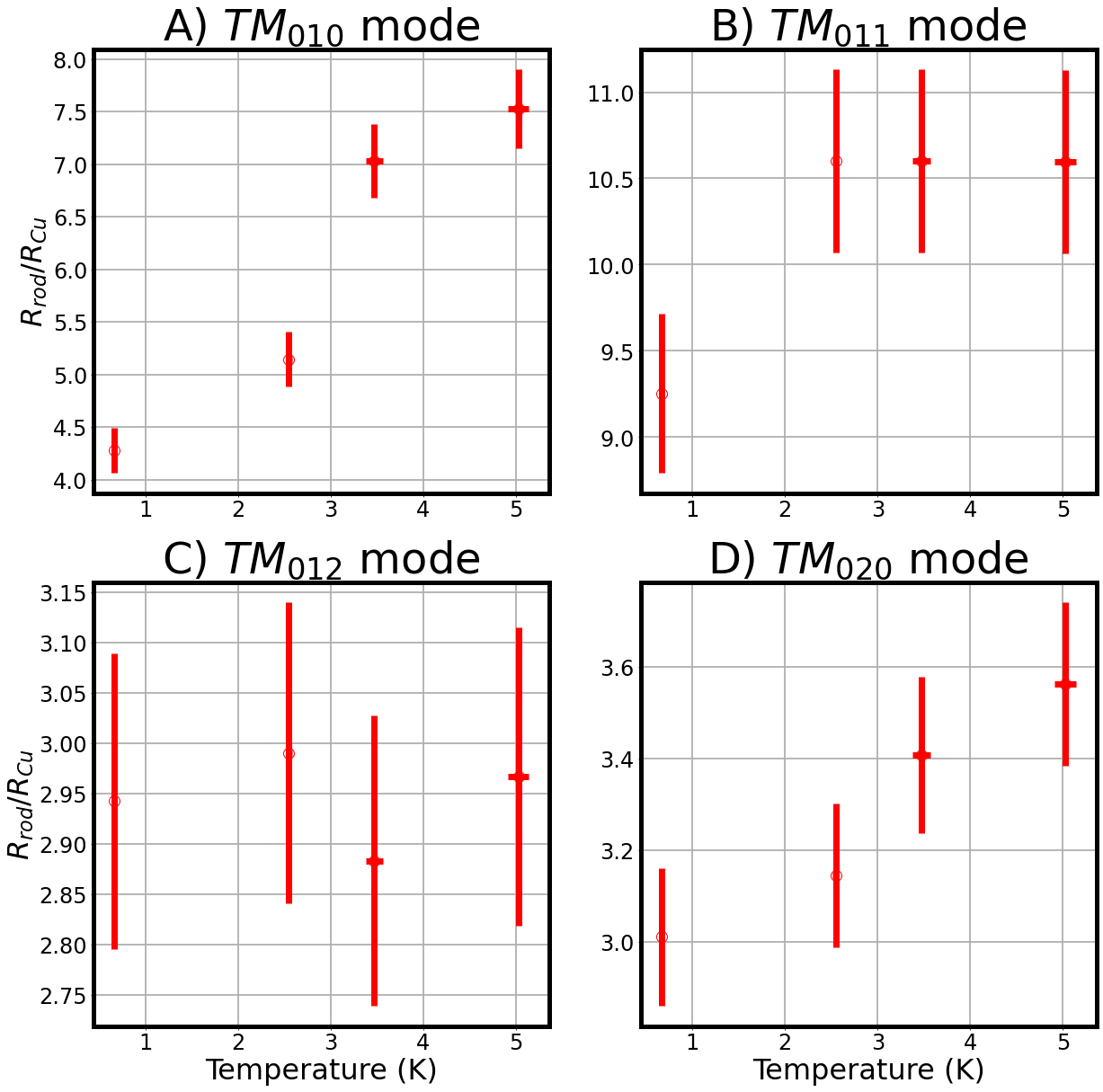}
    \caption{The ratio of the tuning rod resistance to copper during the Sidecar cool down. This is calculated using Equation \ref{eqn:hybridkappa}, with the experimentally measured quality factors, and the expected quality factors for an all copper cavity. It doesn't account for any RF leakage, or sub-quality copper resistance within the cavity. This method will also produce different values than the implied resistance values in Figure \ref{fig:sidecar_cooldown_Qmodes}.}
    \label{fig:sidecar_cooldown_kappa}
    \end{figure*}

 Another simple way of getting at the behavior of the rod in the cool down, without a full multi-mode decomposition, is to follow the analysis done in section \ref{simplehybridcavity}, and solve for the ratio of the rod resistance to $R_{Cu}$ using Equation \ref{eqn:hybridkappa}. From the ratio of the rod and cavity geometric factors, one can see that the ideal hybrid Q will only be $\approx1.20$ that of the all copper cavity. Substituting the experimental values of quality factor to solve for $\kappa$ as is done in Figure \ref{fig:sidecar_cooldown_kappa}, one can see the rod resistance is still much higher than copper. Its important to note here that this will produce slightly different ratios than \ref{fig:sidecar_cooldown_Qmodes} because it doesn't account for the geometric factor of the axle assemblies and only looks at $G_{cavity}$ and $G_{rod}$. This also assumes that the copper resistance is exactly as expected from theory and that there is no leakage. Nonetheless it tells the same story; the resistances are much higher than expected.
    \par
\begin{figure*}[htb!]
    \centering
    \includegraphics[width=0.92\textwidth]{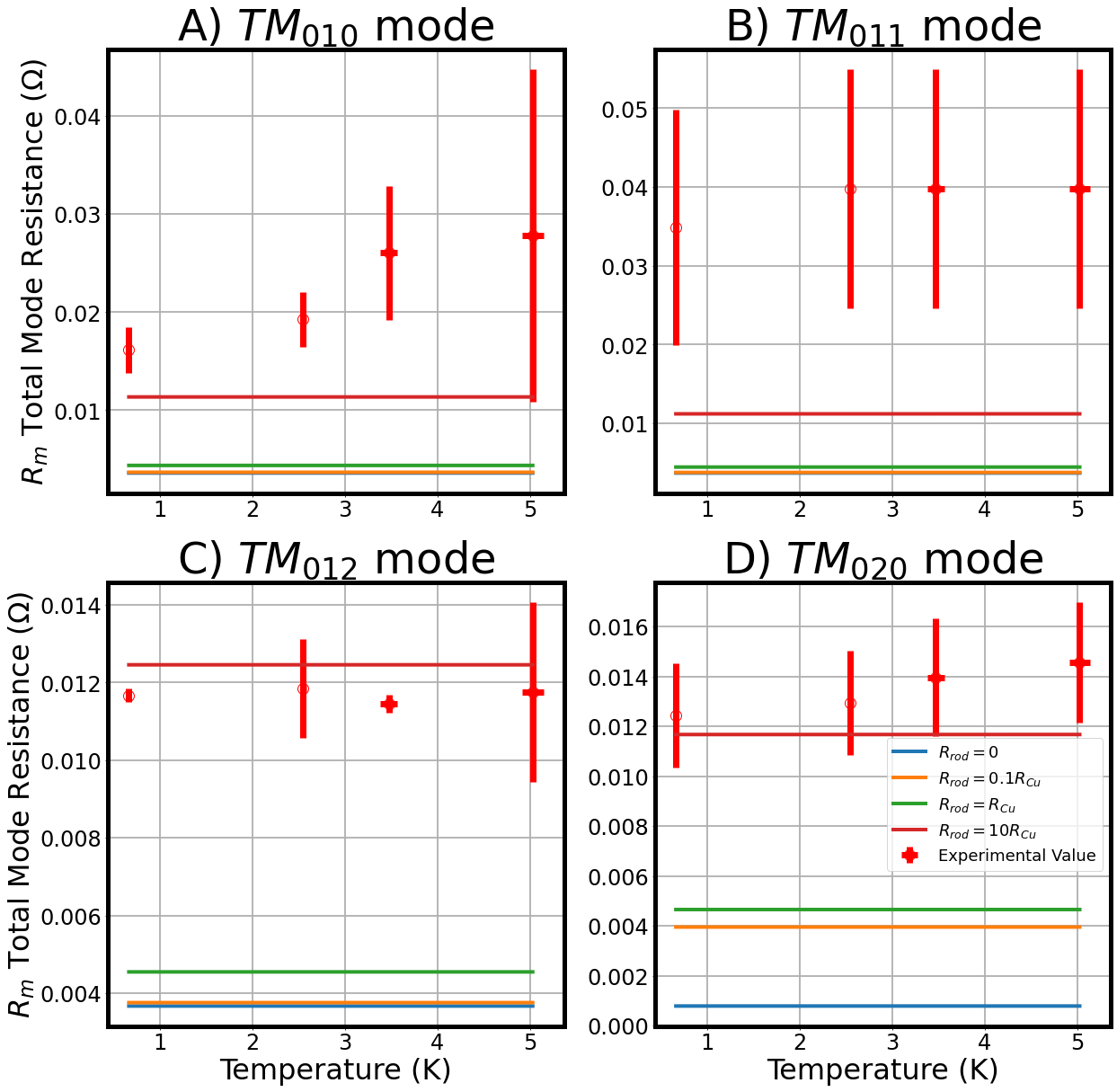}
    \caption{The total mode resistance versus Sidecar cavity temperature for each mode of interest. The horizontal lines represent the expected total mode resistance for possible resistance values of the tuning rod; the total mode resistance is a weighted average based on the geometric factors of the superconducting surfaces and copper surfaces.}
    \label{fig:sidecar_cooldown_Rms}
    \end{figure*}

 Finally, we can translate this quality factor information into total mode resistances using the total geometric factor for each mode, as shown in Figure \ref{fig:sidecar_cooldown_Rms}. This would be done in preparation of performing a multi-mode decomposition, but several factors prevented that analysis from going further. As outlined in section \ref{mulitmodedecomposition}, once a set of mode resistances had been constructed, the $C_{ms}$ matrices would be constructed and their determinants calculated. For simplicity, we stuck with just 2x2 matrix decompositions. With 4 modes, there are 6 possible 2-mode combinations. In this case, there were 10 possible 2-surface combinations from the 11 geometric factors that were simulated. The base decomposition would be just the tuning rod versus the exterior cavity. The subsequent ones would combine the other geometric factors with the exterior cavity versus a single surface of interest: the rod walls, the rod end caps, top end cap, bottom end cap, and the 5 wall segments top to bottom. The geometric factors would be selected based on a chosen tuning rod angle that best corresponded to the system state. As outlined with Figure \ref{fig:cooldowntheta}, we were not very confident in this value because of the anomaly in the $TM_{012}$ mode; for simplicity, the angle corresponding to the $TM_{010}$ mode was chosen, 26.7 degrees. Once these 60 matrices were constructed, their determinants were calculated, and only 18 of them were statistically non-zero, meaning they qualified for being inverted and use in a decomposition. Of the remaining 18, only 1 of them was a mode combination that didn't include the anomalous $TM_{012}$ mode that we were not confident about. All of them produced non-physical, negative values for the surface resistances, which were not usable. The best solution to this would be to tune the cavity to another frequency, further away from mode crossings for all modes of interest, and converge on a better value for the tuning rod angle; one wants the cavity modes to be in region where they are distinct from one another to minimize the number of matrices with zero determinants. Unfortunately, at the time of writing, the Sidecar cavity has been unable to be tuned from its starting frequency using the piezoelectric motors, so this is not possible. I'll also note that it won't be possible to get new cool down data until the system is warmed up, which is most likely not to occur until December 2024.
    \subsection{Magnet ramp}
    Although the Quality factor measurements during the cool down were far less than anticipated, the magnet ramp might still shed some light on an important unanswered question: Is the rod actually superconducting? If it was still in a normal state, one would expect there to be no degradation in the quality factor as the magnetic field was increased. If it was superconducting, but there was some unaccounted for RF leakage, impurities, etc. that were degrading the overall base Q, then one still expects the quality factor to degrade in field.
    \par
\begin{figure*}[htb!]
        \centering
        \includegraphics[width=0.6\textwidth]{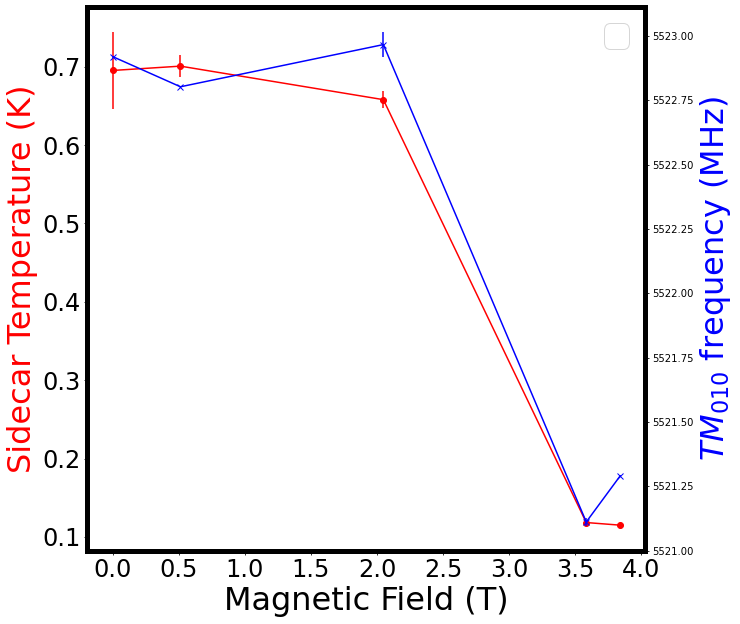}
        \caption{The Sidecar cavity temperature versus the average Sidecar magnetic field during the ramp. The initial ramp was done before the system had fully cooled to base temperature, whereas the later two points the insert had fully cooled to its base value at around 115 mK. This temperature shift was well correlated with a shift in the Sidecar mode frequencies with the $TM_{010}$ frequency displayed on a separate y-axis here. This is also seen as an increase in Q within most of the subsequent ramp plots.}
        \label{fig:sidecar_ramp_temps}
    \end{figure*}

 The plots in this section will pretty much be the same as the last section, but replace the X-axis with magnetic field instead of temperature. Measurements were made at 0 T, 0.5 T, 2 T, 3.6 T, and 3.8 T, where the field strength is the average magnetic field within the Sidecar cavity for the given magnet current. As alluded to earlier, when the magnet current is being increased, the insert is heated significantly, and it takes significant time for things to cool off. The magnet was initially ramped at a starting temperature of about 700 mK to 200 amps of current with two short pauses at the 0.5 T and 2 T points; the system then cooled completely before the 3.6 T data point was taken. Similarly, when the magnet was finally ramped to 225 amps several weeks later, the system was cooled until the 3.8 T data was taken. Unfortunately, this created a dataset that had different temperatures and fields, which makes deciphering the behavior difficult. The first three data points were at about 700 mK whereas the last two were at 115 mK. Figure \ref{fig:sidecar_ramp_temps} shows this relation, and how it was well correlated with a shift in resonant frequency for all the modes.
    \par
\begin{figure*}[htb!]
        \centering
        \includegraphics[width=0.92\textwidth]{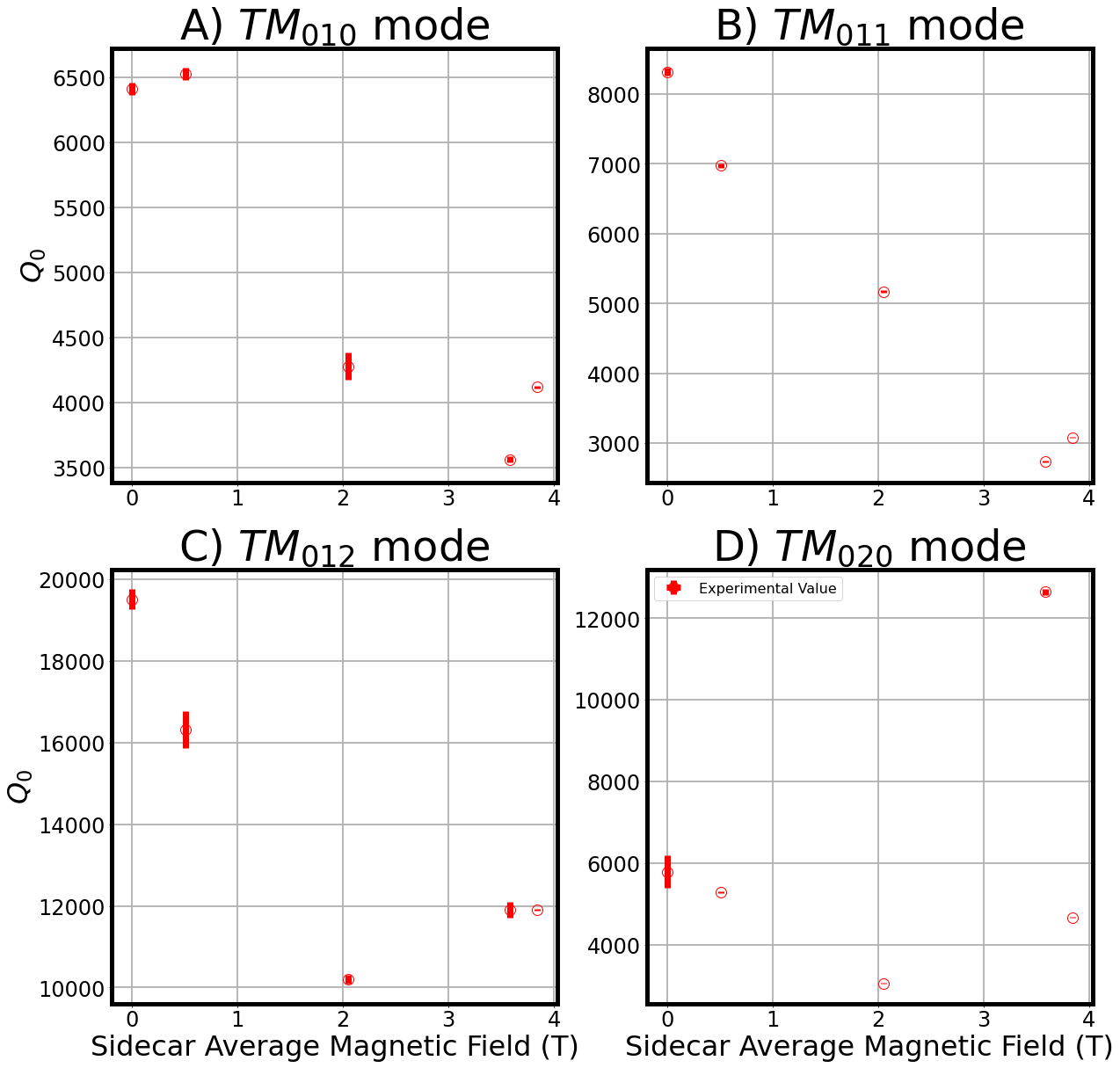}
        \caption{The unloaded quality factors for the modes during the magnet ramp. All of them seem to show signs of Q-degradation, indicative of superconducting behavior. The small rises in Q for the 3.6 T and 3.8 T values could be because of the much lower temperature of the insert when they were taken.}
        \label{fig:sidecar_ramp_Qs}
    \end{figure*}

 This was slightly corrected by adjusting the resonant frequency of all the points to be the same for all Q values within a given mode, but this didn't capture the decrease in $\Delta f$ as the cavity cooled to 115 mK. This information is entangled with the possible increase in $\Delta f$ due to Q-degradation from the magnetic field. Even after correcting for the resonant frequency shift, some of these points still showed an increase in Q, as shown in Figure \ref{fig:sidecar_ramp_Qs}, which implies the cooling dominated over any degradation due to the field. To add to this, the last two points were also taken after the Sidecar antenna had been critically coupled for axion data-taking; this should be fine as it is accounted for in $\beta$ during the calculation of $Q_0$, but the relation does break down for $\beta >> 1$, and the $TM_{020}$ did register the highest $\beta$ of 10.8 and 3.7 for its 3.6 and 3.8 T values respectively.
    \par
\begin{figure*}[htb!]
        \centering
        \includegraphics[width=0.92\textwidth]{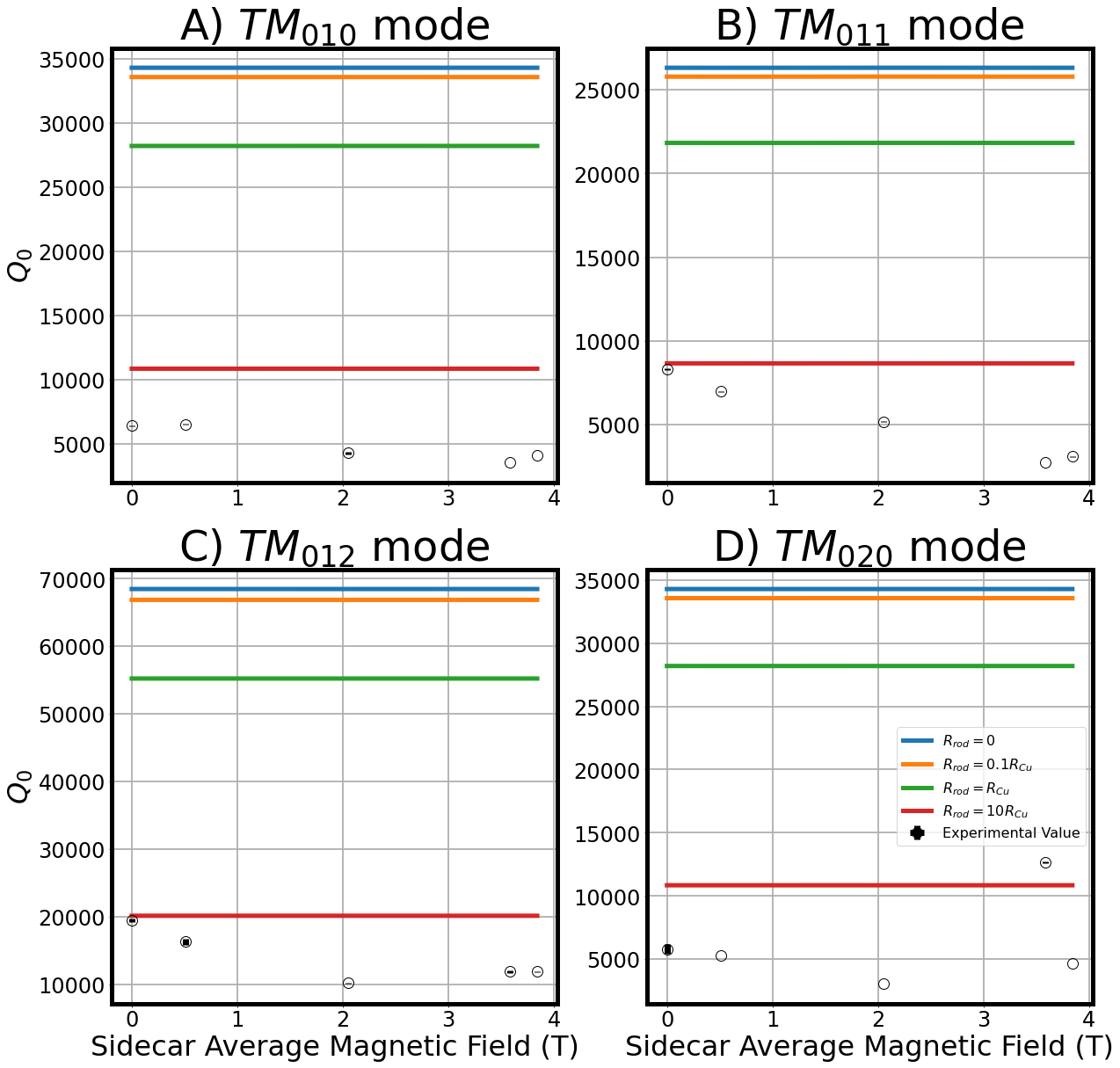}
        \caption{The unloaded quality factors for the modes during the magnet ramp compared to expected hybrid cavity Q values.}
        \label{fig:sidecar_ramp_Qsline}
    \end{figure*}

\begin{figure*}[htb!]
        \centering
        \includegraphics[width=0.76\textwidth]{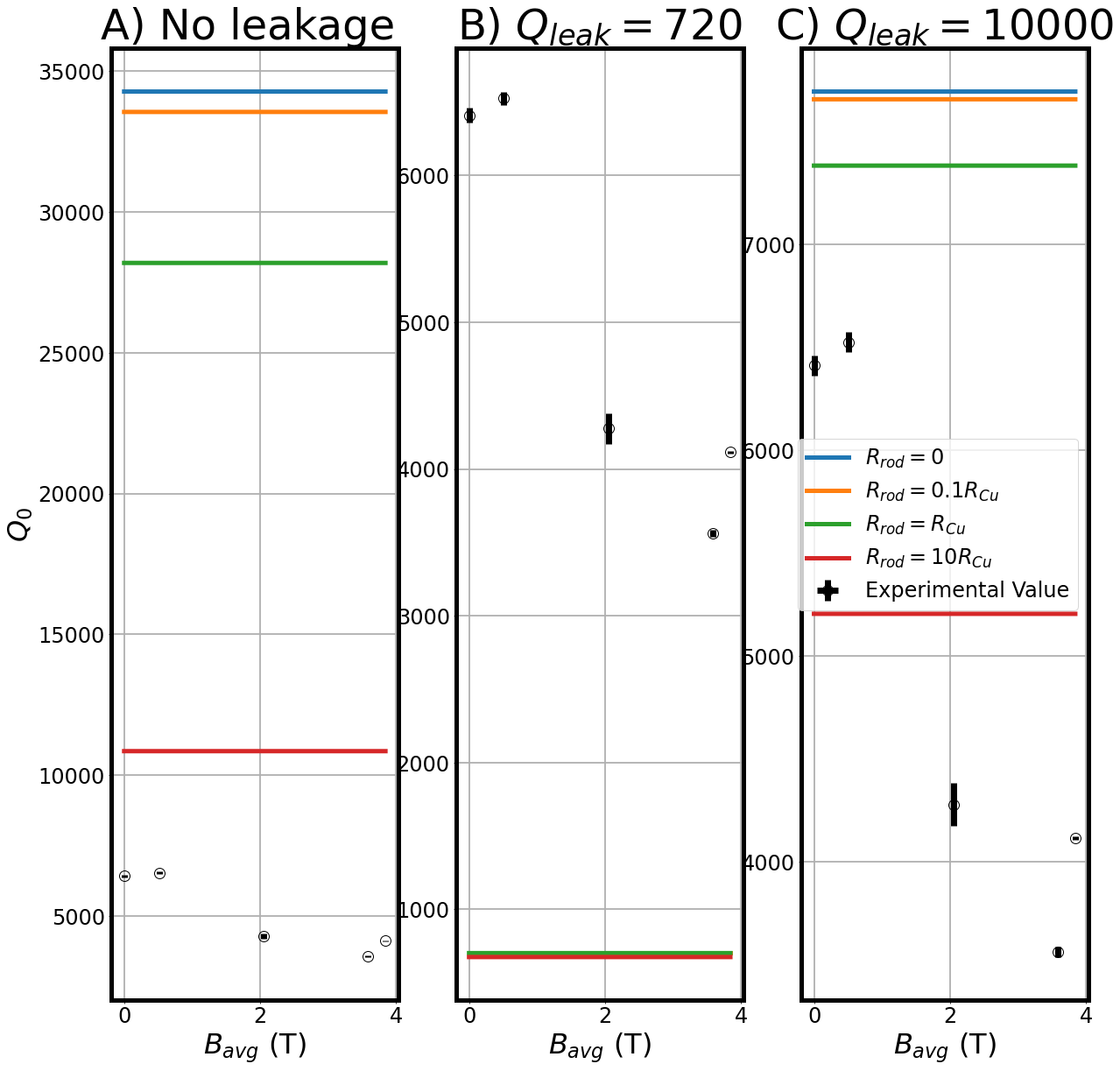}
        \caption{The $TM_{010}$ mode unloaded quality factor versus magnetic field strength compared to the expected quality factor for varying rod resistances and different leakage Qs. A) shows the expected values for a cavity with no leakage ($Q_{leak}\rightarrow\infty$). B) shows a very low $Q_{leak}=720$ that was estimated in the previous Sidecar run 1C. As one can see, this value is too low, as it implies the experimental data exceeds the ideal hybrid cavity case, which is nonphysical. C) shows a $Q_{leak}=10000$, this was a prescribed lower bound estimate for $Q_{leak}$ in the cool down data (see Figure \ref{fig:sidecar_cooldown_Qleak}) because it set the lowest temperature experimental value near the ideal hybrid cavity expected Q. In this case, the zero field data point does not fall near the $R_{rod}=0$ value but below the $R_{rod}=R_{Cu}$ this might be because the cavity temperature was slightly warmer (700 mK versus 666 mK), changing the resistance of the rod, but one doesn't expect $Q_{leak}$ to change with temperature at this regime because copper has already fully thermally contracted.}
        \label{fig:sidecar_ramp_Qleaky}
    \end{figure*}

 When one overlays the expected quality factor lines to the ramp data, as shown in Figure \ref{fig:sidecar_ramp_Qsline} it is still clear that the cavity is behaving very poorly. Nonetheless, following the same $Q_{leak}$ analysis done on the cool down data, as shown in Figure \ref{fig:sidecar_ramp_Qleaky}, one can see that the cavity behavior, if the cavity had a $Q_{leak}=10000$, is in line for what one would expect of a hybrid SRF cavity of this construction; the Q degrades with field strength, just not from a high zero field Q. It is important to note the lower zero field Q here as compared to the coldest temperature Q in the cool down which was about 1000 higher; its not entirely clear what changed here, as the temperature is only slightly higher, $\approx 40 mK$, at 700 mK. Another likely reason is the coupling constant increased from about 0.17 in the cool down to 0.45 at the beginning of the ramp for unknown reasons, as the antenna had not been adjusted at all. It is important to note however, that one expects the leakage quality factor to be constant at this temperature because the copper exterior should already be fully thermally contracted.
    \par
\begin{figure*}[htb!]
        \centering
        \includegraphics[width=0.92\textwidth]{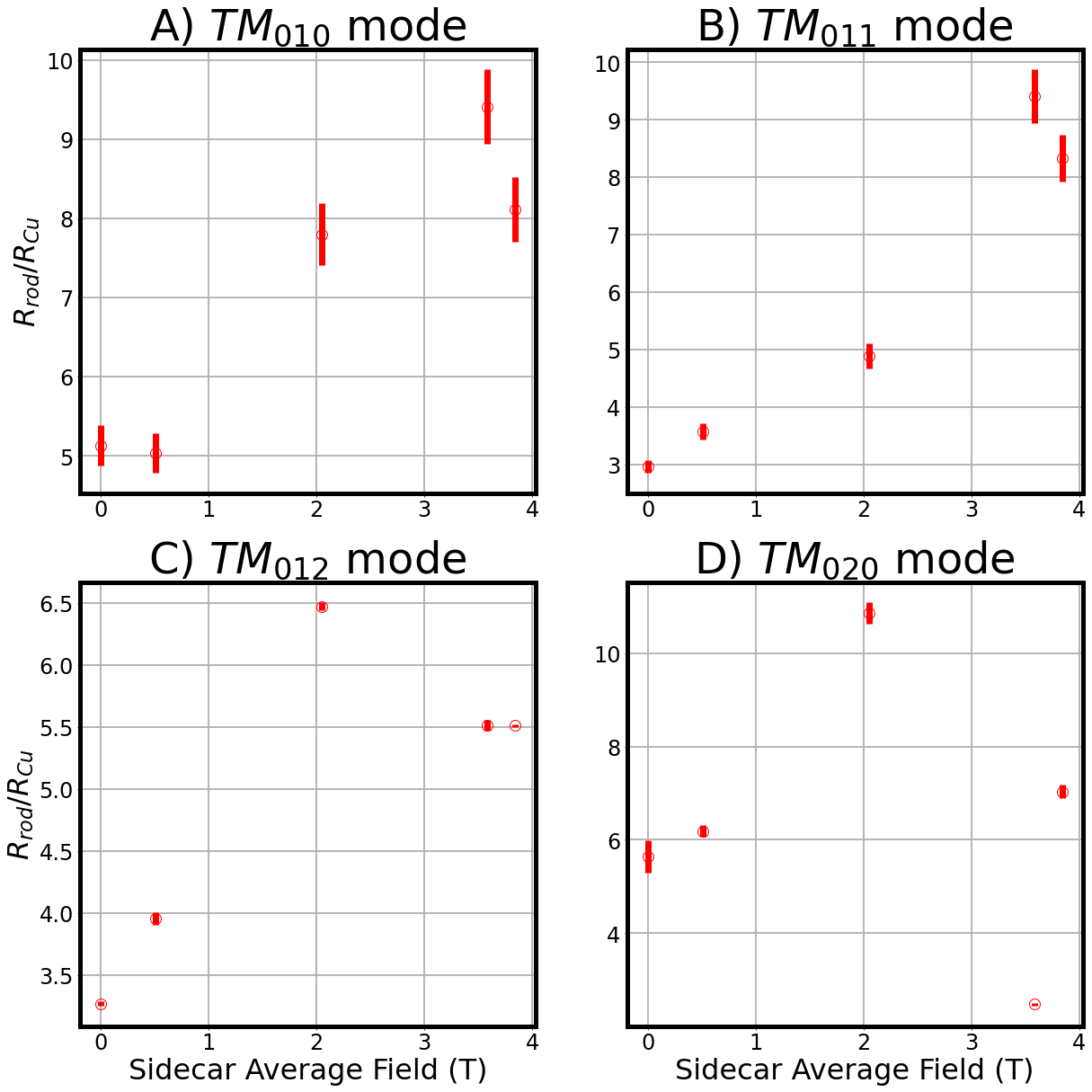}
        \caption{The expected ratio of tuning rod resistance to copper resistance for each mode during the magnet ramp. This value is derived using Equation \ref{eqn:hybridkappa} and the other analysis outlined in section \ref{simplehybridcavity}.}
        \label{fig:sidecar_ramp_kappa}
    \end{figure*}

\begin{figure*}[htb!]
        \centering
        \includegraphics[width=0.92\textwidth]{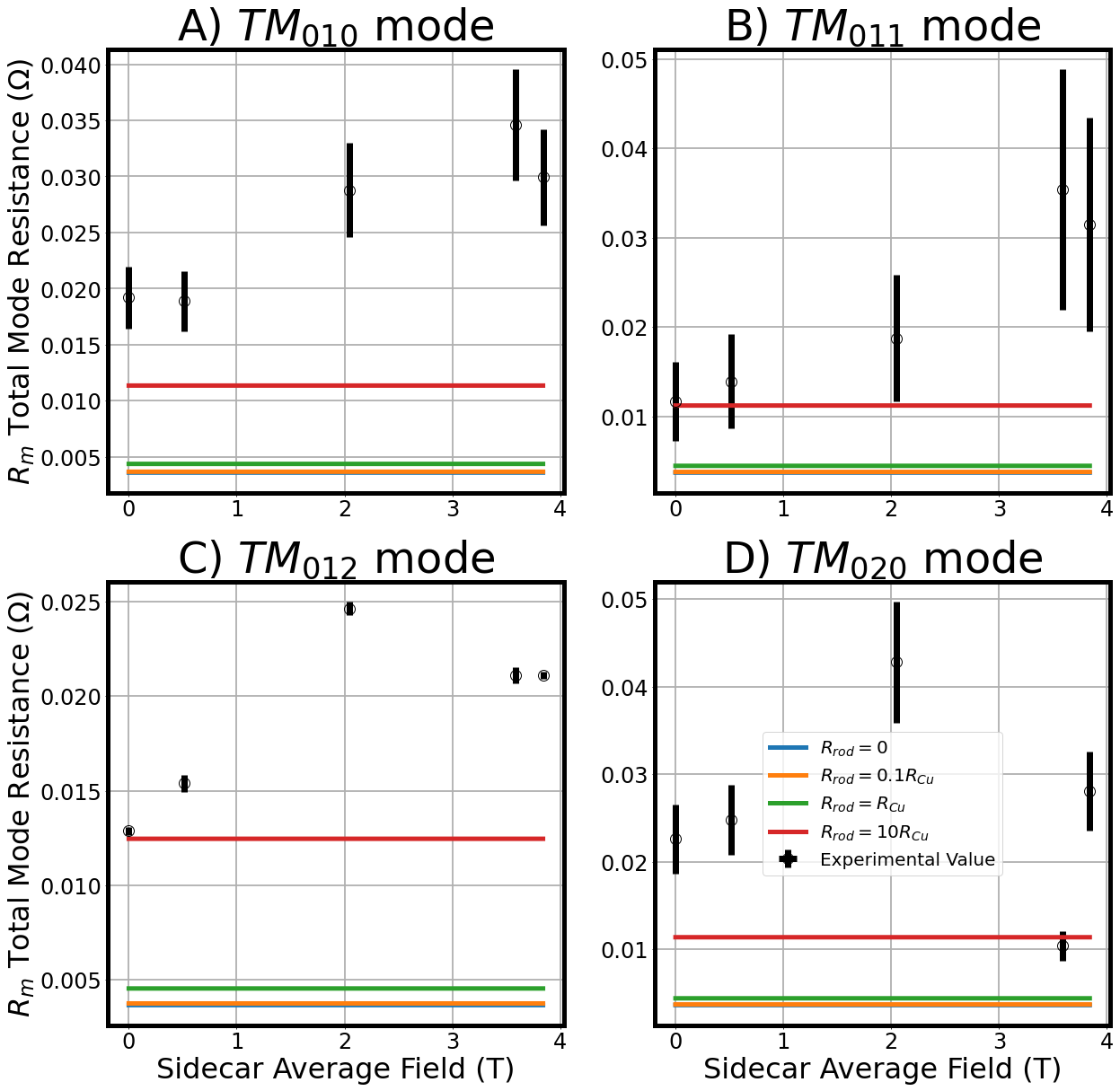}
        \caption{The Sidecar total mode resistances versus the magnetic field strength during the run 1D magnet ramp. The expected total mode resistance for a hybrid cavity with varying tuning rod resistance is also shown as colored horizontal lines.}
        \label{fig:sidecar_ramp_modeR}
    \end{figure*}

 The rod resistance to copper resistance ratio during the ramp is shown in Figure \ref{fig:sidecar_ramp_kappa}. The values are similarly high as they were in Figure \ref{fig:sidecar_cooldown_kappa}, but there is still a clear increase in resistance from the zero to 3.8 T field points, as would be expected if the rod is superconducting. Figure \ref{fig:sidecar_ramp_modeR} shows a similar story for the total mode resistances during the ramp; the total mode resistance increases with field. Similarly to the cool down data, the 60 possible surface decompositions all had statistically zero determinants, included the anomalous $TM_{012}$ mode, or just produced non-physical surface resistance values. One would need to tune to a different tuning rod angle and wait until the magnet is ramped down again, which will not be possible until November 2024 at the earliest. In conclusion, the magnet ramp shows signs that the rod is superconducting, and could just be suffering from some sort of unknown radiative Q. 
\section{Noise temperature calibrations}
\begin{figure*}[htb!]
\centering
    \includegraphics[width=0.7\textwidth]{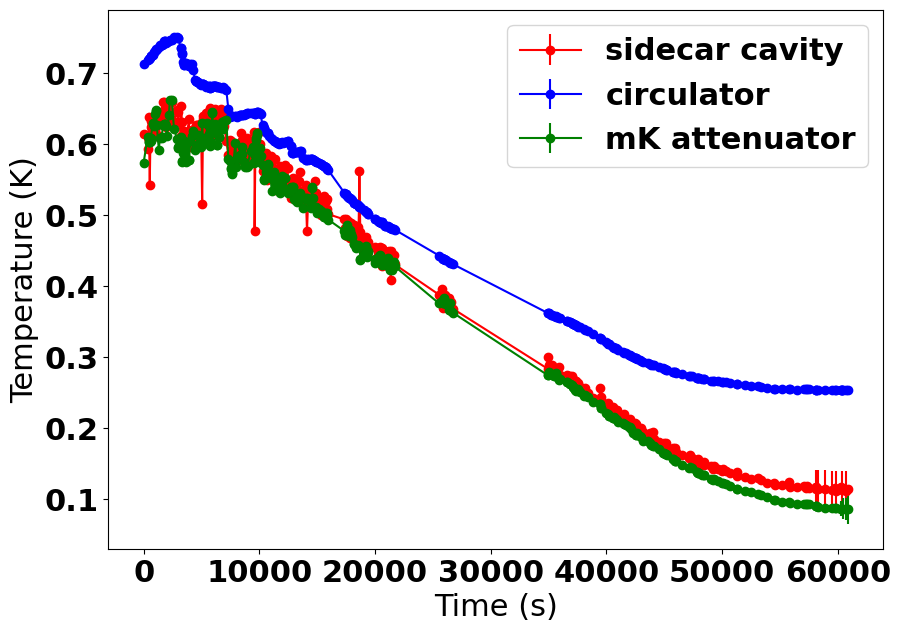}
    \caption{The relevant temperature sensor values versus time during the December magnet ramp cool down noise characterization period. This period was divided into to separate measurements because the frequency steps taken during the wide-power measurement changed part way through. The points that occurred after the gap in data at about 30,000 seconds constitute the temperatures  used for the second measurement.}
    \label{fig:sidecar_noise_DecTempvst}
\end{figure*}

\begin{figure*}[htb!]
\centering
    \includegraphics[width=0.5\textwidth]{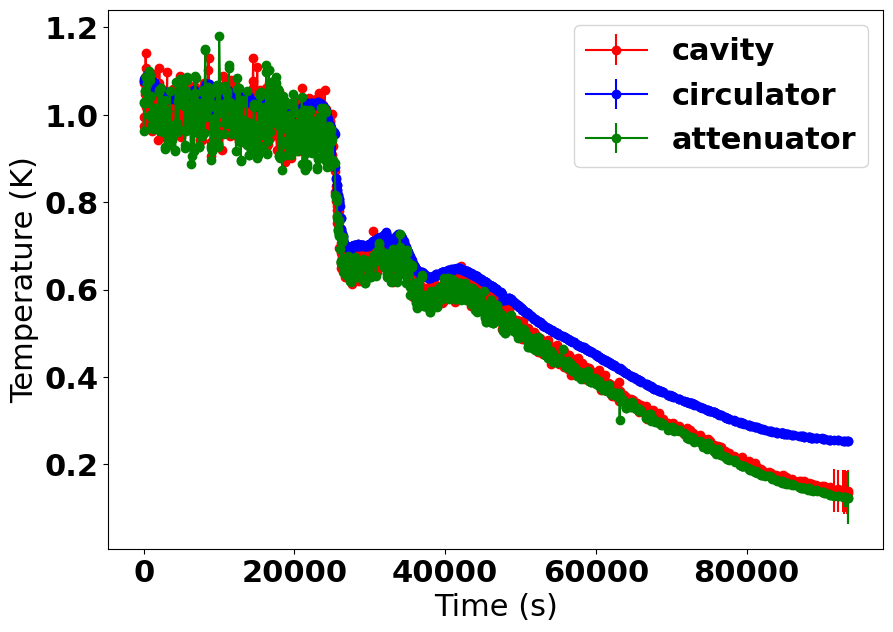}
    \caption{The relevant temperature sensor values versus time during the March anomalous fridge crash cool down noise characterization period. This crash heated the system above 1 kelvin, whereas the previous magnet ramp cool down only reached about 0.75 kelvin. This gave this measurement a much larger swing in temperature, which is better for the fit integrity.}
    \label{fig:sidecar_noise_MarchTempvst}
\end{figure*}

There were two opportunities so far in run 1D to characterize the HFET noise temperature, and therefore the overall system noise temperature of the Sidecar system. Both of these implemented the y-factor method 2, outlined in section \ref{hotload2}, that uses the mK attenuator as a variable temperature noise source. This was because the Radiall coaxial switch was unable to be actuated successfully to point towards the dedicated hot load; plans are in order to bypass the pulsed switching system and attempt a direct DC connection to actuate, but this hasn't been approved or scheduled at the time of writing. Nonetheless, two opportunities arose where the mK attenuator was heated close to 1 Kelvin, enabling the Y-factor method 2 to be performed as it cooled back down: The main magnet ramp in December 2023 (see Figure \ref{fig:sidecar_noise_DecTempvst}), and an anomalous heating of the dilution refrigerator that took place at the end of March 2024 (see Figure \ref{fig:sidecar_noise_MarchTempvst}). 
\par
\begin{figure*}[htb!]
\centering
    \includegraphics[width=0.7\textwidth]{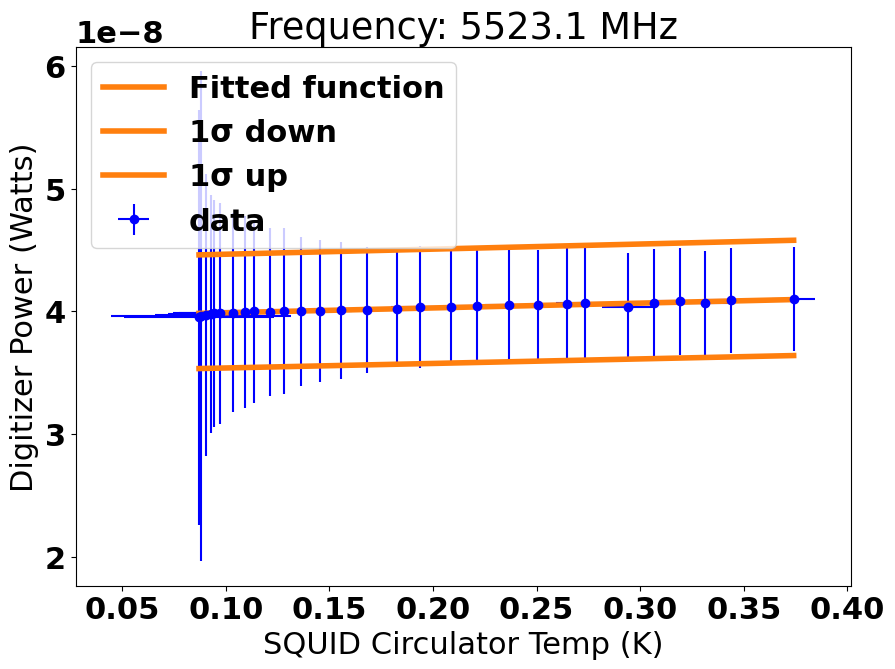}
    \caption{An example fit from the 2nd half of the December magnet ramp cool down. The blue error bars represent the systematic error due to the temperature difference between the attenuator and cavity, which is carried over as an uncertainty in the measured power. Because this temperature uncertainty is relatively constant, the fractional uncertainty increases at lower temperatures, causing the increase in the power uncertainty near 0.1 K. The orange $1\sigma$ lines represent the error in the fit itself. These two errors are then combined when setting the error in the HFET noise temperature.}
    \label{fig:sidecar_noise_DecPt2fits}
\end{figure*}

\begin{figure*}[htb!]
\centering
    \includegraphics[width=0.7\textwidth]{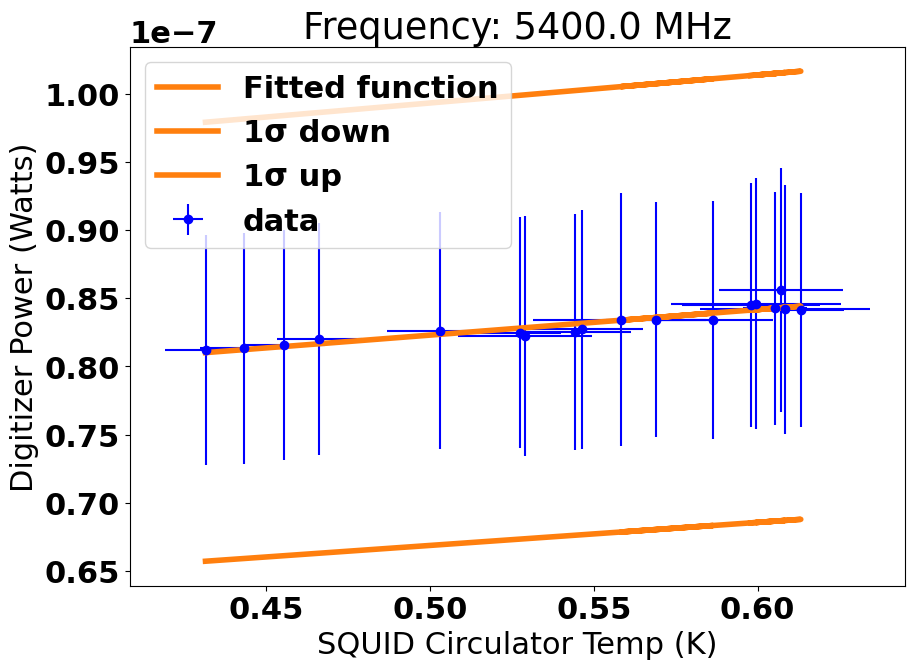}
    \caption{An example fit from the 1st half of the December magnet ramp cool down, which has significantly reduced errors compared to the 2nd half. The blue error bars represent the systematic error due to the temperature difference between the attenuator and cavity, which is carried over as an uncertainty in the measured power. The orange $1\sigma$ lines represent the error in the fit itself. These two errors are then combined when setting the uncertainty in the HFET noise temperature.}
    \label{fig:sidecar_noise_DecPt1fits}
\end{figure*}

\begin{figure*}[htb!]
\centering
    \includegraphics[width=0.7\textwidth]{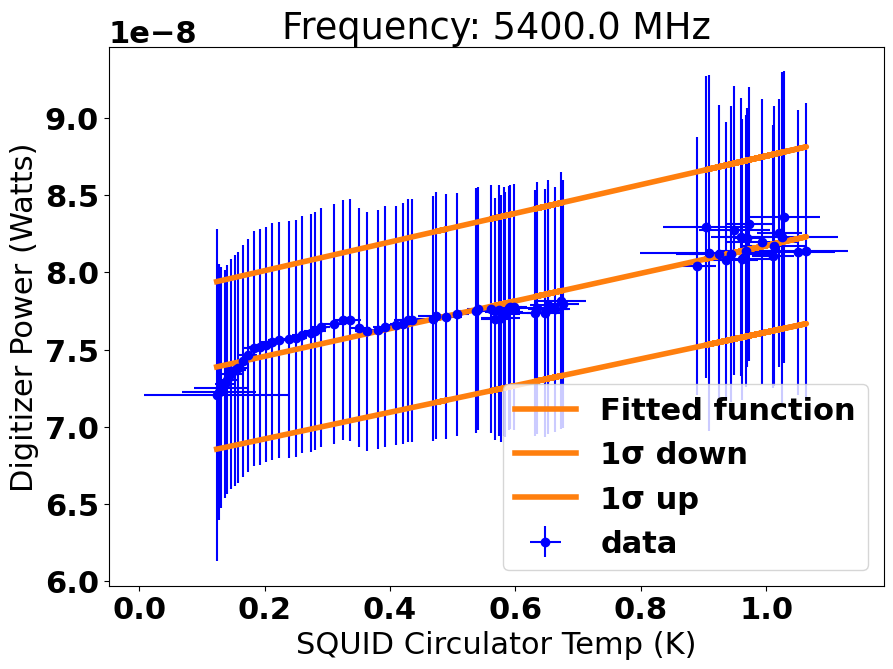}
    \caption{An example fit from the March fridge cool down. The blue error bars represent the systematic error due to the temperature difference between the attenuator and cavity, which is carried over as an uncertainty in the measured power. This measured temperature difference was roughly 20-30 mK higher than in the December cool down, increasing the overall error. The orange $1\sigma$ lines represent the error in the fit itself. These two errors are then combined when setting the error in the HFET noise temperature.}
    \label{fig:sidecar_noise_MarchFit}
\end{figure*}

 During these periods, regular data taking on-resonance digitizations were turned off, and a wide-power measurement was performed over the entire tuning range of the cavity; a 100 second power digitization (same 3 MHz bandwidth as normal data-taking) was taken every 200 MHz from 4-6 GHz. The average power was then calculated for each of these digitizations. Due to complications with the control scripts during the magnet ramp, these predefined frequency steps changed about half-way through the December cool down, and the latter half took steps every 60 MHz from 5223-5823 MHz. Because of how the analysis was set up, it was easier to divide this magnet ramp cool down period into two separate measurements for each set of frequencies. The limitation to doing this, however, is that each individual measurement has roughly half the sweep in temperatures; the first half only took data from 0.75 K to 0.45 K, and the second half from 0.45 K to 0.1 K. Additionally, since the 2nd half of data was taken at colder temperatures, but the uncertainty in temperature stayed constant, it has overall higher fractional uncertainties in the systematic error, which is evident at the lower fit temperatures in Figure \ref{fig:sidecar_noise_DecPt2fits}. In contrast, the fits from the first half of the December cool down (see Figure \ref{fig:sidecar_noise_DecPt1fits}) do not have this effect, and are more dominated by the error in the fit itself. The fits from the march cool down (see Figure \ref{fig:sidecar_noise_MarchFit}), covering the entire temperature swing, does show an increase in systematic error at the low temperatures, but it also had a higher temperature difference registered between the cavity and attenuator, increasing the systematic error.
\par
\begin{figure*}[htb!]
\centering
    \includegraphics[width=0.6\textwidth]{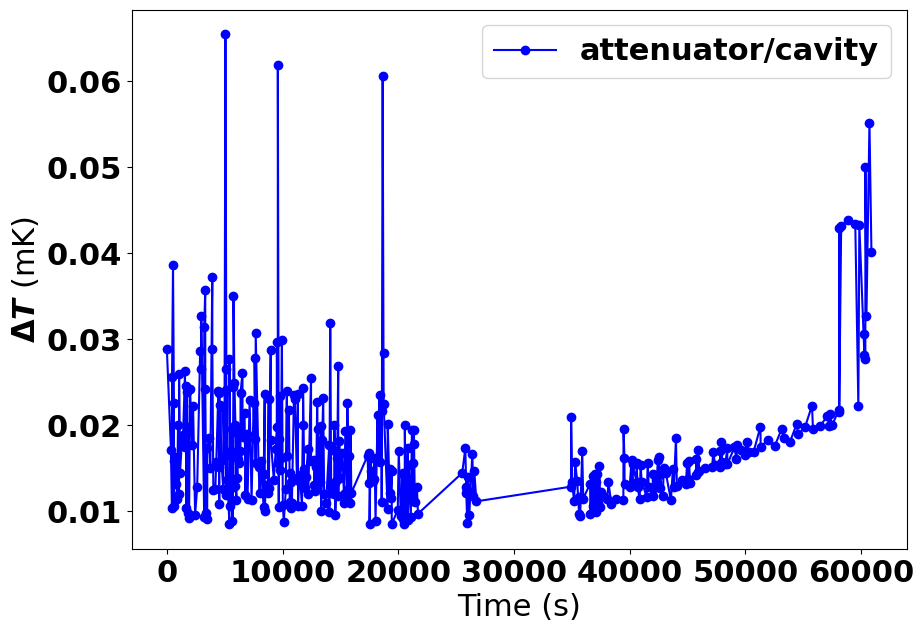}
    \caption{The registered temperature difference between the attenuator and cavity during the December magnet ramp cool down measurement.}
    \label{fig:sidecar_noise_DecdeltaT}
\end{figure*}

\begin{figure*}[htb!]
\centering
    \includegraphics[width=0.6\textwidth]{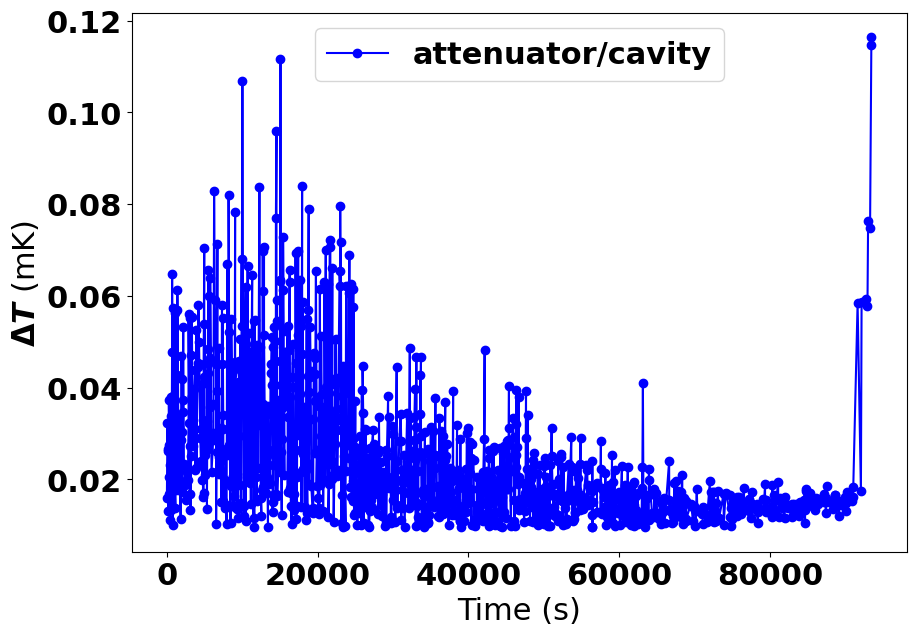}
    \caption{The registered temperature difference between the attenuator and cavity during the March cool down measurement.}
    \label{fig:sidecar_noise_MarchdeltaT}
\end{figure*}

 This is a good time to discuss how the error bars were set in Figs. \ref{fig:sidecar_noise_DecPt1fits}, \ref{fig:sidecar_noise_DecPt2fits}, and \ref{fig:sidecar_noise_MarchFit}. A key assumption when using Equation \ref{eqn:hotload4} for these fits, is that the attenuator and cavity have the same temperature. In the previous run 1C Sidecar run, the temperature sensors used were not fully calibrated from the manufacturer, Lakeshore, and as a result they measured statistically the same temperature within their mediocre uncertainties. In this run, the sensors had been fully calibrated, and had an improved uncertainty $\pm2$ mK, which meant they now registered an observable temperature difference between them. This meant that the temperature uncertainty had to be set not only on the calibration uncertainty of the individual sensors, but the measured difference between the two sensors, which was an order of magnitude higher, at 40-60 mK. Figs. \ref{fig:sidecar_noise_DecdeltaT} and \ref{fig:sidecar_noise_MarchdeltaT} show this measured difference versus time for each cool down period; one can see that the March period registered a noticeably higher difference. These temperature differences set the blue error bars seen in Figs. \ref{fig:sidecar_noise_DecPt1fits}, \ref{fig:sidecar_noise_DecPt2fits}, and \ref{fig:sidecar_noise_MarchFit} directly as the error in the x-axis, and indirectly for the y-axis error. The systematic uncertainty in the power was set by the equation:
\begin{equation}
    \Delta P = |P|\sqrt{(\frac{\Delta T}{T})^2+(\frac{\Delta \epsilon}{\epsilon})^2}
    \label{eqn:hotloaddeltaP}
\end{equation}
where $P$ is the measured average power for the scan, $T$ is the attenuator temperature during that scan, $\Delta T$ is predominantly the temperature difference between the attenuator and cavity during the scan (but it also includes the calibration error in the sensors themselves), and $\epsilon$ and $\Delta \epsilon$ is the emissitivity, or transmission efficiency, between the cavity and JTWPA, which is measured ex-situ. Since this part of the receiver remained pretty much unchanged, it was fair to use the previous ex-situ measurement of $\epsilon$ during run 1C, which was $0.4 \pm 0.04$. It is planned to remeasure this value the next time the insert is removed from the magnet. Both are comparable sources of error to one another, $\approx 10\%$, depending on the temperature at the time of the digitization; warmer digitizations $\epsilon$ error dominates and vice-versa. The fit is then performed which gives an uncertainty in the fitted values; this error was then combined with the systematic error when setting the uncertainty in the receiver temperature.
\par
\begin{figure*}[htb!]
    \includegraphics[width=0.99\textwidth]{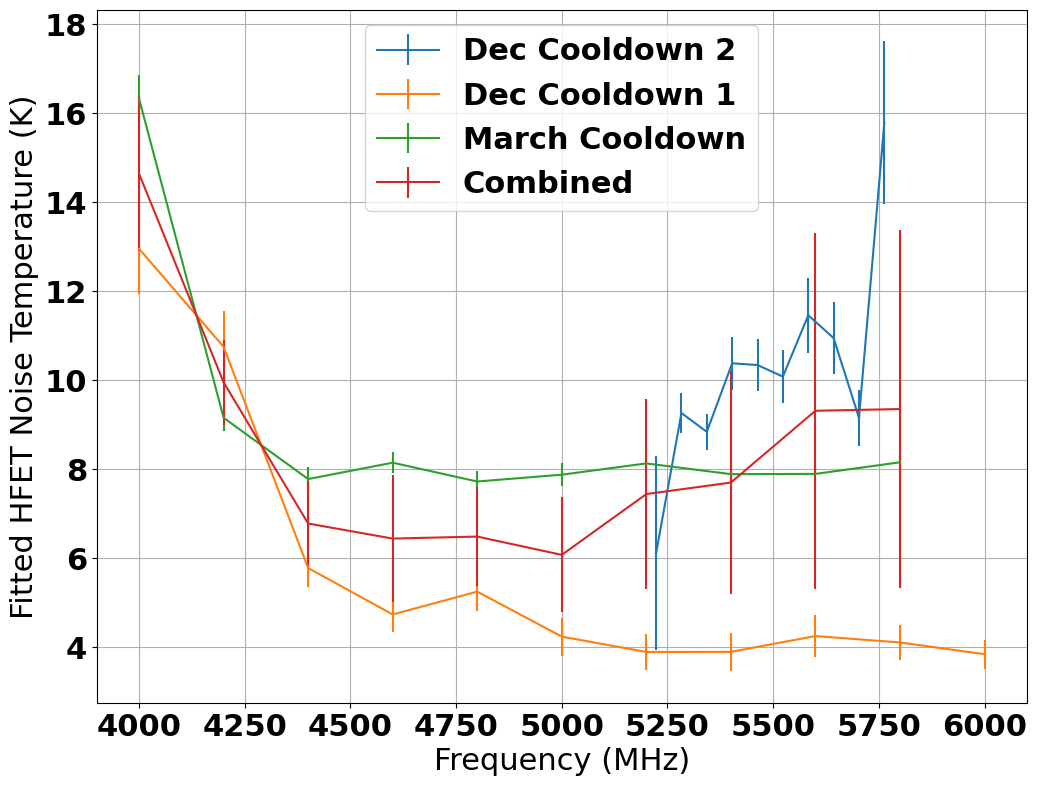}
    \caption{The HFET noise temperature versus frequency for each of the three noise calibration measurements, as well as the combined result for all three. Note that the December cool down part 2 measurement was taken at a different set of frequencies than the other two. Since all three measurements seem to disagree with one another, a conservative combined result was made to be used in the subsequent axion search analysis.}
    \label{fig:sidecar_noise_HFETtemp}
\end{figure*}

 The HFET receiver noise temperature plot for the 3 measurements is shown in Figure \ref{fig:sidecar_noise_HFETtemp}. The first half of the December cool down gave the lowest noise temperature of $4.1 \pm 0.25 K$ at 5521 MHz, the frequency the cavity is currently tuned to. This measurement agrees the best with the run 1C measured temperature, $3.7 \pm 0.2 K$ at 4798 MHz, and the run 1A temperature, $4.0 \pm 0.3 K$, albeit it predicts a warmer temperature at that lower frequency of $5.2 \pm 0.2 K$. The second half of the December cool down has the most bizarre, inconsistent curve, and predicts the highest noise temperature of $10.1 \pm 0.3 K$ at 5521 MHz; because of the constrained set of frequencies, it is also only, roughly, valid from 5.2-5.8 MHz. The March cool down data produced a result much closer in shape to the first half of the December cool down, but that had an overall noise temperature in between the two December results, $7.9 \pm 0.2 K$ at 5521 MHz. Nonetheless, none of these measurements statistically agreed with one another, and a very conservative, combined result had to be set using data from all 3 measurements, as shown in Figure \ref{fig:sidecar_noise_HFETtemp}. This increased the uncertainty in the measurement quite a bit. The final combined noise temperature was $8.7 \pm 1.7 K$ at 5521 MHz. This combined result would be used in the subsequent axion search analysis for calculating the overall system noise temperature. 
\section{Axion search analysis}
The Sidecar's ability to search for axions has been very limited so far in run 1D. The piezoelectric rotary motor has been unable to tune the cavity, making the search limited to the bandwidth of the digitizer. It was for this reason that the digitizer bandwidth was expanded to a maximum width of 3 MHz from its 2 MHz value during run 1C, which had also been expanded for the same reason. A bandwidth larger 3 MHz was not possible because of the presence of a 3 MHz pass-band filter in the warm electronics. Due to complications with programming the digitizer, another quirk of this change was that the width of a single frequency bin was kept at 100 Hz, and the number of bins expanded to 30,000, whereas it is typically only 500-2000 in the main cavity analysis. This significant increase in the number of bins and the fact that the same frequencies were being probed in hundreds of digitizations posed some unique problems to running the analysis. To add to this, a very low loaded quality factor, $\approx 1900$, and higher than normal HFET temperature, $8.7 \pm 1.7 K$, as discussed earlier, lowered the sensitivity of any given individual scan.
\par The Sidecar cavity antenna was critically coupled on December 21st 2023, and the JTWPA enabled and biased for data-taking. Due to complications with the JTWPA re-biasing script not working very well when the cavity isn't tuning, the initial data did not keep a consistent TWPA gain; the script was fixed by January 11th and data-taking commenced in earnest with an average TWPA gain of about 10 dB. It was noticed later in the run that the JTWPA was being compressed when measuring its gain, because of the network analyzer power being set too high; somewhat fortunately, this resulted in the gain being underestimated, and therefore using this data would be fine as it would only produce a more conservative limit.
\par It also became apparent that, when running the analysis on hundreds of digitization scans with the relatively the same frequencies, that co-adding spectra after individually fitting them, compounded otherwise small errors into much larger errors in the grand spectrum; normally there are much fewer overlapping spectra that get added together so this error is mitigated. This resulted in a non-Gaussian SNR spectrum, even for a data set of less than one day. Solving this issue is still in progress at the time of writing. For the purposes of this dissertation, the subsequent sections go through the analysis performed on a single random day of data-taking, January 15th, even though there are months of data at this point. It is a goal of the future Sidecar analysis to resolve this issue and determine the maximum length dataset one could use before further systematic errors arise and limit further axion sensitivity.
    \subsection{Extracting the Background Warm Receiver Shape}
\begin{figure*}[htb!]
    \centering
        \includegraphics[width=0.6\textwidth]{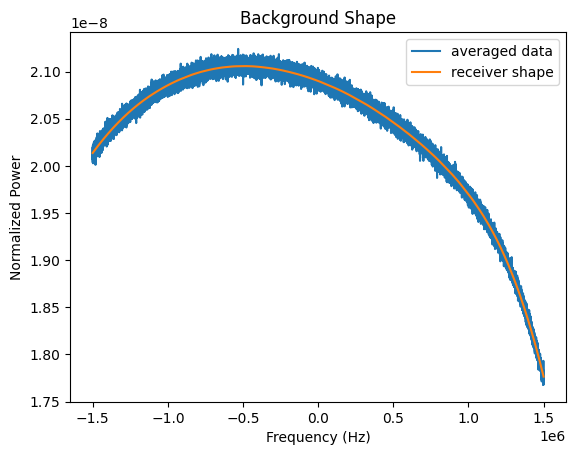}
        \caption{The Sidecar warm receiver shape and its fit for background extraction.}
        \label{fig:sidecar_analysis_backgrounfit}
    \end{figure*}

\begin{figure*}[htb!]
    \centering
        \includegraphics[width=0.6\textwidth]{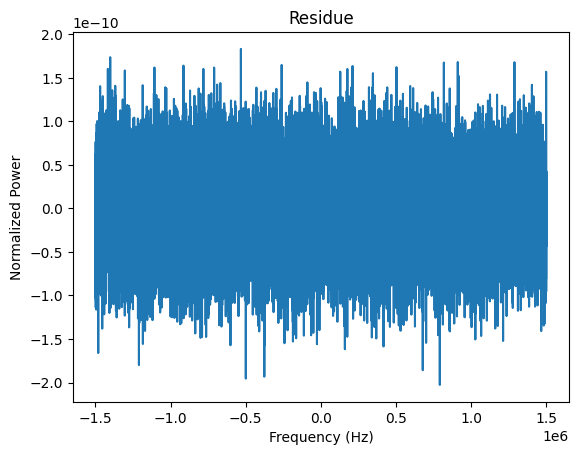}
        \caption{The Sidecar warm receiver residue after the background fit is removed.}
        \label{fig:sidecar_analysis_backgroundresidue}
    \end{figure*}

    The first step in the analysis is to fit the background in the digitizer due to the warm receiver components. This is done by disconnecting the coaxial connections from the insert, terminating them, and performing a wide power measurement for at least 3 cycles. This is the same wide power measurement used in characterizing the HFET noise temperature. These digitizations' timestamps are called out to be used in the background determination, and are cut from the axion dataset. In this case, a Savitisky-Golay filter was used to fit the background with a window size of 7700 points and of 3rd order; the main experiment currently uses the same but with a window size of just 17 points. The window size was increased significantly because the number of bins had been increased by a factor of 60. In theory, keeping this value at 17 should be fine, because, since it's a warm receiver measurement, there is no chance of a narrow axion signal being hidden in this background. Nonetheless, if kept small, it would introduce fine structure when divided out of the individual spectra later in the analysis; this exacerbated the compounding fit error issues we were seeing at the stage of this analysis (more on this later).  Figure  \ref{fig:sidecar_analysis_backgrounfit} shows the warm receiver shape and its fit, and Figure \ref{fig:sidecar_analysis_backgroundresidue} shows the resultant power residue after subtracting the fit; the residue is a normalized noise spectrum.
    \subsection{Run Parameter Extraction}
    \begin{table}
    \centering
    \begin{tabular}{||c|| c||} 
    \hline
    Parameter & Average Value for 14 hour dataset  \\ [0.5ex] 
        \hline\hline
        Reflection frequency& $5521.052\pm0.035$ MHz\\ 
        \hline
        Transmission frequency& $5521.311\pm0.024$ MHz\\ 
        \hline
        Reflection Q& $1919\pm209$\\ 
        \hline
        Transmission Q& $1941\pm40$\\ 
        \hline
        $\beta$ coupling& $1.01\pm0.03$\\ 
        \hline
        transmission $\Delta f$ FWHM& $2.84\pm0.06$ MHz\\ 
        \hline
        JTWPA gain& $12.40\pm0.06$ dB\\ 
        \hline
        JTWPA SNRI& $9.78\pm0.02$ dB\\
        \hline
        $T_{HFET}$ noise temperature& $9.34\pm1.78$ K\\
        \hline
        $T_{sys}$ noise temperature& $1.37\pm0.42$ K\\
        \hline
        $V_{eff}=CV_{cav}$ effective volume& $0.165\pm0.005$ L\\
        \hline
        $B_0$ average field& $3.585\pm0.000003$ T\\
        \hline \hline
        \end{tabular}
        \caption{The average value of run parameters and their uncertainties for a random 14 hour data-set used for initial Sidecar run 1D results.}
        \label{tab:sidecar_analysis_params}
        \end{table}
    The next major step in the analysis pulls relevant sensor and RF calibration values during the time each raw spectra used in the run dataset was taken. This includes all the variables that go into calculating Equations \ref{eqn:axionpower4} and \ref{eqn:DickeRadiometer}: $f_0$, $Q_0$, $\beta$, $C$, $V$, $B_0$, and $T_{sys}$. The $T_{sys}$ is calculated using Equation \ref{eqn:Tsyson}, with values of the $T_{HFET}$ derived in the previous section, $\epsilon$ from ex-situ measurements, and the SNRI of the TWPA measured during the last calibration period of the spectra; the SNRI is calculated from Equation \ref{eqn:SNRI} based on the TWPA gain. The average $B_0$ for Sidecar is calculated based on the current in the main magnet and assuming a constant ratio of field strength to current (3.58 T/220 A). The $C$ and $V$ are coded in a form factor configuration file based on the COMSOL simulations done for the $TM_{010}$ mode; they are actually expressed in the analysis as $V_{eff}=V_{cav}C(f_0)$, an effective volume based on the form factor at the given resonant frequency. The $f_0$, $\beta$, and $Q_L$ are pulled from the periodic network analyzer measurements in the data-taking cycle and associated with the timestamp of each individual spectra; both transmission and reflection values are pulled, but the transmission values of $f_0$ and $Q_L$ are what are used for weighting the spectra later. Table \ref{tab:sidecar_analysis_params} lists the average values of these various parameters for the 14 hour data set that was used to set initial results.
    \par
\begin{figure*}[htb!]
        \centering
        \includegraphics[width=1\textwidth]{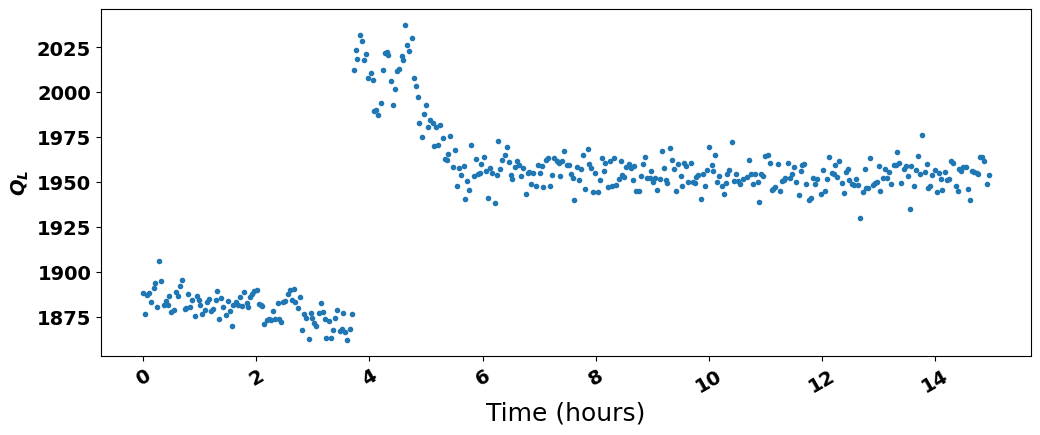}
        \caption{The Sidecar loaded quality factor (transmission) over the course of the 14 hour data set used for analysis. Evident in this plot is a small shift in Q ($\approx 150$) due to a helium reservoir fill that took place roughly at hour 4. The value recovers after about 2 hours, but it is shifted by $\approx 75$; This is probably within the uncertainty in Q for any individual fit, so it is not considered a problem at the moment.}
        \label{fig:sidecar_analysis_Qvst}
    \end{figure*}

 A weakness in the Sidecar data-taking process is that it doesn't stop taking data during helium reservoir fills. These happen approximately once a day, and last only a few minutes. This is a problem because it causes several of these parameters to shift temporarily for about 1-3 hours, predominantly the resonant frequency, coupling, and loaded quality factors; this temporary shift is visible in a time plot of the quality factor for the specific day that this initial analysis used (Figure \ref{fig:sidecar_analysis_Qvst}). As one can see its a relatively small shift compared to the normal variance in the data. No cuts were made on this data, but it is worth noting, as it may play into the systematic errors of a more precise search.
    \par
\begin{figure*}[htb!]
        \centering
        \includegraphics[width=1\textwidth]{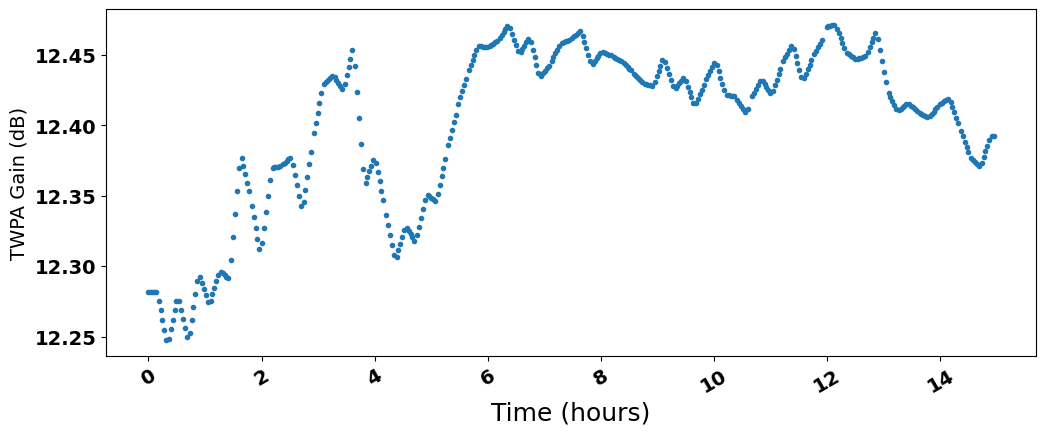}
        \caption{The Sidecar TWPA gain performance over the course of a 14 hour data set.}
        \label{fig:sidecar_analysis_TWPAgain}
    \end{figure*}

 Finally, the JTWPA gain performance over the course of the day used for analysis is plotted in Figure \ref{fig:sidecar_analysis_TWPAgain}; it is more resilient to the helium fills, but also has a larger variance to begin with throughout the day.
    \subsection{Raw Spectrum Preparation}
\begin{figure*}[htb!]
    \centering
        \includegraphics[width=1\textwidth]{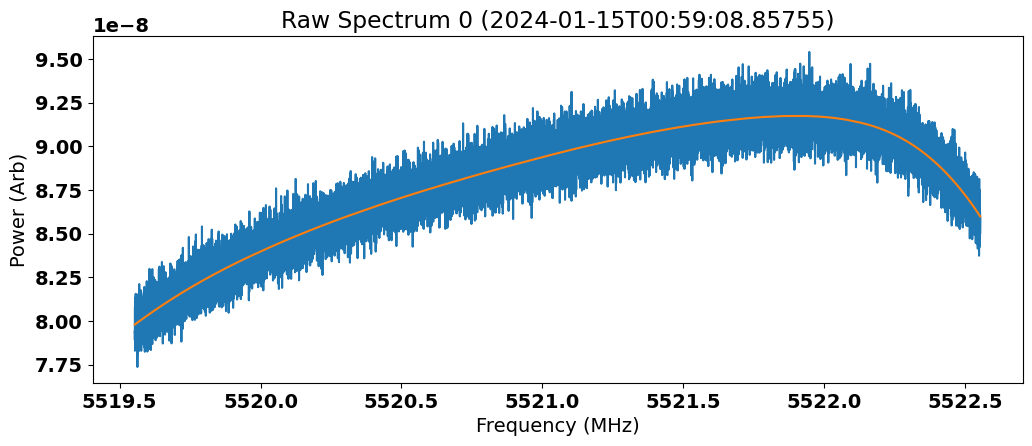}
        \caption{An example raw spectrum shown in blue with its cold receiver Pad\'e fit shown in orange.}
        \label{fig:sidecar_analysis_rawspectra}
    \end{figure*}

\begin{figure*}[htb!]
    \centering
        \includegraphics[width=1\textwidth]{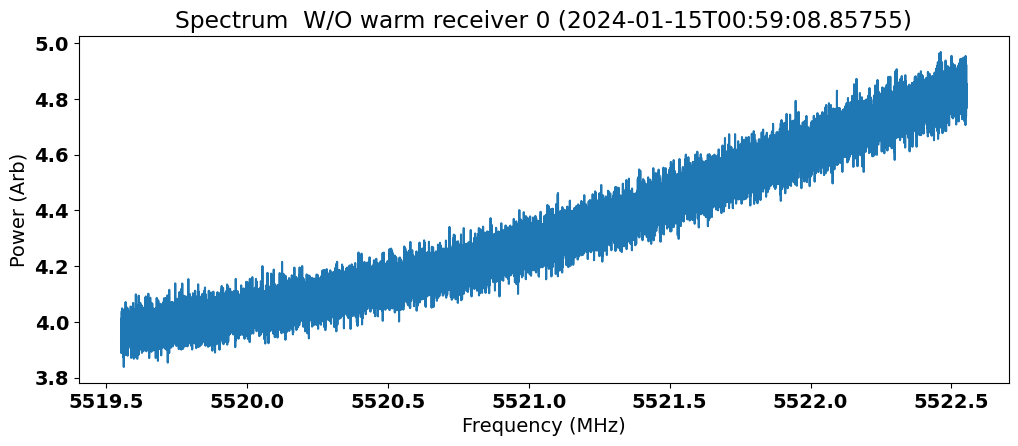}
        \caption{An example raw spectrum with the warm receiver shape removed, but no cold receiver fit applied; there is still residual structure unaccounted for.}
        \label{fig:sidecar_analysis_NowarmBG}
    \end{figure*}

\begin{figure*}[htb!]
    \centering
        \includegraphics[width=1\textwidth]{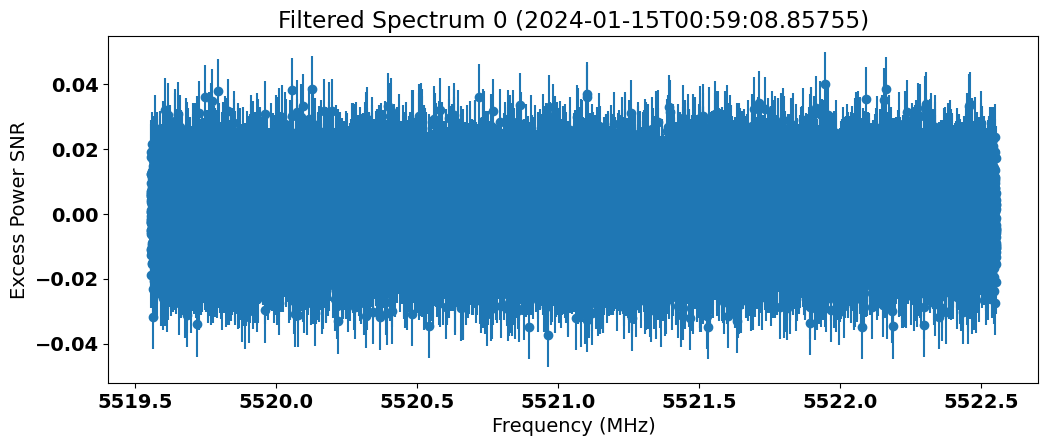}
        \caption{The raw spectrum after the cold receiver fit being removed. It is now normalized, and should be flat and centered about zero power.}
        \label{fig:sidecar_analysis_filtered}
    \end{figure*}

\begin{figure*}[htb!]
    \centering
        \includegraphics[width=1\textwidth]{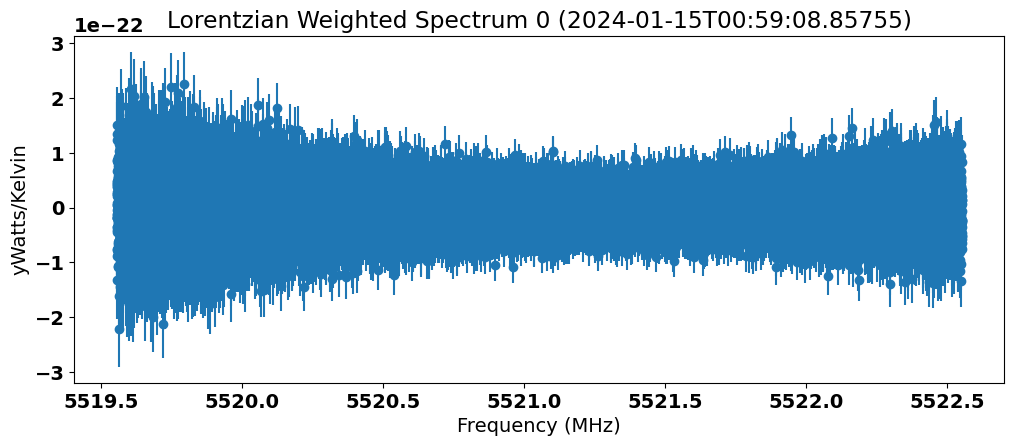}
        \caption{The final Lorentzian weighted spectrum sent for co-adding into a grand spectrum. The Lorentzian weighting applied matches the associated resonant frequency and quality factor at the time of the digitization.}
        \label{fig:sidecar_analysis_lorentzspectrum}
    \end{figure*}

    The next step in a typical axion analysis would be to prepare all of the raw spectra taken in the data-taking time period. The typical steps in this process for each spectra would be:
    \begin{enumerate}
        \item Check to see if the spectra should be cut from the dataset for several reasons (time cuts. parameter cuts, etc.). This isn't very relevant in this set because there were no cuts during the 14-hour period used for this analysis.
        \item Subtract/divide out the warm receiver fit shape from the raw spectra that was made earlier. This is shown in Figure \ref{fig:sidecar_analysis_NowarmBG}.
        \item Perform a "cold-receiver" filter fit to the individual spectra and divide it out from the individual spectra. Figure \ref{fig:sidecar_analysis_rawspectra} shows a raw spectrum and its Pad\'e fit at this stage of the process. Special care has to be taken here not to filter out potential axion signals. A Pad\'e polynomial fit was used in this analysis. This produces a normalized filtered spectra that is shown in Figure \ref{fig:sidecar_analysis_filtered}.
        \item Apply a Lorentzian weighting to each bin of the spectra according to the quality factor and resonant frequency associated with the individual spectra. This ensures that when co-adding spectra later that bins further off-resonance do not get counted equal to bins closer to resonance. It is expressed in yocto-watts/Kelvin because a system noise temperature has not been applied to the spectrum yet; this will be done in the grand assembly. This is shown in Figure \ref{fig:sidecar_analysis_lorentzspectrum}.
    \end{enumerate}
    \par
\begin{figure*}[htb!]
    \centering
        \includegraphics[width=0.97\textwidth]{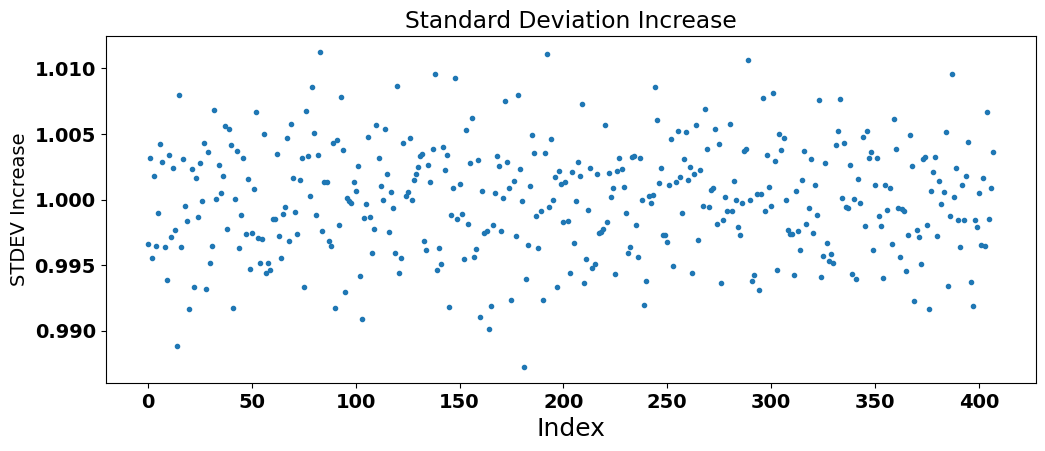}
        \caption{The standard deviation increase versus the spectrum index of the 408 spectra included in the data set. The data is centered about the expected mean with a $\pm 0.01$ variance as expected.}
        \label{fig:sidecar_analysis_stdevincrease}
    \end{figure*}

\begin{figure*}[htb!]
    \centering
        \includegraphics[width=0.97\textwidth]{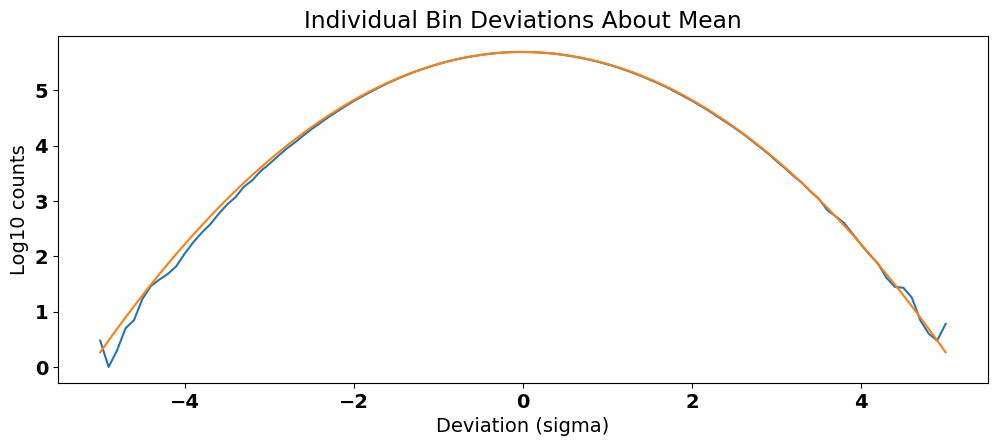}
        \caption{The individual bin deviations of raw spectra about the mean fitted to a Gaussian distribution.}
        \label{fig:sidecar_analysis_deviation}
    \end{figure*}

 This process was done for the 408 spectra that were in the random 14-hour data set chosen. The final statistics of the process agreed well with the ideal. The distribution of the standard deviations was Gaussian about the expected value with the expected variance. This variance should be 0.01 about a mean value of 1. The standard deviation increase, which is the ratio of the individual spectrum standard deviation to the expected value, is plotted as a function of digitization index in Figure \ref{fig:sidecar_analysis_stdevincrease}. The deviation of individual bins about the mean is plotted in Figure \ref{fig:sidecar_analysis_deviation}. This was also very Gaussian, indicating this process successfully normalized the noise for each individual spectra.
    \par We will run into problems with what we did here in the next step however; the cold receiver fit isn't perfect and there is some receiver structure in each bin that wasn't captured. This means when we add many of these spectra together that error will start to compound into a larger error. This isn't an issue in a normal analysis with cavity tuning because only on the order of 10 spectra get added into a single grand spectrum assembly bin, and the fit error is uncorrelated with the grand spectrum bins across spectra because there is tuning between each spectra. In this case, because all the spectra cover the relatively same frequency range, this error will compound on the order of the number of spectra included in the set, 408 scans in this case. More on this next.
    \subsection{Grand Spectrum Assembly}
\begin{figure*}[htb!]
    \centering
        \includegraphics[width=1\textwidth]{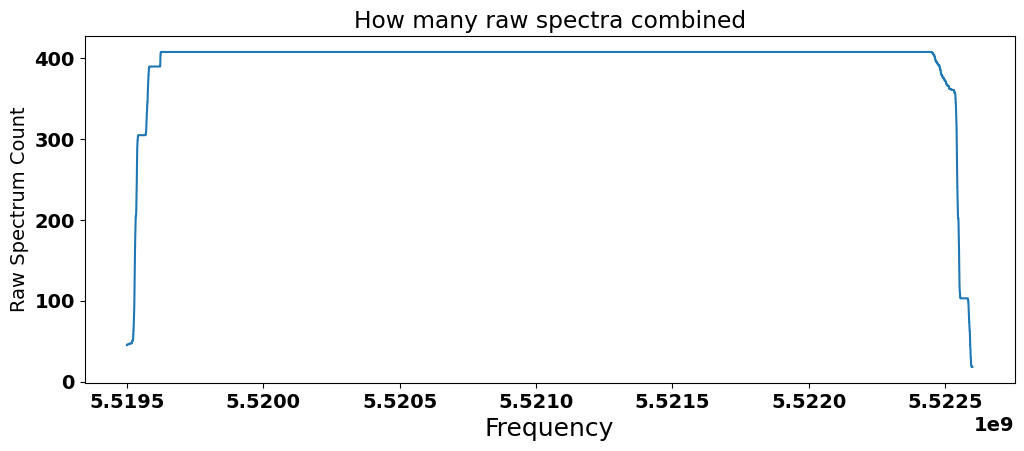}
        \caption{The number of raw spectra that contributed to each frequency bin in the grand spectrum.}
        \label{fig:sidecar_analysis_rawspectracombo}
    \end{figure*}

    The next step in the analysis is to co-add all of these prepared spectra into a grand spectrum as outlined in section \ref{GrandSpectrum}. Fig \ref{fig:sidecar_analysis_rawspectracombo} shows the number of spectra being added in any given frequency bin of what will become the grand spectrum. Because there is no tuning, almost every spectra in the sample set, 408 total, contributes to every bin in the grand spectrum, with the exception of the fringe frequencies, because there are slight differences in the center frequency for each spectrum.
    \par
\begin{figure*}[htb!]
    \centering
        \includegraphics[width=1\textwidth]{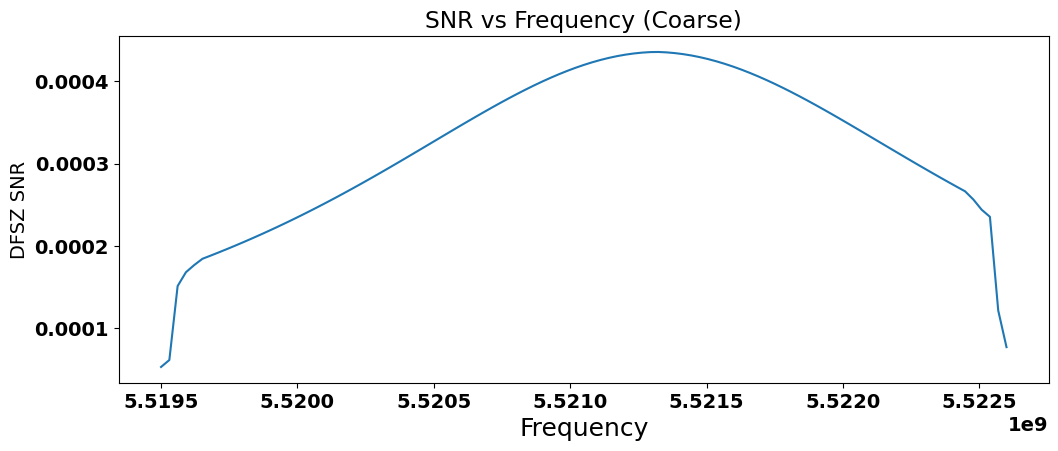}
        \caption{The DFSZ SNR of the grand spectrum versus frequency. Its Lorentzian shape is a product of the Lorentzian weighting of the individual raw spectra. The off-center peak is due to a mismatch between the spectra centers and the measured resonant frequency of the cavity.}
        \label{fig:sidecar_analysis_DFSZSNR}
    \end{figure*}

 The DFSZ SNR for each frequency in the grand spectrum is shown in Figure \ref{fig:sidecar_analysis_DFSZSNR}; remember that Sidecar is not meant to be sensitive to DFSZ axions, hence the very low SNR. These DFSZ SNR sensitivity values are in agreement with the expected sensitivity of Sidecar for a set of 408 spectra. Again, because this data set has no tuning, a unique expected feature of this 'static' spectrum is that it should produce a Lorentzian shape about the resonant frequency as seen in the plot. The reason it isn't centered in the plot is because of a slight mismatch between the target frequency that the DAQ instructed the digitizer to center spectra to and the measured resonant frequency of the cavity; it is a difference of about 200 kHz.
    \par
\begin{figure*}[htb!]
    \centering
        \includegraphics[width=0.9\textwidth]{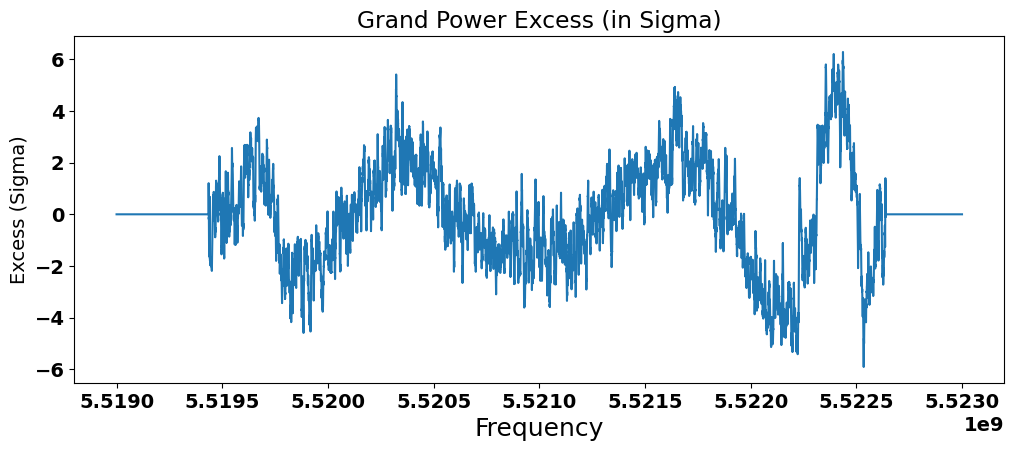}
        \caption{The grand power excess spectrum versus frequency for the Sidecar run 1D sample dataset. It is apparent that there is significant noise from the receiver leaking through due to the compounding error from individual spectra fits. Ideally, this spectrum should be Gaussian for a static frequency dataset.}
        \label{fig:sidecar_analysis_Grandspectrum}
    \end{figure*}

\begin{figure*}[htb!]
    \centering
        \includegraphics[width=0.9\textwidth]{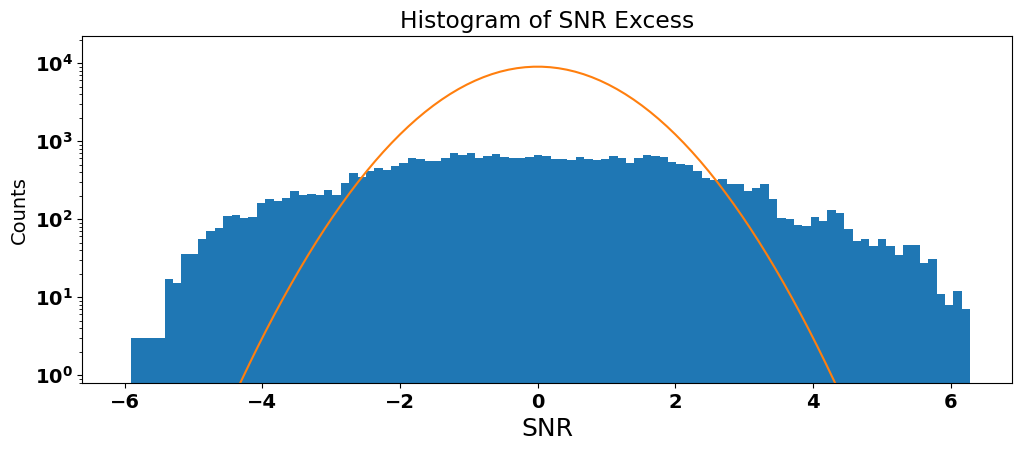}
        \caption{A histogram of the SNR excess in the grand spectrum.}
        \label{fig:sidecar_analysis_SNRhistogram}
    \end{figure*}

 The final spectrum is expressed as power excess in each bin, and its associated standard deviation. In this way, any spikes in power above $3\sigma$ in the grand spectrum would be flagged as potential axion candidates; this wasn't the case in this Sidecar dataset. Furthermore, one expects a Gaussian distribution about the center frequency for a set of data all taken about the same center frequency; in the case of a data set where there was tuning at a constant rate, one expects this shape to be averaged out and the grand spectrum flattened, as is the case for normal ADMX grand spectra. Figure \ref{fig:sidecar_analysis_Grandspectrum} shows the grand spectrum for the sidecar dataset; it is filled with excess structure and not ideal. Figure \ref{fig:sidecar_analysis_SNRhistogram} shows the histogram of SNR excesses, which does not fit well to its expected Gaussian, indicating unknown structure in the combined dataset. 
    \par This excess structure in the grand spectrum is presumed to be because of the error alluded to earlier; the individual error in each cold receiver fit is being compounded as the raw spectra are being added together. Normally this error would not grow very large if there is tuning, because the raw spectra bins would not all align with each other in the same grand spectra bin very well; each spectra would be offset by a tuning step. This error only grows for the number of spectra included in the dataset, which is why only a 14 hour data set is used here. The solution to this would be to restructure the analysis to perform the raw spectra adding procedure before the cold receiver fit, and then perform a single cold receiver fit on the combined spectrum to normalize it. This, unfortunately, requires some significant changes to the analysis code to get working and is outside the scope of this dissertation. However, the plan is to solve this eventually and be able to expand the size of the dataset to improve the search limits. Nonetheless, with the less than ideal grand spectrum shown here, some axion limits can be set.
    \subsection{Axion limits}
\begin{figure*}[htb!]
        \centering
        \includegraphics[width=1\textwidth]{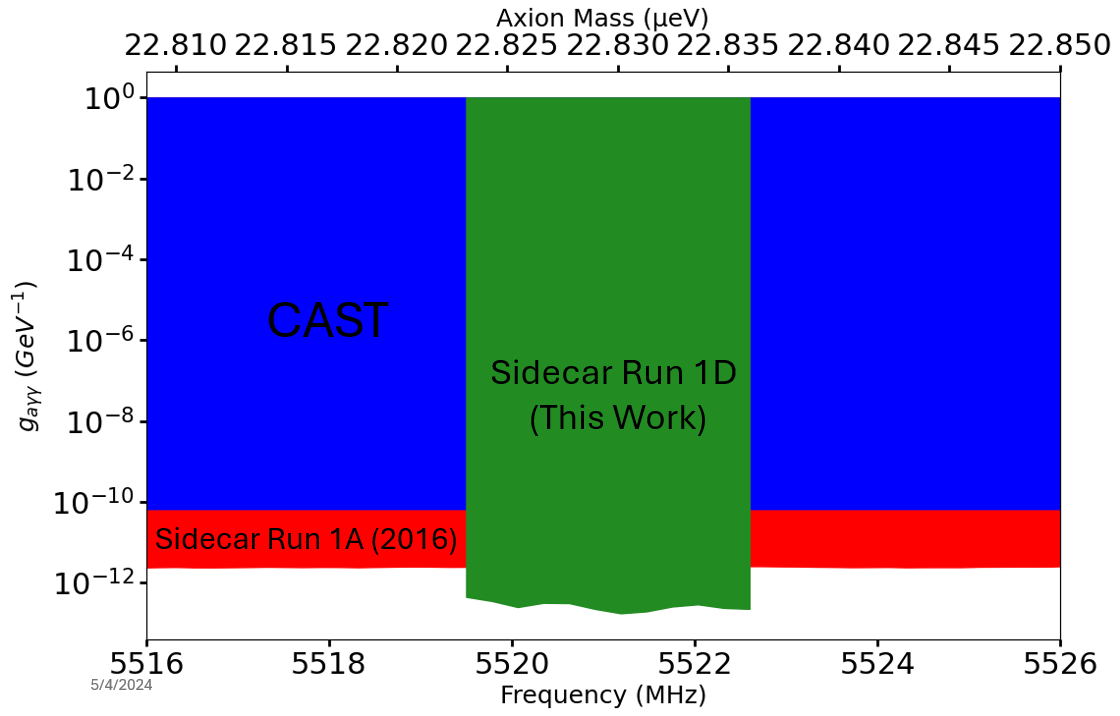}
        \caption{Preliminary axion exclusion limits for Sidecar run 1D, based on a 14 hour dataset. This work is shown in green against the previous Sidecar run 1A limits in this region (red), and the CAST broadband limit in this region (blue).}
        \label{fig:sidecar_analysis_limits}
    \end{figure*}

\begin{figure*}[htb!]
    \centering
        \includegraphics[width=0.9\textwidth]{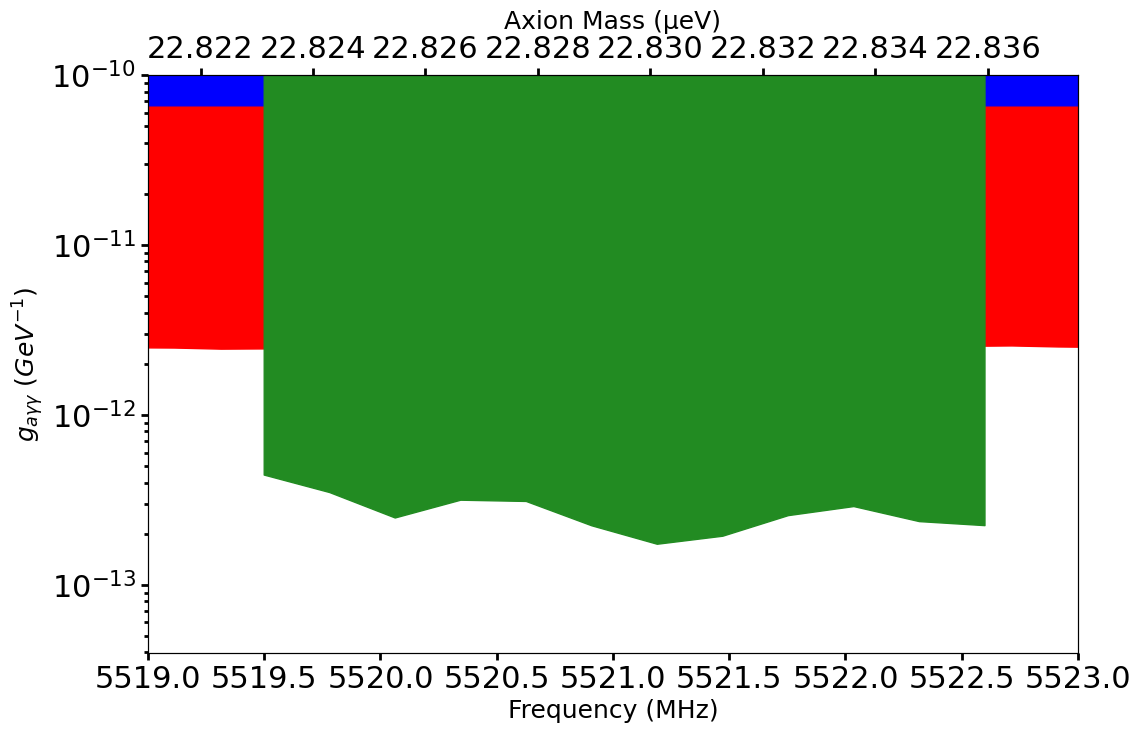}
        \caption{A zoomed in exclusion limit plot of the Sidecar run 1D preliminary dataset shown in green. The previous Sidecar run 1A limit is shown in red, and the CAST experimental limit shown in blue.}
        \label{fig:sidecar_analysis_limitszoom}
    \end{figure*}

    Although the traditional analysis process would require one to temper the grand spectrum to make sure there are no axion candidates hidden within the residual structure of Figure \ref{fig:sidecar_analysis_Grandspectrum}. Considering Sidecar is not even close to KSVZ or DFSZ sensitivity, one can go forward and set a limit under the assumption that all candidates were eliminated within the sensitivity of Sidecar. This is done by setting a minimum value of $g_{a\gamma\gamma}$ based on a 90\% confidence limit. As shown in Figs. \ref{fig:sidecar_analysis_limits}, \ref{fig:sidecar_analysis_limitszoom}, for the 14 hour sample data set used throughout this section, the average minimum $g_{a\gamma\gamma}$ set was $2.71 \pm 0.70 \times 10^{-13} GeV^{-1}$ between 5519.5 and 5522.6 MHz (22.823-22.836 $\mu$eV). The previous best limit was Sidecar run 1A with $g_{a\gamma\gamma}=2.46 \pm 0.06 \times 10^{-12} GeV^{-1}$, which tuned through this 3 MHz region in 2016. By just increasing the number of spectra at this very specific frequency to the order of hours, one is able to get a limit 10x better. Assuming this analysis could be run for a longer data set with a leveled grand spectrum, one could theoretically use the 2+ months of Sidecar data to drill down further; this should follow a $\sqrt{N}$ relation with the number of spectra. Expanding from a 14 hour data set to 30 days would then theoretically improve the limit by a factor 7.7. However one will eventually run up against further systematic issues as the data set is expanded to be longer and longer. It is a goal of the future Sidecar paper to find what this maximum set length would be, given that Sidecar is potentially stuck at this frequency for many months more before an extraction can occur.
\section{Future prospects for Sidecar}
As should be apparent from the previous sections, there are still quite a few mysteries and problems to solve on the current ADMX Sidecar; even though new, better limits could be set in the small frequency range the cavity was stuck taking data on.
\par The performance of the superconducting tuning rod remains unclear as it stands; there maybe RF power leaking through the apertures in the clamshell that are causing the very low baseline Q, but one can't know for sure. This degraded performance from external factors is entangled with the performance of the tuning rod. Plans are being made to test the second, solid $\mathrm{Nb}_{3}\mathrm{Sn}$ tuning rod in another copy of the Sidecar cavity within a dilution refrigerator at LLNL. This will give a new baseline zero field Q. Additionally, because the amplification will only be done through LNF HFETs, cavity data can hopefully be taken at higher temperatures during the cool down, opening the possibility of observing the superconducting transition. This fridge can be opened on a more regular basis, and therefore a variety of tests can be performed quantifying any potential leakage from the clam shell. In addition there is some work to be done on optimizing the geometric factor calculations and determining optimal tuning rod angles for performing a surface decomposition on Sidecar successfully. \par In terms of data-taking and analysis, Sidecar will continue to take axion data until approximately December 2024. In that time, it may be possible to attempt to tune the rotor piezoelectric motor again; even a small change could move the $TM_{020}$ mode away from its mode crossing, enabling axion data taking on that mode as well. There is still the possibility of successfully actuating the coaxial switch by directly connecting it to its power supply; this would enable a dedicated hot load measurement, which could improve the HFET noise calibration. As it stands, the HFET noise temperature is significantly degraded compared to previous runs, potentially because its recent repairs were not successful; any further noise measurements would only help. Finally, the filtering error issues present in the power excess spectrum must be resolved to get a publishable limit. After this is resolved, increasing the size of the dataset further will allow the limits to be improved by an order of magnitude potentially. 
\par All of this work will eventually inform the future direction of ADMX. Superconducting cavities have gained a lot of interest in my time studying them as an ADMX graduate. The CAPP institute has continued to push the limits of their YBCO HTS superconducting cavities, with many anticipated results coming soon. With more axion experiments appearing in the field, competition for the immediate higher frequencies above 1 GHz is possible. Being able to tune at a competitive rate is crucial for ADMX to remain relevant; superconducting cavities may be a necessity to do so. To implement this technology successfully will involve taking this graduate prototype experiment into an engineering development phase for ADMX-EFR, ensuring the cavities will meet the quality factor standard to operate. There is still much work to be done for anyone interested in either axions or superconducting cavities!

%
\nocite{*}   
\bibliographystyle{plain}
\bibliography{uwthesis}
%
%
\appendix
\raggedbottom\sloppy
 

\vita{Tom Braine is now a postdoctoral researcher located at Pacific Northwest National Laboratory (PNNL) in Richland, WA as of July 2024. He welcomes any comments or inquiries at {\tt thomas.braine@PNNL.GOV}.
}

\end{document}